\documentclass[12pt,a4paper,oneside]{book}
\usepackage{preamble}
\usepackage{microtype}
\usepackage{aas_macros}
\usepackage[utf8]{inputenc}
\usepackage[english]{babel}
\usepackage{csquotes}
\usepackage{graphicx} 
\usepackage{amsmath,amssymb,bm}
\usepackage{slashed}
\usepackage{xcolor,comment}
\usepackage{soul}
\DeclareUnicodeCharacter{2212}{\ensuremath{-}}

\setulcolor{thesislilac}   
\usepackage{float}
\usepackage{subcaption}
\usepackage{cleveref}
\usepackage{moresize}
\usepackage[style=numeric-comp,sorting = none]{biblatex}

\begin{document}

\thispagestyle{empty}
\begin{center} 
{\Large \sc PhD thesis} \vspace{2.5 cm}\\
\Huge Primordial black holes \vspace{-0.5 cm}
\Huge in braneworld gravity
\\
\vspace{0.1cm} 
\Large \it A gateway to hidden dimensions and the dark Universe 
\vspace{1cm}

\end{center}
\vspace{4 cm}
\begin{center}
    {\LARGE Itziar Aldecoa Tamayo} \vspace{0.5 cm} \\
    {\large Submitted for the degree of Doctor of Philosophy \\
Astronomy Centre, University of Sussex \\
June 2026} \vspace{3 cm} \\

    
    \vfill
    \includegraphics[width = 0.3\textwidth]{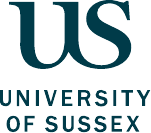}

    \end{center}
\newpage

\clearpage

\chapter*{Abstract}
\thispagestyle{empty}
\pagestyle{empty}
This thesis investigates the cosmological dynamics of primordial black holes (PBHs) within the Randall--Sundrum Type II braneworld scenario, examining gravitational collapse, accretion, evaporation, and their cosmological signatures, with the aim of assessing their viability as dark matter candidates. The work comprises a preliminary discussion on gravitational collapse in the small-black-hole limit at early cosmological epochs, and two principal studies.

The first study presents a fully relativistic treatment of PBH accretion in the early Universe, employing Michel accretion theory within the covariant Shiromizu--Maeda--Sasaki formalism. Special attention is given to the propagation of information in the cosmological fluid and to the role of braneworld corrections at high energies. The analysis reveals that higher-dimensional effects substantially enhance early-time accretion onto light PBHs, with discrepancies between our black hole masses and previous estimates growing with cosmic time as $t^{0.33}$ --- a difference that reaches several orders of magnitude by the end of the strong-accretion epoch. This finding fundamentally alters the mapping between the initial PBH mass spectrum and present-day populations.
The second study reassesses the observational and cosmological constraints on higher-dimensional PBHs in Randall--Sundrum cosmologies. Constraints from direct Hawking radiation, unresolved diffuse gamma-ray backgrounds, microlensing surveys, and energy injection into the cosmic microwave background are reanalysed in light of the modified higher-dimensional evaporation dynamics and updated observational surveys.

Taken together, these results identify viable regions of parameter space in which PBHs can account for the entirety of dark matter while remaining consistent with current observational bounds. Notably, due to the enhanced accretion epoch, PBHs capable of constituting a significant fraction of dark matter may originate from masses below the effective four-dimensional Planck scale. This thesis thus puts forward a dark matter candidate with a distinct evolutionary history and observational signatures today.

\clearpage


%



%


\small
\frontmatter 
\minitoc
\setstretch{1.8}
\begingroup
\hypersetup{linkcolor=black}
\tableofcontents
\endgroup
\thispagestyle{plain}
\newpage
\normalsize

\mainmatter 
\setstretch{1.8}

\chapter*{List of abbreviations}
\thispagestyle{fancy}
Here follow the most repeated abbreviations within the thesis.

\begin{table}[h]
    \renewcommand\arraystretch{1.8}
    \begin{tabular}{ll}
        \textbf{\textcolor{thesislilac}{ADD}} & Arkani-Hamed--Dimopoulos--Dvali\\
        \textbf{\textcolor{thesislilac}{AdS}} & Anti-de Sitter\\
        \textbf{\textcolor{thesislilac}{BBN}} & Big Bang Nucleosynthesis
        \\
        \textbf{\textcolor{thesislilac}{BDL}} & Bin\'{e}truy--Deffayet--Langlois\\
        \textbf{\textcolor{thesislilac}{CFT}} & Conformal field theory\\
        \textbf{\textcolor{thesislilac}{CMB}} & Cosmic microwave background\\
        \textbf{\textcolor{thesislilac}{DM}} & Dark matter\\
        \textbf{\textcolor{thesislilac}{DMPR}} & Dadhich--Maartens--Papadopoulos--Rezania\\
        \textbf{\textcolor{thesislilac}{FLRW}} & Friedmann--Lema\^{i}tre--Robertson--Walker\\
        \textbf{\textcolor{thesislilac}{FSR}} & Final state radiation\\
        \textbf{\textcolor{thesislilac}{GR}} & General relativity\\
        \textbf{\textcolor{thesislilac}{IA}} & Inflight annihilation\\
        \textbf{\textcolor{thesislilac}{ISM}} & Interstellar medium\\
        \textbf{\textcolor{thesislilac}{KK}} & Kaluza--Klein\\
        \textbf{\textcolor{thesislilac}{NR}} & Non-relativistic\\
        \textbf{\textcolor{thesislilac}{PBH}} & Primordial Black Hole\\
        \textbf{\textcolor{thesislilac}{RS-I}} & Randall--Sundrum 2-brane model\\
        \textbf{\textcolor{thesislilac}{RS-II}} & Randall--Sundrum 1-brane model\\
        \textbf{\textcolor{thesislilac}{SMS}} & Shiromizu--Maeda--Sasaki\\
    \end{tabular}
\end{table}

\addcontentsline{toc}{chapter}{\protect\numberline{}List of abbreviations}
\clearpage

\chapter{Introduction to this thesis}
\thispagestyle{fancy}
\pagestyle{fancy}
\setlength{\epigraphwidth}{0.61\textwidth}
\epigraph{{\it Perhaps the open questions of cosmology speak the same language --- what if the answer is geometry?}}{} 
\setlength{\epigraphwidth}{0.6\textwidth}
Primordial black holes are among the most active areas of research in cosmology today. Braneworlds, once equally vibrant, have largely receded from the spotlight. Yet the two are connected by a striking common theme. Although their periods of greatest activity were separated by a quarter of a century, both rose to prominence by proposing answers to two of the deepest problems in modern physics through a remarkably similar idea: that geometry itself might hold the key.

On one hand, dark matter remains one of the greatest unsolved problems in cosmology. The primordial black hole hypothesis offers a conceptually simple alternative to the introduction of new particle species, suggesting instead that the missing matter may consist of compact gravitational objects described by a theory we already possess and have directly observed. On the other hand, braneworld theories addressed a different mystery: the enormous disparity between the scales associated with gravity and the other fundamental interactions. While contemporary approaches introduced new particle content to reconcile these scales, braneworld models proposed that the hierarchy could emerge as a consequence of the geometry of a higher-dimensional spacetime, with extra dimensions hidden from direct observation. They further provide a geometrical interpretation of the cosmological constant problem. 

There is, moreover, a deeper reason why this parallel is worth pursuing. The regimes in which braneworld models differ most strongly from conventional gravity are precisely those in which primordial black holes naturally reside: small scales, strong gravitational fields, and high energies. Primordial black holes therefore provide a natural setting in which to explore the consequences of higher-dimensional gravity. Their formation, their characteristic scales, and their very nature place them squarely within the domain where such effects are expected to be most pronounced.

In this thesis, we begin with braneworld gravity and gradually introduce cosmology, exploring the various interpretations and dual descriptions that emerge within this framework (\href{part:braneworlds}{\textbf{\textcolor{thesislilac}{Part I}}}). We then turn to black holes, examining the spacetime structure and physical properties they acquire in higher-dimensional settings (\href{part:higher-d}{\textbf{\textcolor{thesislilac}{Part II}}}). Finally, we place five-dimensional black holes in a cosmological context and follow the evolution of a population of these objects in a universe with hidden dimensions (\href{part:pbhs}{\textbf{\textcolor{thesislilac}{Part III}}}). 

While a substantial portion of the work presented here is original, the first two parts of this thesis also contain a significant review of the development of braneworld theories. This is not merely for completeness. Many of the ideas employed throughout this work are best appreciated in the context from which they emerged, and understanding the development of the field provides insight into the structure of the theory itself. Accordingly, we follow the literature closely, tracing how the modern picture was assembled piece by piece.

As a consequence, a unified notation has not always been maintained. This choice is deliberate. Retaining the notation of the original works not only facilitates comparison with the source material, but also preserves the distinct perspectives through which different authors approached the same underlying physics. In a subject where equivalent descriptions coexist, notation is often intertwined with interpretation. The symbols chosen by an author frequently reflect the perspective from which they approach the underlying physics. Notes have been included wherever conventions change, and the reader is encouraged, where possible, to approach the thesis in a linear fashion.
\\
\\
\noindent\textbf{Notation}. We use natural units $c=\hbar=k_B=1$ and the sign convention (-,+,+,+,+).

\part{Braneworld cosmology}\label{part:braneworlds}

    \chapter[Intuitions beyond the visible dimensions]{Intuitions beyond\\ the visible dimensions}\label{chap1:intuitions}
    \thispagestyle{fancy}
\pagestyle{fancy}

The coexistence of beings, realms, or dimensions beyond the limits of human perception has occupied human thought for millennia. From spirits and mythological creatures to hidden worlds and transcendent planes of existence, human cultures have long imagined realities that exceed the visible and tangible world. Such beliefs reveal a persistent dissatisfaction with the assumption that observable reality exhausts existence. Long before higher dimensions became a mathematical or physical concept, the idea that the world available to human perception might be only a partial aspect of a deeper reality had already become a recurring theme in philosophy, mythology, and religion.

The modern concept of higher-dimensional space emerged through mathematics. During the nineteenth century, developments in geometry made it possible to formalise spaces with more than three dimensions~\cite{Riemann1854,Kragh2012}, separating the idea of dimensionality from immediate physical intuition. Higher-dimensional spaces could then be treated rigorously as consistent mathematical structures, rather than imagined only symbolically or philosophically.

Although higher-dimensional spaces may initially appear to belong to pure mathematic rather than to the physical world, they eventually became central to some of the most ambitious attempts in theoretical physics. Over the past century, physicists began to consider the possibility that hidden dimensions might form part of physical reality, even if they are not directly accessible to human observation. Over the past century, higher-dimensional theories have played an important role in attempts to address some of the deepest problems in fundamental physics, including the unification of forces, quantum gravity, and the hierarchy problem. In this context, extra dimensions are not introduced merely as speculative extensions of space, but as possible structural features of a deeper physical theory.

The introduction of higher dimensions into physics occurred most famously through the work of Kaluza in 1921~\cite{Kaluza1921} and marked a decisive shift. By extending general relativity to five dimensions, Kaluza showed that, under certain assumptions, Einstein’s gravitational theory could incorporate electromagnetism, making higher dimensions attractive as a possible route toward unification. Klein later proposed that the additional spatial dimension could be compactified --- curled up at scales too small to be observed directly~\cite{Klein1926} --- explaining why only three spatial dimensions are apparent at everyday energies. The resulting Kaluza--Klein framework established a pattern that would recur throughout later theoretical physics: extra dimensions introduced as hidden structures whose effects might become visible only indirectly~\cite{OverduinWesson1997}. 

A further conceptual shift occurred with the emergence of localisation mechanisms in higher-dimensional field theory. Rubakov and Shaposhnikov proposed that ordinary matter could be confined to a lower-dimensional region within a higher-dimensional spacetime~\cite{RubakovShaposhnikov1983}. In their domain-wall picture, particles associated with ordinary four-dimensional physics are localised near a defect, while the higher-dimensional bulk remains largely inaccessible at low energies. This idea anticipated a central feature of later brane-world scenarios: our observed universe need not fill the entire higher-dimensional space, but may instead be a lower-dimensional submanifold embedded within it~\cite{Rubakov2001}.

This localisation perspective became especially influential toward the end of the twentieth century, when higher-dimensional models acquired a new motivation through the hierarchy problem: the question of why gravity appears so much weaker than the other fundamental interactions, or equivalently why the electroweak scale is so small compared with the Planck scale. In the ADD model proposed by Arkani-Hamed, Dimopoulos, and Dvali, the weakness of gravity is explained by allowing gravity to propagate through large extra dimensions while Standard Model fields remain confined to a lower-dimensional brane, diluting gravitational strength across the higher-dimensional volume~\cite{ArkaniHamedDimopoulosDvali1998}.

The Randall--Sundrum models~\cite{RandallSundrum1999a,RandallSundrum1999b} instead introduced a warped five-dimensional geometry in which physical scales vary exponentially along the extra dimension. In the first Randall--Sundrum model, this warping provides a possible explanation of the large hierarchy between the Planck and electroweak scales. Thus, while earlier higher-dimensional theories such as Kaluza--Klein theory were primarily motivated by unification, later braneworld scenarios often used extra dimensions to address naturalness, localisation, and the apparent weakness of gravity.

A further modern development is provided by holography. In contrast to models where higher dimensions are treated as literal hidden components of spacetime, the holographic principle --- anticipated by ’t Hooft and Susskind~\cite{tHooft1993,Susskind1995} and given its most influential realisation in Maldacena's AdS/CFT correspondence~\cite{Maldacena1998} --- suggests that a gravitational theory in a higher-dimensional bulk may be equivalently described by a lower-dimensional theory without gravity defined on its boundary. In this framework, certain quantum field theories in four dimensions can be related to gravitational or string-theoretic dynamics in a higher-dimensional anti-de Sitter spacetime~\cite{Maldacena1998,Witten1998}. Higher dimensions therefore acquire a dual role: they may be interpreted either as possible physical extensions of spacetime or as mathematical tools that encode strongly coupled four-dimensional physics in a geometrical language. Even when the extra dimension is not regarded as a directly observable part of our universe, it can provide a powerful framework for studying gravity, quantum field theory, and the relation between them.

Higher-dimensional frameworks therefore occupy a distinctive position in modern theoretical physics. They connect mathematical abstraction, historical speculation, and physical model-building, but their role has changed over time. In Kaluza--Klein theory, extra dimensions served mainly as a route toward unifying known forces. In string theory, they became a condition for theoretical consistency~\cite{GreenSchwarzWitten1987,Polchinski1998}. In domain-wall and braneworld models, they allowed the possibility that ordinary matter is confined to a lower-dimensional world embedded in a higher-dimensional bulk. In ADD and Randall--Sundrum models, they were used to reformulate the hierarchy problem and the weakness of gravity. In holographic dualities, higher-dimensional geometries can also serve as alternative descriptions of lower-dimensional quantum theories. Across these developments, the central idea remains the same: the observed four-dimensional world may be either an effective description of a deeper higher-dimensional structure, or a theory that admits a higher-dimensional dual representation.

In this thesis, we focus on braneworld models of the Randall--Sundrum type, and we exploit precisely the regimes where deviations from standard four-dimensional gravity are most pronounced. Braneworld corrections to the gravitational equations are largest at high energies, where the effective Friedmann equation acquires modifications quadratic in the energy density, and in the vicinity of black holes, where the geometry of the bulk imprints itself on the brane in ways that have no four-dimensional analogue. These are also the regimes that are hardest to probe observationally in the late Universe --- but primordial black holes, formed in the early Universe when energy densities were extreme, offer a natural window into both simultaneously. They are sensitive to the high-energy modification of the expansion rate at formation, and their subsequent evolution is governed by black hole physics in a braneworld setting. Primordial black holes therefore serve in this work not merely as astrophysical objects, but as a theoretical probe of extra-dimensional gravity in the regimes where it departs most sharply from the four-dimensional expectation.

    \chapter{Randall--Sundrum braneworlds}\label{chap2:RS}
    \thispagestyle{fancy}
\pagestyle{fancy}

    Having introduced the general idea of braneworld theories, we now turn to one of the most well-known realisations: the Randall--Sundrum models. In contrast with the flat extra dimensions of the ADD scenario, the Randall--Sundrum construction is based on a curved five-dimensional bulk. The brane is not merely a passive hypersurface embedded in the bulk, but a gravitating object actively coupled through its energy content to the surrounding geometry. Even in the absence of ordinary matter, it carries localised vacuum energy, parametrised by its tension $\lambda$, and therefore gravitationally backreacts on the surrounding bulk geometry. Remarkably, for an appropriate tuning between the bulk cosmological constant and the brane tension, the induced geometry on the brane remains four-dimensional Minkowski spacetime. These features make the RS models particularly interesting from a general relativistic point of view, and a particularly rich arena for exploring gravitational physics beyond four dimensions.
\section{Notation and conventions}
Before proceeding to the models themselves, we fix our notation and conventions to avoid ambiguity. We work with signature $(-,+,+,+,+)$ and consider a five-dimensional bulk spacetime ($\mathcal{M},\,^{(5)}g_{AB}$) containing four-dimensional hypersurfaces $\Sigma_i$ which we identify as \textit{branes}. Capital Latin indices ($A, B, C,\dots$) are bulk indices and run over all five coordinates $\{0,1,2,3,5\}$, where we label the extra dimension $5$ rather than $4$ to avoid confusion with four-dimensional labels. Greek indices ($\mu, \nu, \alpha,\dots$) are brane indices and run over $\{0,1,2,3\}$ only. Latin indices $(i,j,k)$ are spatial brane indices and run over $\{1,2,3\}$ only. These appear when specialising to a cosmological setting.
\begin{itemize}
    \item A left superscript $^{(5)}$ labels geometric objects intrinsic to the bulk, such as the bulk metric $^{(5)}g_{AB}$, the bulk Riemann tensor $^{(5)}R_{ABCD}$ and the bulk Einstein tensor $^{(5)}G_{AB}$. These objects carry capital Latin indices and are defined with respect to the bulk Levi-Civita connection. 
    \item A left superscript $^{(4)}$ labels geometric objects intrinsic to the brane, such as the induced metric $^{(4)}g_{\mu\nu}$,  the brane Riemann tensor $^{(4)}R_{\mu\nu\rho\sigma}$ and the brane Einstein tensor $^{(4)}G_{\mu\nu}$. These carry Greek indices and are defined with respect to the Levi-Civita connection of the \textit{induced} metric on the brane. When there is no risk of confusion, the superscript $^{(4)}$ is dropped. 
    \item For actions we write $\mathcal{S}_5$ and $\mathcal{S}_4$ to avoid cumbersome notation.
    \item Occasionally, one needs to embed a brane tensor into the bulk algebra. In this case it is written with capital Latin indices but understood to be \textit{tangentially projected}: ${}^{(4)}R_{AB}$ denotes the Ricci tensor induced on the brane viewed as a five-dimensional tensor with all components involving the normal direction set to zero, ${}^{(4)}R_{AB}n^B = 0$, where $n^A$ is the unit spacelike normal to $\Sigma$  with $n^A n_A = +1$. 
 \end{itemize}
 As an illustrative example, note that the induced metric on the brane, should obey: 
 \begin{equation}
     ^{(4)}g_{AB}n^B=0\,\qquad ^{(4)}g_{AB}=^{(5)}g_{AB}-n_An_B\,,
 \end{equation}
with this notation and sign conventions.
\section{Randall--Sundrum 2-brane model}
The first Randall--Sundrum model~\cite{RandallSundrum1999a}, usually denoted RS-I, consists of a five-dimensional bulk bounded by two branes with tensions $\lambda_1$ and $\lambda_2$, separated by a proper distance $L$. The fifth dimension is compact and has the topology of an $S^1/Z_2$ orbifold. Denoting the coordinate along the fifth dimension by $y$, this space may be constructed by imposing the periodicity $y\sim y+2L$ together with the reflection symmetry $y\leftrightarrow -y$. This identification folds the circle $S^1$ into an interval whose endpoints, located at $y=0$ and $y=L$, are the fixed points of the $Z_2$ symmetry. The two branes are placed at these fixed points and form the boundaries of the physical spacetime.

We now choose Gaussian normal coordinates $x^A=(x^\mu,y)$, where $x^\mu$ are coordinates along the brane and the transverse coordinate $y$ measures proper distance along geodesics orthogonal to the branes. 

If $n^A$ denotes the unit normal to hypersurfaces of constant $y$, then
\begin{equation}
    n_A\dd x^A=\dd y\,.
\end{equation}
In these coordinates the metric contains no mixed $\dd x^\mu\dd y$ terms and can be written in the form
\begin{equation}
    \dd s^2=g_{\mu\nu}\dd x^\mu \dd x^\nu+\dd y^2\,,
\end{equation}
where $g_{\mu\nu}$ is the metric induced on the hypersurfaces of constant $y$. 

For the RS-I background, let us take the most general 4D Poincaré-invariant metric within this coordinate choice,
\begin{equation}\label{metric-rs-1}
    \dd s^2=e^{2A(y)}\,\eta_{\mu\nu}\dd x^\mu \dd x^\nu+\dd y^2\,, \qquad 0\leq y\leq L\,,
\end{equation}
where $\eta_{\mu\nu}$ is the four-dimensional Minkowski metric, so that each constant-$y$ hypersurface is four-dimensional Minkowski spacetime up to the warp factor $A(y)$. Since the $S^1/Z_2$ orbifold identifies $y\leftrightarrow -y$, the warp factor must be even, $A(y)=A(-y)$. It therefore suffices to solve the Einstein equations on the interval $y\in[0,L]$ and extend the solution by reflection.

We follow the normalisation $\kappa_5^2=8\pi/ M_5^{3}$ for the fundamental gravitational scale, so that the five-dimensional Einstein equations take the form 
\begin{equation}\label{EFE-5d}
    \,{}^{(5)}G_{AB}=-\Lambda_5 {}^{(5)}g_{AB}+\kappa_5^2{}^{(5)}T_{AB}.
\end{equation}

The five-dimensional Einstein field equations are therefore
\begin{equation}\label{EFE-5d-ext}
    \left(^{(5)}R_{AB}-\frac{1}{2}\,^{(5)}R\, g_{AB} \right)=-\Lambda_5 \,g_{AB}-\kappa_5^2\sum_i
\lambda_i g_{\mu\nu}(x,y_i)\delta_A^\mu\delta_B^\nu\delta(y-y_i)\,.
\end{equation}
Note that we have introduced a bulk cosmological constant
$\Lambda_5$, and localised brane tension sources on the right-hand side in eq.~\eqref{EFE-5d-ext}. 

In the present coordinates, the non-vanishing components of the Einstein tensor
for the above ansatz are
\begin{align}
G_{\mu\nu}
&=
3\left(A''+2A'^2\right)e^{2A}\eta_{\mu\nu},
\\
G_{55}
&=
6A'^2\,,
\end{align}
where the prime denotes partial differentiation with respect to $y$. Away from the branes, the delta-function terms vanish. The \(55\)-component of
the Einstein equations therefore gives
\begin{equation}
A'^2=-\frac{\Lambda_5}{6}.
\end{equation}
Since $A'(y)$ must be real, a flat-sliced warped solution of this form exists only if
\begin{equation}
\Lambda_5<0.
\end{equation}
Thus, the requirement of four-dimensional Poincaré invariance --- equivalently, flat constant-$y$ hypersurfaces --- forces the bulk cosmological constant to be negative. Defining
\begin{equation}
\ell^2
=
-\frac{6}{\Lambda_5},
\end{equation}
with \(\ell>0\), the solution in the bulk is
\begin{equation}
A(y)=-|y|/\ell\,.
\end{equation}
Hence the metric becomes
\begin{equation}\label{metric-rs-2}
    \dd s^2=e^{-2|y|/\ell}\,\eta_{\mu\nu}\dd x^\mu \dd x^\nu+\dd y^2\,, \qquad -L\leq y\leq L\,,
\end{equation}
where $\ell$, the AdS radius, sets the AdS curvature scale of the bulk.

The remaining conditions arise from the singular (delta-function) contributions localised on each brane. To extract them, we use the Israel junction conditions~\cite{Israel:1966rt}: a standard result in general relativity stating that a thin shell of matter --- here, each brane --- sources a discontinuity in the extrinsic curvature of the hypersurface it occupies, proportional to its surface energy-momentum tensor. In our setting, where the bulk metric has $Z_2$ symmetry across each brane, this discontinuity is directly encoded in the jump of $A'(y)$.

Concretely, we integrate the $\mu\nu$-component of the Einstein equations across an infinitesimal interval $(-\epsilon, +\epsilon)$ straddling the brane at $y=0$, and take the limit $\epsilon \to 0$. The smooth bulk contributions vanish in this limit, leaving only the distributional part. Since $A(y) = -|y|/\ell$ has a kink at $y=0$, its derivative $A'$ jumps discontinuously there, and therefore $A''$ contains a delta-function contribution proportional to $\delta(y)$. Only this term survives the integration:
\begin{equation}
    \int_{-\epsilon}^{+\epsilon}G_{\mu\nu}\,\dd y
=3\int_{-\epsilon}^{+\epsilon}\left(A''+2A'^2\right)e^{2A}\,\eta_{\mu\nu}\dd y\xrightarrow{\epsilon\rightarrow 0}3\left[A'\right]^{0^+}_{0^-}\,\eta_{\mu\nu}\,,
\end{equation}
where $\left[A'\right]^{0^+}_{0^-}$ denotes the jump across the brane, and the $2A'^2$ is smooth and contributes nothing. Equating this to the right-hand side of the Einstein equations integrated across the same interval gives
\begin{equation}
3\big[A'\big]_{0^-}^{0^+}\,\eta_{\mu\nu}
=
-\kappa_5^2\lambda_1\eta_{\mu\nu}\,.
\end{equation}
Since $A(y)=-|y|/\ell$, we have $A'(0^+)=-1/\ell$ and $A'(0^-)=+1/\ell$, so the jump is 
\begin{equation}
\big[A'\big]_{0^-}^{0^+}=-2/\ell\,,
\end{equation}
and therefore
\begin{equation}
\lambda_1=\frac{6}{\kappa_5^2\ell}\,.
\end{equation}
The same procedure at $y=L$, where the orbifold identification $L+y\leftrightarrow L-y$ reverses the orientation and hence the sign of the jump, gives
\begin{equation}
\big[A'\big]_{L^-}^{L^+}=+\frac{2}{\ell}\,.
\end{equation}
The second junction condition then gives
\begin{equation}
\lambda_2=-\frac{6}{\kappa_5^2\ell}\,.
\end{equation}
Thus, starting from a general bulk cosmological constant and general brane
tensions, the requirement of flat four-dimensional slices yields
\begin{equation}
\Lambda_5=-\frac{6}{\ell^2}\,,
\qquad
\lambda_1=\frac{6}{\kappa_5^2\ell}\,,
\qquad
\lambda_2=-\frac{6}{\kappa_5 ^2\ell}\,.
\end{equation}
Equivalently,
\begin{equation}
\lambda_2=-\lambda_1.
\end{equation}
Note that this constitutes two independent fine-tuning conditions: one fixing the bulk cosmological constant in terms of the AdS radius, and one relating the two brane tensions. This sensitivity to the precise values of $\Lambda_5$, $\lambda_1$ and $\lambda_2$ constitutes a fine-tuning problem: no dynamical mechanism enforces these two relations, and any deviation destroys the flat four-dimensional geometry on the brane. This is a known theoretical drawback of the RS-I model, though it reduces to one relation in the RS-II scenario on which this work focuses. 

In eq.~\eqref{metric-rs-2}, the exponential factor multiplies the four-dimensional part of the line element and therefore acts as a position-dependent redshift factor along the extra dimension. As one moves from the positive-tension brane to the negative-tension brane, physical mass scales are exponentially suppressed, so that a mass parameter of order $M_5$ on the Planck brane corresponds to a physical mass of order $M_5e^{-L/l}$ on the TeV brane. This directional redshift along the extra dimension is precisely the mechanism by which RS-I resolves the hierarchy problem. Remarkably, the sixteen orders of magnitude separating the Planck scale $M_4\sim 10^{18}\,$GeV from the electroweak scale $\sim 10^2\,$GeV can be reproduced with only a modest separation $L/\ell\approx 37$, requiring no intermediate energy scale or unnatural parameter. 

The relation between the \textit{fundamental} five-dimensional and \textit{effective} four-dimensional gravitational scales ($\kappa_5^2$ and $\kappa_4^{2}$, respectively) is obtained by starting from the five-dimensional Einstein--Hilbert action,
\begin{equation}\label{hil}
        \mathcal{S}_5=\frac{1}{2\kappa_5^2}\int\dd^4x\,\dd y\sqrt{-^{(5)}g}\,^{(5)}R\,,
\end{equation}
where $^{(5)}R$ is the Ricci scalar and $^{(5)}g$ the determinant of the five-dimensional metric in eq.~\eqref{metric-rs-2}, and comparing it with the four-dimensional Einstein--Hilbert action $\mathcal{S}_4$, where $^{(4)}g_{\mu\nu}=\eta_{\mu\nu}$. This results in the relation
\begin{equation}\label{kappas}
    \frac{1}{\kappa_4^2}=\frac{1}{\kappa_5^2}\int_{-L}^L e^{-2|y|/\ell}\dd y\,.
\end{equation}
Evaluating this integral gives
\begin{equation}\label{m4-m5-rs-1}
    M_4^2=M_5^3\ell\left(1-e^{-2L/\ell}\right)\,,
\end{equation}
where we have used the relations $\kappa_4^{2}=8\pi/M_4^2$ and $\kappa_5^{2}=8\pi/M_5^3$.
Thus, the observed four-dimensional Planck mass is obtained from the five-dimensional gravitational scale weighted by the warped volume of the extra dimension. In RS-I, the hierarchy problem is not solved by making this volume large, as in ADD, but by the exponential redshift of mass scales between the two branes. 

In the original hierarchy interpretation, the positive-tension brane at $y=0$ is the Planck brane, while the Standard Model is usually placed on the negative-tension brane at $y=L$, where mass parameters are redshifted by the factor $e^{-L/\ell}$.

Although RS-I provides an elegant warped explanation of the hierarchy problem, the model also requires the separation between the two branes to be stabilised. This interbrane distance is a modulus, and without an additional stabilisation mechanism the hierarchy generated by the warp factor is not dynamically fixed. A standard mechanism to address this is the Goldberger--Wise proposal~\cite{Goldberger:1999uk}, in which a bulk scalar field with different boundary values on each brane generates a potential that dynamically fixes the interbrane separation; we do not discuss this further as the present work concerns the single-brane RS-II limit.
\section{Randall--Sundrum 1-brane model}
The second Randall--Sundrum model~\cite{RandallSundrum1999b}, usually denoted RS-II, is obtained by taking the limit in which the negative-tension brane is sent to infinity. The remaining setup consists of a single positive-tension brane embedded in an infinite five-dimensional AdS bulk. In this limit the extra dimension is no longer compact, but the metric retains the same warped form,
\begin{equation}\label{static-RS}
  \dd s^2 = e^{-2|y|/\ell}\,\eta_{\mu\nu}\dd x^\mu \dd x^\nu + \dd y^2\,,  \qquad -\infty<y<\infty\,.
\end{equation}
The brane is located at $y=0$, and the $Z_2$ symmetry identifies the two sides of the bulk across the brane. The warp factor is maximal on the brane and decreases exponentially as one moves away from it into the bulk.

The RS-II bulk metric~\eqref{static-RS} admits an equivalent 
description in terms of a Poincar\'e radial coordinate $r$, defined by
\begin{equation}
r = \ell\,e^{-|y|/\ell}\,,
\label{eq:poincare_coord}
\end{equation}
under which $\ell^2 r^{-2}\,\dd r^2 = \dd y^2$ and the warp factor 
becomes $e^{-2|y|/\ell} = r^2/\ell^2$. After rescaling $x^i \to 
\ell\,x^i$, the metric~\eqref{static-RS} takes the form
\begin{equation}\label{rs-poincare}
\dd s^2 = -\frac{r^2}{\ell^2}\,\dd t^2 + \frac{\ell^2}{r^2}\,\dd r^2 
+ r^2\,\delta_{ij}\,\dd x^i\dd x^j,
\end{equation}
where the brane sits at $r = \ell$ and the deep bulk corresponds to 
$r \to 0$. This coordinate system, in which the bulk geometry is 
manifestly static, will prove useful when we discuss the cosmological 
dynamics on the brane in Chapter~\ref{ch:Kraus}.

The relation between the four- and five-dimensional Planck scales is obtained by taking the limit $L\to \infty$ in the RS-I expression~\eqref{m4-m5-rs-1}. This gives
\begin{equation}\label{M4-M5-rel}
    M_4^2 = M_5^3\ell\,.
\end{equation}
Thus the effective four-dimensional Planck mass remains finite even though the extra dimension is infinite. This is one of the central results of RS-II: the warped geometry makes the effective volume of the extra dimension finite, despite the non-compact fifth dimension. This is further supported by the perturbative analysis of gravitational fluctuations: the massless graviton zero mode is normalisable and localised near the positive-tension brane. This zero mode reproduces four-dimensional gravity at large distances. In addition, there is a continuum of massive Kaluza--Klein modes propagating in the bulk; their effect on the brane is suppressed at low energies and appears only as small corrections to four-dimensional gravity. Since our focus will be on the background cosmological dynamics rather than the perturbative gravitational spectrum, we will not analyse these modes explicitly here, though their effect on this theory will be revisited in Section~\ref{sec:garriga}.

    \chapter[The brane as a cosmological universe]{The brane as a\\ cosmological universe}\label{ch:Kraus}
    \thispagestyle{fancy}
\pagestyle{fancy}

    The Randall--Sundrum models, as originally formulated, describe a static configuration: a brane of fixed tension sitting at a definite position in a time-independent bulk. How does this static brane become a dynamical universe --- one that expands, cools, and evolves? The natural question is how to promote the brane from a static hypersurface to a genuine cosmological spacetime, one admitting a Friedmann--Lemaître--Robertson--Walker (FLRW) geometry on its world-volume and an expansion history compatible with observation. 
%
%
\section{A non-conventional Friedmann equation}
The first systematic derivation of the cosmological equations on a brane universe was obtained by Binétruy, Deffayet and Langlois, in a paper submitted in May 1999~\cite{Binetruy:1999ut} (hereafter BDL) --- before the RS models had been fully assimilated into the cosmological literature. Working directly from the five-dimensional Einstein equations and Israel junction conditions, and allowing for a general bulk cosmological constant $\Lambda_5$, they showed that the Friedmann equation governing expansion on the brane takes a form fundamentally different from its four-dimensional counterpart, with the square of the Hubble parameter $H^2$ depending quadratically rather than linearly on the  energy density of the matter localised on the brane. The result was immediately recognised as deeply problematic: such a modification of the expansion rate appeared to persist even at low energies and was in sharp conflict with Big Bang Nucleosynthesis, which constrains deviations from the standard Friedmann equation at MeV temperatures to be negligibly small~\cite{Walker:1991ap}. It was initially unclear whether brane cosmology was physically viable at all.

The derivation proceeds as follows. Consider a five-dimensional bulk with metric 
\begin{equation}\label{eq:bulk_metric}
\dd s^2 = -N^2(t,y)\,\dd t^2 + a^2(t,y)\,\gamma_{ij}\,\dd x^i \dd x^j +\,\dd y^2,
\end{equation}
where $\gamma_{ij}$ is the maximally symmetric three-metric with curvature 
$k \in \{-1,0,+1\}$. Note that $y$ is the Gaussian normal coordinate discussed in the previous section. Unlike the static RS ansatz, the metric functions 
$N$ and $a$ are now allowed to depend on both $t$ and $y$, encoding the 
dynamical evolution of the brane universe. Matter on the brane is modelled as a perfect 
fluid, so the brane energy-momentum tensor takes the form 
\begin{equation}
    ^{(4)}T_{AB} = \mathrm{diag}(-\rho,p,p,p,0)\,\delta(y)\,.
\end{equation}
At this stage, $\rho$ and $p$ denote the total energy density and pressure localised on the brane.
To extract the junction conditions, we follow the 
same procedure as in Chapter~\ref{chap2:RS}: we integrate each component of the 
Einstein equations across an infinitesimal interval $(-\epsilon, +\epsilon)$ 
straddling the brane and take the limit $\epsilon \to 0$. The smooth bulk 
contributions vanish in this limit, and only the distributional parts sourced 
by the delta-function matter survive. Imposing $Z_2$ symmetry across 
the brane, so that $a'(0^+) = -a'(0^-)$, and working in the gauge where $N|_{y=0} = 1$, the $(ij)$ and $(tt)$ components yield, respectively,
\begin{equation}\label{eq:junction_ij}
    \left.\frac{a'}{a}\right|_{0^+} = -\frac{\kappa_5^2}{6}\rho\,,\qquad 
\left.\frac{N'}{N}\right|_{0^+} = \frac{\kappa_5^2}{6}(2\rho + 3p)\,.
\end{equation}
These are the dynamical junction conditions: they relate the normal derivatives 
of the metric functions at the brane to the matter content on it.

Separately, the $(yy)$ component of the bulk Einstein equations, evaluated 
away from the brane, provides a first integral --- a constraint equation that must 
hold throughout the bulk. Evaluated at the brane position $y = 0$ in the same 
gauge, it takes the form
\begin{equation}\label{eq:bulk_constraint}
\left(\frac{\dot{a}}{a}\right)^2 - \left(\frac{a'}{a}\right)^2 
= -\frac{k}{a^2} + \frac{\Lambda_5}{6} + \frac{\mathcal{C}}{a^4}\,,
\end{equation}
where $\mathcal{C}$ is an integration constant and a dot denotes the partial derivative with respect to $t$. 
Substituting the junction condition eq.~\eqref{eq:junction_ij} into 
eq.~\eqref{eq:bulk_constraint}, and identifying $H = \dot{a}/a\big|_{y=0}$, one 
obtains the generalised Friedmann equation on the brane,
\begin{equation}
H^2 = \frac{\kappa_5^4}{36}\rho^2 - \frac{k}{a^2} + \frac{\Lambda_5}{6} 
+ \frac{\mathcal{C}}{a^4}\,.
\label{eq:BDL_Friedmann}
\end{equation}
The $\rho^2$ dependence arises because the junction conditions~\eqref{eq:junction_ij} relate the extrinsic curvature of the brane, encoded in the normal derivative $a'/a$, linearly to the brane energy density. Since the bulk constraint equation depends quadratically on these geometric quantities, the resulting Friedmann equation is quadratic in $\rho$. The integration 
constant $\mathcal{C}$, which enters as a term scaling as $a^{-4}$, behaves on 
the brane like a radiation component of purely geometric origin.

The resolution of the apparent conflict with nucleosynthesis came quickly and 
from two independent groups, Cline, Grojean and Servant~\cite{Cline_1999} and 
Cs\'aki, Graesser, Kolb and Terning~\cite{Cs_ki_1999}, within weeks of the BDL 
paper. The key observation was that in the RS setup, the brane carries not only matter but also a bare tension $\lambda$, and the bulk contains a negative 
cosmological constant $\Lambda_5 < 0$ fine-tuned against $\lambda$ through 
the RS consistency condition $\lambda^2 = -6\Lambda_5/\kappa_5^4$, which ensures that the junction conditions are satisfied in the static vacuum where the brane carries only tension and $H = 0$. Separating the constant brane tension from the dynamical matter contribution according to $\rho_{\scriptscriptstyle\mathrm{total}} = \lambda + \rho$, the Friedmann equation~\eqref{eq:BDL_Friedmann} becomes
\begin{equation}
H^2 = \frac{\Lambda_4}{3} + \frac{8\pi}{3M_4^2}\rho
\left(1 + \frac{\rho}{2\lambda}\right) 
+ \frac{\mathcal{C}}{a^4} - \frac{k}{a^2},
\label{eq:BDL_Friedmann_expanded}
\end{equation}
where the effective four-dimensional cosmological constant is identified as
\begin{equation}\label{Lambda-4}
\Lambda_4=\frac{1}{2}\left(\frac{\kappa_5^4\lambda^2}{6}+\Lambda_5\right)\,,
\end{equation}
and where
\begin{equation}\label{kappas-lambda}
    \kappa_4^2=\frac{8\pi}{M_4^2}=\frac{1}{6}\lambda \kappa_5^4
\end{equation}
has been introduced in order to match the linear term of the conventional Friedmann equation. 
In general, $\Lambda_4$ is a free parameter determined by the bulk-brane 
configuration. The RS-II fine-tuning condition $\lambda^2 = -6\Lambda_5/\kappa_5^4$ corresponds to the special case $\Lambda_4 = 0$, in 
which the vacuum brane ($\rho = 0$) has vanishing expansion rate 
and flat induced geometry. When matter is present, the brane expands as a 
standard FLRW universe at low energies $\rho \ll \lambda$, with 
$\Lambda_4 = 0$ meaning that the expansion is driven purely by matter rather 
than by any residual bulk-brane contribution. Thus, at low energies $\rho\ll\lambda$, standard four-dimensional cosmology is recovered, while the high-energy correction term $\rho^2/\lambda$ becomes important only in the early Universe when $\rho\gtrsim\lambda$.

It is worth pausing to appreciate what this means. The RS-II fine-tuning 
condition $\lambda^2 = -6\Lambda_5/\kappa_5^4$, which in the static setting 
appears as a precise relation between bulk and brane parameters, is in the 
cosmological setting simply the condition that the effective four-dimensional 
cosmological constant vanish, $\Lambda_4 = 0$. The question of why 
$\Lambda_4$ is anomalously small is therefore not unique to braneworld 
cosmology, but rather a higher-dimensional reformulation of the usual 
cosmological constant problem already present in the conventional scenario. 
From this perspective, RS-II does not introduce a fundamentally new 
fine-tuning; instead, it makes explicit the geometric origin of the 
effective vacuum energy in terms of bulk and brane contributions. 
\section{Motion or expansion: a dual vision of the Universe}
The BDL approach treats the bulk geometry as dynamical and derives the cosmological equations by solving for the metric functions $N(t,y)$ and $a(t,y)$ in the presence of brane matter. At first glance, however, the very premise of the BDL approach seems to be in tension with a fundamental result in five-dimensional gravity. The five-dimensional generalisation of Birkhoff's theorem~\cite{Bowcock_2000} 
states that any vacuum solution of the five-dimensional Einstein equations with a negative bulk cosmological constant $\Lambda_5 < 0$ that admits a foliation by homogeneous and isotropic co-dimension 1 surfaces is necessarily static, and is locally the Schwarzschild--$\mathrm{AdS}_5$ spacetime. Although the brane itself carries matter, the bulk Einstein equations are vacuum everywhere away from the brane, with the brane appearing only as a distributional source in the junction conditions. The theorem therefore applies independently in the two bulk regions on either side of the brane. This spacetime has the metric
\begin{equation}\label{eq-ads-schwarzschild}
\dd s^2 = -f(r)\,\dd T^2 + f(r)^{-1}\,\dd r^2 + r^2\,\gamma_{ij}\,\dd x^i\dd x^j,
\end{equation}
where\footnote{Equation~\eqref{eq-ads-schwarzschild} is not the most general Schwarzschild--AdS$_5$ metric, which contains additional terms depending on the spatial curvature. We have set this curvature to zero to simplify the calculations, since it will vanish anyway once we tie this to a physical universe on the brane.}
\begin{equation}
f(r) = \frac{r^2}{\ell^2} - \frac{\mu}{r^2},
\label{eq:f_SAdS}
\end{equation}
with $T$ the bulk time coordinate, and $\mu$ a mass parameter 
proportional to the mass of a black hole located at $r = 0$ in the bulk (when $\mu \neq 0$). 

We can see how this connects to RS-II:  eq.~\eqref{eq-ads-schwarzschild} is the generalisation of the RS-II metric in Poincar\'e coordinates provided in eq.~\eqref{rs-poincare}, where the choice $\mu = 0$ was made. However, the time dependence introduced by hand in the bulk metric functions $N(t,y)$ and $a(t,y)$ in the BDL approach seems to be forbidden by 
Birkhoff's theorem --- and with it, any cosmological evolution on the 
brane. 

It turns out that it is compatible, and that the apparent contradiction is resolved by a change of perspective. Kraus~\cite{Kraus_1999} and, independently, Ida~\cite{Ida_2000} observed that the explicit time dependence of the bulk metric functions $N(t,y)$ and $a(t,y)$ in Gaussian normal coordinates is a coordinate artefact: it is simply what 
a static Schwarzschild--$\mathrm{AdS}_5$ bulk looks like from the frame of reference tied to the brane. What BDL described as a time-dependent bulk with a fixed brane can equally be described as a static bulk with a moving brane --- two coordinate representations of the same spacetime, related by the explicit transformation derived in Ref.~\cite{Mukohyama:1999wi}. Crucially, this brane motion is not introduced by hand: as we shall see, in order for the junction conditions to be satisfied, a non-vanishing stress-energy tensor on the brane immediately results in a time-dependent position of the brane with respect to the static bulk. 

Let us therefore consider the brane moving radially through the static bulk~\eqref{eq-ads-schwarzschild}, with its trajectory given by
\begin{equation}\label{brane-trajectory}
T = T(\tau)\,,\qquad r = R(\tau)\,,
\end{equation}
where $\tau$ is the induced proper time measured by observers on the brane (i.e.~cosmic time)\footnote{Note that the relative motion between the brane and the bulk gives rise to two distinct time coordinates, $T$ and $\tau$, rather than a single common time $t$. We keep this notation throughout this section in order to remain consistent with the literature, although when we proceed to cosmology we will relabel $\tau$ as the cosmic time $t$.}. Substituting into the bulk metric~\eqref{eq-ads-schwarzschild} 
and evaluating at $r=R(\tau)$, the induced metric on the brane is
\begin{equation}\label{induced-metric}
\dd s_{\scriptscriptstyle\mathrm{brane}}^2 = \left(-f\dot{T}^2 + f^{-1}\dot{R}^2\right)\dd\tau^2 
+ R^2(\tau)\,\gamma_{ij}\,\dd x^i\dd x^j\,.
\end{equation}
The condition that $\tau$ is proper time on the brane is equivalent to the normalisation of the brane's five-velocity $u^a=(\dot{T},\dot{R},0,0,0)$, namely $g_{ab}u^a u^b=-1$. Evaluating this in the bulk metric~\eqref{eq-ads-schwarzschild} gives
\begin{equation}
-f\dot{T}^2 + f^{-1}\dot{R}^2 = -1,
\label{eq:normalisation}
\end{equation}
so that the induced metric~\eqref{induced-metric} reduces to
\begin{equation}
\dd s^2_{\scriptscriptstyle\mathrm{brane}} = -\dd\tau^2 + R^2(\tau)\,\gamma_{ij}\,\dd x^i\dd x^j\,,
\end{equation}
which is precisely a spatially flat FLRW metric with scale factor $a(\tau) = R(\tau)$. The cosmological expansion of the brane universe is therefore nothing but the radial motion of the brane through the static bulk --- what observers on the brane interpret as the expansion of space is, from the bulk perspective, simply the brane moving through the five-dimensional geometry.

To complete our understanding of the moving brane picture, we need to study the junction conditions for a moving hypersurface. In the previous sections, we obtained the junction conditions by integrating the Einstein equations across the brane at a fixed position $y = 0$, concluding that the discontinuity in the normal derivatives $a'/a$ and $N'/N$ is sourced by the brane matter. In the present setting, we cannot simply integrate around a fixed $y$, since the brane's position is changing. In order to obtain the relation between 
the discontinuity in the normal derivatives and the brane content for an arbitrary moving hypersurface, we use the generalised Israel junction conditions~\cite{Israel:1966rt, Maartens_2004}, which make use of the extrinsic curvature $K_{\mu\nu}$,
\begin{equation}\label{extrinsic-curvature}
K_{\mu\nu} = e_\mu^{\ A}\,e_\nu^{\ B}\,\nabla_A n_B,
\end{equation}
with $e_\mu^{\ A} = \partial x^A/\partial\sigma^\mu$ the tangent vectors to the brane\footnote{Here $\sigma^\mu$ are intrinsic coordinates on the brane worldvolume 
and $x^A$ are coordinates in the bulk, so that $e_\mu^{\ A}$ are 
the tangent vectors to the brane embedded in the bulk --- they map 
displacements along the brane into displacements in the bulk.}. $K_{\mu\nu}$ is the covariant object that captures the discontinuity in the normal derivatives of the metric across the brane, regardless of the choice of coordinates. The Israel junction conditions then state 
that this discontinuity is sourced by the brane energy-momentum tensor~\cite{Israel:1966rt, Maartens_2004}:
\begin{equation}
\left[K_{\mu\nu}\right] = -\kappa_5^2\left(T_{\mu\nu} 
- \frac{1}{3}T\,g_{\mu\nu}\right),
\label{eq:Israel_covariant}
\end{equation}
where $[K_{\mu\nu}] = K_{\mu\nu}|_{0^+} - K_{\mu\nu}|_{0^-}$ is the jump across the brane and $T = g^{\mu\nu}T_{\mu\nu}$ is the trace of the brane stress-energy tensor. Under $Z_2$ symmetry, $[K_{\mu\nu}] = 2K_{\mu\nu}|_{0^+}$.

To evaluate $K_{\mu\nu}$ explicitly, we need the unit normal $n^A$ to the brane. It must satisfy $^{(5)}g_{AB}\,n^A u^B = 0$ and the normalisation $^{(5)}g_{AB}n^An^B = 1$. Solving these two conditions using the bulk metric~\eqref{eq-ads-schwarzschild} gives
\begin{equation}
n^A = \left(f(R)^{-1}\dot{R},\ -f(R)\dot{T},\ 0,0,0\right),
\label{eq:normal}
\end{equation}
where the sign is chosen so that $n^A$ points away from the brane into the bulk. Since the spatial tangent vectors are $e_i^{\ A} = \delta_i^{\ A}$, the spatial extrinsic curvature reduces to
\begin{equation}
K_{ij} = -\Gamma^r_{ij}\,n_r = -\Gamma^r_{ij}\,g_{rr}\,n^r,
\end{equation}
where the relevant Christoffel symbol is $\Gamma^r_{ij} = 
-r\,f(R)\,\gamma_{ij}$ and the radial component is $n_r = g_{rr}n^r = 
f(R)^{-1}\cdot(-f(R)\dot{T}) = -\dot{T}$. Substituting and evaluating at $r = R(\tau)$,
\begin{equation}
K_{ij} = -R\,f(R)\,\dot{T}\,\gamma_{ij} = -R\,f(R)\,\dot{T}\,\frac{g_{ij}}{R^2} 
= -\frac{f(R)\dot{T}}{R}\,g_{ij}\,.
\label{eq:K_spatial}
\end{equation}
Using the normalisation constraint~\eqref{eq:normalisation}, one can show that $f(R)\dot{T} = \sqrt{\dot{R}^2 + f(R)}$, so that
\begin{equation}
K_{ij} = -\frac{\sqrt{\dot{R}^2 + f(R)}}{R}\,g_{ij}\,.
\label{eq:K_spatial_final}
\end{equation}
The spatial component of the Israel conditions~\eqref{eq:Israel_covariant}, with $Z_2$ symmetry, gives
\begin{equation}\label{eq:junction_Kraus}
\frac{2\sqrt{\dot{R}^2 + f(R)}}{R} = \frac{\kappa_5^2}{3}\rho\,,
\end{equation}
where we have used $T_{\mu\nu} - \frac{1}{3}Tg_{\mu\nu} = 
\frac{\rho_{\scriptscriptstyle\mathrm{total}}}{3}g_{\mu\nu}$ for a perfect fluid with energy density $\rho_{\scriptscriptstyle\mathrm{total}}$ evaluated on the spatial components. Squaring 
eq.~\eqref{eq:junction_Kraus} and rearranging,
\begin{equation}
\frac{\dot{R}^2}{R^2} = \frac{\kappa_5^4}{36}\rho_{\scriptscriptstyle\mathrm{total}}^2 - \frac{f(R)}{R^2}.
\end{equation}
We now insert $f(R) = R^2/\ell^2 - \mu/R^2$ and identify 
$H = \dot{R}/R$, and obtain
\begin{equation}\label{h2-kraus}
H^2 = \frac{\kappa_5^4}{36}\rho_{\scriptscriptstyle\mathrm{total}}^2 + \frac{\Lambda_5}{6} + \frac{\mu}{R^4}\,,
\end{equation}
where we have used $\Lambda_5 = -6/\ell^2$. This is precisely the BDL Friedmann equation~\eqref{eq:BDL_Friedmann} with the integration constant identified as $\mathcal{C} = \mu$. The Kraus picture therefore not only leads to the same metric and Friedmann equation, but also provides a geometric interpretation of the 
otherwise unexplained integration constant $\mathcal{C}$ in the BDL derivation: $\mathcal{C}$ is not a free parameter to be fixed by initial conditions on the brane, but encodes the mass of a black hole in the bulk~\cite{Kraus_1999, Gregory_2002}. Since it enters the Friedmann equation as a term scaling as $a^{-4}$, it mimics a radiation fluid on the brane --- hence the name dark radiation. 

In the simplest case, one sets $\mu = 0$, reducing the bulk to pure $\mathrm{AdS}_5$. The dark radiation term vanishes, and the Friedmann equation~\eqref{h2-kraus} retains only the $\rho^2$ and $\Lambda_5$ terms. None of the geometric interpretation is lost: the bulk remains static, the brane still moves radially through it, and observers on the brane still interpret that motion as cosmological 
expansion driven entirely by their matter content. 

This duality has a striking conceptual consequence. Let us examine eq.~\eqref{h2-kraus}. For $\mu=0$, a vacuum brane ($\rho_\mathrm{total} = \lambda$) sits at rest at $r = \ell$ in the bulk when the RS-II tuning condition $\lambda^2 = -6\Lambda_5/\kappa_5^4$ is satisfied, 
recovering the static RS-II configuration of eq.~\eqref{rs-poincare}. A non-zero bulk black hole mass $\mu \neq 0$ sets the brane in motion even in the absence of matter, through the gravitational influence of the bulk black hole on the brane trajectory. Placing matter on the brane further perturbs the junction conditions and drives additional radial motion through the $\mathrm{AdS}_5$ bulk. What observers on 
the brane call the Hubble rate $H = \dot{a}/a$ is literally the radial velocity $\dot{R}/R$ of the brane, and the entire cosmological history of the brane universe --- its expansion rate, its matter content, its epoch of high-energy braneworld corrections --- is encoded in that single trajectory.

Crucially, however, Birkhoff's theorem applies only because the bulk is vacuum and the brane is perfectly homogeneous and isotropic. The moment either condition is broken --- by matter propagating into the bulk, or by inhomogeneous or anisotropic matter distributions on the brane --- the theorem no longer holds, and the bulk need no longer be Schwarzschild--$\mathrm{AdS}_5$. It is in the special case of homogeneity and isotropy where the influence of the bulk geometry on the brane reduces to the single scalar parameter $\mu$.

Beyond this symmetry, the bulk imprints additional degrees of freedom onto the brane that cannot be determined from brane data alone, and the projected gravitational equations on the brane cease to form a closed system. It is precisely to make this structure explicit, and to identify which bulk quantities appear in the brane equations and in what form, that Shiromizu, Maeda and Sasaki developed the covariant formalism we turn to next.

\section{Covariant brane gravity}\label{sec_sms}
The BDL and Kraus--Ida approaches, while powerful for cosmological 
backgrounds, both rely on the specific symmetries of a homogeneous and 
isotropic brane. As argued at the end of the previous section, the 
moment these symmetries are broken --- by inhomogeneous or anisotropic matter distributions on the brane, or by matter propagating into the bulk --- a fully covariant framework is needed. This was provided by Shiromizu, Maeda and Sasaki~\cite{Shiromizu_2000} (hereafter SMS), who derived the effective gravitational equations on the brane by projecting the five-dimensional Einstein equations onto the brane using the Gauss--Codazzi formalism.
The starting point is the most general five-dimensional bulk metric in 
Gaussian normal coordinates,
\begin{equation}
\dd s^2 = g_{\mu\nu}(x,y)\,\dd x^\mu \dd x^\nu + \dd y^2,
\label{eq:bulk_GN_general}
\end{equation}
where $g_{\mu\nu}(x,y)$ is a general metric depending on both the brane 
coordinates $x^\mu$ and the extra-dimensional coordinate $y$.

The key tool for projecting the five-dimensional curvature onto the brane is provided by the Gauss equation~\cite{wald1984general, Israel:1966rt}, which is a standard result from the differential geometry of embedded hypersurfaces. It relates the five-dimensional Riemann tensor ${}^{(5)}R_{ABCD}$ to the 
four-dimensional Riemann tensor $R_{\mu\nu\rho\sigma}$ of the induced metric $g_{\mu\nu}$ and the extrinsic curvature $K_{\mu\nu}$,
\begin{equation}\label{eq:Gauss}
{}^{(4)}R_{ABCD}= {}^{(5)}R_{EFGH} \,g_A^{\ E}g_B^{\ F}g_C^{\ G}g_D^{\ H} + 2K_{A[C}K_{D]B}\,,
\end{equation}
where square brackets denote antisymmetrisation. 

The bulk Riemann tensor ${}^{(5)}R_{EFGH}$ appearing on the 
right-hand side of eq.~\eqref{eq:Gauss} admits a natural decomposition into a part 
determined by the local matter content via the Einstein equations --- encoded in the Ricci tensor ${}^{(5)}R_{AB}$ and Ricci scalar ${}^{(5)}R$ --- and a remainder that is not locally determined by matter: the Weyl tensor ${}^{(5)}C_{ABCD}$. In conventional general relativity, the Weyl tensor is the traceless, fully antisymmetric part of the Riemann tensor and represents the free gravitational field --- tidal forces and gravitational waves --- that can propagate in the absence of any local matter source. In five-dimensional theories, it plays the same role, encoding the non-local gravitational degrees of freedom of the bulk that are not determined by the bulk energy-momentum tensor alone.

Specifically, the five-dimensional Riemann tensor decomposes as
\begin{equation}
{}^{(5)}R_{ABCD} = {}^{(5)}C_{ACBD} 
+ \frac{2}{3}\left(g_{A[C}{}^{(5)}R_{D]B} 
- g_{B[C}{}^{(5)}R_{D]A}\right) 
- \frac{1}{6}{}^{(5)}R\,g_{A[C}g_{D]B},
\label{eq:Riemann_decomp}
\end{equation}
where the factors of $2/3$ and $1/6$ are specific to five dimensions~\cite{wald1984general}. Projecting the bulk Weyl tensor onto the brane yields the symmetric traceless tensor $\mathcal{E}_{\mu\nu}$, defined as
\begin{equation}\label{eq:Weyl_proj}
\mathcal{E}_{\mu\nu} = {}^{(5)}C_{ACBD}\,n^C\,n^D\,g_{\mu}^{\ A}\,g_{\nu}^{\ B}\,,
\end{equation}
commonly referred to as the electric part of the bulk Weyl tensor, by analogy with the electric part of the four-dimensional Weyl tensor in the $(3+1)$-decomposition of general relativity~\cite{Maartens_2004}. It is symmetric ($\mathcal{E}_{[AB]}=0$), traceless ($g^{\mu\nu}\mathcal{E}_{\mu\nu} = 0$) and carries information about the bulk gravitational field \textit{at the brane} ($\mathcal{E}_{AB}n^B=0$).
Consider the five-dimensional Einstein equations in the bulk given in eq.~\eqref{EFE-5d}, where 
\begin{equation}\label{EFE-complete}
   {}^{(5)}G_{AB}=-\Lambda_5\, {}^{(5)}g_{AB}+\kappa_5^2\left( {}^{(5)}T_{AB}^{\scriptscriptstyle\mathrm{bulk}}+T_{AB}^{\scriptscriptstyle\mathrm{brane}}\delta(y)\right)
\end{equation}
with
\begin{equation}\label{tbrane}
  T_{\mu\nu}^{\scriptscriptstyle\mathrm{brane}}=T_{\mu\nu}-\lambda g_{\mu\nu}\,.
\end{equation}
Projecting ${}^{(5)}G_{AB}$ onto the brane and using the Gauss equation~\eqref{eq:Gauss}, gives
\begin{equation}
   {}^{(5)}G_{AB}\,g^A_{\ \mu}\,g^B_{\ \nu}={}^{(5)}R_{AB}\,g^A_{\ \mu}\,g^B_{\ \nu}-\frac{1}{2}{}g_{\mu\nu}{}^{(5)}R\,.
\end{equation}
Contracting the Gauss equation~\eqref{eq:Gauss}, using the decomposition of the Riemann tensor~\eqref{eq:Riemann_decomp}, the definition of the Weyl tensor on the brane~\eqref{eq:Weyl_proj}, and collecting $G_{\mu\nu}=R_{\mu\nu}-\tfrac{1}{2}Rg_{\mu\nu}$, one obtains
\begin{equation}
     {}^{(5)}G_{AB}\,g^A_{\ \mu}\,g^B_{\ \nu}={}^{(4)}G_{\mu\nu}-KK_{\mu\nu}+K_{\mu\rho}K^\rho_{\ \nu}+\frac{1}{2}g_{\mu\nu}\left(K^2-K_{\rho\sigma}K^{\rho\sigma}\right)+\mathcal{E}_{\mu\nu}\,.
\end{equation}
By projecting the right-hand side of eq.~\eqref{EFE-complete}
and given that the extrinsic curvature for the stress-energy tensor~\eqref{tbrane} is
\begin{equation}\label{Kmunu}
    K_{\mu\nu}=-\frac{1}{2}\kappa_5^2\left[T_{\mu\nu}+\frac{1}{3}(\lambda-T)g_{\mu\nu}\right]\,,
\end{equation}
we combine both sides of the projected Einstein equation and obtain
\begin{equation}\label{EFE-1}
    G_{\mu\nu}=-\Lambda_4g_{\mu\nu}+\kappa_4^2T_{\mu\nu}+6\frac{\kappa_4^2}{\lambda}S_{\mu\nu}-\mathcal{E}_{\mu\nu}+4\frac{\kappa^2_4}{\lambda}\mathcal{F}_{\mu\nu}\,,
\end{equation}
where we have used the relation~\eqref{Lambda-4} between $\Lambda_4$, $\Lambda_5$ and $\lambda$ introduced in the BDL calculation, and the relation~\eqref{kappas-lambda} between $\kappa_4$ and $\kappa_5$ characteristic of RS-II. We have also grouped terms into two new tensors, defined as
\begin{equation}
    \mathcal{F}_{\mu\nu}={}^{(5)}T_{AB}^{\scriptscriptstyle\mathrm{bulk}}\,g_\mu^{\ A}g_{\nu}^{\ B}+\left({}^{(5)}T_{AB}^{\scriptscriptstyle\mathrm{bulk}}\,n^An^B-\frac{1}{4}{}^{(5)}T^{\scriptscriptstyle\mathrm{bulk}}\right)\,g_{\mu\nu}\,,
\end{equation}
which is zero when the bulk is empty, ${}^{(5)}T_{AB}^{\scriptscriptstyle\mathrm{bulk}}=0$, and 
\begin{equation}\label{Smunu-general}
    S_{\mu\nu}=\frac{1}{12}T T_{\mu\nu}-\frac{1}{4}T_{\mu\alpha}T^\alpha_\nu+\frac{1}{24}g_{\mu\nu}\left(3T_{\alpha\beta}T^{\alpha\beta}-T^2\right)\,,
\end{equation}
which gathers all the quadratic terms of the stress-energy tensor arising from the extrinsic curvature terms. 

Finally, the projected Weyl tensor $\mathcal{E}_{\mu\nu}$, defined in eq.~\eqref{eq:Weyl_proj}, appears as an additional source term on the right-hand side of the effective Einstein equations, encoding the gravitational influence of the bulk on the brane.
\subsection{The brane viewpoint}
We will consider an empty bulk, $\mathcal{F}_{\mu\nu}=0$ and a perfect fluid of matter fields on the brane,
\begin{equation}\label{Tmunu}
    T_{\mu\nu}=(\rho+p)\,u_\mu u_\nu +p \, g_{\mu\nu}\,,
\end{equation}
with $u_\mu$, $\rho$ and $p$ being the four-velocity, energy density and pressure of the matter content confined to the brane, respectively, from this moment onwards throughout this entire work. For a stress-energy tensor of this kind, eq.~\eqref{Smunu-general} reduces to
\begin{equation}\label{Smunu}
    S_{\mu\nu}=\frac{1}{12}\rho\left[\rho\, u_\mu u_\nu + (\rho+2p)\, h_{\mu\nu}\right]\,,
\end{equation}
where $h_{\mu\nu}=g_{\mu\nu}+u_\mu u_\nu$ is the spatial part of the brane's metric. In the regime $\rho \gg \lambda$, the tensor $S_{\mu\nu}$ dominates over the matter stress-energy tensor in eq.~\eqref{EFE-1}. Consequently, although $T_{\mu\nu}$ describes the physical matter content on the brane, it is $S_{\mu\nu}$ that primarily governs the spacetime geometry encoded in $^{(4)}G_{\mu\nu}$ in this limit.
On the other hand, the projected Weyl tensor $\mathcal{E}_{\mu\nu}$ in eq.~\eqref{EFE-1} encodes contributions from five-dimensional graviton effects. From the perspective of a brane observer, these appear as non-local corrections. A non-vanishing Weyl tensor may be sourced by bulk matter (bulk black holes and/or $\mathcal{F}_{\mu\nu}\neq 0$) or arise as a backreaction to inhomogeneities on the brane, as we shall see in the next section, encoded via the Codazzi equation. 

One can interpret the Weyl term as a dark stress-energy tensor and take the fluid ansatz:
\begin{equation}\label{Emunu}
    \mathcal{E}_{\mu\nu}=-\kappa_4^2\big(\rho_\mathcal{E}\,u_\mu u_\nu +\frac{1}{3}\rho_\mathcal{E\,}h
    _{\mu\nu}+q_\mu^\mathcal{E}u_\nu+q_\nu^\mathcal{E}u_\mu+\pi_{\mu\nu}^{\mathcal{E}}\big)\,,
\end{equation}
where the scalar $\rho_{\mathcal{E}}$ can be interpreted as the energy density of an effective dark fluid residing in the bulk, while $q_\mu^\mathcal{E}$ and $\pi_{\mu\nu}^{\mathcal{E}}$ represent a momentum density and anisotropic stress, respectively.
\section{Conservation equations}\label{sec:conservation-equations}
The effective Einstein equations~\eqref{EFE-1} are not sufficient by themselves to determine the dynamics on the brane. Additional constraints arise from the Codazzi equation, which relates the divergence of the extrinsic curvature to the bulk geometry, together with the four-dimensional Bianchi identity,
\begin{equation}\label{eq:bianchi}
    \nabla^\mu G_{\mu\nu}=0\,.
\end{equation}
The Codazzi (or Mainardi--Codazzi) equation is the second fundamental relation from the differential geometry of embedded hypersurfaces~\cite{wald1984general}, alongside the Gauss equation. It describes how the extrinsic curvature $K_{\mu\nu}$ varies along the brane:
\begin{equation}\label{eq:Codazzi}
\nabla_\nu K^\nu_{\ \mu}-\nabla_\mu K= \,^{(5)}R_{AB}\,n^A\,g_\mu^{\ B}\,.
\end{equation}
The right-hand side couples the evolution of $K_{\mu\nu}$ to the bulk Ricci tensor, and hence to the bulk matter content via the bulk Einstein equations. In the absence of bulk matter, ${}^{(5)}T_{AB}=0$, the right-hand side reduces to a contribution from the bulk cosmological constant only.

%
\subsection{Local conservation equations}
Using the four-dimensional Bianchi identity together with the Codazzi equation~\eqref{eq:Codazzi} and the 
Israel junction conditions with $Z_2$ symmetry, one obtains the set of local conservation equations on the brane:
\begin{equation}\label{eq:local_conservation}
    \nabla^\mu T_{\mu\nu} = 0\,,
\end{equation}
for the special case $^{(5)}T^{\scriptscriptstyle\mathrm{bulk}}_{AB}=0$. Otherwise, the right-hand side of eq.~\eqref{eq:local_conservation} would be non-zero and the equation would reflect an exchange between the brane and bulk (see e.g.~\cite{van_de_Bruck_2000,PhysRevD.62.124009}). Equation~\eqref{eq:local_conservation} states that in the absence of bulk fields, the brane energy-momentum tensor is conserved on the brane. In this case, one recovers the standard conservation equations of general relativity. If the matter on the brane has a perfect-fluid stress-energy tensor, the corresponding conservation equations are~\cite{Maartens_2004}
\begin{equation}\label{LCE-1}
    u^\nu\nabla_\nu\rho+\Theta(\rho+p)=0\,,
\end{equation}
\begin{equation}\label{LCE-2}
   \vec{\nabla}_\mu
p+(\rho+p)A_\mu=0\,,
\end{equation}
where $\Theta=\nabla^\mu u_\mu$ is the volume expansion rate of the $u^\mu$ worldlines, $A_\mu=u^\nu\nabla_\nu u_\mu$ the 4-acceleration and $\vec{\nabla}$ the spatially projected derivative,
\begin{equation}
    \vec{\nabla}_\mu F^{\alpha\dots}_{\ \dots\beta}=h_\mu^{\ \nu}h^{\alpha}_{\ \gamma}\dots h_\beta^{\ \delta}\nabla_\nu F^{\gamma\dots}_{\ \dots\delta}\,.
\end{equation}
On a FLRW brane in the local inertial frame comoving with $u^\mu$, i.e.~$u^\mu=(1,\vec{0})$, $A_\mu=0$ and $\Theta=3H=3\dot{a}/a$, eq.~\eqref{LCE-1} becomes
\begin{equation}\label{LCE-1-a}
    \dot{\rho}+3H(\rho+p)=0\,,
\end{equation}
where the dot is the derivative with respect to cosmic time $t$. One thus recovers the conservation equation of conventional cosmology on the brane as long as the bulk is empty.
\subsection{Non-local conservation equations}
The evolution of the projected Weyl tensor $\mathcal{E}_{\mu\nu}$ for $^{(5)}T^{\scriptscriptstyle\mathrm{bulk}}_{AB}=0$ is governed by the non-local conservation equations, obtained by combining the Bianchi identity~\eqref{eq:bianchi} with the local conservation equation~\eqref{eq:local_conservation},
\begin{equation}\label{eq:nonlocal_conservation}
    \nabla^\mu \mathcal{E}_{\mu\nu} = 
    \frac{6\kappa_4^2}{\lambda}\nabla^\mu S_{\mu\nu}\,.
\end{equation}
For a perfect fluid on the brane, we use eqs.~\eqref{Smunu} and~\eqref{Emunu} and obtain 
\begin{equation}\label{NLCE-1}
    u^\nu\nabla_\nu\rho_\mathcal{E}+\frac{4}{3}\Theta\rho_{\mathcal{E}}+\vec{\nabla}^\mu q_\mu^{\mathcal{E}}+2A^\mu q_\mu^\mathcal{E}
    +\sigma^{\mu\nu}\pi_{\mu\nu}^{\mathcal{E}}=0
\end{equation}
and
\begin{equation}\label{NLCE-2}
   \dot{q}^{\mathcal{E}}_{\langle\mu\rangle} 
+ \frac{4}{3}\Theta q^{\mathcal{E}}_\mu 
+ \frac{1}{3}\vec{\nabla}_\mu\rho_{\mathcal{E}} 
+ \frac{4}{3}\rho_{\mathcal{E}}A_\mu 
+ \vec{\nabla}^\nu\pi^{\mathcal{E}}_{\mu\nu} 
+ A^\nu\pi^{\mathcal{E}}_{\mu\nu} 
+ \sigma_\mu{}^{\nu}q^{\mathcal{E}}_\nu 
- \varepsilon_\mu{}^{\nu\alpha}\omega_\nu q^{\mathcal{E}}_\alpha 
= -\frac{(\rho+p)}{\lambda}\vec{\nabla}_\mu\rho\,,
\end{equation}
where $\sigma_{\mu\nu}=\vec{\nabla}_{\langle \mu} u_{\nu\rangle}$ is the shear rate and $\omega_\mu=-\frac{1}{2}\varepsilon_{\mu\alpha\beta}\vec{\nabla}^\alpha\,u^\beta$ the vorticity rate, with $\varepsilon_{\mu\alpha\beta}$ the projection orthogonal to $u^\mu$ of the four-dimensional brane alternating tensor. 
Equation~\eqref{NLCE-2} shows how the components of $\mathcal{E}_{\mu\nu}$ (left-hand side) are sourced by gradients in the brane matter (right-hand side). In FLRW branes and comoving coordinates, $\sigma_{\mu\nu}=0$ and $q^{\mathcal{E}}_\mu=0$~\cite{Maartens_2004}, and eq.~\eqref{NLCE-1} becomes
\begin{equation}
    \dot{\rho_{\mathcal{E}}}+4H\rho_{\mathcal{E}}=0\,,
\end{equation}
which has the dark radiation solution,
\begin{equation}
    \rho_\mathcal{E}=\rho_{\mathcal{E},i}\left(\frac{a_i}{a}\right)^4\,,
\end{equation}
where $\rho_{\mathcal{E},i}$ and $a_i$ are the dark radiation and scale factor evaluated at some cosmic time $t_i$, respectively. 

Note that in this limit we can identify $\mathcal{C}=\rho_{\mathcal{E}}$, connecting the integration 
constant appearing in the BDL and Kraus formalisms to the SMS covariant approach. In the SMS framework, $\rho_{\mathcal{E}}$ is the scalar component of the projected Weyl tensor $\mathcal{E}_{\mu\nu}$, defined via eq.~\eqref{Emunu}, which encodes the bulk gravitational degrees of freedom projected onto the brane. It only reduces 
to the dark radiation term $\mathcal{C}/a^4$ of the BDL approach in the homogeneous and isotropic limit.

The precise form of all these conservation equations depends on the spacetime configuration under consideration. For highly symmetric configurations --- such as static or spherically symmetric spacetimes --- many of the terms vanish identically by symmetry, and the system simplifies considerably.

Note that, even though eq.~\eqref{LCE-1} takes the usual form of stress-energy conservation in general relativity, the covariant derivatives depend on the underlying metric, which is itself modified by the new terms in the Einstein field equations. This effect becomes particularly pronounced for $\rho\gg\lambda$, where $S_{\mu\nu}$ dominates and braneworld corrections to the geometry are no longer negligible. 
Furthermore, the non-local conservation equations ~\eqref{NLCE-1} and~\eqref{NLCE-2}, which constrain the evolution of $\mathcal{E}_{\mu\nu}$, have no analogue in four-dimensional general relativity. 

These conservation equations will play a central role in the remainder of this work. They are the foundation of the gravitational collapse and Michel accretion calculations. In both cases, the interplay between the local and non-local equations is what distinguishes braneworld dynamics from standard general relativity.
\section{Cosmological epochs}
The RS-II generalisation of the first Friedmann equation~\eqref{eq:BDL_Friedmann_expanded} 
allows us to study the early Universe, governing the background evolution on the brane. 
For early epochs after the inflationary era, we neglect the brane cosmological constant 
$\Lambda_4$ and consider a flat Friedmann brane ($k=0$), so that 
eq.~\eqref{eq:BDL_Friedmann_expanded} reduces to
\begin{equation}\label{mod-Friedmann}
    H^2\simeq\frac{8\pi}{3M_4^2}\left(\rho+\frac{\rho^2}{2\lambda}+\rho_{\mathcal{E}}\right)\,,
\end{equation}
where we have identified $\mathcal{C}=\rho_{\mathcal{E}}$ following the covariant approach 
of Section~\ref{sec_sms}. In principle, the conservation equations~\eqref{NLCE-1} and~\eqref{NLCE-2} permit a background with a homogeneous and isotropic yet non-vanishing $\rho_{\mathcal{E}}$, even when the brane matter content is a perfect fluid with $\nabla_\mu \rho = 0$. However, cosmological constraints require $\rho_{\mathcal{E}}$ to be subdominant at the background level; for instance, $\rho_{\mathcal{E}} / \rho_{\scriptscriptstyle\mathrm{rad}} \lesssim 0.062$ at the epoch of Big Bang Nucleosynthesis (BBN)~\cite{Sasankan_2017-1,Sasankan_2017-2}. We therefore set 
$\rho_{\mathcal{E}}/\rho \to 0$, yielding
\begin{equation}\label{mod-Friedmann-1}
    H^2\simeq\frac{8\pi}{3M_4^2}\left(\rho+\frac{\rho^2}{2\lambda}\right)\,.
\end{equation}
Equation~\eqref{mod-Friedmann-1} reveals two distinct expansion regimes depending on the magnitude of $\rho$ relative to $\lambda$. For $\rho\ll\lambda$, expansion follows the standard linear relation $H^2\sim\rho$, while for $\rho\gg\lambda$ it obeys the non-standard quadratic scaling $H^2\sim\rho^2$. The crossover between regimes occurs when $\rho \sim \lambda$, defining a cosmic crossover time
\begin{equation}
    t_c\equiv \frac{\ell}{2}\,.
\end{equation}
Assuming a radiation-dominated background with $\rho \propto a^{-4}$, the two regimes give
\begin{subequations}
\begin{equation}\label{quad-reg}
    a\propto t^{1/4}\,,\qquad    
    \rho=\frac{3M_4^2}{32\pi t_c t}\,,\qquad  
    H^{-1}=4t \qquad (t\ll t_c)\,,
\end{equation}
\begin{equation}\label{linear-reg}
    a\propto t^{1/2}\,,\qquad  
    \rho=\frac{3M_4^2}{32\pi t^2}\,,\qquad     H^{-1}=2t \qquad (t\gg t_c)\,.
\end{equation}
\end{subequations}
For $t \gg t_c$, standard cosmology is recovered and the conventional expansion history is reinstated. 
\section{The effective prescription: a word of caution}\label{sec:eff-pres}
In the case of a perfect fluid, one common approach is to rewrite 
eq.~\eqref{EFE-1} with an effective stress-energy tensor,
\begin{equation}
 ^{(4)}G_{\mu\nu}=\kappa_4^2\,\mathcal{T}_{\mu\nu}\,,
\end{equation}
such that
\begin{equation}\label{Tmunu-eff}
    \mathcal{T}_{\mu\nu}=(\rho_{\scriptscriptstyle\mathrm{eff}}+p_{\scriptscriptstyle\mathrm{eff}})\,u_\mu u_\nu + p_{\scriptscriptstyle\mathrm{eff}}\,g_{\mu\nu}+\pi_{\mu\nu}^{\mathcal{E}}\,,
\end{equation}
where the effective quantities are~\cite{Maartens_2004} 
\begin{subequations}\label{effective-quantities}
\begin{equation}\label{rho-eff}  
\rho_{\scriptscriptstyle\mathrm{eff}}=\rho\left(1+\frac{\rho}{2\lambda}+\frac{\rho_\mathcal{E}}{\rho}\right),
\end{equation}
\begin{equation}\label{p-eff}
    p_{\scriptscriptstyle\mathrm{eff}}=p+\frac{\rho}{2\lambda}\left(2p+\rho\right)+\frac{\rho_\mathcal{E}}{3}\,.
\end{equation}
\end{subequations}

Depending on the properties of $\pi_{\mu\nu}^{\mathcal{E}}$, the effective description may or may not correspond to that of a perfect fluid. We assume that the effective fluid shares the same unit four-velocity $u_\mu$ as in eq.~\eqref{Tmunu}, implying the absence of any relative velocity between the dark radiation and the matter fluid. This assumption is valid for an exact FLRW brane with an empty bulk~\cite{Maartens_2004}, where symmetry arguments ensure that the dark component is comoving with $u^\mu$.

We emphasise that the effective fluid interpretation can be misleading and should be treated with care. As shown in eq.~\eqref{EFE-1}, in braneworld scenarios --- particularly in the high-energy regime $\rho \gg \lambda$ --- the spacetime curvature sourced by a fluid with energy density $\rho$ and pressure $p$ differs from that in the standard four-dimensional case. One may introduce an effective stress-energy tensor $\mathcal{T}_{\mu\nu}$ which, within conventional $(3+1)$-dimensional general relativity, reproduces the same spacetime geometry; however, this correspondence is purely formal. Consequently, quantities such as the sound speed or the equation of state must be interpreted with caution. It is tempting to introduce an effective equation of state and sound speed based on eqs.~\eqref{effective-quantities}. However, although the modified geometry influences the propagation of pressure-supported disturbances, these perturbations propagate within the physical matter fluid rather than any effective medium. Consequently, an effective sound speed lacks a clear physical meaning: the effective fluid is not a real medium supporting information transfer, but merely a formal construct reproducing the same geometry in standard gravity.

Moreover, the equation of state is intrinsically a thermodynamic quantity, encoding short-range microscopic interactions within the fluid, and should therefore be independent of the spacetime geometry. Introducing effective quantities that do not correspond to physical degrees of freedom can thus be misleading. While dimensional reduction provides a useful description from the perspective of a brane observer, we refrain from adopting an effective-fluid interpretation in most of this work for these reasons. In Section~\ref{ch:perturbations}, we instead show how the additional terms modify the propagation of scalar perturbations and hence the speed at which information propagates within the matter fluid.

    \chapter[Perturbations and the flow of information]{Perturbations and the flow\\ of information}\label{ch:perturbations}
    \thispagestyle{fancy}
\pagestyle{fancy}

    In compressible fluid dynamics, a distinguished role is played by the information propagation speed associated with pressure-supported disturbances. Since pressure gradients provide the only mechanism by which spatially separated fluid elements can influence one another, the propagation of small-amplitude compressive perturbations determines the causal structure of the flow. A paradigmatic example is black hole 
accretion: comparing the fluid velocity to the information 
propagation speed determines whether perturbations can propagate upstream or are irreversibly advected downstream, a distinction that underlies the definition of the sonic point and the global structure of Bondi--Michel solutions.

A general characterisation of information transport is obtained by analysing linear perturbations of the flow. For a given background configuration, perturbations obey a dispersion relation $\omega(k)$ and the propagation of localised disturbances is governed by the group velocity $v_g=\dd \omega/\dd k$. In the conventional scenario, this group velocity coincides with the characteristic speeds of the first-order Euler equations and with the square-root of the thermodynamic derivative $(\dd p/\dd \rho)_s$ evaluated at constant entropy $s$, known as the sound speed, $c_s$. However, in more general situations --- such as dispersive effective theories, higher-derivative models, or fluids with additional dynamical degrees of freedom --- the perturbation equations need not be first order, the dispersion relation may be non-linear in $k$, and the information propagation speed is no longer determined by the thermodynamic quantity $c_s$. In such cases, the relevant propagation scale must be computed directly from the perturbation spectrum, with the group velocity (and, where appropriate, its high-frequency limit) providing the appropriate measure of information transport.\footnote{See Brillouin's discussion (Ref.~\cite{Brillouin1960}) on information/perturbation propagation speeds.} 

To investigate the propagation of information in a fluid in braneworld scenarios, we study the dynamics of scalar perturbations around the background solution by means of the perturbed Einstein equations and junction conditions. A crucial aspect of perturbation theory in braneworld models is that consistency requires perturbing both the metric and the stress-energy tensor simultaneously. In conventional cosmology, perturbing only the matter sector over a fixed background is often a valid approximation, since the coupling between matter and metric perturbations is weak enough to be neglected at leading order. However, in braneworld scenarios the intrinsic coupling between matter fluctuations on the brane and five-dimensional gravitational perturbations in the bulk geometry demands greater care: neglecting either sector leads to an incomplete description, as the perturbed Einstein equations enforce their mutual backreaction. In fact, restricting the analysis to matter perturbations alone artificially suppresses this coupling and recovers the standard result $v_g=c_s$, as if the extra dimension were absent. Carrying out the full perturbation analysis, as we do in this section, yields instead a third-order wave equation that encodes the brane–bulk coupling. Nevertheless, in the quadratic-regime and sub-horizon limits, the solutions of this richer equation reduce to the conventional result, confirming that the effect of the bulk is suppressed at small scales. 
\section{Scalar perturbations}
Scalar perturbations in braneworld scenarios have been studied extensively in the literature (see the references collected in Ref.~\cite{Maartens_2010}). The central difficulty is that, unlike at the background level --- where the projected Weyl tensor can be absorbed into an effective dark-radiation term and the dynamics reduce to a closed set of equations on the brane --- there is, in general, no consistent physical justification for setting $\delta \mathcal{E}_{\mu\nu}$ to zero, as these perturbations encode non-local gravitational degrees of freedom in the bulk. Consequently, scalar perturbations on the brane couple to bulk perturbations, and the resulting system is generally not closed. In certain regimes, however, suitable approximations allow the system to be closed, yielding a complete description in terms of brane quantities alone (e.g.~\cite{Langlois_2001}).

The parametrisation employed in this work follows the notation and definitions introduced by Kodama and Sasaki in Ref.~\cite{Kodama:1984ziu}, who developed a gauge-invariant formalism for cosmological perturbations in arbitrary spacetime dimensions. This framework was later adapted to braneworld cosmology by Kodama, Ishibashi and Seto in Ref.~\cite{Kodama_2000}. They derived master equations for scalar, vector and tensor perturbations in the bulk and related bulk and brane perturbations through the Israel junction conditions, working in Kraus's perspective of a moving brane in a static bulk. The scalar sector was subsequently investigated numerically by Cardoso et al.~\cite{Cardoso_2007}, who analysed the high-energy quadratic regime during the radiation era. Their results showed that the matter power spectrum on small scales exceeds the corresponding prediction of general relativity by at least an order of magnitude, while confirming agreement with general relativity on sufficiently large scales.

We follow Refs.~\cite{Kodama_2000, 2002PhRvD} in deriving the coupled evolution equations for the gauge-invariant scalar perturbations. To facilitate an analytic treatment, we first specify the matter content and subsequently introduce approximations under which the system becomes tractable. In all three applications considered in this thesis --- background cosmology, spherical gravitational collapse and Michel accretion --- the flow is assumed to be adiabatic, with entropy conserved along streamlines. The thermodynamic closure relation is therefore $p = p(\rho)$. To obtain analytic solutions in the high-energy quadratic regime, we further assume that the equation-of-state parameter $w$ remains constant, adopting the barotropic relation
\begin{equation}
    p=w\rho\,.
\end{equation}
Following the notation in~\cite{Kodama_2000,Cardoso_2007}, we parametrise the perturbed brane metric as
\begin{equation}\label{perturbed-brane-metric}
    \dd s^2_{\scriptscriptstyle\mathrm{brane}}=-(1+2\alpha Y)\,\dd t^2-2a\beta Y_i\dd t\dd x^i+a^2\left[\,(1+2h_L)\delta_{ij}+2h_TY_{ij}\,\right]\dd x^i\dd x^j
\end{equation}
where $a$ is the scale factor. The gauge-dependent variables $(\alpha,\beta,h_L,h_T)$ can be combined to construct the gauge-invariant quantities,
\begin{equation}
    \Phi=h_L+\frac{1}{3}h_T-\frac{Ha}{k}\sigma_g\,,\qquad \Psi=\alpha-\frac{1}{k}\frac{\dd}{\dd t}\left(a\sigma_g\right)\,,\qquad \sigma_g=\frac{a}{k}\frac{\dd h_T}{\dd t}-\beta\,,
\end{equation}
and $Y$, $Y_i$ and $Y_{ij}$ in eq.~\eqref{perturbed-brane-metric} denote the scalar harmonic and its associated vector and tensor harmonics,
\begin{equation}
    Y=e^{i\mathbf{k\cdot x}}\,,\qquad Y_i=-\frac{1}{k}\partial_i Y\,,\qquad Y_{ij}=\frac{1}{k^2}\partial_i\partial_j Y+\frac{1}{3}\delta_{ij}Y\,.
\end{equation}
For the right-hand side of the Einstein field equations~\eqref{EFE-1} we parametrise the perturbations of the matter sector,
\begin{equation}\label{deltaTs}
    \delta T_0^{\ 0}=-\delta\rho Y \,,\qquad \delta T_i^{\ 0}=-kY_i\delta q\,,\qquad \delta T_i^{\ j}=\delta pY\delta_i^{\ j}\,,
\end{equation}
where $\delta q$ denotes the scalar momentum potential and the matter anisotropic stress has been set to zero. For a perfect fluid this follows from the constitutive assumption itself --- $\Pi_{\mu\nu} = 0$ exactly at all orders --- and is therefore not an approximation. We also parametrise the perturbations of the Weyl fluid,
\begin{equation}\label{delta-Es}
    \delta \mathcal{E}_0^{\ 0}=\kappa_4^2\delta\rho_{\mathcal{E}} Y\,, \qquad \delta \mathcal{E}_i^{\ 0}=\kappa_4^2k Y_i\delta q_{\mathcal{E}}\,, \qquad \delta \mathcal{E}_i^{\ j}=-\kappa_4^2\left(\frac{1}{3}\delta\rho_{\mathcal{E}} Y\delta_i^{\ j}+k^2\delta\pi_{\mathcal{E}}Y_i^{\ j}\right)\,,
\end{equation}
where we adopt the fluid ansatz for $\mathcal{E}_{\mu\nu}$ given in eq.~\eqref{Emunu}. The variable $\delta\pi_{\mathcal{E}}$, which quantifies the perturbation of the Weyl anisotropic stress, is defined by
\begin{equation}
\delta\pi_{ij}^{\mathcal{E}}=\delta\pi_{\mathcal{E}}\left(\partial_i\partial_j-\frac{1}{3}\delta_{ij}\nabla^2\right)Y\,.
\end{equation}
By the Stewart--Walker lemma, $\delta\pi_{\mathcal{E}}$ is gauge-invariant at linear order because $\pi^{\mathcal{E}}_{\mu\nu}$ vanishes in the background spacetime. However, unlike the matter anisotropic stress, this is a geometric coincidence of the conformally flat FLRW background rather than a microphysical property, and $\delta\pi_{\mathcal{E}}$ is in general a non-trivial propagating degree of freedom sourced by bulk gravitational fields.

On the other hand, the perturbations in eqs.~\eqref{deltaTs} and~\eqref{delta-Es} are not individually gauge-invariant, but can be combined into the following gauge-invariant quantities:
\begin{subequations}
\begin{align}
\label{scalar-perturb-matter}
    \rho\Delta &= \delta\rho - 3H\delta q\,, \quad 
    & a(\rho+p)V &= -k\delta q - (\rho+p)a\sigma_g\,, \quad 
    & \Gamma &= \delta p - c_s^2\delta\rho\,,\\[10pt] \label{scalar-perturb-weyl}
    \rho\Delta_{\mathcal{E}} &= \delta\rho_{\mathcal{E}} - 3H\delta q_{\mathcal{E}}\,, \quad 
    & a(\rho+p)V_{\mathcal{E}} &= -k\delta q_{\mathcal{E}} - (\rho+p)a\sigma_g\,, \quad 
    & 0 &= \delta p_{\mathcal{E}} - \frac{1}{3}\delta\rho_{\mathcal{E}}\,.
\end{align}
\end{subequations}
In eq.~\eqref{scalar-perturb-matter} we introduce the thermodynamic quantity $c_s^2=\partial p/\partial\rho\big|_s$ evaluated at constant entropy, i.e.~$\Gamma=0$ for adiabatic processes. Note that there is no analogous perturbation $\Gamma_\mathcal{E}$ relating $\delta\rho_\mathcal{E}$ and $\delta p_{\mathcal{E}}$. Since $\mathcal{E}_{\mu\nu}$ is traceless, the Weyl energy density has a radiation equation of state $\rho_{\mathcal{E}}\propto a^{-4}$ both in the background and at linear order. Consequently, $\delta p_{\mathcal{E}}=\frac{1}{3}\delta\rho_\mathcal{E}$ exactly. The equation of state of the Weyl fluid is fixed geometrically and therefore admits no intrinsic entropy perturbation. 

Using the Einstein field equations, we relate the metric gauge-invariant quantities to the matter and Weyl perturbations,
\begin{subequations}\label{pert-EFE}
\begin{align}
    \label{Poisson}
    \Phi&=\frac{\rho a^2}{M_4^2k^2}\left[\left(1+\frac{\rho}{\lambda}\right)\Delta+\Delta_{\mathcal{E}}\right]\,,
\\[10pt] \label{velocity-metric}
    H\Psi-\dot{\Phi}&=\frac{(1+w)\rho a}{2M_4^2 k}\left[\left(1+\frac{\rho}{\lambda}\right)V+V_\mathcal{E}\right]\,,
\\[10pt] \label{slip-eq}
    \Phi+\Psi&=-\frac{a^2}{M_4^2}\delta\pi_{\mathcal{E}}\,,
    \end{align}
\end{subequations}
where an overdot denotes a derivative with respect to cosmic time $t$ defined in eq.~\eqref{perturbed-brane-metric}. Perturbing the spatial components of the local conservation equation~\eqref{eq:local_conservation}, we also obtain the Euler equation\footnote{Equation~\eqref{pert-nablaT} assumes $\rho+p\neq 0$, since it is obtained by dividing the momentum conservation equation by $\rho+p$. Consequently, it does not apply to a pure cosmological constant ($w=-1$), for which the momentum density vanishes identically and no fluid velocity perturbation is defined.}, following Ref.~\cite{Cardoso_2007},
    \begin{equation}\label{pert-nablaT}
       \dot{V}=-HV+\frac{k}{a}\Psi+\frac{k}{a}\frac{\Gamma+c_s^2\rho\Delta}{\rho(1+w)}\,.
    \end{equation}
We now identify the braneworld counterparts of the standard gauge-invariant equations. Equation~\eqref{Poisson} is the Poisson equation on the brane, with a high-energy correction factor $1+\rho/\lambda$ arising from $S_{\mu\nu}$ and $\Delta_\mathcal{E}$ by the projected Weyl tensor. Equation~\eqref{velocity-metric}, sourced by the $(0i)$ Einstein equation, is the momentum constraint, with high-energy and Weyl corrections analogous to those appearing in the Poisson equation. Equation~\eqref{slip-eq} is the slip equation. In the conventional scenario the right-hand side vanishes for a perfect fluid, giving $\Phi=-\Psi$, but here the Weyl anisotropic stress sources a non-zero gravitational slip even in the absence of matter anisotropic stress. Finally, eq.~\eqref{pert-nablaT} is the Euler equation, encoding momentum conservation of the fluid. The right-hand side describes acceleration due to gravity and pressure gradients, while the $HV$ term is the Hubble friction term, a kinematic consequence of the expanding background. The quantities $\Psi$ and $H$ are influenced by five-dimensional dynamics. The potential $\Psi$ receives contributions from the Weyl sector through the modified slip relation, while the background expansion rate $H$ is altered by the braneworld corrections appearing in the modified Friedmann equation~\eqref{mod-Friedmann}. 

To obtain a closed system of equations, we perturb the remaining equations relating the bulk geometry to the brane matter content. Specifically, we perturb the extrinsic curvature equation~\eqref{Kmunu}, where $K_{\mu\nu}$ is defined geometrically in eq.~\eqref{extrinsic-curvature}, following Refs.~\cite{2002PhRvD, Cardoso_2007}. For this, we need to define gauge-invariant variables that are combinations of five-dimensional perturbations in the bulk. Following Kodama et al.'s formalism~\cite{Kodama_2000}, the perturbed bulk metric reads
\begin{equation}
    {}^{(5)}\dd s^2=(q_{\alpha\beta}+f_{\alpha\beta})\dd x^\alpha \dd x^\beta+\frac{2r}{l}f_\alpha Y_i\dd x^\alpha\dd x^i+\frac{r^2}{l^2}\left[\,\left(1+2H_L\right)\delta_{ij}+2H_TY_{ij}\,\right]\,\dd x^i\dd x^j\,,
\end{equation}
where 
\begin{equation}
    q_{\alpha\beta}\dd x^\alpha\dd x^\beta=\frac{\ell^2}{z^2}(-\dd T^2+\dd z^2)
\end{equation}
with $z=\ell^2/r$ is the 2-metric in Poincaré coordinates, and $T$ is the bulk time, as in Chapter~\ref{ch:Kraus}. From these, one constructs the gauge-invariant combinations,
\begin{equation}\label{eqs-F}
    F=H_L+\frac{1}{3}H_T+\frac{X^\alpha D_\alpha r}{r}\,,\quad F_{\alpha\beta}=f_{\alpha\beta}+D_\alpha X_\beta+D_\beta X_\alpha\,,\quad X_\alpha=\frac{r}{kl}\left(f_\alpha+\frac{r}{kl}D_\alpha H_T\right)\,,
\end{equation}
where $D_\alpha$ is the covariant derivative with respect to the 2-metric $q_{\alpha\beta}$. For an empty maximally symmetric bulk, Mukohyama introduced a master variable $\Omega$~\cite{Mukohyama_2000}, in terms of which $F$ and $F_{\alpha\beta}$ can be expressed as
\begin{equation}
    F=\frac{1}{6r}\left(D^\mu D_\mu-\frac{2}{\ell^2}\right)\Omega\,,\qquad F_{\alpha\beta}=\frac{1}{r}\left[D_\alpha D_\beta-\frac{1}{3}q_{\alpha\beta}\left(2D^\mu D_\mu-\frac{1}{\ell^2}\right)\right]\Omega\,,
\end{equation}
and the bulk master variable satisfies
\begin{equation}
    \left(-\frac{\partial^2}{\partial T^2}+\frac{\partial^2}{\partial z^2}+\frac{3}{z}\frac{\partial}{\partial z}+\frac{1}{z^2}-k^2\right)\Omega=0\,.
\end{equation}
The perturbed junction conditions provide a boundary condition for $\Omega$ on the brane and determine the Weyl-fluid perturbations ($\Delta_\mathcal{E},V_{\mathcal{E}},\delta\pi_{\mathcal{E}}$) in terms of $\Omega$,
\begin{subequations}\label{epsilon-eqs}
    \begin{align}
    \label{omega-ev}
           0&= (\partial_n\Omega)|_b+\frac{1}{\ell}\left(1+\frac{\rho}{\lambda}\right)\Omega_b+\frac{6\rho a^3}{\lambda k^2}\Delta\,,
           \\[10pt]
\label{eqs-e-1}
    \kappa_4^2\rho\Delta_\mathcal{E}&=\frac{k^2}{\ell a^3}\left(\frac{k^2+3H^2a^2}{3a^2}\Omega_b-H(\partial_u \Omega|_b)\right)\,,
\\[10pt]\label{eqs-e-2}
    \kappa_4^2(1+w)\rho V_\mathcal{E}&=\frac{k^3}{3\ell a^2}\left(H\Omega_b-(\partial_u\Omega)|_b\right)\,,
\\[10pt]\label{eqs-e-3}
    \kappa_4^2\delta\pi_\mathcal{E}&=\frac{1}{2\ell a^3}\left(\frac{1}{3}k^2\Omega_b-H(\partial_u\Omega)|_b-\frac{3\rho(1+w)}{\ell \lambda}(\partial_n\Omega)|_b+(\partial_u^2\Omega)|_b\right)\,,
    \end{align}
\end{subequations}
where the subscript $b$ denotes evaluation at the brane, and
\begin{subequations}
\begin{align}
\label{partial-n}
     \partial_n=n^A\partial_A &=\frac{1}{a}\left(-H\ell\frac{\partial}{\partial T}+\sqrt{1+H^2\ell^2}\frac{\partial}{\partial z}\right)\,,
    \\[10pt]
    \label{partial-u}
    \partial_u=u^A\partial_A &=\frac{1}{a}\left(-H\ell\frac{\partial}{\partial z}+\sqrt{1+H^2\ell^2}\frac{\partial}{\partial T}\right)=\frac{1}{a}\frac{\dd}{\dd t}
\end{align}
\end{subequations}
are the directional derivatives orthogonal and parallel to the brane, respectively. Thus, the three Weyl-fluid perturbations are not independent; they are determined by a single bulk master variable evaluated on the brane.

Substituting eqs.~\eqref{eqs-e-1}--\eqref{eqs-e-3} into eqs.~\eqref{pert-EFE} allows us to express the gauge-invariant quantities $\Phi$, $\Psi$ and $V$ in terms of the density contrast $\Delta$ and the master variable $\Omega_b$,
\begin{subequations}
\begin{align}
    \Phi&=\frac{3a^2\rho(\rho+\lambda)}{k^2\ell^2\lambda^2}\Delta+\left(\frac{3H^2a^2+k^2}{6\ell a^3}\right)\Omega_b-\frac{H}{2\ell a^2}\Omega_b'\,,
\\[10pt]\label{psi}
    \Psi&=-\frac{3\rho a^2(3w\rho+4\rho+\lambda)}{k^2\ell^2\lambda^2}\Delta-\left[\frac{(3w+4)\rho^2}{2\ell^3a\lambda^2}+\frac{(5+3w)\rho}{2\ell^3a\lambda}+\frac{k^2}{3\ell a^3}\right]\Omega_b+\frac{3H}{2\ell a^2}\Omega'_b-\frac{1}{2\ell a^3}\Omega_b''\,,
\\[10pt]\label{v}
    V&=\frac{3wHa}{k(1+w)}\Delta-\frac{1}{k(1+w)}\Delta'+\frac{kH}{2\ell a^2}\Omega_b-\frac{k}{2\ell a^3}\Omega_b'\,,
    \end{align}
\end{subequations}
where the prime denotes derivative with respect to conformal time $a\partial_t=\partial_\eta$. Substituting~\eqref{psi} and~\eqref{v} into the Euler equation~\eqref{pert-nablaT} gives a wave equation for $\Delta$,
\begin{subequations}\label{wave-eq}
    \begin{equation}
\Delta''+\left(1-3w\right)aH\Delta'+\left(k^2c_s^2+\frac{3\rho a^2}{\ell^2\lambda}A+\frac{3\rho^2a^2}{\ell^2\lambda^2}B\right)\Delta=-\frac{k^2\Gamma}{\rho}+\left(\frac{k^4(1+w)}{3\ell a^3}\right)\Omega_b
 \end{equation}
 \begin{equation}
    A=3w^2-2w-1\qquad B=-6w-4\,.
 \end{equation}
\end{subequations} 
This equation closely resembles the standard density-contrast equation of general relativity, but contains both the high-energy corrections proportional to $\rho/\lambda$ and a non-local source term $\Omega_b$ coupling the brane dynamics to the bulk. 

Since the expressions above mix cosmic time and conformal time derivatives, we now establish the relevant scaling relations. Combining the energy-momentum conservation of ordinary matter~\eqref{LCE-1-a} with the modified Friedmann equation~\eqref{mod-Friedmann} allows us to find, in the $\rho/\lambda\gg1$ approximation,
\begin{equation}
    a\propto \eta^{1/(2+3w)}\,,\qquad H=\frac{1}{(2+3w)a\eta}\,.
\end{equation}
In the high-energy regime $H\ell\gg1$, eqs.~\eqref{partial-n} and~\eqref{partial-u} imply $\partial_n\simeq-\partial_u$. Since $\partial_u=a^{-1}\dd/\dd t$, eq.~\eqref{omega-ev} reduces to the boundary condition
\begin{equation}\label{boundary-cond-1}
    \frac{\dd\Omega_b}{\dd\eta}\simeq\frac{1}{3\eta}\left(1+\frac{3a\eta}{\ell}\right)\Omega_b+\frac{2\ell a^3}{k^2\eta}\Delta\,.
\end{equation}
\section{Bulk-brane coupling and the propagation of information}
In the previous section we derived the coupled evolution equations for the gauge-invariant quantities $\Delta$ and $\Omega_b$. These equations make explicit the mutual coupling between matter pertubations on the brane and scalar perturbations in the bulk. The quantity $\Omega_b$ originates from the bulk master variable and therefore encodes non-local gravitational effects. However, in the high-energy regime, the junction condition~\eqref{omega-ev} reduces to the effective brane equation~\eqref{boundary-cond-1}, allowing the bulk influence to be expressed entirely in terms of quantities evaluated on the brane. Equations~\eqref{wave-eq} and~\eqref{boundary-cond-1} therefore form a closed system. 

Moreover, these can be further combined into a single third-order differential equation of $\Delta(\eta)$, independent of the quantity $\Omega_b$. The appearance of a third-order equation  reflects the additional dynamical degree of freedom introduced by the bulk geometry.
We now plug in the plane wave ansatz: 
\begin{equation}
    \Delta=e^{i(kx-\omega\eta)}\,,
\end{equation}
and solve for $\omega$. The group velocity is then defined by
\begin{equation}
    v_g=\frac{\dd \omega}{\dd k}\,.
\end{equation}
We obtain three branches: the first two are equal but opposite in sign, $\pm\omega^{(1)}(k)$, and with all imaginary terms being suppressed on sub-horizon scales, $(k\eta)^{-1}\to 0$; the third solution $\omega^{(3)}(k)$ is purely imaginary (i.e.~not oscillatory) and fully suppressed in sub-horizon scales. Since the imaginary part of $\omega^{(1)}(k)$ is suppressed,
\begin{equation}
    \frac{\Im(\omega^{(1)}(k))}{\Re(\omega^{(1)}(k))}\ll1\,,
\end{equation}
the solution is a well-defined wavepacket. 

We compute the group velocity by differentiating the real part of $\omega^{(1)}(k)$ with respect to $k$ and obtain
\begin{equation}
    v_g=\frac{\dd \omega^{(1)}}{\dd k}\simeq \sqrt{w}+\left[f_1(w)+f_2(w,\eta)\left(\frac{a\eta}{\ell}\right)\right]\left(\frac{1}{k\eta}\right)^2\,,
\end{equation}
up to second order in the expansion around $(k\eta)^{-1}\to0$, with
\begin{equation}
f_1=-\frac{315w^4-318w^3-313w^2+152w+48}{72w^{5/2}\,(2+3w)^2},
\end{equation}
\begin{equation}
    f_2=-\frac{27w^3-12w^2+w+4}{6w^{3/2}\,(2+3w)}\,\eta^{\frac{1}{2+3w}-\frac{1}{3}}.
\end{equation}
We highlight that the dimensionless quantity $(a\eta/\ell)$ is a suppression factor for any $\eta$ during the quadratic regime, imposed by the condition $t/t_c\ll 1$ in cosmic time. Due to both suppression factors, we see that 
\begin{equation}
    v_g\simeq \sqrt{w}
\end{equation}
for sub-horizon scales in the adiabatic scenario. 

During the quadratic regime, the appropriate choice of equation of state is $w=1/3$, given an isotropic plasma dominated by ultra-relativistic degrees of freedom. In that case, the group velocity is
\begin{equation}
    v_g\simeq \frac{1}{\sqrt{3}}-\left[\frac{7}{3\sqrt{3}}+\frac{2}{\sqrt{3}}\left(\frac{a\eta}{\ell}\right)\right]\left(\frac{1}{k\eta}\right)^2.
\end{equation}
We conclude that for a barotropic fluid in the quadratic regime, the approximation $v_g\simeq c_s$ is reasonable in sub-horizon scales.

In previous literature \cite{Maartens_2004}, an effective sound speed, defined as $c_{s,\scriptscriptstyle\mathrm{eff}}^2=\partial p_{\scriptscriptstyle\mathrm{eff}}/\partial \rho_{\scriptscriptstyle\mathrm{eff}}$, was introduced. In the quadratic regime, this quantity evaluates to $c_{s,\scriptscriptstyle\mathrm{eff}}^2 = 5/3$ (corresponding to $w = 1/3$), which may naively suggest a superluminal propagation and thus a potential violation of causality. However, this interpretation is misleading: the quantity does not correspond to a physical signal propagation speed. 

\begin{mdframed}[style = mybbox]
We conclude that perturbations within the fluid propagate subluminally in the quadratic regime, and causality remains preserved. Even though the underlying geometry differs from the conventional scenario, the corrections to the group velocity in the adiabatic case are negligible in sub-horizon scales, and $v_g\approx c_s$. 
\end{mdframed}

\part{Higher-dimensional black holes}\label{part:higher-d}

 \chapter[Introduction]{Introduction}\label{intro-part2}
    \thispagestyle{fancy}
\pagestyle{fancy}

    Black holes in spacetimes with more than four dimensions have a surprisingly long theoretical history. The first exact higher-dimensional black hole solution was obtained by Tangherlini in 1963 \cite{Tangherlini:1963bw}, who generalised the Schwarzschild metric to an arbitrary number $(d+1)$ of spacetime dimensions. The rotating counterpart came much later: Myers and Perry (1986) \cite{Myers:1986un} found exact solutions for black holes rotating in all possible independent rotation planes in $d>3$ dimensions, providing the higher-dimensional analogue of the Kerr solution.

For several decades these results were regarded primarily as formal extensions of general relativity, of mathematical rather than physical interest. The situation changed dramatically in the late 1990s, when braneworld proposals gave concrete phenomenological motivation for studying gravity in higher dimensions. In particular, since models with large or warped extra dimensions can lower the effective fundamental Planck scale, they opened up the possibility of producing microscopic higher-dimensional black holes in particle collisions~\cite{Giddings_2002,Dimopoulos_2001,Dimopoulos_2002,Cheung_2002-a,Cheung_2002,Park:2001xc,Giudice_2002,CASADIO_2002,Mocioiu:2003gi}, making their properties directly relevant to high-energy physics. This spurred a large body of work on the geometry, thermodynamics, and Hawking emission of such objects, comprehensively reviewed by Kanti~\cite{Kanti:2004nr} and, from a more formal standpoint, by Emparan and Reall~\cite{Emparan_2008}.

Although the prospect of black hole production at colliders has largely receded in the absence of experimental evidence~\cite{2016,2018}, the study of braneworld black holes remains an active area today, pursued along two complementary lines. On one hand, higher-dimensional black holes continue to serve as probes of extra dimensions in specific phenomenological frameworks: Friedlander et al.~\cite{Friedlander:2022ttk,Friedlander:2023qmc} have studied the viability of higher-dimensional PBHs as dark matter candidates in ADD through their evaporation products and cosmological signatures, while Vitale et al.~\cite{vitale2026microscopicprimordialblackholes} have studied their coupling to the cosmological background and accretion history. In our works~\cite{Aldecoa-Tamayo_2026,aldecoatamayo2026cosmologicalaccretionbraneworldblack}, we have focused on warped geometries and studied overdensity collapse, accretion, evaporation history and signatures, while Ghoshal et al.~have studied PBH formation produced by the supercooled radion phase transition~\cite{ghoshal2025complementaryprobeswarpedextra}. 

On the other hand, braneworld black holes play a central role in the holographic description of quantum gravity effects. Within the holographic interpretation of AdS braneworlds, a classical bulk black hole localised on the brane is expected to be dual to a quantum-corrected black hole on the brane, with the CFT stress-energy tensor encoding the back-reaction of Hawking radiation~\cite{Emparan_2002}.

In Chapter~\ref{chap5:st-geometries}, we review the higher-dimensional black hole geometries most relevant to braneworld cosmology and discuss their applicability and limitations. In Chapter~\ref{chap6:hawking-evap}, we study Hawking radiation in the RS-II braneworld, discuss how the emission spectrum is modified with respect to the four-dimensional case, and present the tools employed to compute the different particle spectra in our framework.

Throughout this chapter, $h_{\mu\nu}$ denotes the metric perturbation around a background metric, $g_{\mu\nu} \rightarrow g_{\mu\nu} + h_{\mu\nu}$, following the universal convention in linearised gravity. This differs from \href{part:braneworlds}{\textbf{\textcolor{thesislilac}{Part I}}}, where $h_{\mu\nu} = g_{\mu\nu} + u_\mu u_\nu$ denoted the spatial projection tensor orthogonal to the fluid four-velocity, following the conventions of covariant braneworld gravity. We adopt the redefinition here because the notation $h_{\mu\nu}$ for the metric perturbation is completely standard in the linearised gravity and Hawking radiation literature, and departing from it would make comparison with existing work unnecessarily cumbersome.

    \chapter[Spacetime geometries]{Spacetime geometries}\label{chap5:st-geometries}
    \thispagestyle{fancy}
\pagestyle{fancy}

    \section{From four to higher dimensions}\label{sec:ST}
\subsection{The Schwarzschild--Tangherlini solution}
\begin{figure}
    \centering
    \includegraphics[width=0.58\linewidth]{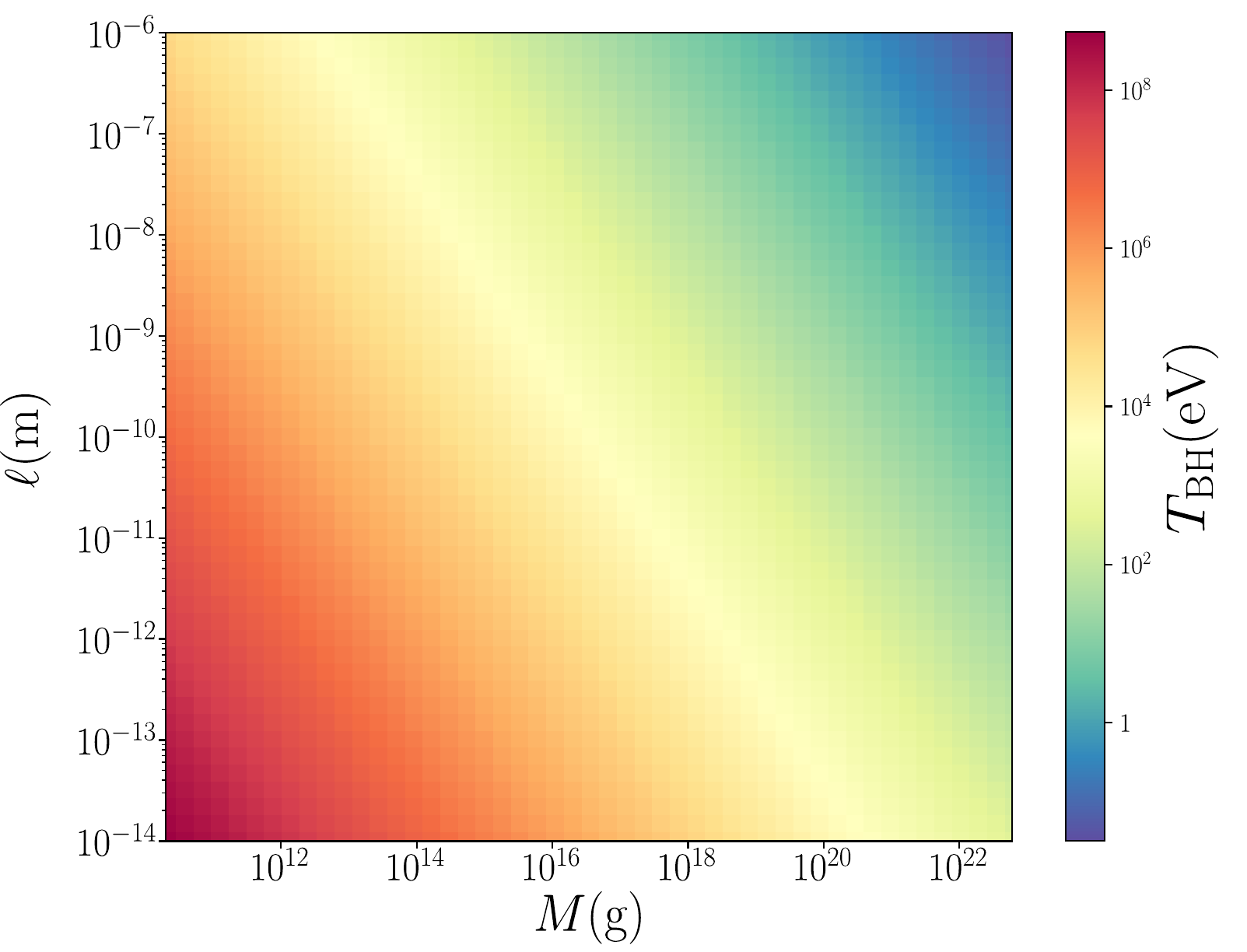}
    \caption[Black hole temperature vs.\ mass and AdS radius]{Black hole temperature dependence on the black hole mass $M$ (in grams) and the bulk AdS radius $\ell$ (in metres)
    for Schwarzschild--Tangherlini black holes in Randall--Sundrum Type-II universes, following \eqref{horizonradius} and \eqref{st-temp}.}
    \label{fig:tBH-5d}
\end{figure}

The simplest higher-dimensional generalisation of the Schwarzschild metric is the Schwarzschild--Tangherlini solution \cite{Tangherlini:1963bw}, which describes a static, spherically symmetric vacuum black hole in $d+1$ spacetime dimensions. We begin with the most general spherically symmetric metric in $d+1$ dimensions, which can be written as
\begin{equation}
    \dd s^2 = -e^{2\alpha(t,r)}\dd t^2 + e^{2\beta(t,r)}\dd r^2 + r^2\,\dd\Omega_{d-1}^2\,,
\end{equation}
where $\dd\Omega_{d-1}^2$ is the line element on the unit $(d-1)$-sphere and $\alpha,
\beta$ are arbitrary functions of $t$ and $r$. Substituting into the $(d+1)$-dimensional vacuum Einstein equations 
\begin{equation}
    R_{AB}=0\,,
\end{equation} 
the off-diagonal $(tr)$-component yields
\begin{equation}
    R_{tr} = \frac{d-1}{r}\,\partial_t\beta = 0\,,
\end{equation}
which immediately implies that $\beta = \beta(r)$ is time-independent. The $(tt)$ and $(rr)$ components then give, respectively,
\begin{align}\label{rr-eq}
    R_{tt} &= 0 \quad \longrightarrow \quad \partial_r(\alpha + \beta) = 0\,,\\
    R_{rr} &= 0 \quad \longrightarrow \quad \frac{2}{r}\partial_r\alpha + (d-2)\frac{e^{2\beta}-1}{r^2} = 0\,.
\end{align}
The first equation tells us that $\alpha + \beta$ has no $r$-dependence, and must therefore be a function of time only. Since $\beta$ is purely a function of $r$, this means 
\begin{equation}
    \alpha(t,r) + \beta(r) = \chi(t)\,.
\end{equation}
The function $\chi(t)$ can always be removed by redefining the time coordinate: setting 
\begin{equation}
    \dd\tilde{t}=e^{\chi(t)}\dd t
\end{equation}
absorbs $\chi(t)$ into the time coordinate, leaving 
\begin{equation}
   \alpha = -\beta\,. 
\end{equation}
This establishes that any spherically symmetric vacuum solution is necessarily static, providing the higher-dimensional analogue of Birkhoff's theorem. It allows us to write both metric functions in terms of a single function $f(r) \equiv e^{2\alpha(r)}$, so that $g_{tt} = -f(r)$ and $g_{rr} = f(r)^{-1}$. Substituting into the $(rr)$ equation~\eqref{rr-eq} then gives
\begin{equation}
    f'r + (d-2)(f-1) = 0\,,
\end{equation}
which is a first-order linear ODE in $f(r)$ and integrates directly to give
\begin{equation}
    f(r) = 1 - \frac{c}{r^{d-2}}\,,
\end{equation}
where $c$ is an integration constant. The constant $c$ is fixed by requiring that in the weak-field, large-$r$ limit the metric reproduces linearised gravity sourced by a point mass $M$~\cite{Emparan_2008}. Writing the metric as a small perturbation around flat Minkowski space,
\begin{equation}
    g_{AB} = \eta_{AB} + h_{AB}\,,\qquad |h_{AB}| \ll 1\,,
\end{equation}
where now $A,B=0,1,...,d$, 
and defining the trace-reversed perturbation
\begin{equation}
    \bar{h}_{AB} = h_{AB} - \frac{1}{d-1}\eta_{AB}\,h\,,\qquad h \equiv \eta^{AB}h_{AB}\,.
\end{equation}
With this choice, Einstein’s equations are simply written to leading order as
\begin{equation}\label{lin-EE}
    \square\,\bar{h}_{AB} = -16\pi G_{d+1}\,T_{AB}\,.
\end{equation}
For a static point mass $M$ at the origin, $T_{tt} = M\delta^{(d)}(\mathbf{x})$ with all other components vanishing, so only the $(tt)$ component of~\eqref{lin-EE} is sourced. For a static configuration $\square \to -\nabla^2$, giving the Poisson equation
\begin{equation}\label{poisson-hd}
    \nabla^2\bar{h}_{tt} = -16\pi G_{d+1}\,M\,\delta^{(d)}(\mathbf{x})\,.
\end{equation}
In spherical symmetry and away from the source, eq.~\eqref{lin-EE} takes the form
\begin{equation}
    \frac{1}{r^{d-2}}\frac{\dd}{\dd r}\left(r^{d-2}\frac{\dd\bar{h}_{tt}}{\dd r}\right) = 0\,,
\end{equation}
with decaying solution $\bar{h}_{tt} = \mu/r^{d-2}$. To fix $\mu$, we integrate~\eqref{poisson-hd} over a $d$-ball of radius $R$ and apply the divergence theorem:
\begin{equation}
    \oint_{S^{d-1}}\partial_r\bar{h}_{tt}\,\dd A = -16\pi G_{d+1}M\,.
\end{equation}
Substituting $\bar{h}_{tt} = \mu/r^{d-2}$ and evaluating the surface integral gives
\begin{equation}
    -(d-2)\frac{\mu}{R^{d-1}}\cdot R^{d-1}\Omega_{d-1} = -16\pi G_{d+1}M\,,
\end{equation}
and therefore
\begin{equation}\label{mu-param}
    \mu = \frac{16\pi G_{d+1}M}{(d-2)\,\Omega_{d-1}}\,.
\end{equation}
To recover $h_{tt}$ from $\bar{h}_{tt}$, we invert the trace-reversal. Since the only non-vanishing component of $\bar{h}_{\mu\nu}$ is $\bar{h}_{tt}$, the trace is $\bar{h} = \eta^{tt}\bar{h}_{tt} = -\bar{h}_{tt}$, and the inverse relation gives
\begin{equation}
    h_{tt} = \bar{h}_{tt} - \frac{1}{d-1}\,\eta_{tt}\,\bar{h} = \bar{h}_{tt}\left(1 - \frac{1}{d-1}\right) = \frac{d-2}{d-1}\,\bar{h}_{tt}\,.
\end{equation}
Substituting~\eqref{mu-param}:
\begin{equation}
    h_{tt} = \frac{16\pi G_{d+1}M}{(d-1)\,\Omega_{d-1}}\frac{1}{r^{d-2}}\,.
\end{equation}
Since $g_{tt} = -(1 - h_{tt})$ in the weak-field limit, identifying with $g_{tt} = -(1 - c/r^{d-2})$ gives
\begin{equation}\label{c-fixed}
    c = \frac{16\pi G_{d+1}M}{(d-1)\,\Omega_{d-1}}\,,
\end{equation}
with $\Omega_{d-1} = 2\pi^{d/2}/\Gamma(d/2)$ the area of the unit $(d-1)$-sphere. Writing $c \equiv r_0^{d-2}$ where $r_0$ is the horizon radius defined by $f(r_0) = 0$, the mass-radius relation reads
\begin{equation}
    M = \frac{(d-1)\,\Omega_{d-1}}{16\pi G_{d+1}}\,r_0^{d-2}\,.
\end{equation}
The unique asymptotically flat, static, spherically symmetric solution of the vacuum Einstein equations is therefore
\begin{equation}\label{ST-metric}
    \dd s^2 = -\left(1-\left(\frac{r_0}{r}\right)^{d-2}\right)\dd t^2 + \left(1-\left(\frac{r_0}{r}\right)^{d-2}\right)^{-1}\dd r^2 + r^2\,\dd\Omega_{d-1}^2\,.
\end{equation}
The Hawking temperature of the Schwarzschild--Tangherlini black hole is obtained from the surface gravity of the horizon. The metric~\eqref{ST-metric} admits a timelike Killing vector $\xi = \partial_t$, normalised so that $\xi^A\xi_A \to -1$ at spatial infinity. The event horizon at $r = r_0$ is a Killing horizon --- a null hypersurface on which $\xi^A\xi_A = 0$ --- and the surface gravity $\kappa_{\scriptscriptstyle\mathrm{S}}$ is defined by~\cite{wald1984general}
\begin{equation}
    \xi^A\nabla_A\xi^B\big|_{\mathcal{H}} = \kappa_{\scriptscriptstyle\mathrm{S}}\,\xi^B\,.
\end{equation}
For a metric of the form 
\begin{equation}
    \dd s^2 = -f(r)\dd t^2 + f(r)^{-1}dr^2 + r^2\dd\Omega_{d-1}^2\,,
\end{equation}
\begin{equation}
    \kappa_{\scriptscriptstyle\mathrm{S}} = \frac{1}{2}\partial_rf(r_0)\,.
\end{equation}
For the Schwarzschild--Tangherlini metric,
\begin{equation}
    \partial_rf(r_0) = \frac{d-2}{r_0}\,,
\end{equation}
giving $\kappa_{\scriptscriptstyle\mathrm{S}} = (d-2)/(2r_0)$. The Hawking temperature then follows from $T_{\scriptscriptstyle\mathrm{BH}}=\kappa _{\scriptscriptstyle\mathrm{S}}/2\pi$~\cite{Hawking:1975vcx},
\begin{equation}
    T_{\scriptscriptstyle\mathrm{BH}} = \frac{d-2}{4\pi r_0}\,,
\end{equation}
or, expressed in terms of the black hole mass,
\begin{equation}
    T_{\scriptscriptstyle\mathrm{BH}} = \frac{d-2}{4\pi}\left(\frac{(d-1)\,\Omega_{d-1}}{16\pi G_{d+1}\,M}\right)^{1/(d-2)}\,,
\end{equation}
so that $T_{\scriptscriptstyle\mathrm{BH}} \propto M^{-1/(d-2)}$. For fixed mass $M$, the temperature therefore decreases as the number of dimensions increases: higher-dimensional black holes are cooler than their four-dimensional counterparts of the same mass. 
In the five-dimensional case ($d=4$), the Hawking temperature is
\begin{equation}\label{st-temp}
    T_{\scriptscriptstyle\mathrm{BH}} = \frac{1}{2\pi r_0}\,,
\end{equation}
and the mass--radius relation gives
\begin{equation}\label{horizonradius}
    r_0 = \sqrt{\frac{8}{3\pi}}\frac{\sqrt{\ell M}}{M_4}\,,
\end{equation}
where we have used eq.~\eqref{M4-M5-rel}. The dependence of the Hawking temperature on the AdS warping and black hole mass can be seen in Figure~\ref{fig:tBH-5d}.

\subsection{The Myers--Perry solution and rotating black holes}
The rotating generalisation of the Schwarzschild--Tangherlini metric is the Myers--Perry solution \cite{Myers:1986un}, which describes a stationary, asymptotically flat vacuum black hole in $d+1 $ dimensions with $\lfloor d/2 \rfloor$ independent angular momenta, corresponding to rotation in each of the available independent planes~\footnote{The floor function $\lfloor x \rfloor$
denotes the greatest integer less than or equal to $x$.}. In four dimensions only one such plane exists and the solution reduces to the Kerr metric; in five and higher dimensions multiple independent angular momenta are possible and qualitatively new features emerge. Most notably, for $d \geq 5$ there is no upper bound on the spin parameter for a given mass, giving rise to ultraspinning configurations with no four-dimensional analogue \cite{Emparan_2008}.

Whether rotation must be included depends on the formation mechanism under consideration. For a black hole forming on the brane from the collapse of brane matter, the $Z_2$ symmetry of the RS-II construction and the absence of bulk matter forbid rotation in any plane involving the extra dimension; angular momentum can only be acquired in planes parallel to the brane. Although the collapse dynamics are expected to differ from the four-dimensional case due to the modified gravitational equations on the brane, the underlying symmetries of the FLRW universe persist: overdense regions in radiation-dominated early epochs are statistically nearly spherical, and the angular momentum acquired through tidal torques during spherical collapse vanishes by symmetry. Primordial black holes forming in this scenario are therefore expected to carry negligible angular momentum. By contrast, black holes produced in trans-Planckian particle collisions generically acquire angular momentum from the non-zero impact parameter of the collision \cite{Frost_2009} and are naturally described by the full Myers--Perry geometry.

In this work we restrict attention to non-rotating black holes. This approximation is well motivated for the primordial formation scenario considered throughout this thesis, and it allows us to isolate the effects associated with the extra dimension without introducing the additional parameters characterising rotation.
\subsection[Beyond \texorpdfstring{$d$}{d}-spherical horizons]{Beyond \texorpdfstring{$\boldsymbol{d}$}{d}-spherical horizons}
It is worth emphasising that the uniqueness result established above --- that Schwarzschild--Tangherlini is the unique asymptotically flat, static, spherically symmetric bulk solution --- rests crucially on the assumption of spherical symmetry encoded in the $r^2 \dd \Omega_{d-1}^2$ term of the ansatz. Relaxing this assumption opens up a much richer landscape of solutions. In contrast to four dimensions, where black-hole uniqueness and horizon-topology theorems imply that stationary asymptotically flat vacuum black holes possess spherical horizon cross-sections 
$S^2$, higher-dimensional vacuum gravity admits stationary black objects with extended or non-spherical horizons~\cite{Emparan_2008}. The most celebrated example is the black ring found by Emparan and Reall \cite{Emparan_2002} in five dimensions. Higher-dimensional gravity also admits black strings, black branes, and a variety of other extended black objects, with an increasingly rich phase structure as the number of dimensions grows \cite{Emparan_2008}; extended objects are moreover often subject to the Gregory--Laflamme instability \cite{Gregory_1993}, which can lead to the fragmentation of sufficiently thin configurations into localised black holes. Crucially, all of these non-spherical solutions rely on rotation or extended structure for their existence. In the spherically symmetric, non-rotating collapse of brane matter that characterises the primordial formation scenario we consider, such configurations are not expected to arise dynamically, and we therefore restrict attention to the Schwarzschild--Tangherlini geometry throughout.
\section{Brane black holes in Randall--Sundrum frameworks}\label{brane-bhs}
The Schwarzschild--Tangherlini solution derived above describes a black hole in an otherwise empty, flat higher-dimensional spacetime. This is, however, far from the situation encountered in the RS-II framework. In reality, a braneworld black hole is embedded in a curved AdS$_5$ bulk and localised on a brane that carries its own self-gravity. One might reasonably ask whether these features alter the black hole geometry in any significant way --- and if so, how. 

Note that in the formation scenario where the bulk is otherwise empty and the black hole arises from gravitational collapse induced by brane matter processes, spherical symmetry along the brane directions is preserved, and the $Z_2$ orbifold symmetry of the RS-II construction implies that any horizon extending into the bulk does so symmetrically on either side of the brane. With this geometry in mind, the rest of this section examines how the AdS bulk and the brane tension modify the picture relative to the flat vacuum case.

\begin{figure}
    \centering
    \includegraphics[width=0.6\linewidth]{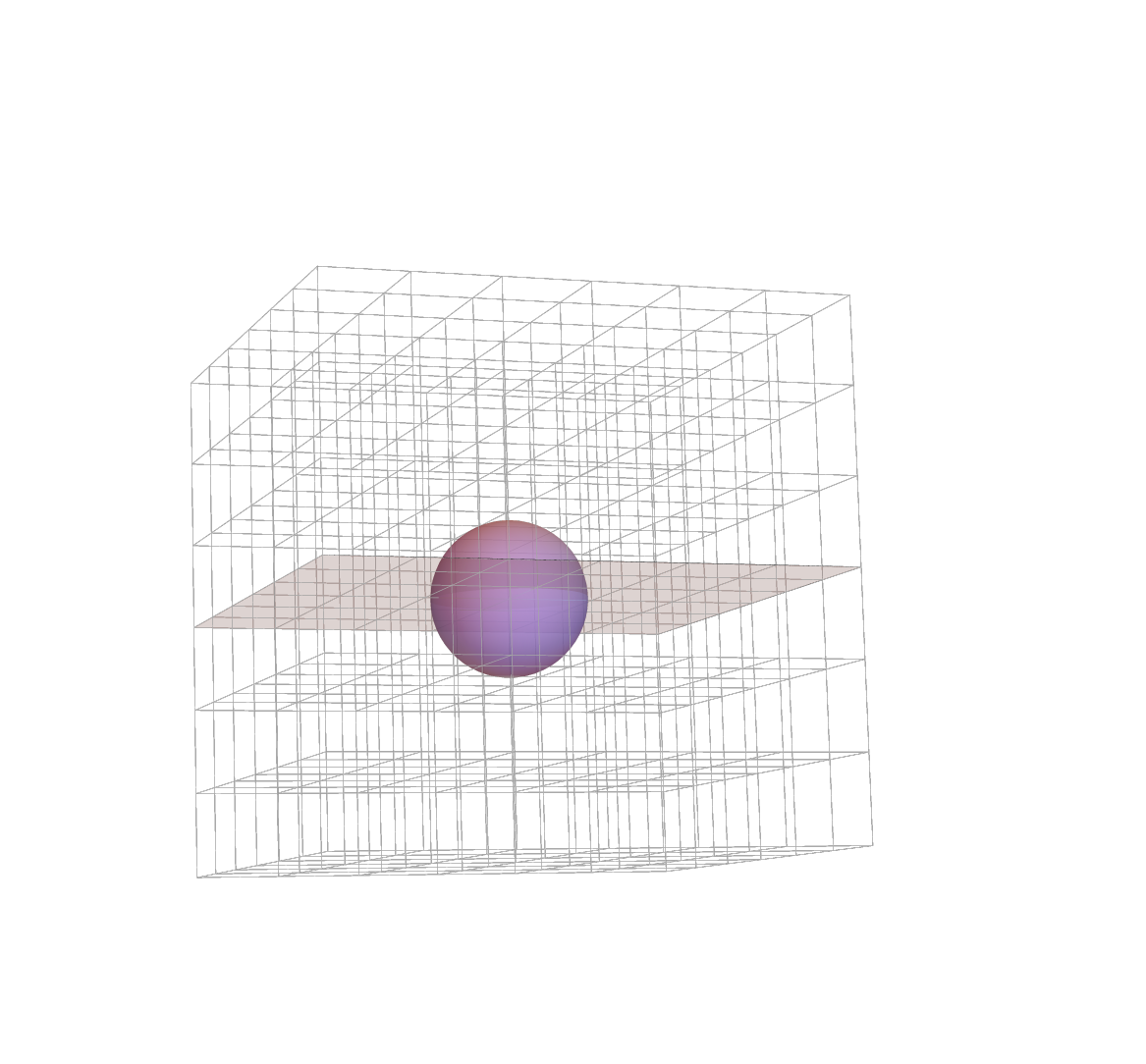}
    \caption[Spherically symmetric horizon intersected by a codimension-1 brane]{Spherically symmetric horizon equatorially intersected by a codimension-1 brane, embedded in a locally flat five-dimensional spacetime. }
    \label{fig:bh-5d}
\end{figure}

Let us first consider scales sufficiently small near the brane ($n_A\dd x^A=\dd y$) so that locally we can assume flat five-dimensional spacetime and ignore the global AdS warping (see Figure~\ref{fig:bh-5d}). In this limit, compact objects are not sensitive to the curvature of the extra dimension, and the Schwarzschild--Tangherlini solution applies. The AdS radius $\ell$ is the scale that parametrises this approximation: it is valid when $r_0 \ll \ell$, and black holes satisfying this condition have relatively low masses. We refer to them as \textbf{\emph{small} black holes}.

One might wonder whether the brane tension introduces any additional modification to the geometry of the black hole even if the bulk geometry is locally flat. However, the brane tension is already expressed in the warping of the spacetime in which the black hole is embedded --- it is precisely the fine-tuning condition between the brane tension and the bulk cosmological constant that produces the AdS$_5$ bulk \cite{RandallSundrum1999b, Gregory} --- and does not introduce any additional geometric deformation of the black hole metric. This is in contrast to higher-codimension scenarios, where the bulk geometry required to support the brane tension is no longer AdS$_5$, and the tension introduces a conical deficit in the transverse angular space \cite{Olasagasti_2000} --- a local geometric effect that persists even in the near-brane region. For example, in codimension-2 scenarios, embedded small black holes retain the local radial structure of the Schwarzschild--Tangherlini solution, but their total horizon area is reduced in proportion to the missing angular measure~\cite{Emparan_2002}, with corresponding modifications to their thermodynamic properties.

Now imagine a black hole forming on the brane with horizon radius $r_0$ \textit{much} larger than $\ell$. Recall from \href{part:braneworlds}{\textbf{\textcolor{thesislilac}{Part I}}} that, even though the extra dimension is infinite, the AdS warping endows the bulk with an effective finite volume, with an extension into the extra dimension of order $\ell$. This means that the horizon is expected to extend only a distance of order $\ell$ into the bulk. If $ r_0 \gg \ell$, the transverse extension of the horizon into the extra dimension is therefore negligible in comparison to its radial extension along the brane directions, and the geometry on the brane is well approximated by the four-dimensional Schwarzschild metric. We refer to these as \textbf{\emph{large} black holes}.

For $r_0\sim \ell$, the horizon can extend into the fifth dimension over distances comparable to its extent along the brane, but the background is no longer approximately flat in that direction. The effective confinement of the bulk to scales of order $\ell$ prevents the horizon from expanding isotropically as it would in flat five-dimensional spacetime, while the AdS warping makes distances transverse to the brane geometrically inequivalent to distances along it. As a result, neither the Schwarzschild--Tangherlini geometry nor the four-dimensional Schwarzschild solution provide an adequate description, making this intermediate regime the least understood of the three. The full description of a brane-localised black hole is expected to have a \emph{flattened pancake} shape~\cite{Emparan_2000-b,Emparan_2008}, and no exact analytical solution is known to date. However, the geometry is expected to
transition between the ``large'' and ``small'' regimes smoothly. 

\subsection{Binding to the brane}
While the brane tension does not modify the geometry of small black holes, it does affect their stability against escape into the bulk. Fraser and Eardley \cite{Fraser_2015} showed that small black holes on a positive-tension RS-II brane exhibit substantial gravitational binding energy, making them stable against escape into the bulk. This is not the case in ADD-type models, where the brane carries no tension and small black holes are not protected against recoil into the bulk from asymmetric Hawking emission into bulk modes \cite{Frolov_2002,Frolov_2002-b, Flachi_2005}. Large black holes, on the other hand, cannot escape into the bulk regardless of the framework, since their horizon size exceeds the characteristic size of the extra dimension.

This binding property is particularly important for the Hawking-radiation calculations discussed later, since it ensures that black holes formed on the brane remain localised there throughout their lifetime.
\section{From bulk to brane: a Weyl cloud and non-uniqueness}
The Schwarzschild--Tangherlini solution, as derived above, is a manifestly five-dimensional object. It is instructive, however, to examine what a brane-bound observer would perceive. Working in Gaussian normal coordinates and restricting the five-dimensional Schwarzschild--Tangherlini geometry to the brane hypersurface $y=0$, one obtains the induced metric
\begin{equation}\label{ST-metric-4d}
    {}^{(4)}\dd s^2 = -\left(1-\frac{r_0^2}{r^2}\right)\dd t^2 + \left(1-\frac{r_0^2}{r^2}\right)^{-1}\dd r^2 + r^2\,\dd\Omega_2^2\,.
\end{equation}
Computing the effective Ricci tensor ${}^{(4)}R_{\mu\nu}$ of this induced metric, one finds that it does not vanish. The induced metric is therefore not a vacuum solution of the four-dimensional Einstein equations; rather, it satisfies the trace-free equation
\begin{equation}\label{nonvanish-Rmunu}
    {}^{(4)}G_{\mu\nu} = {}^{(4)}R_{\mu\nu} \neq 0\,.
\end{equation}
Since there is no matter on the brane, this non-vanishing Ricci tensor is purely geometric in origin, arising from dimensional reduction of the bulk geometry. In the covariant framework, we identify ${}^{(4)}R_{\mu\nu} = -\mathcal{E}_{\mu\nu}$ and adopt the fluid decomposition of eq.~\eqref{Emunu}. The black hole is therefore surrounded by a \emph{Weyl cloud} --- a distribution of dark radiation sourced entirely by the bulk gravitational field and invisible to an observer who has no knowledge of the extra dimension.
To characterise the morphology of the gravitational “fluid” around the black hole as seen by a brane observer, we impose the traceless condition for a Schwarzschild--Tangherlini spacetime, yielding
\begin{equation}\label{rhoE-asr0}
    \rho_\mathcal{E}^{\scriptscriptstyle\mathrm{BH}}=-\frac{1}{\kappa_4^2}\frac{r_0^2}{r^4}\,,
\end{equation}
or, expressed in terms of the mass,
\begin{equation}\label{rhoE-vacuum}
    \rho_\mathcal{E}^{\scriptscriptstyle\mathrm{BH}}=-\frac{\ell M}{3\pi^2r^4}\,,
\end{equation}
and 
\begin{equation}\label{Pimunu}
    \pi^{\mathcal{E}}_{\mu\nu}=\text{diag}\bigg[0,-\frac{1}{3}\rho_\mathcal{E}^{\scriptscriptstyle\mathrm{BH}}g_{rr}+\bigg(\frac{r_0}{r}g^{rr}\bigg)^2,-\frac{1}{3}\rho_\mathcal{E}^{\scriptscriptstyle\mathrm{BH}}g_{\theta\theta},-\frac{1}{3}\rho_\mathcal{E}^{\scriptscriptstyle\mathrm{BH}}g_{\phi\phi}\bigg]\,,
\end{equation}
where we have taken $\pi^{\mathcal{E}}_{\mu\nu}$ to be diagonal due to the symmetries of the underlying metric. 
We find that, if $\pi^{\mathcal{E}}_{\mu\nu}$ is written in the form
\begin{equation}
\pi^{\mathcal{E}}_{\mu\nu}=\pi_{\mathcal{E}}\bigg(\frac{1}{3}\left(g_{\mu\nu}+u_\mu u_\nu\right)-r_\mu r_\nu\bigg)
\end{equation}
for some function $\pi_{\mathcal{E}}(r)$ and where $r_\mu$ denotes the unit radial vector, one obtains the relation
\begin{equation}\label{eos-weyl}
    \pi_{\mathcal{E}}=-\rho_\mathcal{E}^{\scriptscriptstyle\mathrm{BH}}\,.
\end{equation} 
This can be interpreted as an effective equation of state for the Weyl term with $w_\mathcal{E}=-1$, as done by Gregory et al.~\cite{Gregory_2004}. In general, $\rho_{\mathcal{E}}^{\scriptscriptstyle\mathrm{BH}}$ need not satisfy this relation in realistic scenarios. However, for an isolated black hole in an empty five-dimensional spacetime, eq.~\eqref{eos-weyl} is expected to hold, while alternative equations of state may characterise other simplified configurations.

We can now re-express the metric~\eqref{ST-metric-4d} in terms of the dark radiation energy density,
\begin{equation}\label{metric-cloud}
   \dd s^2=-\left(1+\frac{8\pi}{M_4^2}\rho_\mathcal{E}^{\scriptscriptstyle\mathrm{BH}}\,r^2\right)\dd t^2+\left(1+\frac{8\pi}{M_4^2}\rho_\mathcal{E}^{\scriptscriptstyle\mathrm{BH}}\,r^2\right)^{-1}\dd r^2+r^2\dd \Omega^2_2\,,
\end{equation}
where we emphasise that $|\rho_\mathcal{E}^{\scriptscriptstyle\mathrm{BH}}|$ decreases radially, and that $\rho_\mathcal{E}^{\scriptscriptstyle\mathrm{BH}} < 0$, as required for an attractive gravitational potential.

Equation~\eqref{metric-cloud} clarifies the interpretation of the Schwarzschild--Tangherlini solution for a brane-bound observer. In this picture, one observes a spherically symmetric cloud of dark radiation, with energy density profile $\rho_\mathcal{E}^{\scriptscriptstyle\mathrm{BH}}(r)$ determined by the five-dimensional parameter $M$ and given by eq.~\eqref{rhoE-vacuum}. The profile decreases rapidly with radius, featuring an event horizon at $r = r_0$ and a singularity at $r = 0$.
%
%
\subsection{Non-uniqueness and the tidal Reissner--Nordstr\"{o}m solution}
The analysis above raises an important subtlety. In four-dimensional general relativity, Birkhoff's theorem guarantees that the Schwarzschild metric is the unique static, spherically symmetric vacuum exterior solution. On the brane, this uniqueness is lost. The effective Einstein equations~\eqref{EFE-1} contain the projected Weyl tensor $\mathcal{E}_{\mu\nu}$ as an undetermined source term, and different bulk configurations give rise to different $\mathcal{E}_{\mu\nu}$
 and hence to different induced geometries on the brane, all of which may appear to a brane observer as a static, spherically symmetric black hole of the same mass $M$, yet with distinct causal structures and horizon properties. The brane field equations alone do not form a closed system, and without additional information about the bulk, the exterior solution is not uniquely determined.
This freedom was used by Dadhich, Maartens, Papadopoulos and Rezania \cite{Dadhich_2000} (hereafter DMPR), who observed that the algebraic symmetries of $\mathcal{E}_{\mu\nu}$ --- symmetric, traceless, and with the same index structure as an electromagnetic stress-energy tensor --- imply that any Einstein--Maxwell geometry can be reinterpreted as a vacuum brane solution by identifying the Maxwell stress-energy tensor with the projected Weyl tensor. This observation yields the \emph{tidal Reissner--Nordstr\"{o}m} solution:
\begin{equation}\label{DMPR}
    \dd s^2 = -\left(1 - \frac{2M}{r} + \frac{Q}{r^2}\right)\dd t^2 + \left(1 - \frac{2M}{r} + \frac{Q}{r^2}\right)^{-1}\dd r^2 + r^2\,\dd\Omega_2^2\,,
\end{equation}
where $Q$ is a \emph{tidal charge} parameter arising from the bulk Weyl field rather than from any electric charge on the brane. Unlike in the Reissner--Nordstr\"{o}m solution of general relativity, $Q$ is not constrained to be positive: $Q<0$ is not only permitted but physically natural, since a negative tidal charge strengthens the gravitational field on the brane, consistent with the intuition that the bulk gravitational field of the source mass $M$ is ``reflected back" onto the brane by the negative bulk cosmological constant \cite{Maartens_2004}. For $Q<0$ there is a single horizon located outside the Schwarzschild radius, and the central singularity remains spacelike. Matching to the expected five-dimensional gravitational potential at scales $r\ll\ell$ suggests $Q\simeq-2M\ell$ \cite{Maartens_2004}, so that the tidal charge is set by the mass and the AdS scale. Unlike the electric charge of the RN solution, a negative tidal charge increases rather than decreases the horizon radius.
The Weyl fluid associated with the DMPR solution satisfies a different effective equation of state from the ST-induced cloud. One finds \cite{Gregory_2004, Maartens_2004}:
\begin{equation}
    \pi_\mathcal{E} = -2\rho_\mathcal{E}^{\scriptscriptstyle\mathrm{BH}}\,,
\end{equation}
corresponding to $w_\mathcal{E} = -2$, in contrast to the $w_\mathcal{E} = -1$ of the Schwarzschild--Tangherlini-induced Weyl cloud. The two solutions therefore differ not only in their metric but in the equation of state of the bulk gravitational fluid they imply.
Which of these --- or indeed any other solution consistent with the brane field equations --- correctly describes a braneworld black hole depends on the bulk geometry, which remains unknown in the absence of an exact five-dimensional solution. This non-uniqueness is a feature of the bottom-up approach. The DMPR solution, for example, is well-motivated and offers a physically appealing interpretation of the Weyl cloud, but the bulk geometry corresponding to this induced metric is not known. The Schwarzschild--Tangherlini induced metric~\eqref{ST-metric-4d}, by contrast, has a transparent top-down origin: it is the projection onto the brane of the known exact five-dimensional Schwarzschild solution, valid in the regime $r_0 \ll \ell$ where the bulk is well approximated as locally flat. It is this clear bulk provenance that motivates our use of the Schwarzschild--Tangherlini metric throughout this work. Since the central aim of this thesis is to understand how bulk physics manifests itself in brane observables, we favour a geometry whose higher-dimensional origin is known, even if this comes at the expense of adopting the simplest small-black-hole approximation.
%
%
\section{The far-field regime in Randall--Sundrum braneworlds}\label{sec:garriga}
In the derivation of the Schwarzschild--Tangherlini solution in Section~\ref{sec:ST}, after imposing spherical symmetry and vacuum, the local Einstein equations determined the metric up to a single integration constant. The physical meaning of this constant emerged only after studying the gravitational field far from the source. There, we considered linearised perturbations of five-dimensional Minkowski spacetime generated by a point mass and used the resulting weak-field solution to fix the remaining freedom in the metric. In this sense, the Schwarzschild--Tangherlini geometry is not determined solely by its local vacuum Einstein equations; its interpretation relies crucially on its asymptotic behaviour.

Recall that on sufficiently small scales the RS-II bulk is locally Minkowskian. If the black hole horizon is much smaller than the AdS radius, then every region relevant to the local determination of the metric is effectively indistinguishable from five-dimensional Minkowski space.

The situation changes as one moves away from the source. At distances comparable to or larger than the AdS radius, the warping of the bulk can no longer be neglected and gravitational perturbations propagate in the full RS-II geometry, eq.~\eqref{static-RS}, rather than in flat five-dimensional spacetime. In what follows we study how a point mass perturbs the AdS bulk, locally deforms the brane, and how the Israel junction conditions relate these bulk-brane effects to the induced metric measured by a brane observer. It is precisely in this regime that the recovery of four-dimensional gravity becomes a non-trivial question. Solar-system observations require that the gravitational field of any localised source on the brane approach that predicted by four-dimensional general relativity at sufficiently large distances (evidently far beyond the flat regime) and this behaviour must emerge from the full five-dimensional dynamics of the RS-II spacetime. This problem was addressed by Garriga and Tanaka \cite{Garriga_2000}, who studied linear perturbations of the RS-II geometry sourced by matter confined to the brane. Their analysis, which we review below, provides the asymptotic behaviour that complements the Schwarzschild--Tangherlini description of the near-horizon region. Together, the two results capture the gravitational field of a localised source on the brane across the regimes in which each approximation is valid.

It is worth noting that the Garriga--Tanaka calculation makes no assumptions about the geometry close to the horizon. The analysis relies only on linear perturbations of the RS-II background and is therefore valid in the weak-field regime far from the source, irrespective of the detailed structure of the near-horizon region. In this sense, the asymptotic corrections derived by Garriga and Tanaka apply equally to the gravitational field of large black holes. However, as discussed in Section~\ref{brane-bhs}, when $r_0 \gg \ell$ the near-horizon geometry is already well approximated by the four-dimensional Schwarzschild solution. At the same time, as we shall see the Garriga--Tanaka corrections scale as $(\ell/r)^2$ and become negligible at the distances where the weak-field approximation is valid. Consequently, there is no analogous two-regime description for large black holes: both the near-horizon and asymptotic geometries are effectively four-dimensional and connect smoothly. The distinction between a locally five-dimensional Schwarzschild--Tangherlini geometry and a four-dimensional Garriga--Tanaka asymptotic regime arises only for small black holes with $r_0 \ll \ell$.
%
\subsection{Metric perturbations in the RS gauge}
Consider metric perturbations $h_{\mu\nu}$ about the RS-II background~\eqref{static-RS},
\begin{equation}
    \dd s^2=\left(g_{AB}+h_{AB}\right)\dd x^A\dd x^B\,,
\end{equation}
where $g_{AB}$ is the static RS-II metric in eq.~\eqref{static-RS}. We now consider the transverse traceless RS gauge introduced in~\cite{Garriga_2000}, where one chooses the hypersurface-orthogonal coordinates with $h_{A5}=0$, reducing the fifteen independent components of the symmetric tensor $h_{AB}$ to the ten components $h_{\mu\nu}$. We further impose 
\begin{equation}\label{RS-gauge}
    \partial^\nu h_{\mu\nu} = 0\,,\qquad h \equiv h^\mu_{\ \mu} = 0\,.
\end{equation}
This gauge has no remaining freedom, and the five polarisations of the five-dimensional graviton are encoded in the five independent components of $h_{\mu\nu}$. In this gauge, the perturbed metric takes the form
\begin{equation}\label{pert-metric}
    \dd s^2 = e^{-2|y|/\ell}\left(\eta_{\mu\nu} + h_{\mu\nu}\right)\dd x^\mu \dd x^\nu + \dd y^2\,.
\end{equation}
Linearising the bulk Einstein equations in $h_{\mu\nu}$, the perturbations satisfy the decoupled field equation \cite{Garriga_2000}
\begin{equation}\label{field-eq-pert}
    \left(e^{2|y|/\ell}\,\Box^{(4)} + \partial_y^2 - \frac{4}{\ell^2}\right)h_{\mu\nu} = 0\,,
\end{equation}
where $\square^{(4)} \equiv \eta^{\mu\nu}\partial_\mu\partial_\nu$ is the four-dimensional d'Alembertian. Boundary conditions for~\eqref{field-eq-pert} must be imposed via the Israel junction conditions on the brane. A subtlety arises here: in the RS gauge, the brane is located at $y = 0$ only in the absence of matter. When matter is present on the brane, it acts as a source that displaces the brane to $y = -\xi^5\left(x^A\right)$. The junction conditions are therefore most conveniently imposed after a coordinate transformation to a frame $\bar{x}^A$ in which the brane is always located at $\bar{y} = 0$ by construction. The transformation between the two gauges will be of the form
\begin{equation}
    \bar{y}=y+\xi^5(x^\rho)\,,\qquad \bar{x}^\mu=x^\mu+\xi^\mu(x^A)\,,
\end{equation}
where $\xi^A$ is expected to be small. In this frame, we also have $\bar{h}_{A5}=0$. We can decompose the vector components $\xi^A$ into a $y$-independent and dependent part,
\begin{equation}
    \xi^5=\hat{\xi}^5(x^\rho)
\end{equation}
\begin{equation}
    \xi^\mu=-\frac{\ell}{2}e^{-2|y|/\ell}\,\eta^{\mu\nu}\,\partial_\nu\hat{\xi}^5(x^\nu)+\hat{\xi}^\mu(x^\rho)
\end{equation}
where the terms $\hat{\xi}^A$ do not depend on $y$. Note that this specific form of the transformation is the most general one consistent with both gauges satisfying $h_{A5}=0$ and $\bar{h}_{A5}=0$, i.e.~maintaining the form of the metric~\eqref{pert-metric}.

Performing this transformation and imposing $Z_2$ symmetry $\bar{y}\rightarrow-\bar{y}$, the Israel junction conditions give \cite{Garriga_2000}
\begin{equation}\label{jc-GN}
    \left(\partial_{\bar{y}} + \frac{2}{\ell}\right)\bar{h}_{\mu\nu} = -\kappa_5^2\left(T_{\mu\nu} - \frac{1}{3}\eta_{\mu\nu}T\right)
\end{equation}
at $\bar{y} = 0^+$. Transforming back to the RS gauge, 
\begin{equation}\label{trans-hs}
    h_{\mu\nu}=\bar{h}_{\mu\nu}-\ell\partial_\mu\partial_\nu\hat{\xi}^5-\frac{2}{\ell}e^{-2|y|/\ell}\eta_{\mu\nu}\hat{\xi}^5+e^{-2|y|/\ell}\eta_{\rho(\mu}\partial_{\nu)}\hat{\xi}^\rho\,,
\end{equation} 
the junction condition for the perturbation $h_{\mu\nu}$ satisfies
\begin{equation}\label{jc-RS}
    \left(\partial_y + \frac{2}{\ell}\right)h_{\mu\nu} = -\kappa_5^2\Sigma_{\mu\nu}\
\end{equation}
at $y = 0^+$, where the effective source term in the RS gauge is
\begin{equation}\label{Sigma}
    \Sigma_{\mu\nu} = \left(T_{\mu\nu} - \frac{1}{3}e^{-2|y|/\ell}\eta_{\mu\nu}T\right) + \frac{2}{\kappa_5^2}\partial_\mu\partial_\nu\hat{\xi}^5\,.
\end{equation}
The second term in~\eqref{Sigma} encodes the brane-bending contribution. Requiring the trace $h^\mu_{\ \mu} = 0$
to hold in the RS gauge imposes $\Sigma^\mu_{\ \mu} = 0$, which determines $\hat{\xi}^5$ as the solution to 
\begin{equation}\label{brane-bend}
    \square^{(4)}\hat{\xi}^5 = \frac{\kappa_5^2}{6}\,T\,,
\end{equation}
confirming explicitly that the brane is bent by the matter content residing on it. Combining~\eqref{field-eq-pert} and~\eqref{jc-RS} by including delta-function terms at the discontinuity, the equations of motion for the perturbations take the unified form 
\begin{equation}\label{eom-RS}
    \left(e^{2|y|/\ell}\,\square^{(4)} + \partial_y^2 - \frac{4}{\ell^2} + \frac{4}{\ell}\delta(y)\right)h_{\mu\nu} = -2\,\kappa_5^2\,\Sigma_{\mu\nu}\,\delta(y)\,,
\end{equation}
which reduces to the vacuum equation~\eqref{field-eq-pert} away from the brane and reproduces~\eqref{jc-RS} upon integration across $y = 0$. 
%
%
\subsection{The Green's function and the KK spectrum}
Since the RS-II background~\eqref{static-RS} is translation-invariant along the brane directions $x^\mu$ but varies in $y$, we can Fourier transform~\eqref{field-eq-pert} in $x^\mu$ only:
\begin{equation}
    h_{\mu\nu}(x^\alpha, y) = \int \frac{\dd^4k}{(2\pi)^4}\,e^{ik_\sigma x^\sigma}\,\tilde{h}_{\mu\nu}(k^\alpha, y)\,,
\end{equation}
where $k^\alpha = (\omega, \mathbf{k})$ is the four-dimensional momentum. Under this transformation $\square^{(4)} \to -k^2 =-k^\mu k_\mu\equiv \omega^2 - |\mathbf{k}|^2$, and the field equation~\eqref{field-eq-pert} reduces to an ODE in $y$ for each mode:
\begin{equation}\label{mode-eq}
    \left(\partial_y^2 - e^{2|y|/\ell}k^2 - \frac{4}{\ell^2}\right)\tilde{h}_{\mu\nu}(k^\alpha,y) = 0\,.
\end{equation}
The equation admits a mode decomposition labelled by the four-dimensional invariant $k^2$, and its solutions naturally separate into two classes depending on the value of $k^2$.
For a mode to be massless from the four-dimensional perspective, it must satisfy $k^2 = 0$, i.e.~ $\omega^2 = |\mathbf{k}|^2$. When $k^2 = 0$, the $e^{2|y|/\ell}k^2$ term vanishes and the ODE reduces to
\begin{equation}
    \left(\partial_y^2 - \frac{4}{\ell^2}\right)\psi_0(y) = 0\,,
\end{equation}
which admits only the normalisable decaying solution $\psi_0(y) \propto e^{-2|y|/\ell}$, after imposing regularity as $y\rightarrow\infty$ and the junction condition $\partial_y\psi_0=-\tfrac{2}{\ell}\psi_0$ at $y=0$. This is the massless graviton zero mode, localised near the brane by the AdS warping and responsible for four-dimensional gravity at large distances.

For $k^2 = -m^2 < 0$, corresponding to a mode with four-dimensional mass $m > 0$, i.e.~$\omega^2 = |\mathbf{k}|^2 + m^2$, the equation becomes
\begin{equation}
    \left(\partial_y^2 + m^2e^{2|y|/\ell} - \frac{4}{\ell^2}\right)\psi_m(y) = 0\,,
\end{equation}
which can be transformed into a Bessel equation whose solutions are oscillatory in the bulk. These modes are not localised near the brane and extend throughout the bulk, forming a continuum labelled by $m \in [0,\infty)$ --- the massive Kaluza--Klein (KK) modes. From the four-dimensional perspective each KK mode appears as a massive graviton whose effect on brane physics is suppressed at large distances. 
Together, the zero mode and the KK continuum form a complete set of eigenstates of the operator in~\eqref{field-eq-pert}. The solution to the sourced equation~\eqref{eom-RS} can therefore be written as a superposition of these eigenstates weighted by the source, which is precisely what the Green's function $G_R(x,y;x',y')$ encodes--- it gives the response of the system at the point $(x^\mu,y)$ due to a perturbation at $(x'^\mu,y')$ in the past. It is defined by
\begin{equation}
    \left(e^{2|y|/\ell}\,\square^{(4)} + \partial_y^2 - \frac{4}{\ell^2} + \frac{4}{\ell}\delta(y)\right)G_R(x,y;x',y') = \delta^{(4)}(x-x')\delta(y-y')\,,
\end{equation}
so that the formal solution to~\eqref{eom-RS} is
\begin{equation}\label{formal-solution}
    h_{\mu\nu}(x^\mu,y) = -2\kappa_5^2\int \dd ^4x'\,G_R(x,y;x',0)\,\Sigma_{\mu\nu}(x')\,,
\end{equation}
where the integration is over the brane at $y'=0$. Subject to the boundary condition of boundedness as $y \to \infty$, the Green's function may be constructed from the complete set of eigenstates as~\cite{Garriga_2000}
\begin{equation}\label{Green-fn-full}
    G_R(x,y;x',y') = -\int\frac{\dd^4k}{(2\pi)^4}\,e^{ik_\mu(x^\mu-x'^\mu)}\left[\frac{\psi_0(y)\psi_0(y')/\ell}{|\boldsymbol{k}|^2-(\omega+i\varepsilon)^2} + \int_0^\infty \dd m\,\frac{\psi_m(y)\psi_m(y')}{m^2+|\boldsymbol{k}|^2-(\omega+i\varepsilon)^2}\right]\,,
\end{equation}
where the first term is the zero mode contribution, with the factor $1/\ell$ following from the normalisation of the zero mode wavefunction $\psi_0(y) \propto e^{-2|y|/\ell}$, and the integral over $m$ sums over the KK continuum. The $i\varepsilon$ prescription specifies the retarded boundary conditions, ensuring that the response receives contributions only from sources in its causal past. 

Since we are interested in the gravitational field of a static source, we work with the time-integrated Green's function 
\begin{equation}
    G(\mathbf{x},y;\mathbf{x}',y') = \int_{-\infty}^{\infty}\dd t'\,G_R(x,y;x',y')\,,
\end{equation}
which depends only on the spatial coordinates $\mathbf{x},\mathbf{x}'$ and $y$,$y'$. In Fourier space, this time integration sets $\omega=0$, reducing the denominators in~\eqref{Green-fn-full}, consistent with solving a static Poisson-like problem.

With suitable expansions of the Bessel functions, and taking both points on the brane ($y=y'=0$), the far-field behaviour on the brane ($r\gg \ell$) is given by~\cite{Garriga_2000}
\begin{equation}\label{Green-fn-brane}
G(\mathbf{x},0;\mathbf{x}',0) \approx -\frac{1}{4\pi\ell r}\left(1 + \frac{\ell^2}{2r^2} + \cdots\right)\,.
\end{equation}
The $1/r$ term in \eqref{Green-fn-brane} arises entirely from the localised graviton zero mode and reproduces four-dimensional Newtonian gravity, while the $\ell^2/r^3$  correction originates from the exchange of the continuum of KK gravitons. The recovery of four-dimensional gravity at large distances is therefore a consequence of the localisation of the zero mode near the brane and the suppressed contribution of the KK tower.

In the short-distance limit $r\ll\ell$ one instead finds $G\propto 1/r^2$, reflecting the locally five-dimensional nature of gravity.

For large separations in \textit{any} direction and one point on the brane and the other in the bulk, a separate leading-order approximation gives~\cite{Garriga_2000}
\begin{equation}\label{Green-fn-off}
G(\mathbf{x},y;\mathbf{x}',0) \approx -\frac{e^{-3|y|/\ell}}{8\pi\ell}\,\frac{2e^{-2|y|/\ell}r^2 + 3\ell^2}{\left(e^{-2|y|/\ell}r^2+\ell^2\right)^{3/2}}\,,
\end{equation}
showing that the gravitational response sourced on the brane decays steeply as one moves into the bulk, $y\rightarrow \infty$. 
%
%
\subsection{The graviton propagator}
%

In order to find the metric on the physical brane we first decompose the formal solution~\eqref{formal-solution} into contributions sourced directly by the matter stress-energy tensor and by the brane-bending term, writing $h_{\mu\nu}=h_{\mu\nu}^{(\rm m)}+h_{\mu\nu}^{(\rm \xi)}$, where $h_{\mu\nu}^{(\rm m)}$ accounts for the first term in $\Sigma_{\mu\nu}$~\eqref{Sigma} and $h_{\mu\nu}^{(\xi)}$ for the second:
\begin{equation}
  h^{(\rm m)}_{\mu\nu} =- 2\kappa_5^2\int \dd^4x'\,G_R(x,y;x',0)\left(T_{\mu\nu}(x') - \frac{1}{3}e^{-2|y|/\ell}\eta_{\mu\nu}T(x')\right)\,,
\end{equation}
\begin{equation}
      h^{(\xi)}_{\mu\nu} =- 4\int \dd^4x'\,G_R(x,y;x',0)\partial_\mu'\partial_\nu'\,\hat{\xi}^5(x')\,.
\end{equation}
With this decomposition, the gauge transformation~\eqref{trans-hs} becomes
\begin{equation}
\bar{h}_{\mu\nu}=\left(h^{(\rm m)}_{\mu\nu} +h^{(\rm \xi)}_{\mu\nu} \right)+\ell\partial_\mu\partial_\nu\hat{\xi}^5+\frac{2}{\ell}e^{-2|y|/\ell}\eta_{\mu\nu}\hat{\xi}^5-e^{-2|y|/\ell}\eta_{\rho(\mu}\partial_{\nu)}\hat{\xi}^\rho\,.
\end{equation} 
Evaluating at $y=0$, since the brane is located there to linear order, sets $e^{-2|y|/\ell}=1$. The remaining terms proportional to derivatives of $\hat{\xi}^5$ and $\hat{\xi}^\mu$ are gauge-dependent and may be absorbed into the residual four-dimensional coordinate freedom encoded in $\hat{\xi}^\mu$. One can verify that, with an appropiate choice of this residual gauge, the derivative contributions cancel against the brane-bending piece $h_{\mu\nu}^{(\xi)}$, leaving
\begin{equation}\label{pert-brane}
    \bar{h}_{\mu\nu} = -2\kappa_5^2\int \dd^4x'\,G(\mathbf{x},0;\mathbf{x}',0)\left(T_{\mu\nu} - \frac{1}{3}\eta_{\mu\nu}T\right) + \frac{2}{\ell}\,\eta_{\mu\nu}\,\hat{\xi}^5\,.
\end{equation}
This gives the metric perturbation induced on the brane. The first term represents the direct gravitational response to the matter source propagated through the bulk Green's function, while the second arises from the displacement of the brane itself within the bulk geometry. As we shall see below, this brane-bending contribution plays a crucial role in recovering the correct four-dimensional Einstein limit at large distances. 
%
%
\subsection{The Newtonian potential and KK corrections}
To make contact with observation, we now compute the metric perturbation for a point mass $M$ at rest on the brane, with stress-energy tensor
\begin{equation}
    T_{\mu\nu} = M\delta^{(3)}(\mathbf{x})\delta^0_\mu\delta^0_\nu\,,
\end{equation}
so that $T = T^\mu_{\ \mu} = -M\delta^{(3)}(\mathbf{x})$. The brane-bending function is determined by eq.~\eqref{brane-bend}. For a static source $\square^{(4)} \to -\nabla^2$, giving
\begin{equation}
    \nabla^2\hat{\xi}^5 = -\frac{\kappa_5^2}{6}M\delta^{(3)}(\mathbf{x})
\end{equation}
with solution
\begin{equation}\label{xi5-point}
    \hat{\xi}^5 = \frac{\kappa_5^2 M}{24\pi r} = \frac{G_5 M}{3r} = \frac{G_4\ell M}{3r}\,,
\end{equation}
where $r=|\mathbf{x}|$ and we used $\kappa_5^2 = 8\pi G_5$
and $G_4 = G_5/\ell$. The brane-bending contribution to eq.~\eqref{pert-brane} is therefore
\begin{equation}\label{brane-bending-term}
    \frac{2}{\ell}\eta_{\mu\nu}\,\hat{\xi}^5 = \frac{2G_4 M}{3r}\eta_{\mu\nu}\,.
\end{equation}
\begin{figure}
    \centering
    \includegraphics[width=0.7\linewidth]{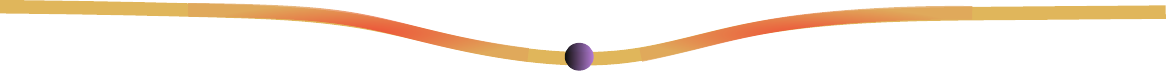}
    \caption[Illustration: Brane bending in the RS gauge]{Illustration of brane bending sourced by a mass on the brane in the RS gauge. Flatness is recovered in the asymptotic limit; in the intermediate regime, brane bending corrections to the induced metric must be included. In the small black hole limit, however, brane bending effects can be neglected in the near-horizon regime.}
    \label{fig:placeholder}
\end{figure}

This term is responsible for correcting the tensor structure of the graviton propagator, as we now see explicitly.
The matter part of the perturbation, evaluated at $y=0$ to linear order as justified above, follows from eq.~\eqref{formal-solution} with the stationary far-field Green's function~\eqref{Green-fn-brane}:
\begin{align}\label{hmunu-m}
    h^{(\rm m)}_{\mu\nu}\big|_{y=0} &= -2\kappa_5^2\int \dd^3x'\,G(\mathbf{x},0;\mathbf{x}',0)\left(T_{\mu\nu}(\mathbf{x}') - \frac{1}{3}\eta_{\mu\nu}T(\mathbf{x}')\right)\notag\\[10pt] 
    &\approx \frac{4G_4 M}{r}\left(1 + \frac{\ell^2}{2r^2}\right)\left(\delta^0_\mu\delta^0_\nu + \frac{1}{3}\eta_{\mu\nu}\right)\,.
\end{align}
Substituting eqs.~\eqref{brane-bending-term} and~\eqref{hmunu-m} into eq.~\eqref{pert-brane}, the full metric perturbation on the brane is
\begin{equation}
    \bar{h}_{\mu\nu} = \frac{4G_4 M}{r}\left(1+\frac{\ell^2}{2r^2}\right)\left(\delta^0_\mu\delta^0_\nu+\frac{1}{3}\eta_{\mu\nu}\right) + \frac{2G_4 M}{3r}\eta_{\mu\nu}+\mathcal{O}\!\left(\frac{\ell^4}{r^4}\right)\,.
\end{equation}
Extracting the components yields
\begin{equation}\label{GT-components}
    \bar{h}_{tt} = \frac{2G_4 M}{r}\left(1 + \frac{2\ell^2}{3r^2}\right)\,,\qquad \bar{h}_{ij} = \frac{2G_4 M}{r}\left(1 + \frac{\ell^2}{3r^2}\right)\delta_{ij}\,.
\end{equation}
The asymmetry between the coefficients of the $\ell^2/r^2$ correction in $\bar{h}_{tt}$ and $\bar{h}_{ij}$ is a direct consequence of the brane-bending contribution: the KK contribution by itself does not reproduce the tensor structure expected for four-dimensional Einstein gravity on the brane, and it is the brane-bending term that modifies this structure, leading to the characteristic coefficients $2/3$ and $1/3$~\cite{Garriga_2000, whisker2008braneworldblackholes}.
The gravitational potential is defined via $g_{tt} = -(1+2\Phi)$, giving $\Phi = -\bar{h}_{tt}/2$:
\begin{equation}\label{GT-potential}
    \Phi(r) = -\frac{G_4 M}{r}\left(1 + \frac{2\ell^2}{3r^2} + \mathcal{O}\!\left(\frac{\ell^4}{r^4}\right)\right)\,,
\end{equation}
recovering the standard $1/r$ Newtonian potential at large distances, with a subleading correction of relative order $(\ell/r)^2$. The $1/r$ term arises predominantly from the localised graviton zero mode and reproduces four-dimensional Newtonian gravity, while the $\ell^2/r^3$ correction originates from the exchange of the continuum of KK gravitons. The recovery of four-dimensional gravity at large distances is therefore a consequence of the localisation of the zero mode near the brane and the suppressed contribution of the KK tower. The corresponding linearised metric, \textit{valid only for} $r \gg \ell$, is the Garriga--Tanaka solution \cite{Garriga_2000},
\begin{equation}\label{GT-metric}
    \dd s^2 = -\left(1 - \frac{2G_4 M}{r} - \frac{4G_4 M\ell^2}{3r^3}\right)\dd t^2 + \left(1 + \frac{2G_4 M}{r} + \frac{2G_4 M\ell^2}{3r^3}\right)\dd r^2 + r^2\,\dd\Omega_2^2\,.
\end{equation}
%

%
%
\subsection{Observational constraint on the AdS radius}

The $(\ell/r)^2$ correction in~\eqref{GT-potential} must be consistent with experimental tests of Newton's inverse-square law. Torsion-balance experiments have confirmed the inverse-square law down to separations of $56\,\mu$m, placing the constraint $\ell \lesssim 44\,\mu$m at 95\% confidence~\cite{Kapner:2006si}. 
This constraint ensures that braneworld corrections to gravitational physics are entirely negligible at all solar-system and astrophysical scales, whilst remaining potentially significant for microscopic black holes with $r_0 \lesssim \ell$.
%
%
\subsection{The two-regime problem}
Neither the Schwarzschild--Tangherlini induced metric~\eqref{ST-metric-4d} nor the DMPR solution~\eqref{DMPR} approaches the Garriga--Tanaka metric~\eqref{GT-metric} in the limit $r \gg r_0$. There are therefore separate analytic solutions describing the near-horizon and far-field regimes, with no known solution that smoothly interpolates between the two. In this work, we employ the Schwarzschild--Tangherlini solution for near-horizon processes such as Hawking evaporation and accretion, where the geometry on scales $r \sim r_0$ is the relevant one, and the Garriga--Tanaka solution for far-field processes such as microlensing.

    \chapter[Hawking evaporation in braneworlds]{Hawking evaporation\\ in braneworlds}\label{chap6:hawking-evap}
    \thispagestyle{fancy}
\pagestyle{fancy}

    The classical description of a black hole is remarkably simple. Once formed, a stationary black hole is completely characterised by a small number of macroscopic parameters --- its mass, angular momentum and charge --- and possesses an event horizon from within which no signal can escape to future infinity. This picture changed dramatically with Hawking's discovery that quantum fields propagating on a black-hole spacetime give rise to a flux of particles observed at future infinity~\cite{Hawking:1974rv, Hawking:1975vcx}. Black holes are therefore not truly black: they emit radiation with a quasi-thermal spectrum and behave as thermodynamic objects possessing a temperature.
The origin of this radiation lies in the fact that the notion of a particle is observer-dependent in curved spacetime. Field modes that define the vacuum state at past infinity evolve non-trivially on a spacetime containing a black-hole horizon, so that an observer at future infinity detects a flux of particles. For a stationary black hole this flux is thermal, with a temperature determined solely by the surface gravity $\kappa_{\scriptscriptstyle\mathrm{S}}$ of the horizon,
\begin{equation}
    T_{\scriptscriptstyle\mathrm{BH}} = \frac{\kappa_{\scriptscriptstyle\mathrm{S}}}{2\pi}\,.
\end{equation}
For a Schwarzschild black hole in four dimensions this reduces to $T_{\scriptscriptstyle\mathrm{BH}} = (4\pi r_0)^{-1}=(8\pi G_4M)^{-1}$, implying that smaller black holes are hotter and radiate more intensely than larger ones. More generally, for the Schwarzschild--Tangherlini black hole in 
$(d+1)$ dimensions, the surface gravity calculation of Section~\ref{sec:ST} gives $T_{\scriptscriptstyle\mathrm{BH}} = (d-2)/4\pi r_0$, with the same qualitative behaviour. Since $T_{\scriptscriptstyle\mathrm{BH}} \propto M^{-1/(d-2)}$, the temperature increases as the black hole loses mass. As a consequence, the evaporation rate increases as the black hole loses mass, ultimately leading to complete evaporation within a finite time in the semiclassical description \cite{Page:1976df}. 

Although Hawking radiation is often described as blackbody emission, the spectrum observed at infinity is modified by the curvature surrounding the black hole. Particles produced near the horizon must propagate through an effective potential barrier before reaching distant observers, and a fraction are scattered back towards the black hole. These propagation effects are encoded in frequency-dependent transmission probabilities known as greybody factors, which distort the spectrum away from that of an ideal blackbody.

Following Ref.~\cite{Harris:2003eg}, the emission rate of a particle with spin $s$ in a $(4+n)$-dimensional spacetime\footnote{In this section we follow the conventions of Harris and Kanti \cite{Harris:2003eg} and use $n$ for the number of extra dimensions, related to the notation of Section~\ref{sec:ST} by $n = d-3$.} is obtained by summing over angular momentum modes $j$ and integrating over the available phase space,
\begin{equation}\label{emission-rate}
    \frac{\dd N^{(s)}(\omega)}{\dd t} = \sum_j\int\Gamma^{(s)}_{j,n}(\omega)\frac{1}{e^{\omega/T_{\scriptscriptstyle\mathrm{BH}}} - (-1)^{2s}}\,\frac{\dd^{n+3}k}{(2\pi)^{n+3}}\,,
\end{equation}
\begin{equation}\label{emission-rate-2}
    \frac{\dd E^{(s)}(\omega)}{\dd t} = \sum_j\int\Gamma^{(s)}_{j,n}(\omega)\frac{\omega}{e^{\omega/T_{\scriptscriptstyle\mathrm{BH}}} - (-1)^{2s}}\,\frac{\dd^{n+3}k}{(2\pi)^{n+3}}\,.
\end{equation}
Summing~\eqref{emission-rate-2} over all emitted species gives the total mass loss rate of the black hole. In eqs.~\eqref{emission-rate} and~\eqref{emission-rate-2}, $|\mathbf{k}| = \omega$ for massless particles, whereas $|\mathbf{k}|^2 = \omega^2 - m^2$ for massive particles. 

The quantities $\Gamma^{(s)}_{j,n}(\omega) \in [0,1]$ are the greybody factors, defined as the transmission probabilities through the effective potential barrier surrounding the black hole \cite{wald1984general}. They depend on the spin $s$ and angular momentum $j$ of the emitted mode and on the structure of the surrounding spacetime, including the number of dimensions. They satisfy $\Gamma^{(s)}_{j,n} \to 1$ in the geometric optics limit $\omega r_0 \gg 1$, where the potential barrier is transparent, and $\Gamma^{(s)}_{j,n} \to 0$ for $\omega r_0 \ll 1$. The greybody factors are obtained by solving the radial wave equation for the relevant field in the black hole background and extracting the transmission coefficient through the effective potential. 

The calculation of Hawking evaporation therefore requires both the horizon temperature and the corresponding greybody factors for the fields under consideration. In the RS-II scenario, there is a clear distinction between small and large black holes. \textbf{For small black holes}, the temperature is lower than that of a four-dimensional Schwarzschild black hole of the same mass, entering the emission rates via the exponential suppression in eqs.~\eqref{emission-rate} and~\eqref{emission-rate-2}. Furthermore, the five-dimensionality of the surrounding spacetime modifies the potential barrier and hence the greybody factors $\Gamma^{(s)}_{j,n}(\omega)$. \textbf{For large black holes}, the geometry on the brane is well described by the four-dimensional Schwarzschild metric, so the mass--temperature relation and the greybody factors of Standard Model particles revert to their standard values. In both regimes, however, additional bulk gravitational degrees of freedom become available as emission channels, with no counterpart in the conventional four-dimensional Hawking radiation calculation. Determining how all these effects alter the evaporation process is therefore central to braneworld black-hole phenomenology. One of the central questions is the impact of emission into the bulk: does the invisible bulk channel dominate over the visible brane channel, or vice versa? Could evaporation be mostly unobservable? The relative emission into the two channels depends on two factors: first, the number of available degrees of freedom in each channel—the Standard Model fields and the graviton zero mode on the brane, versus the continuum of KK modes in the bulk—and second, the probability that these modes penetrate the effective potential barrier. The latter is determined by both the structure of the barrier and the properties of the emitted particles, including their spin, angular momentum and frequency. Higher spin and higher angular momentum generally increase the height of the effective potential barrier and therefore tend to suppress the transmission probability of individual modes \cite{Page:1976df, Harris:2003eg, Cardoso:2005mh}. The interplay between these two factors, and the question of which channel dominates, has been the subject of debate spanning several decades. 
\section{The small black hole regime}\label{sec:hawking-hd}
Determining the relative importance of brane and bulk emission requires the calculation of the corresponding greybody factors. For Schwarzschild--Tangherlini black holes, these have been studied extensively in the literature. The greybody factors for brane-localised Standard Model fields propagating on this geometry were studied systematically by Harris and Kanti \cite{Harris:2003eg}, who computed the emission spectra for scalars, fermions and gauge bosons. Bulk scalar emission from non-rotating Schwarzschild--Tangherlini black holes was studied analytically by Kanti and March-Russell \cite{Kanti:2002nr}, who obtained results in the low-energy regime. Bulk graviton emission was subsequently computed independently by Creek, Efthimiou, Kanti and Tamvakis \cite{Creek_2006}, Cardoso, Cavagliá and Gualtieri \cite{Cardoso:2005mh,Cardoso_2006}, and Cornell, Naylor and Sasaki \cite{Cornell_2006}, all for non-rotating black holes. Together, these studies provided quantitative tests of the heuristic argument of Emparan, Horowitz and Myers \cite{Emparan_2000}, who proposed that higher-dimensional black holes radiate predominantly into brane-localised fields despite the presence of bulk channels.

These analyses were subsequently extended to rotating Myers--Perry black holes. Brane-localised scalar emission was studied by Harris and Kanti \cite{Kanti:2002nr}, gauge boson emission by Casals, Kanti and Winstanley \cite{Casals_2006}, and fermion emission by Ida et al.\ \cite{Ida_2006}.
To compute the greybody factors numerically, one exploits the fact that the wave equation for a field of any spin propagating in the Schwarzschild--Tangherlini background is separable due to the symmetries of the metric. Writing the field as a product of time, angular and radial parts,
\begin{equation}
    \Psi_s = e^{-i\omega t}e^{im\varphi}R_s(r)S_{s,j}^m(\theta)\,,
\end{equation}
the angular and radial equations decouple. The angular part yields spin-weighted spherical harmonics labelled by angular momentum quantum number $j$ and magnetic quantum number $m$, with separation constant $\lambda_{sj} = j(j+1) - s(s+1)$~\cite{Harris:2003eg}. Harris and Kanti derived a single master equation valid for fields of arbitrary spin $s$ in the Schwarzschild--Tangherlini background~\eqref{ST-metric-4d}:
\begin{equation}\label{master-eq}
    \Delta^{-s}\frac{\dd}{\dd r}\left(\Delta^{s+1}\frac{\dd R_s}{\dd r}\right) + \left(\frac{\omega^2 r^2}{f} + 2i\omega sr - \frac{is\omega r^2 f'}{f} + s(\Delta''-2) - \lambda_{sj}\right)R_s = 0\,,
\end{equation}
where $\Delta \equiv f(r)r^2$ and a prime denotes differentiation with respect to $r$. For $s=0$ this reduces to the Klein--Gordon equation in the background~\eqref{ST-metric-4d}.

The greybody factor is extracted by solving eq.~\eqref{master-eq} subject to physically motivated boundary conditions. Near the horizon $r \to r_0$, classical physics demands that no radiation escapes from the black hole, so only ingoing modes are permitted:
\begin{equation}
    R_s \sim A_{\scriptscriptstyle\mathrm{in}}^{(h)}\,\Delta^{-s}\,e^{-i\omega r^*}\,,\qquad r \to r_0\,,
\end{equation}
where $r^*$ is the tortoise coordinate defined by $\dd r^*/\dd r = 1/f(r)$, which maps the horizon to $r^* \to -\infty$ and stretches the near-horizon geometry so that the wave equation takes a simple form there. At spatial infinity $r \to \infty$, the solution is a superposition of ingoing and outgoing waves:
\begin{equation}
    R_s \sim A_{\scriptscriptstyle\mathrm{in}}^{(\infty)}\frac{e^{-i\omega r}}{r} + A_{\scriptscriptstyle\mathrm{out}}^{(\infty)}\frac{e^{i\omega r}}{r^{2s+1}}\,,\qquad r \to \infty\,.
\end{equation}
The ingoing wave at infinity represents radiation falling into the black hole from outside, while the outgoing wave represents radiation escaping to the observer. The greybody factor measures what fraction of the radiation generated at the horizon actually reaches infinity rather than being backscattered by the gravitational potential. The corresponding absorption probability is
\begin{equation}\label{absorption-coeff}
    |\mathcal{A}_j^{(s)}|^2 = 1 - \left|\frac{A_{\scriptscriptstyle\mathrm{out}}^{(\infty)}}{A_{\scriptscriptstyle\mathrm{in}}^{(\infty)}}\right|^2\,,
\end{equation}
from which the greybody factor for brane emission summed over all magnetic quantum numbers entering the Hawking spectrum is obtained as
\begin{equation}\label{greybody-brane}
    \Gamma_j^{(s)}(\omega) = \frac{\pi}{\omega^2}(2j+1)\,|\mathcal{A}_j^{(s)}|^2\,,
\end{equation}
where the factor $(2j+1)$ counts the degeneracy of modes with the same angular momentum quantum number $j$. Although the emission of brane-localised modes is kinematically four-dimensional, $|\mathcal{A}_j^{(s)}|^2$ retains a dependence on $n$ through the metric function $f(r) = 1-(r_0/r)^{n+1}$, so the greybody factors carry information about the dimensionality of the bulk spacetime \cite{Harris:2003eg}.

An analogous procedure can be carried out for fields propagating in the full $(4+n)$-dimensional Schwarzschild--Tangherlini background to obtain the bulk emission spectrum \cite{Kanti:2002nr, Creek_2006, Cardoso:2005mh}. However, the situation is qualitatively different.  Gravitons are not confined to the brane and propagate freely throughout the full $(4+n)$-dimensional spacetime, giving them access to a larger phase space than brane-localised fields. The relevant background is now the full Schwarzschild--Tangherlini metric~\eqref{ST-metric}, and the gravitational perturbation equation must be solved in all dimensions.
A graviton in $(4+n)$ dimensions carries more polarisation states than in four dimensions. Under the rotational symmetry of the extra-dimensional sphere, these polarisations decompose into three distinct types --- classified according to how they transform on the (n+2)-sphere of the Schwarzschild--Tangherlini geometry, and conventionally labelled scalar-, vector- and tensor-type --- each satisfying a separate radial equation with a different effective potential \cite{Creek_2006, Cardoso:2005mh}. These are all gravitational degrees of freedom; the labelling refers to their transformation properties in the extra dimensions, not to their spin on the brane. For each type, the same matching procedure as in the brane case applies: one imposes purely ingoing boundary conditions at the horizon and extracts the absorption probability $|\mathcal{A}_j|^2$ from the asymptotic behaviour of the solution at infinity. The angular part of the wave equation now involves hyperspherical harmonics on the $(2+n)$-sphere rather than ordinary spherical harmonics, and each angular momentum mode $j$ carries a multiplicity $N_j$ that counts the number of independent polarisation states on this higher-dimensional sphere \cite{Creek_2006, Cardoso:2005mh}. The total bulk graviton mass loss rate is
\begin{equation}\label{bulk-graviton}
    -\frac{\dd M_{\scriptscriptstyle\mathrm{bulk}}}{\dd t} = \int\sum_j N_j|\mathcal{A}_j|^2\,\frac{\omega}{\exp(\omega/T_{\scriptscriptstyle\mathrm{BH}})-1}\,\frac{\dd k}{2\pi}\,,
\end{equation}
where the sum runs over all angular momentum modes $j$ and all three graviton types, and the integral is over the bulk momentum $k$. At low energies, bulk graviton emission is strongly suppressed because the gravitational perturbation spectrum contains no monopole ($j=0$) or dipole ($j=1$) modes, so the emission begins at higher angular momentum. Consequently, the corresponding absorption probabilities vanish rapidly as $\omega r_0\to 0$ \cite{Creek_2006,Cardoso:2005mh}. The massless zero-mode graviton, being brane-localised, is not subject to this suppression and its emission is already included in the brane channel. From the perspective of a brane observer, these bulk graviton states correspond to excitations of the continuum KK spectrum discussed in Section~\ref{sec:garriga}. Emission into the bulk therefore appears as missing energy on the brane, since the emitted gravitons are free to propagate away through the extra dimension.
\subsection*{Black hole evaporation codes}
Several numerical codes have been developed to compute the semiclassical Hawking emission spectra of higher-dimensional black holes.\footnote{A separate code, \textsc{QBH} \cite{Gingrich:2009hj}, was developed by Gingrich and Undseth specifically for quantum black holes in higher dimensions, where the semiclassical approximation breaks down.} The earliest generators --- \textsc{CHARYBDIS} \cite{CDF:2002nno} and \textsc{CATFISH} \cite{Cavaglia:2006uk} --- were collider-oriented, but the emission spectra they produce are computed via the higher-dimensional greybody factors of \cite{Harris:2003eg, Casals_2006, Ida_2006}, making their emission spectra applicable beyond their original collider context. \textsc{BlackMax} \cite{Dai:2007ki} followed with a more complete implementation: it includes the full set of numerically computed greybody factors for brane-localised fields of all Standard Model spins following the same formalism, as well as spin-2 bulk graviton emission for non-rotating black holes and the effects of rotation for brane-localised fields. It also includes an option to account for recoil into the bulk due to small asymmetries in bulk emission; however, as discussed in Chapter~\ref{chap5:st-geometries}, small black holes on a positive-tension RS-II brane are gravitationally bound and this effect does not apply. \textsc{BlackMax} further includes an option for non-zero brane tension, implemented via a deficit angle parameter that modifies the bulk geometry. One must be careful not to apply this option to RS-II: the formula provided in their documentation is valid only for codimension-2 brane geometries and its use for the five-dimensional warped spacetime of RS-II is not warranted. \textsc{BlackHawk} \cite{Arbey:2019mbc} and its higher-dimensional extension \cite{Arbey:2021mbl} are today the most widely used public codes for PBH phenomenology, computing primary and secondary Hawking emission spectra for general black hole mass and spin distributions and their cosmological evolution. The results from all these numerical tools consistently confirm that brane emission dominates overwhelmingly over bulk emission for small black holes, consistent with heuristic arguments in the literature~\cite{Emparan_2000}.

In this work, we use \textsc{BlackMax} to extract the emission rates and apply them within the RS-II framework. In the non-rotating case considered here, both \textsc{BlackMax} and \textsc{BlackHawk} contain the necessary ingredients to calculate the emission spectra in the Kanti et al.~formalism discussed above; if spin were to be included, \textsc{BlackHawk} would be the more suitable choice as it is more actively maintained. 

We use these results to study the signatures of primordial black holes that would survive today associated with radiation and particle emission in \href{part:pbhs}{\textbf{\textcolor{thesislilac}{Part III}}}. For the study the impact of energy injection into the universe on the cosmic microwave background (CMB) spectrum, however, we do not use \textsc{BlackMax} but the generator incorporated in \textsc{CosmoLED}~\cite{Friedlander:2022ttk}, as we will discuss in Chapter~\ref{chap11:energy-injection}. 

\begin{figure}
    \centering
    \includegraphics[width=0.8\linewidth]{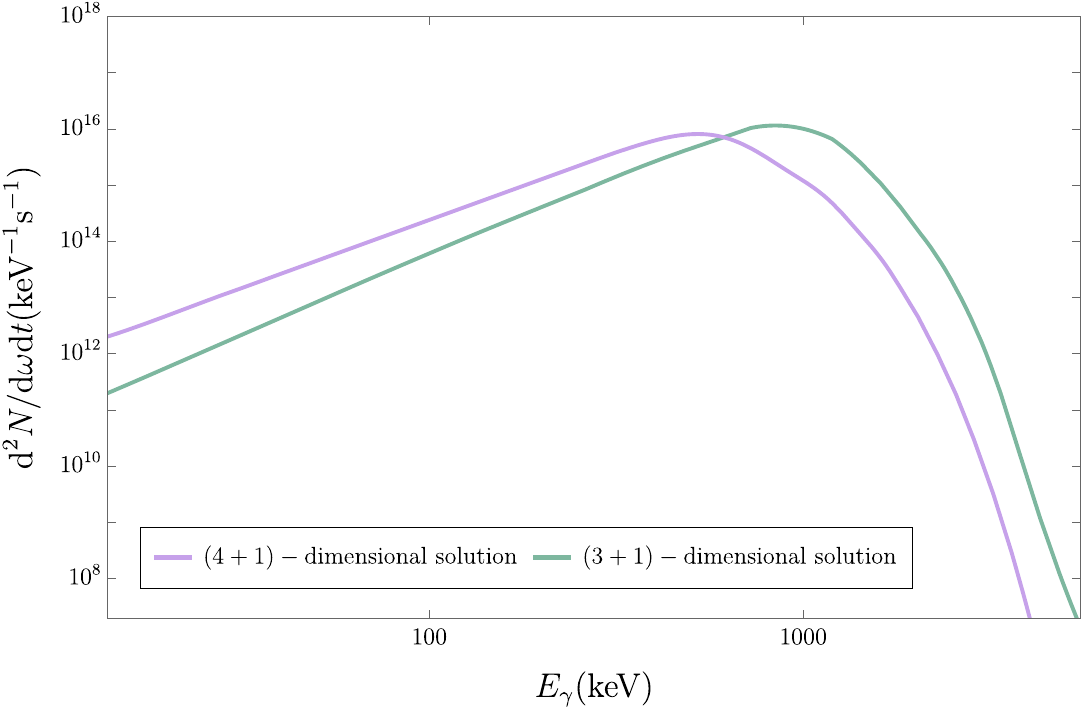}
    \caption[Photon emission spectra in 4D and 5D]{Photon emission spectrum by black holes with Hawking temperature $150$~keV for the $(3+1)$-dimensional (green) and $(4+1)$-dimensional (purple) case.}
    \label{fig:comp-spectra}
\end{figure}

\begin{mdframed}[style = mybbox]
By studying the radiation emission spectra with five-dimensional greybody factors using \textsc{BlackMax}, we find that the spectrum $\dd N/\dd t$ is very similar to that emitted by a four-dimensional black hole with the same Hawking temperature, to within factors of order unity, as illustrated in Figure~\ref{fig:comp-spectra}. The two spectra are shifted relative to each other: the spectral peaks satisfy $E_{\scriptscriptstyle\mathrm{peak}}\approx 3.96\, T_{\scriptscriptstyle\mathrm{BH}}$ in the five-dimensional case, compared to $E_{\scriptscriptstyle\mathrm{peak}}\approx 5.8\, T_{\scriptscriptstyle\mathrm{BH}}$ in the standard four-dimensional scenario~\cite{Carr_2016}, a mild shift attributable to the differences in the greybody factors. The emission spectrum is thus predominantly governed by $T_{\scriptscriptstyle\mathrm{BH}}$, rather than by the dimensionality of spacetime. 
\end{mdframed}

Since the Hawking spectrum of each particle species in a fixed number of dimensions depends solely on $T_{\scriptscriptstyle\mathrm{BH}}$, independently of the parameters of the theory, in RS-II different combinations $(M,\ell)$ yield the same temperature and therefore produce identical energy injection into the surrounding spacetime. This provides a useful translation between different parameter choices in the phenomenological analysis.
%
%
%
\section{The large black hole regime}
The situation for large black holes, $r_0 \gg \ell$, is more subtle and has been the subject of considerable debate. In this regime the near-horizon geometry is well approximated by the four-dimensional Schwarzschild metric, so the brane emission reverts to its standard four-dimensional form. The question of bulk emission, however, is considerably more involved.
In this case, the bulk channel admits a dual holographic interpretation, corresponding to coupling the black hole to a large number of strongly interacting conformal fields on the brane \cite{Emparan_2002}. This would suggest the black hole has access to many additional degrees of freedom, and it was originally proposed that large RS-II black holes would therefore evaporate much more rapidly than their four-dimensional counterparts.\footnote{Note that in ADD frameworks such holographic considerations do not apply, and large black holes are understood to evaporate according to the standard four-dimensional law once their horizon exceeds the compactification scale \cite{Emparan_2000}.} If so, the observed persistence of long-lived astrophysical black holes would impose an upper bound on the AdS radius, since otherwise their enhanced evaporation would have already depleted them.
This is no longer believed to be the case. Wiseman et al.\ \cite{Fitzpatrick_2006} argued that the naïve $N^2$-enhanced Hawking flux was not generic: static, localised brane black holes need not radiate into the full tower of holographic modes. This was later confirmed by the numerical construction of large static RS-II black holes by Figueras and Wiseman \cite{Figueras_2011}, which demonstrated that long-lived solutions indeed exist. Most recently, Emparan et al.\ \cite{Emparan_2023} clarified that enhanced evaporation is restricted to the \emph{connected} (funnel) phase, where the brane horizon merges with a bulk horizon and energy can flow freely into the bulk. In the \emph{disconnected} (droplet) phase, which provides a viable branch of solutions, the black hole is effectively insulated from the bulk and evaporates essentially as in four dimensions, up to small corrections.\footnote{Their argument also applies in the small black hole limit.} In this case, the phenomenology of Hawking evaporation of large RS-II black holes becomes observationally indistinguishable from that in ADD scenarios.
Although this work considers black holes across both mass regimes, the large black hole regime is not phenomenologically relevant for the Hawking evaporation signatures we study. Large black holes in RS-II, described by the four-dimensional Schwarzschild metric on the brane, are significantly colder than their small counterparts of the same mass. In the range of parameter space where small black holes survive until today and produce observable signatures, large black holes are too cold for their Hawking emission to yield any detectable signal. Their evaporation is therefore negligible for the purposes of the phenomenological analysis carried out in \href{part:pbhs}{\textbf{\textcolor{thesislilac}{Part III}}}, and we do not compute their emission spectra nor delve into the intricacies of holography.
%
%
\section{Black hole lifetimes}
An analytical estimate of the black hole lifetime can be obtained by approximating the greybody factors in the high-frequency limit $\omega \gg T_{\scriptscriptstyle\mathrm{BH}}$. As established in Section~\ref{sec:hawking-hd}, greybody factors satisfy $\Gamma^{(s)}_{j,n} \to 1$ in the geometric optics limit $\omega r_0 \gg 1$, where the potential barrier is transparent and all species emit as a perfect blackbody regardless of spin or angular momentum. In this limit, all absorption cross-sections reduce to the same geometric expression \cite{Guedens_2002-1}:
\begin{equation}\label{sigma-geom}
    \Gamma^{(s)}_{j,n} \approx \frac{A_{{\scriptscriptstyle\mathrm{eff}},n}}{4} \equiv \frac{\Omega_{n+2}\,r_{{\scriptscriptstyle\mathrm{eff}},n}^{n+2}}{4}\,,
\end{equation}
where $\Omega_{n+2}$ is the area of the unit $(n+2)$-sphere and $r_{{\scriptscriptstyle\mathrm{eff}},n}$ is an effective radius for blackbody emission defined as \cite{Emparan_2000}
\begin{equation}
    r_{{\scriptscriptstyle\mathrm{eff}},n} = \left(\frac{n+3}{2}\right)^{1/(n+1)}\!\left(\frac{n+3}{n+1}\right)^{1/2}r_0\,.
\end{equation}
Substituting~\eqref{sigma-geom} into the energy loss rate~\eqref{emission-rate-2} and evaluating the resulting integral gives the $(4+n)$-dimensional Stefan--Boltzmann law \cite{Guedens_2002-1}:
\begin{equation}\label{stefan-law}
    \frac{\dd M}{\dd t} \approx -g_{n}\,\tilde{\sigma}_{n}\,A_{{\scriptscriptstyle\mathrm{eff}},n}\,T_{\scriptscriptstyle\mathrm{BH}}^{4+n}\,,
\end{equation}
where $\tilde{\sigma}_n$ counts the effective degrees of freedom at temperature $T_{\scriptscriptstyle\mathrm{BH}}$, and
\begin{equation}
    \tilde{\sigma}_{n} = \frac{\Omega_{n+2}}{4(2\pi)^{n+3}}\Gamma(n+4)\zeta(n+4)
\end{equation}
is the $(4+n)$-dimensional Stefan--Boltzmann constant per degree of freedom \cite{Guedens_2002-1}. In the present setup, the total emitted power receives contributions from both brane and bulk emission,
\begin{equation}\label{total-dMdt}
    \frac{\dd M}{\dd t} \approx -g_{\scriptscriptstyle\mathrm{brane}}\,\tilde{\sigma}_0\,\tilde{A}_{{\scriptscriptstyle\mathrm{eff}},0}\,T_{\scriptscriptstyle\mathrm{BH}}^4 - g_{\scriptscriptstyle\mathrm{bulk}}\,\tilde{\sigma}_1\,A_{{\scriptscriptstyle\mathrm{eff}},1}\,T_{\scriptscriptstyle\mathrm{BH}}^5\,,
\end{equation}
where $\tilde{A}_{{\scriptscriptstyle\mathrm{eff}},0} = 4\pi r_{{\scriptscriptstyle\mathrm{eff}},1}^2$ accounts for the induced brane metric, $g_{\scriptscriptstyle\mathrm{brane}}$ counts brane degrees of freedom with rest mass below $T_{\scriptscriptstyle\mathrm{BH}}$, and $g_{\scriptscriptstyle\mathrm{bulk}}$ counts the polarisation states of the five-dimensional graviton. Since brane emission dominates as established in Section~\ref{sec:hawking-hd}, the bulk term constitutes a small correction.
%
%
\subsection*{Large black holes}
For large black holes, the near-horizon geometry is four-dimensional Schwarzschild with $T_{\scriptscriptstyle\mathrm{BH}} = M_4^2/8\pi M$ and $d=3$. Retaining only the leading brane term in~\eqref{total-dMdt} and integrating gives \cite{Page:1976df, Guedens_2002-1}
\begin{equation}\label{lifetime-4d}
    \tau \propto \frac{M^3}{M_4^4}\,,
\end{equation}
the familiar $\tau \propto M^3$ result of four-dimensional black hole evaporation.
%
%
\subsection*{Small black holes}
For small black holes ($r_0 \ll \ell$), the evaporation proceeds in five dimensions with $T_{\scriptscriptstyle\mathrm{BH}} = 1/2\pi r_0$ and $d=4$. Using $r_0 = \sqrt{8/3\pi}\,\sqrt{\ell M}/M_4$ from Section~\ref{sec:ST} and substituting into~\eqref{total-dMdt}, the dominant brane contribution gives~\cite{Guedens_2002-1}
\begin{equation}\label{dMdt-5d}
    \frac{\dd M}{\dd t} \approx -\frac{\tilde{g}\,M_4^2}{\ell M}\,,
\end{equation}
where $\tilde{g}$ is a numerical factor encoding the effective degrees of freedom. For the massless particle limit $\tilde{g} \approx 0.023$, while including electron-positron pairs gives $\tilde{g} \approx 0.032$~\cite{Guedens_2002-1}. The results are relatively insensitive to the precise value of $\tilde{g}$. Integrating~\eqref{dMdt-5d} gives
\begin{equation}\label{lifetime-5d}
    \frac{M}{M_4} = \tilde{g}^{1/2}\left(\frac{t_{\scriptscriptstyle\mathrm{evap}}}{t_4}\right)^{1/2}\left(\frac{\ell}{\ell_4}\right)^{-1/2}\,,
\end{equation}
where $t_4 = M_4^{-1}$ and $\ell_4 = M_4^{-1}$ are the four-dimensional Planck time and length respectively, following the notation in Ref.~\cite{Guedens_2002-1}. The lifetime therefore scales as $\tau \propto \ell M^2/M_4^2$
, i.e.\ as $M^2$ rather than $M^3$. Since $T_{\scriptscriptstyle\mathrm{BH}} \propto M^{-1/2}$ in five dimensions compared to $T_{\scriptscriptstyle\mathrm{BH}} \propto M^{-1}$ in four dimensions, a small RS-II black hole of fixed mass is colder and longer-lived than its four-dimensional counterpart, with significant consequences for primordial black hole constraints discussed in \href{part:pbhs}{\textbf{\textcolor{thesislilac}{Part III}}}.

\part{Primordial black holes}\label{part:pbhs}

    \chapter{Origins and perspectives}\label{chap7:pbh-intro}
    \thispagestyle{fancy}
\pagestyle{fancy}

    \setlength{\epigraphwidth}{0.61\textwidth}
\epigraph{{\it
Consideration of black holes suggests, not only that God does play dice, but that he sometimes confuses us by throwing them where they can't be seen.}
}{-- S.~Hawking~\cite{hawking1996nature}\\ {\footnotesize 
During a 1994 exchange with Penrose, as a direct riposte to Einstein's famous line.}}
\setlength{\epigraphwidth}{0.6\textwidth}

The two preceding parts of this thesis developed, separately, the two ingredients that this final part brings together. \href{part:braneworlds}{\textbf{\textcolor{thesislilac}{Part I}}} described the expanding Universe evolving under modified Friedmann dynamics. \href{part:braneworlds}{\textbf{\textcolor{thesislilac}{Part II}}} studied higher-dimensional black holes as exact solutions, characterised by their geometry, temperature and emission spectra. Primordial black holes live precisely at the intersection of these two subjects: they are black holes whose formation requires no star, no accretion disk, and no astrophysical environment whatsoever  ---  only the early Universe itself, provided that somewhere, conditions arose extreme enough  ---  be it through large inhomogeneities, phase transitions, or some other process operating in the primordial plasma  ---  to concentrate energy within small enough scales.

The hypothesis that black holes could have formed in the early Universe predates the modern theory of structure formation. It was first put forward by Zeldovich and Novikov in 1967~\cite{Zeldovich:1967lct}, who considered ``cores retarded during expansion" in the hot Big Bang model: regions whose expansion lagged behind the cosmic average and which could recollapse into black holes. Ironically, they concluded that such objects probably did not exist, on the grounds of a Bondi-type accretion argument suggesting that any black hole formed in the radiation era would grow catastrophically, keeping pace with the cosmological horizon and reaching absurdly large masses today. The idea was revived independently by Hawking in 1971~\cite{hawking1971gravitationally}, and the accretion objection was laid to rest by Carr and Hawking in 1974~\cite{Carr_Hawking_1974}, who showed that there is no self-similar solution in which a black hole grows at the same rate as the horizon: a PBH formed in the radiation era retains a mass of the order of its formation value.

The original motivation was therefore largely theoretical: small black holes were the only context in which quantum-gravitational effects on horizons could be physically relevant, and it was Hawking's study of such low-mass objects that led him, in 1974, to the discovery of black hole evaporation~\cite{Hawking:1974rv}. In the conventional scenario, PBHs with initial mass below $M_* \simeq 5\times 10^{14}$ g would have evaporated by today, while heavier ones survive as effectively stable, cold, non-baryonic relics. This last observation prompted Chapline, in 1975, to make the first explicit proposal that a cosmological population of PBHs could constitute the dark matter in the Universe  ---  a remarkably prescient suggestion, made before WIMPs or axions had even been proposed~\cite{Chapline:1975ei}. The same year, Carr derived the first PBH mass spectrum~\cite{Carr1975}, assuming scale-invariant initial conditions, and the formalism for converting a primordial perturbation amplitude into a PBH abundance was born.

\section{Evolution of the field}
The subsequent development of the subject can be read as an alternation between theoretical refinement and new observational input. Through the late 1970s and 1980s, the main activity consisted of constraining evaporating PBHs: Hawking radiation from holes near $M_*$ would contribute to the extragalactic $\gamma$-ray background, alter Big Bang nucleosynthesis, and distort the CMB, leading to the first abundance limits~\cite{Page:1976wp,1991ApJ...371..447M,Carr1998,Boudoul:2003xr}. 

The 1990s brought two decisive inputs. First, inflation matured into the standard theory of primordial perturbations and provided a natural formation mechanism: PBHs form from the collapse of rare, large-amplitude curvature perturbations upon horizon re-entry during radiation domination  ---  the same perturbations that seed galaxies on large scales, but amplified by several orders of magnitude on much smaller ones. Second, the microlensing surveys (MACHO, EROS, later OGLE) turned the question of compact dark matter in the Galactic halo into an experimental one, largely excluding solar-mass compact objects as the dominant dark matter component and pushing interest toward other mass ranges~\cite{Alcock_2000,Tisserand_2007}.

The modern era of the field opened abruptly in 2016 with the LIGO detection of GW150914, a merger of two $\sim 30\,M_\odot$ black holes~\cite{abbott2016observation}. Several groups immediately pointed out that black holes of this mass and merger rate were compatible with a primordial population~\cite{Bird_2016,Sasaki:2016jop,Clesse_2017}, and the suggestion that LIGO might have detected (a component of) the dark matter triggered an explosion of activity that has not subsided. Since then, the field has been driven by gravitational-wave astronomy in a second, indirect way as well: the large scalar perturbations required to form PBHs unavoidably source a stochastic background of scalar-induced gravitational waves (SIGWs) at second order in perturbation theory~\cite{Ananda:2006af,Baumann:2007zm,Saito:2008jc,Domenech_2021}, and the common-spectrum signal reported by the pulsar timing arrays (NANOGrav, EPTA, PPTA, CPTA) in 2023 was widely analysed in terms of such induced backgrounds~\cite{Afzal_2023,Franciolini:2023pbf}  --- though this interpretation is contested, since the required perturbations may overproduce PBHs unless non-Gaussianity is invoked~\cite{Franciolini:2023pbf}, and an astrophysical origin from supermassive black hole binaries remains the default~\cite{Agazie_2023b}. Most recently, JWST observations of unexpectedly massive galaxies and quasars at high redshift have renewed interest in a long-standing side branch of the subject: the possibility that massive PBHs provided the seeds of supermassive black holes~\cite{Bean_2002,Duechting:2004kp,Carr_Silk_2018,Bernal:2018aum}, whose early growth is otherwise difficult to explain~\cite{Inayoshi_2020,Volonteri2010}. However, this explanation is severely challenged by CMB spectral distortion constraints on the amplitude of the primordial perturbations required to produce such massive PBHs~\cite{Pritchard_2025}, which remain tight even when allowing for significant non-Gaussianity.
\section{Main topics and open questions}
As it stands today, the theory of PBHs is organised around a few central problems. 

The first is \ul{\textbf{formation}}: in the standard scenario, a perturbation collapses if its amplitude at horizon crossing exceeds a critical threshold $\delta_c$, and the resulting black hole mass follows the critical-collapse scaling law, producing an extended rather than monochromatic mass function~\cite{Carr1975,Evans1994,PhysRevLett.70.9,Niemeyer_1998,Musco_2013,Harada_2013}. Beyond this minimal mechanism, PBHs may also form at cosmological phase transitions~\cite{Crawford_1982,Jedamzik:1996xz}, from the collapse of cosmic strings~\cite{1989PhLB..231..237H,PhysRevD.43.1106}, domain walls~\cite{Rubin_2001,Deng_2017} or vacuum bubbles~\cite{PhysRevD.26.2681,Deng_2017,Garriga_2016}, or during early matter-dominated epochs~\cite{Khlopov:1980mg,Harada_2016}. 

The second is \ul{\textbf{abundance}}: in essentially every scenario, PBH
formation is a rare event  ---  the collapse of an exceptional fluctuation,
the chance overlap of bubble walls, the occasional closed domain wall or
string loop  ---  so the predicted number density is extremely sensitive
to the parameters of the underlying mechanism; in the standard scenario,
to the collapse threshold, the shape of the power-spectrum peak, and
primordial non-Gaussianity.

The third is \ul{\textbf{evolution}}: accretion, clustering, binary
formation, mergers, and  ---  at the light end  ---  Hawking evaporation. All
of these processes depend strongly on the cosmological epoch under
consideration and on the properties of the population itself (its mass
function, spins, and spatial distribution), so each has generated an
extensive literature of its own. 

The fourth is \ul{\textbf{phenomenology}}: the observational signatures by which a PBH population could be detected or excluded~\cite{Carr_2026}.

Each of these areas contains questions that are genuinely unsettled. The precise value of the collapse threshold and its dependence on the perturbation profile is still the subject of numerical-relativity studies~\cite{Musco_2019,Escriva_2021}; the computation of the abundance suffers from ambiguities in the choice of collapse variable and in the treatment of non-Gaussianity, with different prescriptions yielding abundances differing by many orders of magnitude~\cite{Young:2014wva,Young_2019,Luca_2019,Gow_2021}; the inflationary model-building required to amplify the power spectrum by many orders of magnitude at small scales has raised debated questions about the size of loop corrections and the consistency of such scenarios~\cite{Kristiano_2024,riotto2023primordialblackholeformation,Riotto:2023qkx,Cole_2023}; the clustering of PBHs at formation and its effect on merger rates and on microlensing constraints remains uncertain~\cite{Desjacques_2018,Young_2020,Ballesteros_2018,clesse2018sevenhintsprimordialblack}; and the end state of evaporation  ---  complete disappearance, a Planck-mass relic~\cite{MacGibbon1987,Barrow:1992bk,aljazaeri2025comprehensivelyconstrainingultralightprimordial}, or a slowdown of the process itself, as in recently proposed ``memory burden" scenarios~\cite{dvali2018microscopicmodelholographysurvival,Dvali_2019,dvali2024memoryburdeneffectblack}  ---  is unknown, with direct consequences for the viability of light PBHs as dark matter. Finally, whether any subpopulation of the LIGO-Virgo-KAGRA merger events is primordial remains an open observational question: the data are compatible with a primordial contribution but show no clear statistical preference for one.
\section{Primordial black holes as a dark matter candidate}
The reason for the persistence of the PBH hypothesis across five decades is that it addresses the dark matter problem with remarkable economy. PBHs are cold, effectively collisionless, and  ---  if formed well before nucleosynthesis  ---  non-baryonic, so they evade the BBN and CMB bounds on baryons~\cite{Green_2021}. They require no extension of the Standard Model particle content, although the mechanism that generates the required perturbations (typically a non-trivial inflationary sector) does constitute physics beyond the minimal picture. In the standard picture, any PBH significantly heavier than $M_* \simeq 5 \times 10^{14}$ g is stable on cosmological timescales. 

The population is conventionally described by the fraction of dark matter in PBHs,
\begin{equation}
    f_{\scriptscriptstyle\mathrm{PBH}}=\frac{\Omega_{{\scriptscriptstyle\mathrm{PBH}},0}}{\Omega_{{\scriptscriptstyle\mathrm{DM}},0}}\,,
\end{equation}
where $\Omega_{i,0}$ is the density parameter today of species $i$, together with a mass function $\psi(M)$. The central question of the field  --- can $f_{\rm PBH} = 1$?  --- has,
after decades of constraint work, a surprisingly specific answer. There
is broad consensus that it remains possible in the \emph{asteroid-mass
window}, $10^{17}\,\mathrm{g} \lesssim M \lesssim
10^{22\text{--}23}\,\mathrm{g}$, where PBHs are too heavy to evaporate observably and too light for microlensing surveys to resolve, owing to wave-optics and finite-source-size effects~\cite{Niikura_2019,MonteroCamacho:2019jdg,Smyth_2020}. Whether any other window survives is still under
debate: the standard constraint compilations restrict a single-mass
population at all other masses to a subdominant fraction of the dark
matter~\cite{Carr_2021,Green_2021}, but part of the community argues that the limits in the
solar-to-stellar-mass range carry enough astrophysical model
dependence  --- in the halo models and source sizes assumed by
microlensing surveys, or in the possible clustering of the
population  --- that a dominant PBH component there is not yet
excluded~\cite{carr2024observational,Hawkins_Garcia_Bellido_2025}. In any case, even subdominant populations would have
important consequences --- providing seeds for supermassive black holes
or contributing to the observed merger events.
\section{Searches and the master plot}
The observational status of the field is traditionally summarised in a single figure --- the ``master plot" of constraints on $f_{\rm PBH}$ as a function of PBH mass --- which will serve as the road map for the last two chapters of this part. Its logic is the following. Each physical effect that a PBH population would produce defines, for each mass $M$, a maximum allowed abundance; the plot is the envelope of these upper limits, conventionally computed for a monochromatic mass function. 
\begin{figure}
    \centering
    \includegraphics[width=1\linewidth]{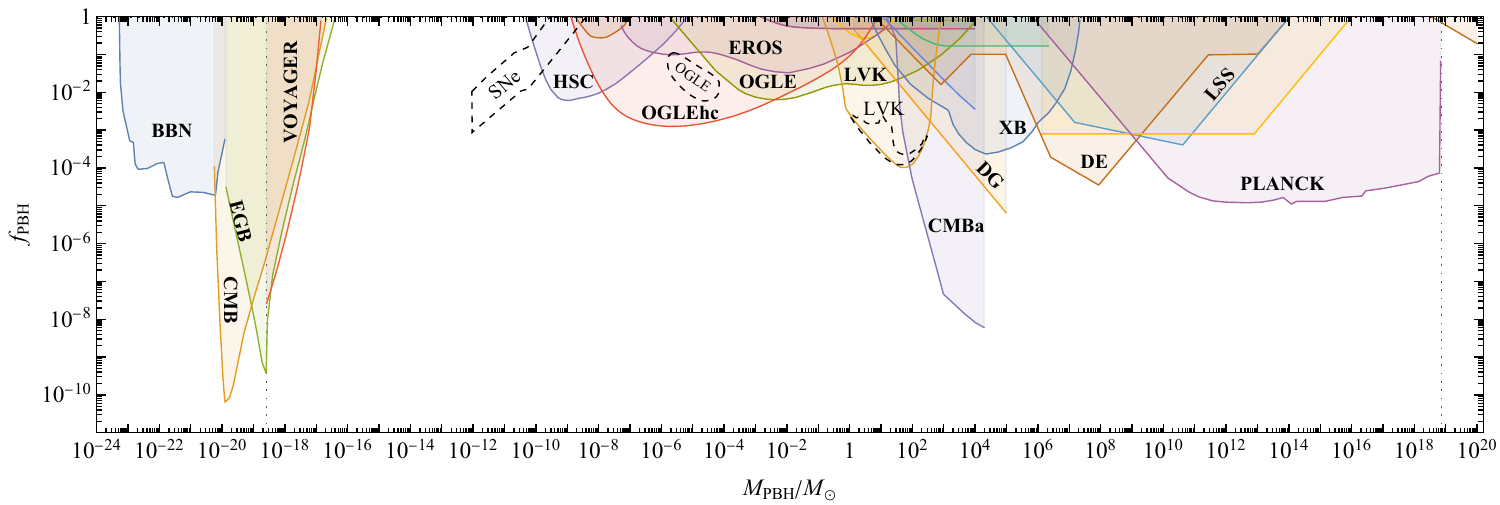}
    \caption[Constraints on the PBH dark matter fraction]{The ``master
plot'': upper limits on the fraction of dark matter in primordial black
holes, $f_{\scriptscriptstyle\mathrm{PBH}} = \Omega_{{\scriptscriptstyle\mathrm{PBH}},0}/\Omega_{{\scriptscriptstyle\mathrm{DM}},0}$~, as a function
of the PBH mass, assuming a monochromatic mass function (solid curves;
only the dominant constraint in each mass range is labelled). From low
to high masses, the limits arise from Hawking evaporation, gravitational
microlensing, gravitational-wave merger rates and accretion, and
dynamical effects, as discussed in the text. The unconstrained region
between the evaporation and microlensing limits is the asteroid-mass
window, where $f_{\scriptscriptstyle\mathrm{PBH}}=1$ remains allowed. Dashed curves indicate
claimed positive signals rather than exclusions (supernova
lensing~\cite{smirnov2022whitedwarfsdwarfspheroidal}, OGLE microlensing events~\cite{Niikura_2019}, LVK
mergers~\cite{Andr_s_Carcasona_2024}), whose interpretation as PBHs is debated. The left
vertical dashed line marks the initial mass $M_* \simeq 5\times
10^{14}\,$g of a black hole completing its evaporation today; the right
one marks the horizon mass at the epoch of CMB formation. Figure
produced with the code accompanying Ref.~\cite{Carr_2026}, where the individual
constraints are referenced.}
    \label{fig:masterplot-4d}
\end{figure}

The most up-to-date constraints, derived within the standard
four-dimensional $\Lambda$CDM framework, are summarised in
Figure~\ref{fig:masterplot-4d}; how this picture is modified in the
braneworld scenario is the subject of the following chapters. Reading Figure~\ref{fig:masterplot-4d} from left to right: at the lightest masses ($M \lesssim 10^{17}$g), the constraints come from evaporation --- Hawking radiation contributing to the extragalactic and Galactic $\gamma$-ray backgrounds, the 511 keV line, CMB anisotropies, and the local positron and electron fluxes measured by Voyager I. Then follows the open asteroid-mass window. Above $\sim 10^{22\text{–}23}$g, gravitational lensing takes over: microlensing of stars by Subaru/HSC, OGLE, EROS and MACHO, and lensing of supernovae at larger masses, excludes $f_{\scriptscriptstyle\mathrm{PBH}}=1$ from sublunar up to and beyond stellar masses. In the stellar-mass range, lensing limits are joined by gravitational-wave constraints from the LVK merger rate, by accretion constraints --- energy injected into the pre-recombination plasma distorts the CMB, while present-day accretion would produce X-ray and radio sources --- and, if the SIGW interpretation of the pulsar timing array (PTA) band is taken at face value, by limits from the induced gravitational-wave background. At still larger masses, dynamical constraints dominate: the survival of wide binaries, of star clusters in ultra-faint dwarfs such as Eridanus II, and of the dwarf galaxies themselves, together with limits from large-scale structure and CMB accretion, exclude a dominant population.

Three caveats accompany any reading of this plot. First, realistic mass functions are extended, and constraints must be reinterpreted accordingly: a broad mass function cannot simply be confronted with
monochromatic limits, and doing so properly tends to narrow, though not necessarily close, the allowed windows~\cite{Carr:2017jsa,Gorton_2024}. Second, individual
constraints carry astrophysical model dependence --- accretion limits
depend on poorly understood accretion physics, microlensing limits on
the assumed halo model and source sizes, dynamical limits on initial
conditions of the perturbed systems --- and several constraints once
considered firm have subsequently been weakened or retracted. Third,
and most importantly for this thesis, the entire compilation is derived
within a definite theoretical framework: four-dimensional general
relativity, a $\Lambda$CDM expansion history, and the Standard Model
particle content. Each constraint inherits these assumptions --- the
evaporation limits, for instance, presuppose the four-dimensional
Hawking spectrum, and the formation masses the standard thermal
history --- so the master plot is not a model-independent statement
about black holes in our Universe, but about black holes in
\emph{this} universe model.

The plot is therefore best understood not
as a static verdict but as a snapshot of an ongoing measurement
programme --- one in which the remaining parameter space is finite,
well charted, and, within a given framework, fully testable.

Chapters~\ref{chap11:energy-injection} and~\ref{chap12:microlensing}
take up precisely the third caveat: we extend this programme to
braneworlds, studying the evolution and observational signatures of
PBHs in the small- and large-black-hole limits of an RS-II brane
universe, and asking how the boundaries of the master plot shift when
the underlying geometry is five-dimensional. In practice, only the
evaporation- and microlensing-related envelopes need to be
re-examined. The reason is that, given the current observational
bounds on the AdS radius $\ell$, the maximum mass attainable by a
five-dimensional black hole only reaches the low-mass tail of the
microlensing envelope. Black holes in the mass ranges probed by the
remaining constraints --- merger rates, accretion, and dynamical
effects --- are therefore necessarily in the large-black-hole limit,
where they are well approximated by their four-dimensional
counterparts and the corresponding envelopes are expected to carry over unchanged from
the conventional scenario.

    \chapter[Gravitational collapse in the early Universe]{Gravitational collapse\\ in the early Universe}\label{chap9:grav-collapse}
    \thispagestyle{fancy}
\pagestyle{fancy}

    Gravitational collapse of overdense regions was the first PBH formation mechanism proposed in the literature~\cite{Zeldovich:1967lct,Hawking:1974rv}, and remains the most thoroughly studied (see references [29-100] in~\cite{Carr_2020}). The dynamics and probability of collapse depend strongly on the symmetries of the system and the properties of the fluid. Deviations from spherical symmetry~\cite{K_hnel_2016,Kokubu_2018,Escriv__2025}, pressure gradients~\cite{Carr1975,Musco:2008hk}, repulsive particle interactions and quantum mechanical effects such as the Pauli exclusion principle~\cite{Flores:2024eyy,Milligan:2025zbu,Parvez:2025wtq,Roberts:2024wup,Padilla:2025bkv} can counteract the gravitational pull, slowing or entirely preventing collapse.

A complete picture of the process requires understanding several interconnected aspects: the origin and shape of the primordial overdensities (e.g.~\cite{Escriv__2025,Musco:2020jjb,Garc_a_Bellido_1996,Randall_1996,Kannike_2017,Byrnes_2018,Green_2016,Dolgov:1992pu,Pattison_2017}), the threshold conditions determining whether a region will collapse (e.g.~\cite{He_2019,Koike_1995,Evans_1994}), the dynamical evolution of the collapse itself (e.g.~\cite{Niemeyer_1999,Escriv__2022}), and the nature of the end product (e.g.~acquired spin~\cite{Luca_2020,Harada_2021,dejong2023spinningprimordialblackholes} or charge~\cite{Alonso-Monsalve:2023brx,Santiago_2025}).

In this chapter, we focus on the spacetime geometry of gravitational collapse in the braneworld setting. We consider spherically symmetric overdensities localised on the brane and an empty AdS$_5$ bulk. This setup was studied in Refs.~\cite{Bruni_2001,Kofinas_2004,Casadio_2005} in the large-black-hole limit; here we follow their procedure and extend the analysis to the small-black-hole limit.

\section{Oppenheimer--Snyder collapse}
Gravitational collapse in braneworld scenarios presents additional complications beyond those of standard general relativity. The fundamental source of difficulty lies in the asymmetry between matter, which is confined to the brane, and gravity, which propagates freely into the extra dimension. This is further compounded by the gravitational interaction between the brane and the bulk, which has no analogue in four-dimensional Einstein gravity.

These intricacies manifest themselves already in the simplest collapse model, the one proposed by Oppenheimer and Snyder~\cite{Oppenheimer:1939zz}. Their model adopts a step-function density profile, in which a homogeneous interior is sharply bounded by a vacuum exterior. In their calculation, the collapsing cloud is composed of pressureless dust ($p=0$). The absence of pressure is precisely what renders this idealisation self-consistent: since dust elements exert no force on one another, there is no dynamical obstruction to maintaining a sharp density discontinuity at the cloud boundary. The interior is described by a closed FLRW metric and the exterior by the Schwarzschild solution, matched across the surface via the Israel junction conditions.

For a fluid with non-vanishing pressure, however, this construction breaks down. The conservation equation~\eqref{LCE-2} shows that $\vec{\nabla}_\mu p\neq 0$ results in the acceleration of the fluid $A_\mu$ to be non-vanishing. A step-function density profile would imply a corresponding discontinuity in the pressure, and hence a delta-function contribution to $\vec{\nabla}_\mu p$ at the cloud boundary. This produces a formally infinite force at the surface, rendering the profile dynamically inconsistent. In standard cosmology, for a radiation fluid, this obstruction is unavoidable. A sharp-boundary profile is therefore not a valid idealisation in this case, and one must instead work with smooth density profiles whose gradients remain finite throughout the evolution. This is the standard approach in modern studies of PBH formation from radiation perturbations~\cite{Escriv__2024}, where the shape of the smooth profile plays a central role in determining the collapse threshold.

Since the local conservation equation~\eqref{LCE-2} applies also in braneworld cosmology with an empty bulk, as discussed in \href{part:braneworlds}{\textbf{\textcolor{thesislilac}{Part I}}}, the same arguments apply. This means that collapse during the quadratic regime cannot be studied reducing the energy density to a top-hat profile, since the brane fluid is highly-relativistic during this epoch. 

We consider the gravitational collapse of overdense regions during matter domination, so that the pressure vanishes and we may set $p = 0$. We take the comoving radius of the cloud to satisfy $r_0 \ll \ell$, so that the end product of collapse is expected to be a five-dimensional black hole. In this regime, the cloud lies at deeply sub-horizon scales, $r_0 \ll H^{-1}$, since $H^{-1} \sim \ell$ when
$t \sim t_c$, where $t_c$ is the time at which the crossover between the quadratic and linear gravitational regimes occurs --- well before matter-radiation equality.

A naïve five-dimensional generalisation of the Oppenheimer--Snyder formalism is not applicable here, since brane localisation of matter breaks the spherical symmetry of the bulk. This naturally calls for the SMS covariant treatment introduced in \href{part:braneworlds}{\textbf{\textcolor{thesislilac}{Part I}}}, where the collapse is described entirely from the perspective of a brane observer in terms of quantities induced on the brane. In this framework, spherical symmetry of the induced metric is preserved and can be exploited.
\subsection{Geometry at the boundary}
We write the FLRW metric for the cloud region in isotropic coordinates,
\begin{equation}\label{curv-metric}
    \dd s^2=-\dd \tau^2+\left(\frac{a(\tau)}{1+\frac{1}{4}Kr^2}\right)^2\left(\dd r^2+r^2\dd \Omega_2^2\right)\,.
\end{equation}
For a perfect fluid on the brane, the Einstein equations applied to this metric yield the Friedmann equation
\begin{equation}\label{Fried-curv}
  H^2=\frac{8\pi}{3M_4^2}\left(\rho+\frac{\rho^2}{2\lambda} \right)+\frac{\mathcal{\rho_E}}{a^4}-\frac{K}{a^2} +\frac{\Lambda_4}{3}\,.
\end{equation}
Note that, although the quadratic correction $\rho^2/\lambda$ is negligible for the cosmological background during matter domination, it need not be so within a collapsing overdensity. Even for a non-relativistic fluid, the local energy density inside the cloud can become sufficiently large to render the high-energy correction significant during the collapse process.

The proper radius in this choice of coordinates is
\begin{equation}
    R=\frac{r a(\tau)}{1+\frac{1}{4}Kr^2}\,,
\end{equation}
and at the boundary surface $\Sigma$ between the interior and the exterior metric,
\begin{equation}\label{rsigma}
    R_{\Sigma}(\tau)=\frac{r_0\,a(\tau)}{1+\frac{1}{4}Kr_0^2}\,.
\end{equation}
Since the fluid is pressureless, the boundary surface $\Sigma$ is in free fall and $r=r_0$ is constant in comoving coordinates. Note that in this regime $\rho=\rho_0(a_0/a)^3$ and $\rho_{\mathcal{E}}=\rho_{\mathcal{E},0}(a_0/a)^4$, where the subindex $0$ indicates evaluation at the moment the collapse process begins. With these scaling relations and eq.~\eqref{rsigma}, one can re-express the Friedmann equation~\eqref{Fried-curv} in terms of $R$,
\begin{equation}\label{rdot-1}
    \dot{R}^2=\frac{2m}{M_4^2R}+\frac{3m^2}{4\pi M_4^2\lambda R^4}+\frac{Q}{R^2}+\bar{K}+\frac{\Lambda_4}{3}R^2\,,
\end{equation}
where, adopting the conventions of~\cite{Bruni_2001}, we define
\begin{equation}
    m=\frac{4\pi }{3}\left(\frac{a_0r_0}{1+\frac{1}{4}Kr_0^2}\right)^3\rho_0\,,\qquad Q=\rho_{\mathcal{E}}\left(\frac{r_0}{1+\frac{1}{4}Kr_0^2}\right)^4\,,\qquad \bar{K}=-\frac{Kr_0^2}{(1+\frac{1}{4}Kr_0^2)^2}\,.
\end{equation}
Following the Oppenheimer--Snyder approach, we match the interior metric~\eqref{curv-metric} to a static, spherically symmetric exterior. Israel junction conditions impose that the metric and the extrinsic curvature must be continuous on the brane at these coordinates. The latter is true if the metric and $\dot{R}$ are continuous, as follows from the definition of the extrinsic curvature, see eq.~\eqref{extrinsic-curvature}. Following the conventional procedure, we take the ansatz
\begin{equation}\label{eq:metric-exterior}
    \dd s^2=-F(R)^2A(R)\dd t^2+\frac{1}{A(R)}\dd R^2+R^2\dd \Omega^2_2
\end{equation}
for the exterior metric, where in our case we consider
\begin{equation}\label{gen-met-c}
    A(R)=1-\frac{\tilde{R}(R)^2}{R^2}\,,
\end{equation}
i.e.~a solution with a form that resembles the expected Schwarzschild--Tangherlini but with $R$-dependent functions $F(R)$ and $\tilde{R}(R)$, so that the ansatz encompasses the most general static, spherically symmetric metric on the brane.  

The radial geodesic equation for the junction surface $\Sigma$ is derived from the exterior metric~\eqref{gen-met-c} by choosing the affine parameter of the geodesic equation to be the proper time of the junction surface $\Sigma$, which coincides with the proper time of comoving observers in the collapsing interior. This choice is always valid for timelike geodesics and ensures that the radial geodesic equation derived from the exterior metric can be directly compared with the interior Friedmann equation, both being expressed in terms of the same $\tau$. 

For radial motion ($\dd\Omega = 0$), the normalisation condition for a timelike geodesic $g_{\mu\nu}\dot{x}^\mu\dot{x}^\nu = -1$ gives
\begin{equation}\label{normalisation}
-F(R)^2 A(R)\dot{t}^2 + \frac{\dot{R}^2}{A(R)} = -1\,.
\end{equation}
The exterior metric admits a timelike Killing vector $\partial_t$, which yields a conserved energy along the geodesic:
\begin{equation}\label{Etilde}
E = F(R)^2 A(R)\,\dot{t}\,.
\end{equation}
Substituting $\dot{t}$ from~\eqref{Etilde} into the normalisation condition~\eqref{normalisation} gives
\begin{equation}\label{rdot-2}
\dot{R}^2 = -A(R) + \frac{E^2}{F(R)^2}\,.
\end{equation}
Equations~\eqref{rdot-1} and~\eqref{rdot-2} must be equal at the boundary surface $\Sigma$, as required by the continuity of $\dot{R}$.

Solving for $\tilde{R}_\Sigma$ yields the condition
\begin{equation}\label{mR-1}
   \tilde{R}_\Sigma^2=\frac{\Lambda_4}{3}R_\Sigma^4+\left(\bar{K}+1-\frac{E^2}{F^2}\right)R_\Sigma^2+\frac{2m}{M_4^2}R_\Sigma+Q+\frac{3m^2}{4\pi M_4^2\lambda R_\Sigma^2}\,.
\end{equation}
The second condition --- metric continuity --- can be derived using Eddington--Finkelstein
coordinates. For the interior, we use~\cite{Bruni_2001}
\begin{equation}
    \dd\tau=\partial_v \tau\,\dd v+\left(1+\frac{1}{4}Kr^2\right)\frac{\dd R}{\dot{a}(\tau)r-1+\frac{1}{4}Kr^2 }
\end{equation}
and \eqref{curv-metric} takes the form~\cite{Bruni_2001}
\begin{equation}\label{FRLW-2}
    \dd s^2=-\frac{(\partial_v\tau)^2\left(1-(K+\dot{a}^2)R^2/a^2\right)}{1-KR^2/a^2 }\dd v^2+\frac{2\partial_v\tau}{\sqrt{1-KR^2/a^2}}\dd v\,\dd R+R^2\dd\Omega^2\,.
\end{equation}
For the exterior metric, as follows from the standard definition of the ingoing Eddington--Finkelstein coordinate,
\begin{equation}
    \dd v=\dd t+\frac{\dd R}{F(R)A(R)}
\end{equation}
and
\begin{equation}\label{exterior-2}
    \dd s^2=-F^2(R)A(R)\dd v^2+2F(R)\dd v\,\dd R+R^2\dd\Omega^2\,.
\end{equation}
Matching the $\dd v^2$ and
$\dd v\dd R$
coefficients of~\eqref{FRLW-2} and~\eqref{exterior-2} at $\Sigma$ yields
\begin{equation}\label{condition-2}
  A(R_\Sigma)=1-(K+\dot{a}^2)\frac{R_\Sigma^2}{a^2}\,,
\end{equation}
which imposes
\begin{equation}\label{cond-2}
    \tilde{R}^2_\Sigma=(K+\dot{a}^2)\frac{R_\Sigma^4}{a^2}\,.
\end{equation}
Substituting the scale factor and its derivatives using~\eqref{rsigma} and using the radial geodesic equation~\eqref{rdot-2}, this expression simplifies to
\begin{equation}\label{condition-2-a}
 \tilde{R}^2_\Sigma=R^2_\Sigma\left(\dot{R}^2|_\Sigma-\bar{K}\right)\,.
\end{equation}
Using again~\eqref{rsigma}, and setting eq.~\eqref{condition-2-a} equal to~\eqref{rdot-1}, gives
\begin{equation}\label{mR-2}
    \tilde{R}_\Sigma^2=\frac{\Lambda_4}{3}R_\Sigma^4+\frac{2m}{M_4^2}R_\Sigma+Q+\frac{3m^2}{4\pi M_4^2\lambda R_\Sigma^2}\,,
\end{equation}
which is the second condition the function $\tilde{R}(R)$ must satisfy at the boundary surface. If we compare condition \eqref{mR-1} and condition \eqref{mR-2}, we see that for them to be compatible,
\begin{equation}\label{F-boundary}
F^2_\Sigma=\frac{E^2}{\bar{K}+1}\,.
\end{equation}
Equation~\eqref{F-boundary} determines $F^2(R)$
 \textit{only at the boundary} $\Sigma$; the behaviour of $F(R)$ in the full exterior spacetime requires additional information.
\subsection{Traceless condition}
In \href{part:higher-d}{\textbf{\textcolor{thesislilac}{Part II}}} we showed that, in the absence of matter, solutions on the brane do not satisfy $R_{\mu\nu} = 0$, but instead obey the traceless condition $\mathcal{E}^\mu_{\ \mu} = 0$, which implies
\begin{equation}\label{traceless-eq}
    R^\mu_{\,\mu}=4\Lambda_4 \,.
\end{equation}
Since the exterior spacetime is empty, the metric~\eqref{eq:metric-exterior} should obey the trace condition~\eqref{traceless-eq}. At the sub-horizon scales relevant to small black hole formation, the cosmological constant is negligible and we can safely set $\Lambda_4 = 0$, i.e.~$R^\mu_{\,\mu}=0$, and eq.~\eqref{mR-2} takes the form
\begin{equation}\label{mR-3}
     \tilde{R}_\Sigma^2=\frac{2m}{M_4^2}R_\Sigma+\frac{3m^2}{4\pi M_4^2\lambda R_\Sigma ^2}+Q\,.
\end{equation}
Imposing the traceless condition, we obtain the coupled differential equation
\begin{equation}\label{dif-eq-1}
    F''\left(\tilde{R}^2-R^2\right)+F'\left(3\tilde{R}'\tilde{R}-2R-\frac{\tilde{R}^2}{R}{}\right)+F\left((\tilde{R}')^2+\tilde{R}\tilde{R}''\right)=0\,,
\end{equation}
with boundary conditions~\eqref{mR-3} and~\eqref{F-boundary}. 

First, for simplicity, we consider $F$ to be a constant\footnote{A constant $F$ can be absorbed into a rescaling of the time coordinate, restoring $g_{tt} \to -1$ as $R \to \infty$ without loss of generality.}. Then eq.~\eqref{dif-eq-1} simplifies to the ordinary differential equation
\begin{equation}
    (\tilde{R}')^2+\tilde{R}\tilde{R}''=0\,,
\end{equation}
whose general solution is
\begin{equation}
    \tilde{R}=c_1R+c_2\,.
\end{equation}
Note that if we evaluate it at the boundary,
\begin{equation}
    c_1R_\Sigma+c_2=\frac{2m}{M_4^2}R_\Sigma+\frac{3m^2}{4\pi M_4^2\lambda R_\Sigma ^2}+Q\,.
\end{equation}
It is essential to realise that $R_\Sigma$ is \textit{not constant in time}, $R_\Sigma\propto a(\tau)$, following~\eqref{rsigma}. This identifies $c_2=Q$ and
\begin{equation}
    c_1(\tau)=\frac{2m}{M_4^2}+\frac{3m^2}{4\pi M_4^2\lambda}\,R_\Sigma ^{-3}(\tau)\,,
\end{equation}
which contradicts the assumption that the exterior metric is static. 

On the other hand, setting $\tilde{R}' =0$ in~\eqref{dif-eq-1} reduces the differential equation to one for $F(R)$ alone, with solution
\begin{equation}
    F(R)=-\frac{c_3}{\tilde{R}}\ln\left(\frac{\tilde{R}+\sqrt{\tilde{R}^2-R^2}}{R}\right)+c_4\,.
\end{equation}
This solution becomes imaginary for $R > \tilde{R}$, and therefore cannot describe the exterior of the cloud where $R > R_\Sigma$, unless $c_3 = 0$. In that case $F = c_4 = \text{const}$, which reduces to the previous case and yields no new solutions.

More generally, neither $F(R)$ nor $\tilde{R}(R)$ need be constant, and eq.~\eqref{dif-eq-1} admits a wider class of solutions in which both functions vary with $R$. In all cases, however, the Israel junction conditions impose a dynamical boundary condition through $\tilde{R}_\Sigma(\tau)$, which propagates into the solution for all $R$ and renders the exterior non-static. A static exterior is therefore ruled out generically for $m\neq 0$, independently of the assumed form of $F(R)$ and $\tilde{R}(R)$, in agreement with the no-go theorem of~\cite{Bruni_2001}. 
\section{Misner--Sharp formalism}
Let us now take a different approach, which will allow us to interpret this non-static conclusion from another angle. For a spherically symmetric metric with areal radius $R(t,r)$, we introduce the notation
\begin{equation}
    D_t\equiv \frac{1}{\sqrt{-g_{tt}}}\,\partial_t\,,\qquad D_r\equiv \frac{1}{\sqrt{g_{rr}}}\,\partial_r\,,
\end{equation}
i.e.~the derivatives with respect to proper time and proper radial distance, respectively. We also define the variables
\begin{equation}
    U\equiv D_t R\,,\qquad \Gamma\equiv D_r R\,,
\end{equation}
and
\begin{equation}
    M_{\scriptscriptstyle\mathrm{MS}}(t,r)=\frac{R}{2}\left(1-g^{ab}\partial_a R\,\partial_b R\right)=\frac{R}{2}\left(1-\Gamma^2+U^2\right)\,.
\end{equation}
The quantity $M_{\scriptscriptstyle\mathrm{MS}}(t,r)$ is known in the literature as the Misner--Sharp mass, and it is defined purely in terms of geometrical quantities. The Bianchi identity then imposes conservation equations on the Misner--Sharp variables $(U,\Gamma,M_{\scriptscriptstyle\mathrm{MS}})$.

Using the Einstein field equations in the covariant approach, one obtains the Misner--Sharp equations
\begin{subequations}
\begin{align}
    &D_tU=-\frac{\Gamma}{\rho_{\scriptscriptstyle\mathrm{eff}}+p_{\scriptscriptstyle\mathrm{eff}}}\,D_rp_{\scriptscriptstyle\mathrm{eff}}-\frac{M_{\scriptscriptstyle\mathrm{MS}}(t,r)}{R^2}-4\pi R\,p_{\scriptscriptstyle\mathrm{eff}}\,,\\[10pt]
    &D_t \Gamma=-\frac{U}{\rho_{\scriptscriptstyle\mathrm{eff}}+p_{\scriptscriptstyle\mathrm{eff}}}\,D_rp_{\scriptscriptstyle\mathrm{eff}}\,,
        \\[10pt]\label{Dr-M}
    &D_rM_{\scriptscriptstyle\mathrm{MS}}=4\pi R^2\, \Gamma\,\rho_{\scriptscriptstyle\mathrm{eff}}\,,
    \\[10pt]\label{Dt-M}
    & D_tM_{\scriptscriptstyle\mathrm{MS}}=-4\pi R^2\,U\,p_{\scriptscriptstyle\mathrm{eff}}\,,
\end{align}
\end{subequations}
where the quantities $\rho_{\scriptscriptstyle\mathrm{eff}}$ and $p_{\scriptscriptstyle\mathrm{eff}}$ are the ones introduced in Section~\ref{sec:eff-pres}, eq.~\eqref{effective-quantities}.

Let us first consider eq.~\eqref{Dr-M}. Integrating with respect to $r$ yields
\begin{equation}
    M_{\scriptscriptstyle\mathrm{MS}}=\int 4\pi \rho_{\scriptscriptstyle\mathrm{eff}}\,R^2R'\,\dd r\,.
\end{equation}
Noting that $4\pi R^2R'\dd r=4\pi R^2\dd R=\dd V$ is the volume element associated with the areal radius $R$, we find
\begin{equation}\label{ms1}
     M_{\scriptscriptstyle\mathrm{MS}}=\int  \rho_{\scriptscriptstyle\mathrm{eff}}\,\dd V\,,
\end{equation}
which can be interpreted as the \textit{effective} gravitational energy enclosed within a volume $V$ of areal radius $R$. Expressing~\eqref{ms1} explicitly in terms of the effective quantities,
\begin{equation}
     M_{\scriptscriptstyle\mathrm{MS}}=\int  \left(\rho+\frac{\rho^2}{2\lambda}+\rho_{\mathcal{E}}\right)\dd V\,,
\end{equation}
we see that it is not simply the gravitational mass of the brane fluid: it also includes the geometric braneworld contributions.

Let us now focus on eq.~\eqref{Dt-M}, and specialize to a homogeneous and isotropic dust region, $p=0$. In four-dimensional general relativity this would imply $p_{\scriptscriptstyle\mathrm{eff}}=p=0$, so that the Misner--Sharp mass enclosed by any comoving surface is conserved --- this is precisely the property underlying the Oppenheimer--Snyder construction. On the brane, however, $p_{\scriptscriptstyle\mathrm{eff}}$ does not vanish for $p=0$; according to~\eqref{effective-quantities}, it takes the value
\begin{equation}
    p_{\scriptscriptstyle\mathrm{eff}}=\frac{\rho^2}{2\lambda}+\frac{\rho_\mathcal{E}}{3}\,.
\end{equation}
These non-vanishing terms imply, through eq.~\eqref{Dt-M}, that the Misner--Sharp mass enclosed for a fixed $r$ is not constant in time. Since the Misner--Sharp mass is defined covariantly on the brane, it includes all the effects induced by the higher-dimensional geometry. Therefore, even in comoving coordinates, a fixed surface $r_0$ does not enclose a conserved quantity, in contrast with four-dimensional Oppenheimer--Snyder collapse. It is thus not reasonable to expect a static exterior.

The junction conditions on the brane therefore forbid matching a FLRW patch directly onto a static exterior. The overdensity profile must instead develop a tail: an intermediate region, whose extent is controlled by the bulk, interpolating between the interior FLRW solution and the asymptotic Schwarzschild--Tangherlini bulk geometry.

It is the full five-dimensional geometry that provides the explanation
for this. A discontinuity in $T_{\mu\nu}$ at the edge of the
overdensity translates, through the junction conditions
\eqref{eq:Israel_covariant}, into discontinuities in the extrinsic curvature $K_{\mu\nu}$ tangential to the
brane --- already enough, via the Taylor expansion of
the bulk metric~\cite{Sasaki_2000}, to make $g_{AB}(x,y)$ discontinuous at $y \neq 0$~\cite{Maartens_2004}. The
non-local conservation equation~\eqref{eq:nonlocal_conservation}
points to a second, independent pathology at the next order: it
relates $\nabla^\mu \mathcal{E}_{\mu\nu}$ to $\nabla^\mu S_{\mu\nu}$,
with $S_{\mu\nu}$ quadratic in $T_{\mu\nu}$ ~\eqref{Smunu-general}, so that
the same step in $T_{\mu\nu}$ produces a delta-distribution in
$\nabla^\mu S_{\mu\nu}$, and hence in $\nabla^\mu \mathcal{E}_{\mu\nu}$,
at the boundary. Either effect on its own signals a breakdown of bulk
regularity, as discussed in~\cite{Maartens_2004,Maartens_2010,casadio2013braneworldstarsminimalgeometric}. This obstruction is independent of
the scale and nature of the fluid, and has accordingly been a
long-standing concern in the study of stellar collapse in
braneworlds.  Indeed, the
continuity of five-dimensional geodesics always requires both the
matter stress tensor and the projected Weyl tensor to be smooth across
the surface and along the brane.

The effect of the resulting tail on the collapse threshold has been
studied extensively in the literature (see e.g.~\cite{Bruni_2001,Kofinas_2004,OVALLE_2008,OVALLE_2009,OVALLE_2010a,OVALLE_2010,Casadio:2012pu,Ovalle:2013vna,Ovalle:2013xla,casadio2013braneworldstarsminimalgeometric}), and the time evolution of the
horizon for large black holes has been worked out in~\cite{Casadio_2005}. A
dedicated analysis for small black holes forming in the high-energy
regime, however, does not yet exist, to our knowledge, and would be
particularly interesting: their mass evolution would be more exotic,
since formation would be followed by the accretion epoch described in
the next chapter, and such an analysis would allow initial conditions
at formation to be matched directly to observational signatures today.

      \chapter[Cosmological accretion: a relativistic description]{Cosmological accretion:\\ a relativistic description}\label{chap10:accretion}
      \thispagestyle{fancy}
\pagestyle{fancy}

    In this chapter, we focus on the study of accretion by primordial black holes in RS-II cosmology. This has already been studied  by Guedens, Clancy and Liddle~\cite{Guedens_2002-2}, Majumdar~\cite{Majumdar_2003}, and Tikhomirov and Tsalkou~\cite{Tikhomirov:2005bt}\footnote{The general case of accretion by a $d$-dimensional Schwarzschild black hole was studied in~\cite{John_2013}, where they also consider a $d$-dimensional fluid.}. The first two articles arrive at the same result, except for an undefined parameter $F$ (to which the result is exponentially sensitive) introduced in Ref.~\cite{Guedens_2002-2} and interpreted as a measure of accretion efficiency, with $F\leq 1$. In both calculations, the authors choose the hydrodynamic approach to describe the accretion process and employ Bondi's approximation~\cite{Bondi1952} for the rate of change of the black hole mass. Their calculations predict a very strong accretion rate in the early stages after formation of small black holes, which clearly diverges from the prediction of standard cosmology where accretion by very light PBHs is expected to be negligible. 

\vspace{0.5cm}
\begin{mdframed}[style = mybbox]
This surprising result arises from the combination of two factors: a larger accretion radius in comparison to their $(3+1)$-dimensional counterpart and a slower dilution rate with respect to cosmic time for the energy density of the matter fields fluid in early epochs. 
\end{mdframed}
\vspace{0.5cm}

Later on, Tikhomirov et al. accounted for frequent particle collisions in the cosmological fluid, and argued that this led to a larger value $F\simeq1.54$. The energy density and velocity profiles during the accretion process in braneworlds were also later studied by Harko~\cite{Harko_2009}, who focused on the low-temperature limit and thus do not apply to early epochs.

The central aim of this chapter is to critically examine the applicability within braneworld models of all standard equations and definitions employed to arrive at the accretion rate, which for simplicity we denote as $\dot{M}(t)$ in this chapter. To this end, we complement the discussion in Refs.~\cite{Guedens_2002-2,Majumdar_2003,Tikhomirov:2005bt} by recalculating the accretion rate using Michel's relativistic extension of the Bondi accretion formalism~\cite{Michel1972}, \emph{a priori} essential in non-conventional scenarios where modifications to the underlying spacetime geometry may play a significant role. 

We also address the validity of Michel's approximation in early epochs, both of which rely on the assumptions of a quasi-steady flow (stationarity) and a negligible self-gravity of the surrounding fluid. This has already been done by Ricotti~\cite{Ricotti_2007} for Bondi's result in the case of PBHs in standard cosmology, but not for braneworlds. It was concluded in Ref.~\cite{Ricotti_2007} that it is adequate only for limited ranges of masses in the conventional scenario. In the case of RS-II, we show that it is valid for a much wider parameter range, with the exact domain determined by the collapse dynamics. In the conventional scenario, most black hole masses require time-dependent self-similar solutions and numerical simulations to accurately describe the accretion flow morphology. In RS-II, these restrictions are softened, allowing for a fully analytic description.
%
%
\section{Constants of integration}\label{section-constants-integration}
Just as in gravitational collapse, the dynamics of accretion in any spacetime geometry are governed by the conservation equations. In \href{part:braneworlds}{\textbf{\textcolor{thesislilac}{Part I}}}, we established the conservation equations arising from the Codazzi equation and Bianchi identity in the SMS framework. In RS-II, these split into two distinct sets: local conservation equations from the independent conservation of the brane matter stress-energy tensor, eq.~\eqref{eq:local_conservation}, and non-local conservation equations relating the additional geometric tensors in the SMS covariant field equations, eq.~\eqref{eq:nonlocal_conservation}. This means that we must ensure consistency: ordinary matter must evolve so that energy-momentum is conserved, but it must do so in a way that does not violate geometric constraints imposed by~\eqref{eq:nonlocal_conservation}. 

The Bianchi identities ensure consistency in the evolution of the geometry of the Universe. 

There is an additional conservation equation that is essential for a complete description of the accretion flow. It enforces conservation of energy along the fluid flow and is obtained by projecting the energy–momentum conservation law along the fluid four-velocity,
\begin{equation}\label{proj-cons}
    u_\mu\, \nabla_\nu T^{\mu\nu}=0\,.
\end{equation}

In the remainder of this section, we identify useful constants of integration arising from the above conservation equations. Since our analysis is formulated within the framework of Michel accretion --- which assumes a quasi-steady flow and, consequently, a time-independent spacetime geometry --- we adopt the following general static metric on the brane
\begin{equation}\label{general-metric}
    \dd s^{2}=-f(r)\dd t^2+\frac{1}{f(r)}\dd r^2+r^2\dd\Omega_2
\end{equation}
and fix our coordinate system so that the fluid four-velocity takes the form
\begin{equation}\label{4-velocity}
    u^{\mu}=(u^t,u,0,0)\,,
\end{equation}
where, in the context of accretion, we expect $u<0$, corresponding to inward flow towards the black hole, while we take $u^t>0$. Imposing the normalisation condition $u^\mu u_\mu=-1$, consistent with the sign convention adopted in this work, for the metric~\eqref{general-metric} the four-velocity vector takes the form
\begin{equation}\label{coordinates}
    u^\mu=\bigg(\frac{\sqrt{f(r)+u^2}}{f(r)},u,0,0\bigg)\,.
\end{equation} 
We distinguish between constants of integration associated solely with brane quantities, which we refer to as \emph{local} constants of integration, and those that depend on bulk degrees of freedom, which we term \emph{non-local} constants of integration.
\subsection{Local constants of integration}
To extract the first constant of integration, we evaluate the time component of the conservation equation \eqref{eq:local_conservation},
\begin{equation}\label{partial-A1}
    \partial_r\left(\sqrt{-g}T^r_t\right)=0\,,
\end{equation}
where $g$ is the determinant of the metric, i.e.~$\sqrt{-g}=r^2\sin\theta$. We  define the first constant $A_1$ by integrating~\eqref{partial-A1} over all angles,
\begin{equation}\label{pre-A1}
    A_1=-\oint \sqrt{-g}T^r_t\;\dd\theta\,\dd\phi=-4\pi r^2\,T^r_t\,,
\end{equation}
such that $\partial_r A_1=0$. For a perfect fluid, eq.~\eqref{pre-A1} takes the form
\begin{equation}\label{A1}
A_1=4\pi\big(\rho(r)+p(r)\big)\,u(r)\,r^2\sqrt{f(r)+u^2(r)}\,.
\end{equation}
The second constant of integration is obtained from~\eqref{proj-cons},
\begin{equation}
    \partial_r\left(\ln(ur^2)+\int^\rho\frac{\dd\tilde{\rho}}{\tilde{\rho}+p(\tilde{\rho})}\right)=0\,.
\end{equation}
We make the choice $\rho(r\to\infty)=\rho_\infty$ for the lower boundary in the integral for convenience and define
\begin{equation}\label{A2}
    A_2=ur^2\exp\left(\int_{\rho_\infty}^\rho\frac{\dd\tilde{\rho}}{\tilde{\rho}+p(\tilde{\rho})}\right)\,.
\end{equation}
These constants of integration are the main ingredients to study the velocity and energy density profiles surrounding the black hole during the accretion process, crucial to determine the mass change rate of the black hole.  
%
%
\subsection{Non-local constants of integration}
We now focus on the conservation equation~\eqref{eq:nonlocal_conservation}. Using eqs.~\eqref{Smunu} and~\eqref{Emunu}, we can write~\eqref{eq:nonlocal_conservation} explicitly\footnote{Notation: the tensor $h^\mu_{\ \nu}$ in eq.~\eqref{NLC-1} is the spatial part of the background metric.},
\begin{equation}\label{NLC-1}
    \nabla_\mu\bigg[\frac{\kappa^2}{2\lambda}\rho\big(\rho u^\mu u_\nu + (\rho+2p) \,h^\mu_{\nu}\big)+\kappa^2\big(\rho_\mathcal{E}u^\mu u_\nu +\frac{1}{3}\rho_\mathcal{E}h^\mu_{\,\nu}+\pi^{\mathcal{E}}{}^{\mu}_{\,\nu}\big)\bigg]=0\,.
\end{equation}
Adopting again a stationary configuration in which the energy densities, metric, and velocity fields depend only on $r$, and using $h^r_t = u^r u_t$, the time component ($\nu = t$) of eq.~\eqref{NLC-1}, integrated over all angles yields
\begin{equation}\label{Cons-eq-2-b}
   \partial_r\bigg[4\pi r^2\bigg(\frac{\rho(\rho+p)}{\lambda}+\frac{4}{3}\rho_\mathcal{E}\bigg) u u_t\bigg]=0\,,
\end{equation}
where we have taken $\pi^{\mathcal{E}}{}^\mu_{\,\nu}$ to be diagonal due to the symmetries of the system and $\pi^{\mathcal{E}}_{tt}=0$. Dividing eq.~\eqref{Cons-eq-2-b} by $A_1$, results in a new constant of integration,
\begin{equation}\label{A3}
   A_3=\frac{\rho}{\lambda}+\frac{4}{3}\frac{\rho_\mathcal{E}}{\rho+p}\,.
\end{equation}
%
%
\section{Michel accretion}\label{section-accretion}
We adopt Michel's relativistic formalism to model accretion within RS-II. This formulation relies on several assumptions that restrict its validity to particular regions of parameter space. We discuss the validity of these approximations in Section~\ref{section-validity}.

Michel accretion follows directly from energy conservation --- specifically, eq.~\eqref{eq:local_conservation} --- and the constant of integration $A_1$ acquires the physical interpretation of the black hole mass accretion rate, $A_1\equiv-\dot{M}$. Under the assumptions taken in this accretion model, $\dot{M}$ must remain constant for all values of radial coordinate $r$. 

Beyond this standard identification, the braneworld framework introduces an additional conserved quantity, given by eq.~\eqref{eq:nonlocal_conservation},
\begin{equation}\label{Cons-eq-2-c}
    \partial_r\left(A_3\,\dot{M}\right)=0\,.
\end{equation}
Here, $A_3$, defined in eq.~\eqref{A3}, plays a role analogous to that of $A_1$: just as $A_1\equiv-\dot{M}$
 encodes the transformation of the energy-momentum tensor $T_{\mu\nu}$ of the accreted fluid into the black hole mass growth rate, $A_3$ encodes the transformation of the gravitational energy stored in $S_{\mu\nu}$ and the component of $\mathcal{E}_{\mu\nu}$ sourced by gradients in the energy density profile, into the growing Weyl cloud of the Schwarzschild--Tangherlini black hole present in the covariant four-dimensional SMS description. Crucially, $A_3$ is conserved alongside the mass accretion rate, and in the late-time limit it drives the Weyl tensor $\mathcal{E}_{\mu\nu}$
toward the form~\eqref{Emunu}, with $\rho_{\mathcal{E}}$ and $\pi^\mathcal{E}_{\mu\nu}$ as given in eqs.~\eqref{rhoE-asr0} and~\eqref{Pimunu}, respectively.

\vspace{0.5cm}
\begin{mdframed}[style = mybbox]
This is the distinctive feature of accretion in braneworlds: a conserved quantity that concurrently transforms the gravitational energy content of the brane. Since $S_{\mu\nu}$ and $\mathcal{E}_{\mu\nu}$ are not fundamental degrees of freedom but rather artifacts of the projection of the 5D bulk geometry onto the brane, they cannot drive the growth of the black hole mass, but they must transform consistently, with the Weyl cloud --- itself a purely four-dimensional construct as well --- evolving according to
\begin{equation}
\dot{\rho}_\mathcal{E}^{\scriptscriptstyle\mathrm{BH}}=-\frac{\ell \dot{M}}{3\pi^2r^4}\,,
\end{equation}
as given by eq.~\eqref{rhoE-vacuum}. 
\end{mdframed}
\vspace{0.5cm}

This is required by energy conservation: were these projection artifacts to evolve without constraint, the Weyl cloud would grow in lockstep with $\dot{M}$ without any physical justification. Since $M$ and $\rho$ are physical observables that exist independently of the dimensional reduction and must remain consistent between the bulk and brane descriptions by construction of the SMS framework, the extra terms in the Einstein field equations must transform accordingly --- and it is precisely this transformation that eq.~\eqref{Cons-eq-2-c} governs.

In the following, we close the discussion of the braneworld-specific conservation structure and turn our attention to the physically observable accretion rate $\dot{M}$, which is the quantity of primary interest from a braneworld cosmology standpoint. To that end, we evaluate the energy density and four-velocity profiles in the vicinity of the black hole in order to determine $\dot{M}$ at a scale $r$ where all relevant quantities are known.
\subsection{Critical radius in Schwarzschild--Tangherlini spacetimes}
Let us combine the radial component of the conservation equation~\eqref{eq:local_conservation},
\begin{equation}\label{radial-cons-1}
    uu'+\frac{f'(r)}{2}=-\frac{p'}{\rho+p}\left(f+u^2\right)\,,
\end{equation}
with eq.~\eqref{proj-cons}, \footnote{In the original Michel treatment~\cite{Michel1972}, this equation is replaced
by rest-mass current conservation, $\nabla_\mu (\rho_0 u^\mu) = 0$, with $\rho_0$ the rest-mass density, which is not available for a radiation
fluid with no conserved particle number.}
\begin{equation}\label{cons-der}
    \frac{\rho'}{\rho+p}+\frac{u'}{u}+\frac{2}{r}=0
\end{equation}
in order to isolate $u'/u$ and $\rho'/\rho$,
\begin{equation}
    \frac{\rho'}{\rho}=\frac{\mathcal{D}_1}{\mathcal{D}_2}\,,\qquad \frac{u'}{u}=\frac{\mathcal{D}_3}{\mathcal{D}_4}\,,
\end{equation}
with
\begin{subequations}
   \begin{equation}
    \mathcal{D}_1=\frac{2u^2}{r}-\frac{f'(r)}{2}\,,
\end{equation}
\begin{equation}\label{eq-D-2}
    \mathcal{D}_2=\frac{wf+(w-1)u^2}{w+1}\,,
\end{equation}
\begin{equation}
    \mathcal{D}_3=\frac{2w(f+u^2)}{r}-\frac{f'(r)}{2}\,,
\end{equation}
\begin{equation}\label{eq-D-4}
    \mathcal{D}_4=-\left(wf+(w-1)u^2\right)\,.
\end{equation}
\end{subequations}
Let us study eq.~\eqref{eq-D-2} at the boundaries. We evaluate $\mathcal{D}_2$ at the horizon $r=r_0$,
\begin{equation}\label{lim-r0}
    \mathcal{D}_2(r_0)=\frac{w-1}{w+1}\,u^2\,,
\end{equation}
and in the asymptotic region,
\begin{equation}\label{lim-inf}
    \mathcal{D}_2(r\to\infty)=\frac{w}{w+1}\,.
\end{equation}
For $0< w\leq 1$, eq.~\eqref{lim-r0} is non-positive and~\eqref{lim-inf} is non-negative. Therefore, $\mathcal{D}_2$ must vanish at some critical radius $r_{\rm cr}$. For the flow to be regular, we must impose $\mathcal{D}_1(r_{\rm cr})=\mathcal{D}_2(r_{\rm cr})=0$  (see Ref.~\cite{shapiro-book} for a pedagogical discussion). The same applies for the other two equations, i.e.~$\mathcal{D}_3(r_{\rm cr})=\mathcal{D}_4(r_{\rm cr})=0$. Both conditions provide the equality
  \begin{equation}\label{eqs-csuc}
   u_{\rm cr}^2=\frac{wf_{\rm cr}}{1-w}=
\frac{f_{\rm cr}'r_{\rm cr}}{4}\,,
\end{equation}
that allows us to define quantities at the critical radius. For the Schwarzschild--Tangherlini metric,
    \begin{equation}\label{eqs-as-w}
r_{\rm cr}^2=\frac{(1+w)\,r_0^2}{2w}\,,
\qquad
    u_{\rm cr}^2=\frac{w}{1+w}\,,
\end{equation}
unlike the conventional scenario, where adiabaticity implies $r_{\rm cr}=(1+3w)\,r_0/4w$ and $u_{\rm cr}^2=w/(1+3w)$.

Whereas the critical point has no particular physical interpretation beyond an algebraic requirement for continuity, the sonic point $r_s$ is where the locally measured physical velocity 
\begin{equation}
    v^2(r)=\frac{u^2(r)}{f(r)+u^2(r)}
\end{equation}
matches the group velocity discussed in \href{part:braneworlds}{\textbf{\textcolor{thesislilac}{Part I}}}, $v^2(r_s)=v_g^2\approx c_s^2$, and dictates where the maximum propagation speed of perturbations becomes comparable to the bulk velocity of the infalling flow. In the supersonic region, perturbations cannot propagate outward (upstream), whereas in the subsonic region, information can travel in both directions. In this geometry, we find $v^2(r_{\rm cr})= c_s^2$. Although the critical and sonic points need not coincide in general, they coincide in the present case. This coincidence was demonstrated by Chaverra and Sarbach~\cite{Chaverra_2015} for this class of relativistic accretion flows, i.e.~steady, adiabatic, spherically symmetric accretion of a perfect fluid.
\begin{figure}
    \centering
    \includegraphics[width=0.5\linewidth]{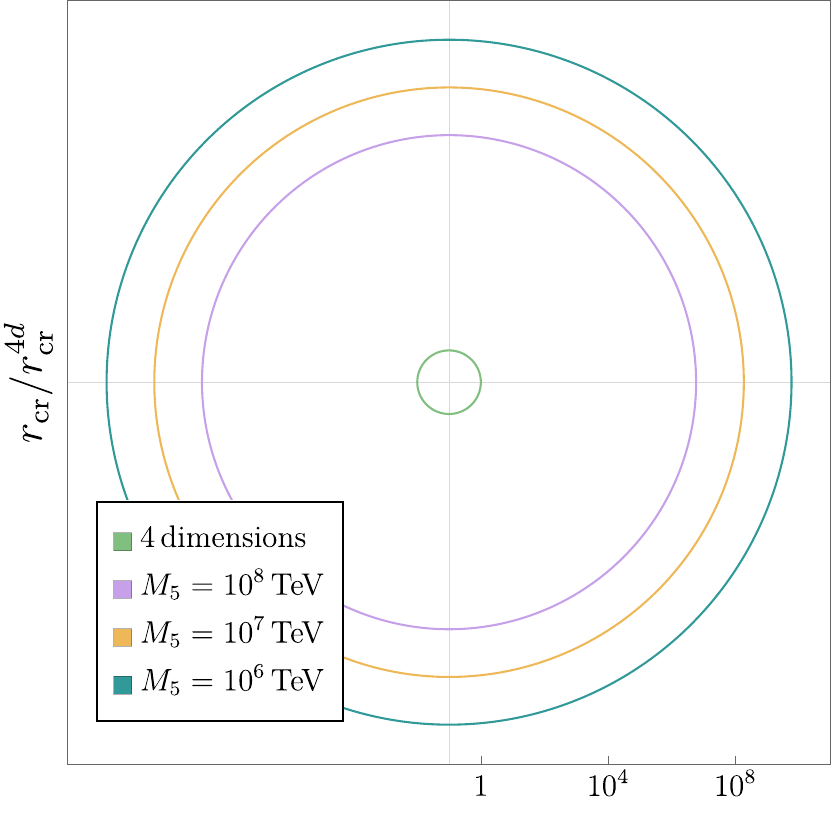}
    \caption[Accretion scale for different fundamental scales]{Accretion scale for a fixed mass ($M=10^{10}M_4$) for different fundamental scales $M_5$ (with $M_5=10^{6}\,\text{TeV}$ being the smallest value allowed by observations~\cite{Kapner:2006si}), compared to the $(3+1)$-dimensional Schwarzschild case where the critical radius takes the value $r_{\rm cr}^{4d}=3M$ \cite{Michel1972}. In both cases, we assume an equation-of-state parameter $w = 1/3$.}
    \label{fig:accretion-scale}
\end{figure}
The critical radius is essential for determining the accretion rate $\dot{M}$ and the overall structure of the inflow. Evaluating the constants of integration~\eqref{A1} and~\eqref{A2} at $r_{\rm cr}$ allows us to define the energy density at the critical point,
\begin{equation}\label{rho-c}
    \rho_{\rm cr}=\left(1+w\right)^{\frac{1+w}{2w}}\rho_\infty\,.
\end{equation}
\subsection{Velocity and energy density profiles}
\begin{figure}
    \centering
    \includegraphics[width=1\linewidth]{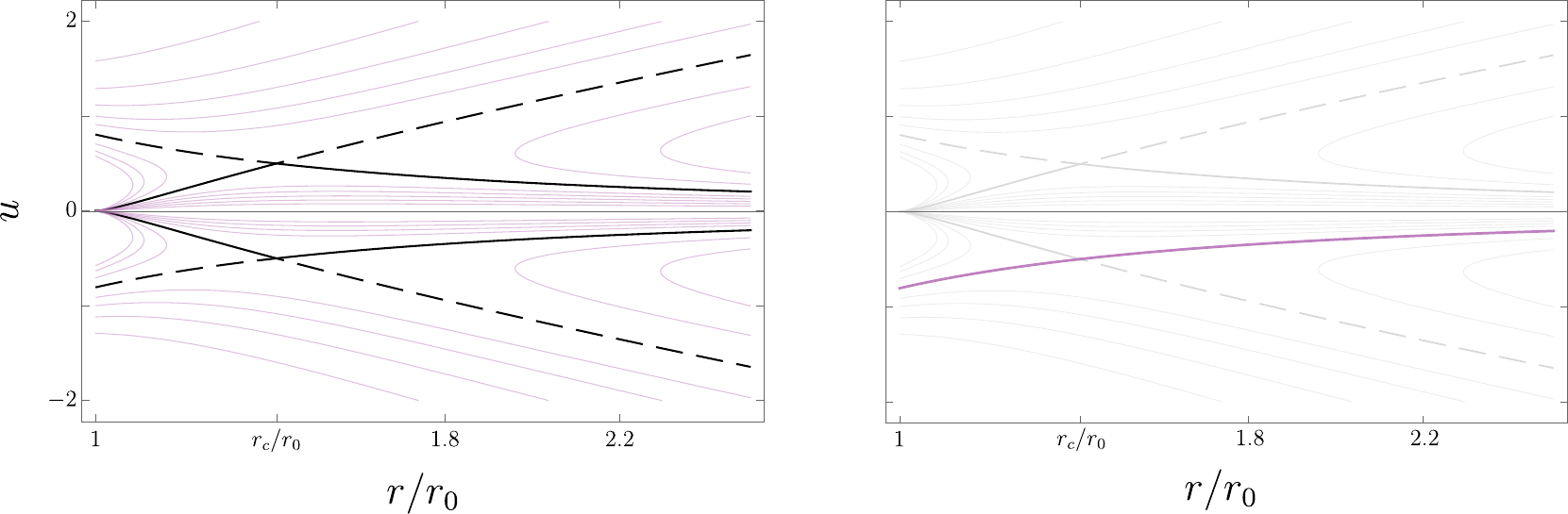}
    \caption[Radial four-velocity: level curves and global solution]{Left: Level curves of the radial component of the four-velocity (purple) as a function of the radial coordinate $r$, expressed in horizon-radius units. The four solution branches are shown in black: $\pm u_1(r)$ (dashed) and $\pm u_2(r)$ (solid), with distinct near-horizon and asymptotic behaviour. Right: Global piecewise solution corresponding to inward accretion matched to a static background.}
    \label{fig:ur}
\end{figure}

For a general radius $r$, eqs.~\eqref{A1} and~\eqref{A2} define a system for the energy density $\rho(r)$ and the radial component of the four-velocity $u(r)$. From this system, one can obtain level curves of $u(r)$ (or $\rho(r)$), shown as contour lines in Figure~\ref{fig:ur} for different values of $A_1$ and $A_2$. Physical solutions must pass through the critical point $(r_{\rm cr}, \pm u_{\rm cr})$ (or $(r_{\rm cr}, \rho_{\rm cr})$).

\begin{figure}
    \centering
    \includegraphics[width=0.75\linewidth]{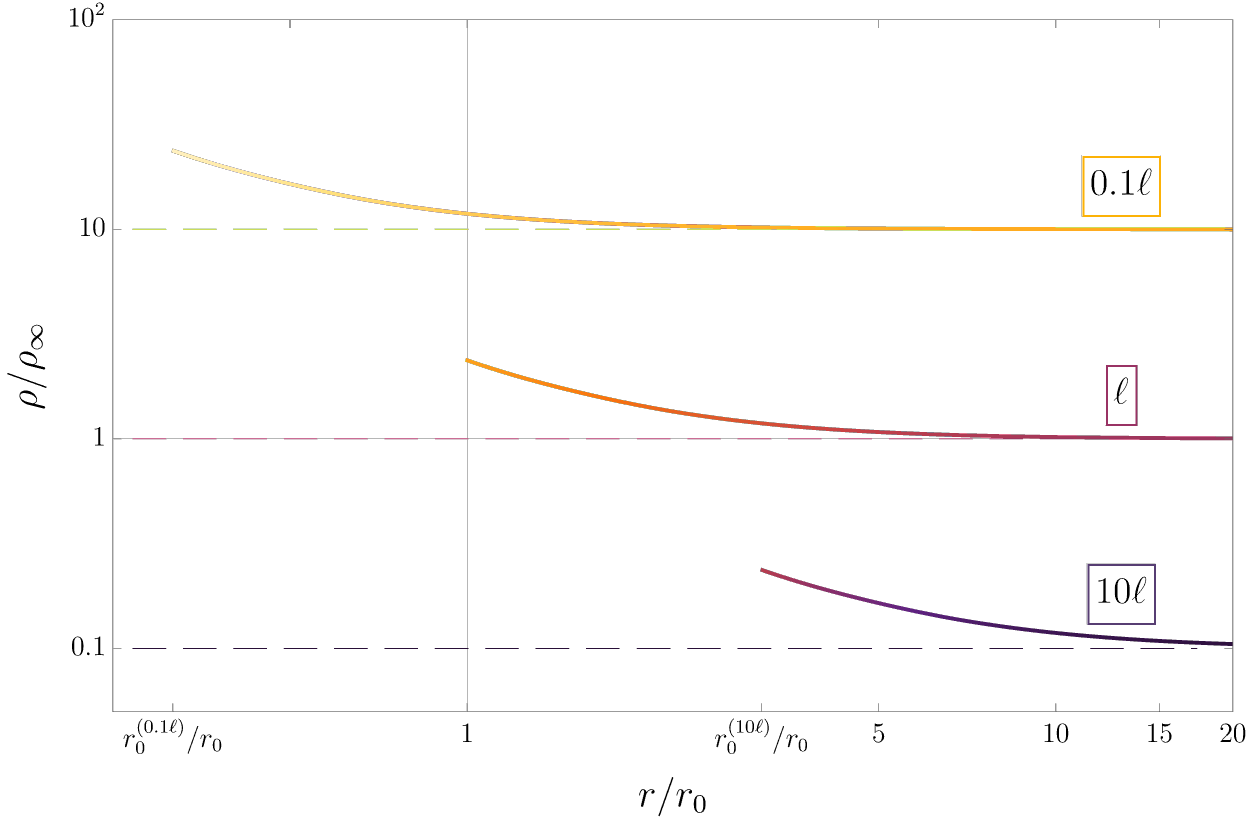}
    \caption[Energy density profiles for different AdS radii]{Energy density profiles for fixed mass and cosmic time, shown for different AdS radii. Both axes are normalised to the background energy density $\rho_\infty$ and horizon radius $r_0$ corresponding to the reference scale $\ell$. The lower and upper curves correspond to $\ell_1 = 10\ell$ and $\ell_2 = 0.1\ell$, respectively.}
    \label{fig:comp-dens}
\end{figure}

We find six branches for $u(r)$ that include the critical point, four of which are real in the domain $r \geq r_0$ and are shown in Figure~\ref{fig:ur} in black. For accretion, we select the inward-flow solutions with $u < 0$ and analyse the near-horizon and asymptotic behaviour of the two relevant branches. For $r < r_{\rm cr}$, we choose the subcritical branch ($u < u_{\rm cr}$), such that $\lim_{r \to r_0} u(r) < 0$, while for $r \geq r_{\rm cr}$ we select the supercritical branch, satisfying $\lim_{r \to \infty} u(r) = 0$. This choice is shown in Figure~\ref{fig:ur}.
The corresponding energy density profile is
\begin{equation}\label{dens-prof}
    \rho(x)=\frac{16}{9}\left(x^2|u(x)|\right)^{-\frac{4}{3}}\rho_\infty\quad,\qquad x=\frac{r}{r_0}.
\end{equation}
Note that, for a fixed mass, the variable $x$ depends on the choice of the AdS scale, as does $\rho_\infty$, following the scaling
\begin{equation}\label{scaling}
    \rho_{\infty}(\ell_1)=\frac{\ell_2}{\ell_1}\rho_\infty(\ell_2)\quad ,\qquad x(\ell_1)=\sqrt{\frac{\ell_1}{\ell_2}}\,x(\ell_2)\,,
\end{equation}
using eqs.~\eqref{quad-reg} and~\eqref{horizonradius}. In Figure~\ref{fig:comp-dens}, we illustrate the impact of AdS warping on the energy density profile. We consider a black hole of mass $M$ at fixed cosmic time $t < t_c$, characterised by a horizon radius $r_0(\ell)$ and background energy density $\rho_\infty$. We compare two AdS radii, $\ell_1 = 10\ell$ and $\ell_2 = 0.1\ell$, with corresponding quantities $(r_{0,i}, \rho_{\infty,i})$. Using the scaling relations~\eqref{scaling}, we express $(r_{0,i}, \rho_{\infty,i})$ in terms of $(r_0, \rho_\infty)$ and compute the resulting energy density profiles. We find that, for fixed mass, the profile becomes more diluted as the warping of the extra dimension increases.
%
%
%
%
\subsection{Black hole mass evolution}
In order to find the accretion rate, we evaluate eq.~\eqref{A1} at the critical point. Using eqs.~\eqref{eqs-as-w} and~\eqref{rho-c}, we find
\begin{equation}\label{mdot1}
    \dot{M}=4\pi F(w)r_0^2\,\rho_\infty
\end{equation}
with
\begin{equation}
    F(w)=\frac{(1+w)^{\frac{1+3w}{2w}}}{2\sqrt{w}}\,,
\end{equation}
which evaluates to $F \simeq 2.05$ for $w = 1/3$. In the regime where Michel’s formalism applies and a barotropic equation of state holds at all radii, $F$ is expected to take this value following from purely geometric considerations, with no thermodynamic input beyond barotropy.

Using eqs.~\eqref{quad-reg} and~\eqref{horizonradius}, we obtain~\eqref{mdot1} as a function of black hole mass and cosmic time,
\begin{equation}
    \dot{M}=\frac{2F(w)}{\pi}\frac{M}{t}\,.
\end{equation}
Integrating from formation time $t_i$ to an arbitrary time $t < t_c$ yields the black hole mass evolution,
\begin{equation}\label{Mt-result}
    M(t)=M_i\left(\frac{t}{t_i}\right)^{\frac{64}{9\sqrt{3}\pi}}
\end{equation}
for $w=1/3$.

To ensure that this strong growth is consistent with causality, we compare the evolution of the cosmological horizon and the Schwarzschild radius of the accreting black hole. We define a parameter $0 < f_H < 1$ that quantifies their relative size at the formation time,
\begin{equation}\label{fH}
    r_0(M_i)=f_H H^{-1}(t_i)\,,
\end{equation}
and, combining eqs.~\eqref{quad-reg},~\eqref{Mt-result} and~\eqref{fH}, we obtain
\begin{equation}\label{r0-H-1}
    \frac{r_0(t)}{H^{-1}(t)}\simeq f_H\left(\frac{t_i}{t}\right)^{0.35}\,.
\end{equation}
Hence, we find that the horizon radius grows more slowly than the cosmological horizon. We impose only the condition $f_H < 1$, and do not assume $f_H \simeq 1$ as in the conventional scenario, since gravitational collapse in braneworld cosmology requires a more careful treatment.

Equation~\eqref{mdot1} remains valid in the regime $t \gg t_c$, provided the energy density is given by eq.~\eqref{linear-reg}. In this case, integration yields
\begin{equation}\label{mass-ev-linear}
    M(t)=M(t_c)\exp\left[\frac{2F(w)}{\pi}\left(1-\frac{t_c}{t}\right)\right]\,,
\end{equation}
in agreement with Ref.~\cite{Guedens_2002-2}. In the regime $t \gg t_c$, the difference between our value of $F$ and the range adopted in the literature is negligible, and the black hole mass remains constant up to a factor of order unity.
\begin{figure}
    \centering
    \includegraphics[width=1\linewidth]{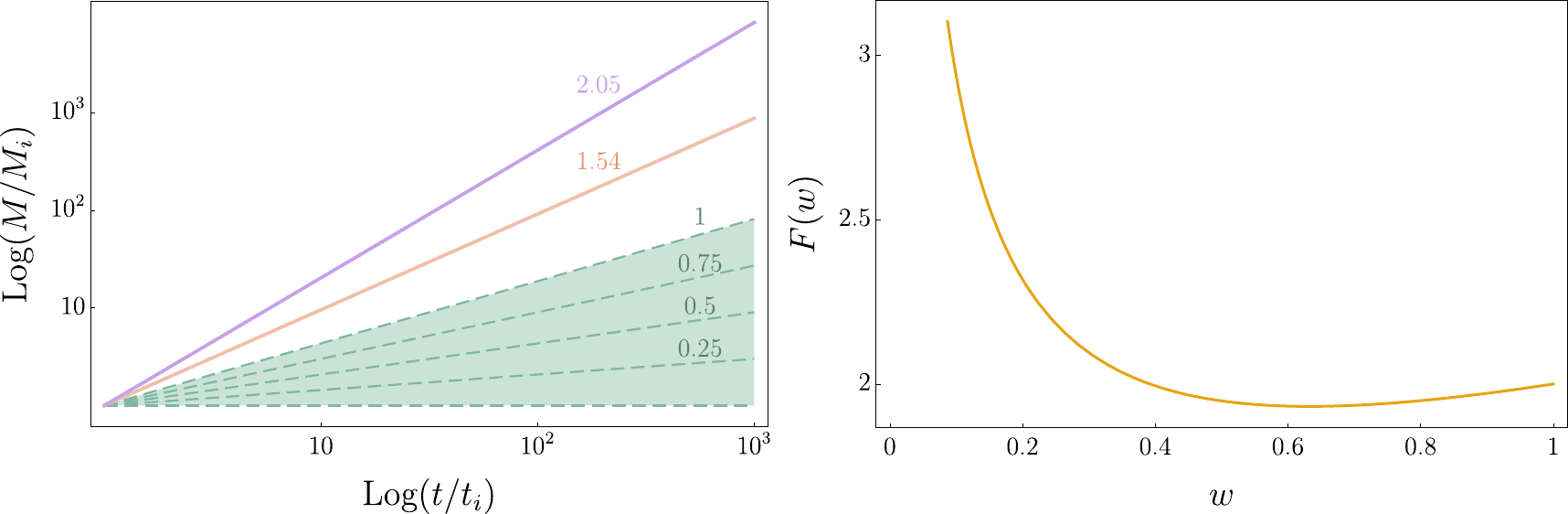}
    \caption[Black hole mass evolution: comparison to previous results]{Left: Black hole mass evolution $M(t)$ as a function of cosmic time for $w=1/3$. The green shaded region shows the range obtained in Refs.~\cite{Guedens_2002-2,Majumdar_2003} for $0<F\leq1$, while the orange line corresponds to the value $F=1.54$ found in Ref.~\cite{Tikhomirov:2005bt}. The purple curve shows our result. See Section~\ref{sec:comparing-results} for a comparison of these results. Both axes are rescaled: the mass is normalised to the initial value $M_i=M(t_i)$, and time is expressed in units of the formation time $t_i$. Right: The factor $F(w)$ as a function of the equation-of-state parameter $w$.}
    \label{fig:comparing-results}
\end{figure}
\subsection{Strong accretion epoch}
To determine the duration of the strong accretion phase, it is necessary to identify the earliest possible formation time, namely the end of inflation in the RS-II scenario. Establishing this initial epoch enables us to relate final black hole masses to their corresponding initial values and to track their evolution in a quantitatively consistent framework. To this end, we make use of the relation between the tensor ($\mathcal{P}_{\scriptscriptstyle\mathrm{T}}$) and scalar ($\mathcal{P}_{\scriptscriptstyle\mathrm{S}}$) power spectra
\begin{equation}
     \mathcal{P}_{\scriptscriptstyle\mathrm{T}}=r\,\mathcal{P}_{\scriptscriptstyle\mathrm{S}}\,,
\end{equation}
with $r$ the tensor-to-scalar ratio. The tensor power spectrum in braneworld scenarios was first derived in~\cite{Langlois_2000}, with subsequent analyses confirming this result~\cite{Seery_2005,Gorbunov_2001}. In these works it is shown that, relative to the four-dimensional case, the tensor power spectrum acquires an additional factor $\mathcal{F}^2$, such that
 \begin{equation}
      \mathcal{P}_{\scriptscriptstyle\mathrm{T}}=\frac{2}{\pi^2}\frac{H^2}{M_4^2} \mathcal{F}^2\quad,\qquad \mathcal{F}^2=\left[\sqrt{1+H^2\ell^2}-H^2\ell^2\ln\left(\frac{1}{H\ell}+\sqrt{1+\frac{1}{H^2\ell^2}}\right)\right]^{-1}\,,
 \end{equation}
 resulting in an enhanced gravitational-wave amplitude in the quadratic regime for fixed $H^2$.\,\footnote{Note that, in contrast to~\cite{Langlois_2000}, we set their parameter $\mu$ to $\mu^{-1} = l$; this identification holds only in the limit $\Lambda_4 \simeq 0$.}

\begin{figure}
    \centering
    \includegraphics[width=0.75\linewidth]{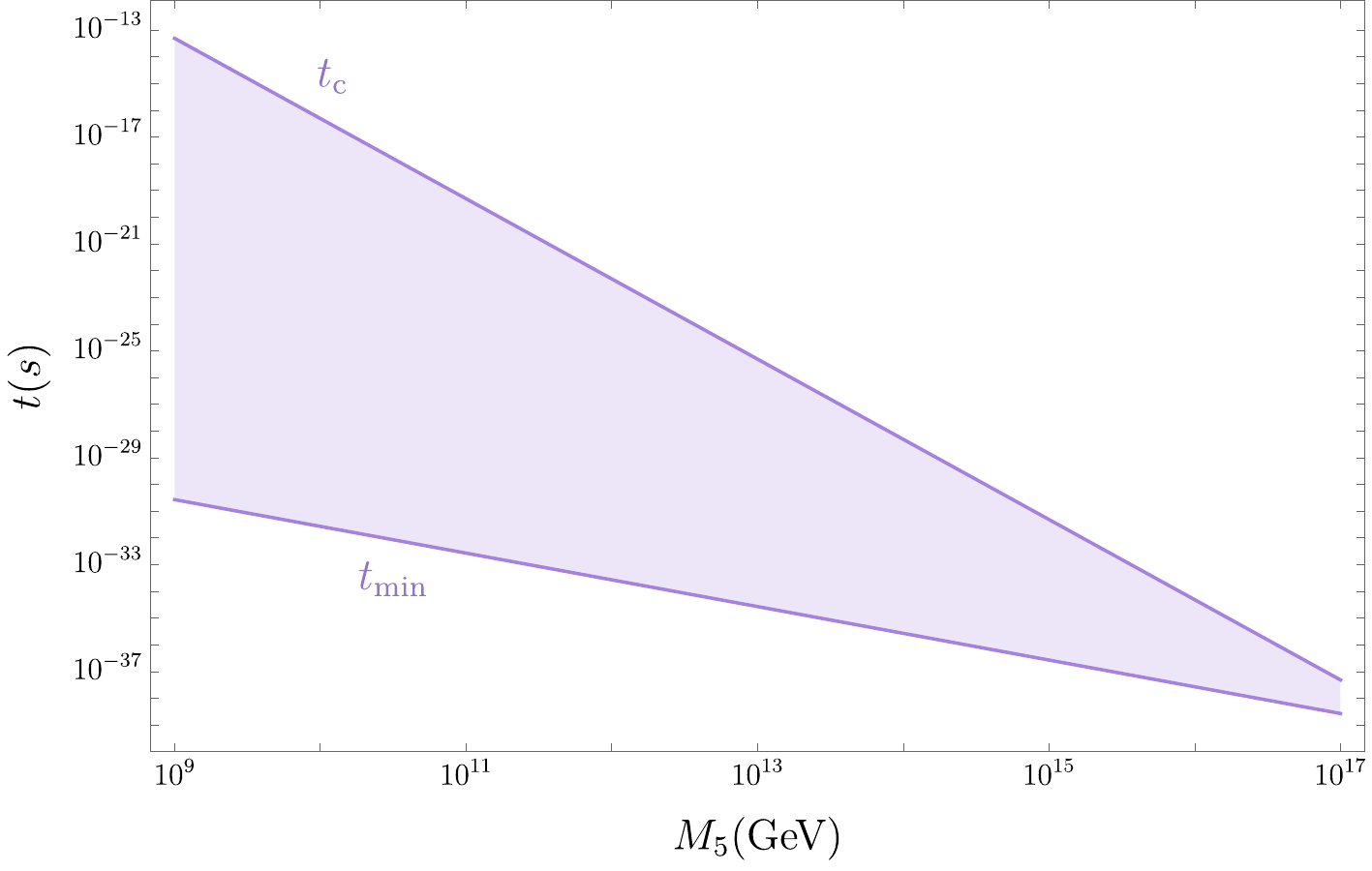}
    \caption[Maximum duration of the strong-accretion epoch]{Maximum duration of the strong-accretion epoch for different $M_5$ values. The upper bound is set by the end of the quadratic regime $t= t_c$ and the lower bound is imposed by the scalar power spectrum and tensor-to-scalar ratio observational constraints.}
    \label{fig:duration}
\end{figure}

The function $\mathcal{F}$ simplifies in the quadratic regime,
\begin{equation}
    \mathcal{F}\simeq \sqrt{\frac{3H\ell}{2}}\,,
 \end{equation}
 and $\mathcal{F} \to 1$ for $\rho \ll \lambda$, recovering the four-dimensional result. Observational constraints on $r$ and $\mathcal{P}_{\scriptscriptstyle\mathrm{S}}$, namely $r < 0.032$ (95\% CL)~\cite{Tristram_2022} and $\mathcal{P}_{\scriptscriptstyle\mathrm{S}} \simeq 2.1 \times 10^{-9}$~\cite{2020}, then allow us to derive a lower bound on $H^{-1}$. This, in turn, sets a corresponding lower bound on $r_0(f_H)$ at formation for different values of $M_5$. The duration of the strong-accretion epoch for different values of $M_5$ can be found in Figure~\ref{fig:duration}. 

\begin{figure}
    \centering
    \includegraphics[width=0.77\linewidth]{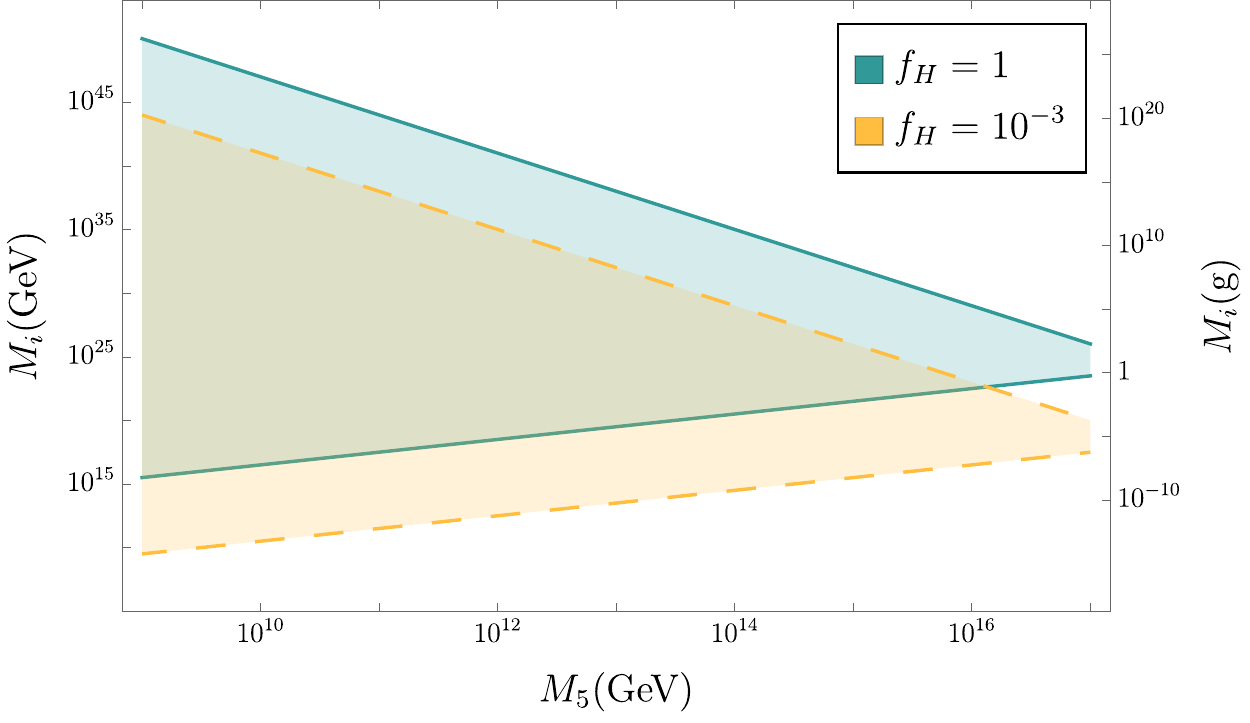}
    \caption[Black-hole formation mass range vs.\ fundamental Planck scale]{Range of black hole masses at formation as a function of the fundamental Planck scale. The blue shaded region corresponds to $f_H = 1$, while the yellow shaded region shows the case $f_H = 0.1$. In both cases the upper mass value corresponds to $t_i\sim t_c$ and the lower mass value to $t_{i,\scriptscriptstyle\mathrm{min}}$ imposed by constraints on the power spectrum. The minimum value of the fundamental Planck scale considered is set by current observational constraints.}
    \label{fig:mass-ranges}
\end{figure}

\begin{figure}
    \centering
    \includegraphics[width=0.77\linewidth]{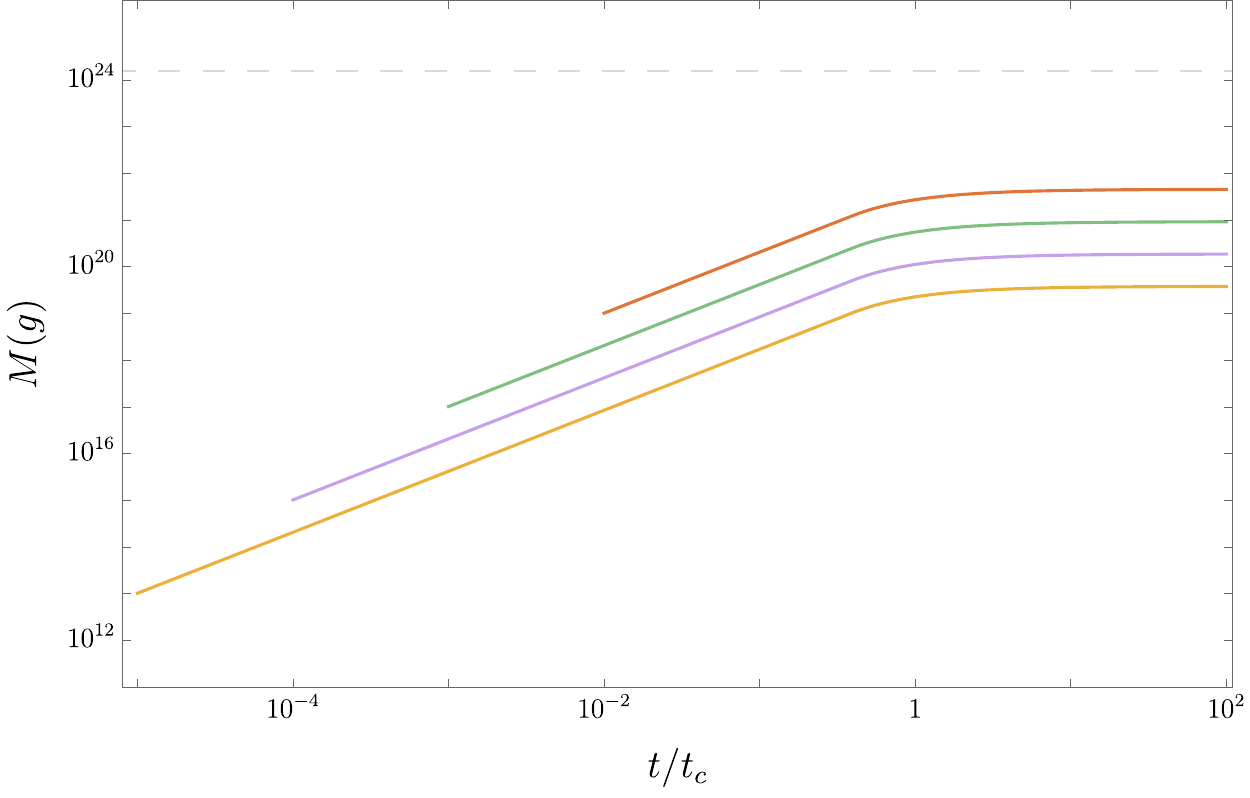}
    \caption[Mass evolution for different formation times]{Mass evolution as a function of cosmic time for $(M_5, f_H) = (M_{5,\rm min}, 0.05)$, shown for four different formation times ($t_i=10^{-x}\,t_c$, $x=\{2,3,4,5\}$). Earlier formation leads to a longer accretion phase and correspondingly relative larger final masses. The dashed line indicates the mass for which $r_0 \sim \ell$, marking the breakdown of the Schwarzschild--Tangherlini description.}
    \label{fig:mass-time}
\end{figure}

 In Figure~\ref{fig:mass-ranges}, we present the range of black hole masses if formation takes place during the quadratic regime for two representative values, $f_H = \{1, 0.1\}$. The parameter $f_H$ depends strongly on the formation mechanism (e.g.~in particle collisions, $f_H \ll 1$). In Figure~\ref{fig:mass-time}, we show the mass evolution as a function of cosmic time for $M_5=10^{6}\,\mathrm{TeV}$, obtained by using eq.~\eqref{Mt-result} for $t<t_c$ and eq.~\eqref{mass-ev-linear} for $t>t_c$, for four formation times with $f_H=0.05$. Earlier formation leads to a longer accretion phase, and hence significantly larger masses. To quantify the magnitude of this enhanced accretion, we consider the limiting case of the minimum observationally allowed five-dimensional Planck scale, with primordial black holes forming immediately after inflation: for $f_H = 0.05$, the initial mass $M_i \sim 10^{-11}\mathrm{g}$ increases to $M(t_c) \sim 10^{11}\mathrm{g}$.

\vspace{0.5cm}
\begin{mdframed}[style = mybbox]
Such an efficient accretion at $t \ll t_c$ implies that in formation scenarios such as primordial overdensity collapse, the relation between the primordial power spectrum and present-day black hole masses is significantly modified relative to the standard picture.
\end{mdframed}
\vspace{0.5cm}

%
%
\subsection{Comparison with previous results}\label{sec:comparing-results}
Note that we recover the same functional form as Guedens et al.~\cite{Guedens_2002-2} and Majumdar~\cite{Majumdar_2003}, although these authors adopt $0 < F \leq 1$ and $F = 1$, respectively. In Figure~\ref{fig:comparing-results}, we compare their results (shown in green) with our prediction (purple). The shaded green region represents the range of possible mass evolution functions associated with different values of $F$, interpreted in Ref.~\cite{Guedens_2002-2} as an accretion efficiency parameter. In contrast, in our treatment $F$ is fixed by the equation of state of the accreted fluid and decreases as the equation of state softens. We find larger values of $F$  which correspond to significantly enhanced accretion during the quadratic regime.

If we compare eq.~\eqref{Mt-result} with the estimates $M_{\scriptscriptstyle\mathrm{GM}}(t)$ obtained by Guedens et al.~\cite{Guedens_2002-2} and Majumdar~\cite{Majumdar_2003},
\begin{equation}
    \frac{M}{M_{\scriptscriptstyle\mathrm{GM}}}=\left(\frac{t}{t_i}\right)^{\frac{2}{\pi}\left(F(w)-F_{\scriptscriptstyle\mathrm{GM}}\right)}\,,
\end{equation}
we find that for the maximal accretion rate $F_{\scriptscriptstyle\mathrm{GM}}=1$, the discrepancy grows with cosmic time as $(t/t_i)^{0.67}$ and can reach several orders of magnitude by $t_c$, depending on the fundamental scale $M_5$ and formation time. 

\vspace{0.5cm}
\begin{mdframed}[style = mybbox]
 For example, for the particular case of the smallest fundamental scale allowed by observations, $M_5\sim10^{6}\,\mathrm{TeV}$, and the corresponding earliest formation time, we find a discrepancy of $(t_c/t_i)^{0.67}\sim 10^{11}$. For $F_{\rm prev}<1$, this effect is even more pronounced. 
\end{mdframed}
\vspace{0.5cm}

Let us also compare our result with the value $F=1.54$ obtained in Ref.~\cite{Tikhomirov:2005bt}, which likewise employs a fully relativistic framework. Their calculations for the critical radius, expressed in terms of entropy conservation and temperature, lead to the same results as ours. However, we disagree with their definition of the accretion rate, which is given in Ref.~\cite{Tikhomirov:2005bt} as $\dot{M}= 4\pi \rho(r) u(r) r^2$. This expression follows from the original Bondi treatment, where relativistic effects are neglected. In a relativistic framework with non-vanishing pressure, it represents a constant of integration only if $\rho$ denotes the proper rest-mass density and the chemical composition is frozen, so that baryon number is strictly conserved. In that case, the expression follows from conservation of the rest-mass current 
\begin{equation}
    \nabla_\mu J^\mu=0\,,\qquad J^\mu=\rho_0 u^\mu\,,
\end{equation} 
where we denote the proper rest-mass density by $\rho_0$ to avoid ambiguity.
However, even when the rest-mass current is conserved, the corresponding rest-mass flux does not necessarily quantify the total energy accreted by the black hole.\,\footnote{See, for example, Refs.~\cite{Babichev_2004,Babichev_2013}, where this distinction is discussed in the context of dark-energy accretion.}

In a fully relativistic treatment, the black-hole mass growth rate is instead obtained from conservation of the stress-energy tensor, with $A_1\equiv-\dot{M}$. The corresponding energy flux includes both internal energy and pressure through the $(\rho+p)$ term in eq.~\eqref{A1}, as well as the relativistic/gravitational factor $\sqrt{f+u^2}$. In the asymptotic regime, $\sqrt{f+u^2}\xrightarrow{r\to\infty}1$. For cold, non-relativistic matter, both the pressure and kinetic-energy contributions can be neglected in eq.~\eqref{A1}, so that $p=0$ and $\rho\simeq\rho_0$. Only in this particular limit do the two expressions coincide in the asymptotic regime. At smaller radial distances, or for relativistic or radiation-like fluids, however, the distinction becomes essential.

The discrepancy is therefore not merely terminological. It is crucial when translating a relativistic fluid solution into an actual black-hole mass-growth rate, and it becomes particularly important in the radiation-dominated regime, where the accreting medium carries a substantial amount of internal and radiation energy.

It is worth noting, however, that in Ref.~\cite{Tikhomirov:2005bt}, $\rho$ is interpreted throughout as the \textit{total} energy density. Consequently, in the regime where $\sqrt{f+u^2}\rightarrow1$, their expression is missing only a factor of $(1+w)$ relative to the appropriate expression for the black-hole mass growth rate. Including this factor restores agreement with the relativistic energy flux. The resulting flux is therefore larger than that quoted in Ref.~\cite{Tikhomirov:2005bt} by a factor of $4/3$ for the equation of state $w=1/3$. Accordingly, the corresponding value of $F$ is enhanced by the same factor, such that the value $F=1.54$ quoted in Ref.~\cite{Tikhomirov:2005bt} becomes $F=(4/3)\times1.54\simeq2.05$.

This difference results in 
\begin{equation}
    \frac{M}{M_{\scriptscriptstyle\mathrm{TT}}}\sim \left(\frac{t}{t_i}\right)^{0.33}
\end{equation}
for $w=1/3$. For the particular case of $M_5\sim10^{6}\,\mathrm{TeV}$ and its corresponding earliest formation time, we find $(t_c/t_i)^{0.33}\sim10^{5}$. 

\vspace{0.8cm}
\begin{mdframed}[style = mybbox]
Although the difference between the two results amounts only to a factor of $(1+w)$, this factor leads to a substantial difference in the resulting mass enhancement.
\end{mdframed}
\vspace{0.5cm}
\section{Validity of Michel's formalism in braneworlds}\label{section-validity}
Both the Bondi and Michel accretion formalisms rely on strong assumptions: an accreting fluid with negligible self-gravity, and a quasi-static flow embedded in an expanding cosmological background. Assessing the validity of these assumptions is therefore crucial in order to identify the region of parameter space in which the prescription applies. This issue was examined by Ricotti~\cite{Ricotti_2007} for Schwarzschild primordial black holes within the Bondi framework. Here, we extend this analysis to the RS-II scenario and to a fully relativistic treatment.
%
%
\subsection{Test fluid approximation}
In Sections~\ref{section-constants-integration} and~\ref{section-accretion}, we assumed that the spacetime geometry is given by the Schwarzschild--Tangherlini solution, thereby neglecting the gravitational backreaction of the surrounding fluid.

Although the energy density profile becomes more dilute in the presence of a warped extra dimension, its contribution to the geometry is governed not only by the stress-energy tensor, but also by the additional terms $S_{\mu\nu}$ and $\mathcal{E}_{\mu\nu}$. To assess the validity of this approximation, we introduce an effective mass function --- following Ref.~\cite{Casadio_2005} in the context of gravitational collapse --- to quantify the enclosed gravitational energy ($r_0<r<r_{\rm cr}$),
\begin{equation}\label{eq-meff}
    M_{\rm eff}=4\pi \int_{r_0}^{r_{\rm cr}} \rho_{\rm eff}(r)\,r^2\dd r\,,
\end{equation}
where we have integrated over the spatial volume $\sqrt{h}\,\dd r\dd \theta\dd\phi$, and compare it to that of the black hole's Weyl cloud,
\begin{equation}\label{eq-mbh}
    M_{\rm \mathcal{E}}=4\pi\int_{r_0}^{r_{\rm cr}}\rho_{\mathcal{E}}^{\scriptscriptstyle\mathrm{BH}}\,r^2\dd r\,,
\end{equation}
with $\rho_{\mathcal{E}}^{\scriptscriptstyle\mathrm{BH}}$ provided by~\eqref{rho-E}. If $|M_{\rm eff}|\ll| M_{\mathcal{E}}|$, the perturbation of the metric induced by the fluid can be neglected, and the procedure used above to compute the accretion rate is justified.

To assess the validity of the test-fluid approximation, one must take into account all gravitational sources that affect the metric in the vicinity of the black hole. We emphasise once more that $\rho_{\rm eff}$ and $\rho_{\mathcal{E}}^{\scriptscriptstyle\mathrm{BH}}$ do not correspond to physical fluids. Rather, they are defined by identifying the relevant contributions to the $00$ component of the effective Einstein equations with an energy-density-like quantity. This is done purely to make the parallel with the conventional cosmological Einstein equations transparent, and to facilitate comparison with the existing literature, and it does not imply the presence of an additional material fluid. In particular, $\rho_{\rm eff}$ contains contributions from both the physical cosmological fluid, with well-defined microscopic properties, and the gravitational backreaction of the bulk on the brane geometry. Similarly, $\rho_{\mathcal{E}}^{\scriptscriptstyle\mathrm{BH}}$ describes the backreaction of the black hole's bulk gravitational field on the brane and should not be interpreted as a physical fluid surrounding the black hole.

To determine $\rho_{\rm eff}(r)$, we consider a spherically symmetric energy density profile that is approximately decoupled from the cosmic expansion, i.e.\ $\dot{\rho}(r) = 0$, and that asymptotically matches the FLRW background. If the static approximation holds, we may use the non-local integration constant~\eqref{A3} and evaluate it in the asymptotic limit during the quadratic regime. This yields
\begin{equation}\label{rho-E}
\rho_\mathcal{E}(r)=  \frac{3(1+w)}{4\lambda}\left( \rho_\infty-\rho(r)\right)\rho(r)\,.
\end{equation} 
This corresponds to the energy density profile of dark radiation sourced by a spherically symmetric distribution of brane matter in the quadratic regime.\,\footnote{This expression does not account for contributions sourced by bulk matter. Throughout, we assume an empty bulk, so that dark radiation is generated exclusively as a backreaction of gradients in the brane energy density.}

\begin{figure}
    \centering
    \includegraphics[width=0.78\linewidth]{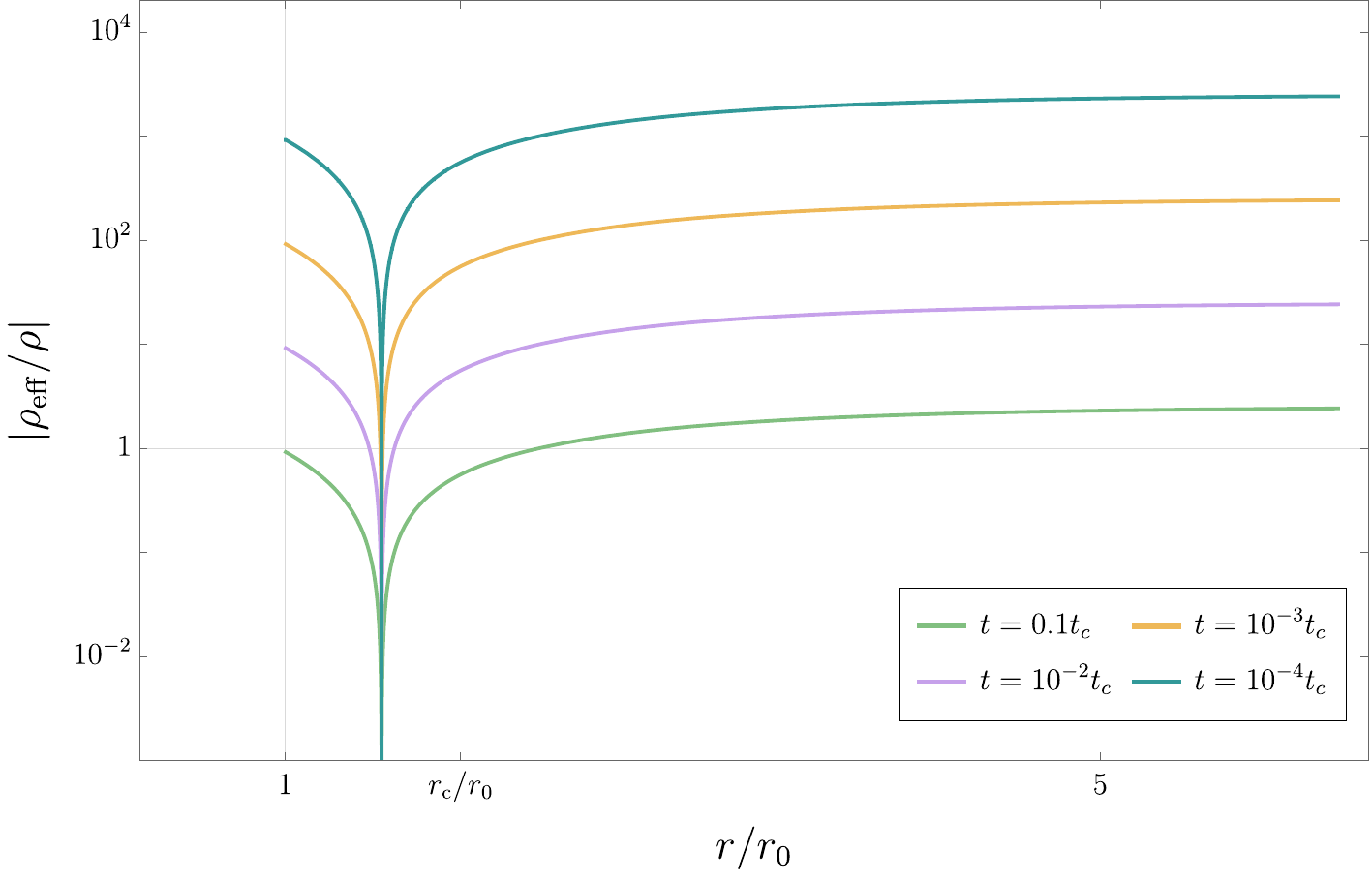}
    \caption[Effective vs.~brane energy density near the horizon]{Absolute value of the ratio of the effective energy density and the brane's fluid energy density in the vicinity of a five-dimensional black hole with horizon radius $r_0$ at different cosmic times.}
    \label{fig:ratio-rhos}
\end{figure}

Therefore, the effective energy density~\eqref{rho-eff} in this region takes the form 
\begin{equation}\label{rho-eff-1}
    \rho_{\rm eff}\simeq \frac{\rho(r)}{\lambda}\left(\rho_\infty-\frac{\rho(r)}{2}\right)
\end{equation}
for $p=\rho/3$ and $\rho_\infty\gg\lambda$. In Figure~\ref{fig:ratio-rhos} one can see the absolute ratio $|\rho_{\rm eff}/\rho|$ for different cosmic times. Allow us to highlight two key features:
\begin{itemize}
    \item The ratio depends strongly on cosmic time, with the contributions from $S_{\mu\nu}$ and $\mathcal{E}_{\mu\nu}$ becoming significant at early times.
    \item The quadratic correction dominates in the asymptotic limit $r \to \infty$. As the density profile steepens, the negative contribution from $\rho_{\mathcal{E}}$ becomes dominant, leading to $\rho_{\rm eff} < 0$, with large negative values $|\rho_{\rm eff}|/\rho \sim t_c/t$.
\end{itemize}
Evaluating the integrals~\eqref{eq-meff} and~\eqref{eq-mbh}, we obtain an estimate of the ratio
\begin{equation}\label{epsilon-M}
   \epsilon_M= \Bigg|\frac{M_{\rm eff}}{M_{\mathcal{E}}}\Bigg|\simeq 0.2\,\left(\frac{r_0}{H^{-1}}\right)^2\,.
\end{equation}
Together with eq.~\eqref{r0-H-1}, this justifies neglecting the backreaction of the surrounding fluid at all cosmic times.

\subsection{Staticity}
In Sections~\ref{section-constants-integration} and~\ref{section-accretion}, we assumed a quasi-static flow, such that the metric, energy densities and four-velocity are time-independent. This approximation is valid when the region near the black hole effectively decouples from the cosmological expansion. We now assess whether this assumption holds in the case under study. To this end, we introduce the parameter
\begin{equation}\label{epsilont}
    \epsilon_t=\frac{t_{\rm acc}}{H^{-1}}
\end{equation}
that allows us to compare the time scales of the system. In~\eqref{epsilont}, $t_{\rm acc}$ is the time scale of accretion, which should be of order
\begin{equation}
    t_{\rm acc}\sim \frac{r_{\rm cr}}{c_{s}}\,,
\end{equation}
since $r_{\rm cr}$ is the scale of the system and $c_s$ is the maximum speed of propagation of information. For this particular case, 
\begin{equation}\label{epsilon-t}
    \epsilon_t=\sqrt{6}\left(\frac{r_0}{H^{-1}}\right)\,.
\end{equation}
Following eq.~\eqref{r0-H-1}, we can conclude that $\epsilon_t<1$ shortly after formation, depending on the parameter $f_H$.

\begin{figure}
    \centering
    \includegraphics[width=0.65\linewidth]{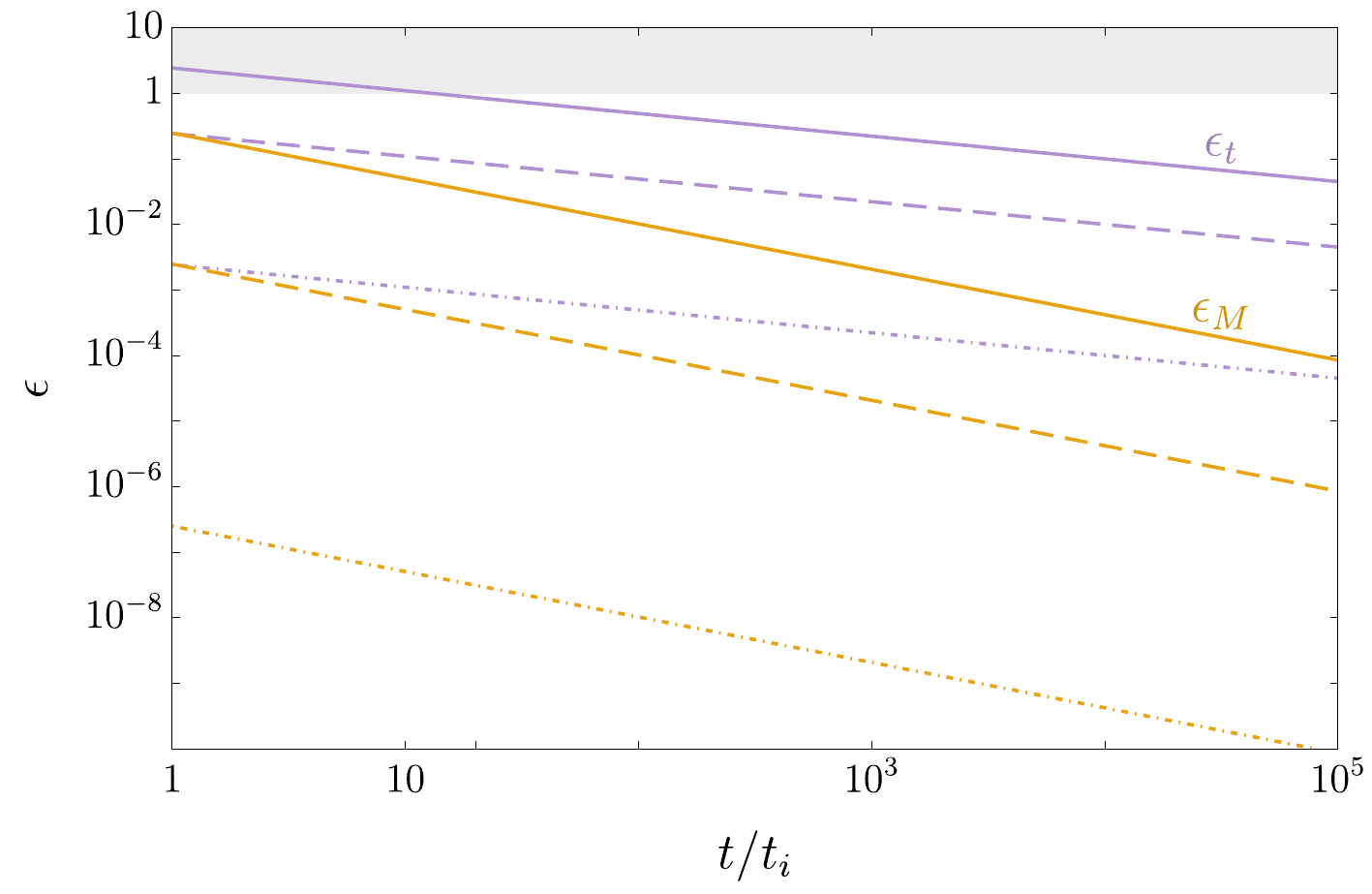}
    \caption[Staticity and self-gravity parameters]{Normalised cosmic time evolution of the staticity and self-gravity parameters, $\epsilon_t$ (in purple) and $\epsilon_M$ (in orange), respectively. for different $f_H$ values ($f_H=1$ solid, $f_H=10^{-1}$ dashed, $f_H=10^{-3}$ dot-dashed). For $\epsilon_t\lesssim 0.1$ and $\epsilon_M\lesssim0.1$, Michel's prescription provides a good approximation.}
    \label{fig:epsilons}
\end{figure}

As shown in Figure~\ref{fig:epsilons}, which tracks the evolution of $\epsilon_{M}$ and $\epsilon_{t}$, $\epsilon_{t}$ exceeds unity at early times for $f_H\simeq1$. Thus, when black holes form with event radii comparable to the cosmological horizon, the Michel approximation may not describe the immediate post-formation accretion accurately. For this reason, in Figure~\ref{fig:mass-time} we show the mass evolution for black holes forming with $f_H=0.05$, for which the Michel formalism provides a reliable description of the accretion throughout the evolution considered. A quantitative treatment of the early-time accretion for $f_H\simeq1$ would require relativistic hydrodynamical simulations of the coupled evolution of the collapsing fluid and newly formed black hole. In four-dimensional general relativity, the corresponding early-time evolution has been studied for radiation~\cite{Niemeyer_1999}, a stiff fluid~\cite{Harada_2005}, and matter domination~\cite{de_Jong_2022}. The first two cases find limited post-formation accretion, whereas the matter-dominated case may exhibit rapid mass growth.

Since no analogous analysis exists for RS-II, the early-time accretion history of PBHs with horizon radii comparable to the cosmological horizon remains undetermined.\,\footnote{Without a complete study of black hole collapse in RS-II, it also cannot be established whether the formation scenario $f_H=1$ is physically realised.} Several possibilities can be envisaged. If accretion is suppressed during the initial non-quasi-static phase, the black hole horizon quickly becomes smaller than the cosmological horizon and $\epsilon_t$ falls into the regime where the Michel framework applies. The mass at the crossover would then be significantly below that obtained by extrapolating the quasi-static Michel evolution back to formation, both because the period of strong accretion is shorter and because the mass-growth rate is steeper at earlier times. Conversely, more efficient non-quasi-static accretion would cause our treatment to underestimate the crossover mass; sufficiently rapid horizon growth could even keep the quasi-static approximation invalid through a substantial fraction, or the entirety, of the high-energy quadratic regime.

These uncertainties, however, do not permit small primordial black holes to depart from the five-dimensional geometry, as argued in \href{part:higher-d}{\textbf{\textcolor{thesislilac}{Part II}}}. Even with very strong accretion, causality limits the cosmological matter incorporated during formation, preventing this transition from occurring during the quadratic regime.

    \chapter[Energy injection into the cosmological background]{Energy injection into\\ the cosmological background}\label{chap11:energy-injection}
    \thispagestyle{fancy}
\pagestyle{fancy}

    Having established the evolution of primordial black holes in the early Universe, we now turn to their cosmological and astrophysical imprint at later times. We focus on primordial black holes that are still evaporating today, continuously injecting energy into the Universe through Hawking radiation. The particles produced interact with the cosmological medium and contribute to a range of observable signals: modifications to the ionisation and thermal history of the Universe, distortions of cosmological backgrounds, and diffuse astrophysical radiation. Confronting these predicted signatures with observations provides a powerful means of constraining the abundance and mass spectrum of a primordial black hole population.

Early studies of RS-II primordial black hole phenomenology were based largely on analytic estimates of the evaporation process and established the qualitative framework for understanding their observational signatures~\cite{Guedens_2002-2,Clancy:2003zd,Majumdar_2003,Majumdar:2003sf,Majumdar:2004mz,Majumdar:2005ba,Inoue:2003di,Sendouda:2003dc,Sendouda:2004hz,Tikhomirov:2005bt,Eiroa:2005vd,Keeton:2006di,Sendouda:2006nu,Sendouda:2006yc}. We revisit the problem using modern numerical tools and updated datasets. We compute the emission spectra of a range of particle species from evaporating braneworld black holes, model the subsequent deposition of this energy into the cosmological background, and compare the resulting signals with current observational constraints. We make use of publicly available computational frameworks, including \textsc{BlackMax}, \textsc{CosmoLed} and \textsc{ExoClass}, which allow a more detailed treatment of particle emission, energy injection and cosmological evolution than was previously possible. Together with the substantial improvements in cosmological and astrophysical observations achieved in recent years, this enables a significantly more robust assessment of the surviving parameter space.

\section{Working framework}
Before turning to the specifics, we fix the framework in which our calculations will take place. This section collects the choices that
define our scenario --- and the reasoning behind them --- together with the quantities through which we will track the PBH population throughout the evolution of the Universe. 
\subsubsection*{Monochromaticity}
First, we comment on the mass spectrum. While most conventional PBH formation scenarios motivate an approximately monochromatic spectrum, we take it to be exactly monochromatic to minimise parameters and facilitate a clearer conceptual analysis. Note, however, that the PBH dark matter fraction $f_{\scriptscriptstyle\mathrm{PBH}}$ is constrained more strongly for extended spectra in the case of standard (3+1)-dimensional Hawking radiation~\cite{Carr:2017jsa,Gorton_2024}.
This is a result of $f_{\scriptscriptstyle\mathrm{PBH}}$ scaling steeply with the black hole mass on the low mass end of the asteroid mass gap, approximately following $f_{\scriptscriptstyle\mathrm{PBH}}\sim M^3$. Even for PBHs originating from a narrow peak in the primordial power spectrum, the asteroid-mass window tightens by roughly half an order of magnitude in terms of the minimum mass~\cite{Gorton_2024}. For five-dimensional black holes, the low-mass tail scaling is less steep but still prominent, close to $\sim M^2$. Therefore, we expect the asteroid mass window to shrink when going beyond the assumption of a monochromatic mass function, albeit by a smaller amount in braneworld models than in the conventional scenario. 
\subsubsection*{Formation time and black hole metrics}
The formation time deserves particular attention. In
\hyperref[part:braneworlds]{\textbf{\textcolor{thesislilac}{Part~I}}}
we introduced the crossover time $t_c$ at which the Hubble horizon
satisfies $H^{-1} \sim \ell$. A black hole formed at $t < t_c$ is
therefore born in the ``small'' regime, and it must remain there until
the crossover: irrespective of the unknown details of collapse and
accretion, its event horizon must fit within the cosmological horizon,
so its radius is bounded by $H^{-1} < \ell$ at all times before
$t_c$. Accretion cannot circumvent this causal bound, and our analysis
in Chapter~\ref{chap10:accretion} confirms that it falls far short of
it in practice. Black holes forming after the crossover, by contrast,
may be four- or five-dimensional depending on the formation process.

Since we are only considering monochromatic mass spectra, we are thus left with two distinct possibilities: a PBH
population of small black holes, or one of large black holes.

\par\noindent\ul{Initial conditions.} We set initial conditions corresponding to a specified mass and initial time $t\geq t_c$, and track their subsequent evolution. This is sufficient for the purposes of the analysis in this work, but would be inadequate if we wished to relate PBH masses to specific $k$-modes in the primordial power spectrum.
\subsubsection*{Mass evolution}
For both ``large'' and ``small'' black holes, we take their mass to be approximately constant throughout their lifetime.
In both cases, accretion is expected to enhance
their masses only by an $\mathcal{O}(1)$ factor for $t>t_c$, as discussed in Chapter~\ref{chap10:accretion}. Given other uncertainties in the calculation, this barely
affects estimates of their lifetime.

We also neglect mass loss
due to emission of Hawking quanta, except in the final
stages of evaporation. This is a standard
approximation for large black holes.
Let us consider why it is also a reasonable approximation for most PBH masses in the small black hole limit.
For a fixed mass, the evaporation lifetime follows~\eqref{lifetime-5d}, i.e.~$t_{\scriptscriptstyle\mathrm{evap}}$ satisfies the proportionality relation
\begin{equation}
    t_{\scriptscriptstyle\mathrm{evap}} =C M_i^2\,,
\end{equation}
with $C$ an approximate constant, assuming that the number of light degrees of freedom available for Hawking emission does not vary significantly over the lifetime of the black hole.
We define $M_*$ as the black hole mass whose lifetime
is equal to the current age of the Universe, i.e.~$t_0=CM_{*}^2$.
Now, consider a black hole with initial mass larger than $M_*$,
i.e., $M_i=\psi M_*$, with $\psi>1$.
The mass of this black hole today $M_0$ satisfies
\begin{equation}
    t_0=C(M_i^2-M_0^2)\,,
\end{equation}
and therefore
\begin{equation}
    M_0=M_*\sqrt{\psi^2-1}\,.
\end{equation}
We conclude that the change in mass, due to evaporation up to the present day for a black hole with initial mass $M_i\geq 3M_*$,
is less than about $6\%$. In what follows, we assume a constant mass for all black holes with $M_i\geq 3M_*$.

\par\noindent\ul{Transition to ``small'' regime.}
Note that as large black holes evaporate,
their surface area shrinks.
Eventually, the horizon radius $r_0$ becomes comparable to the AdS
radius $\ell$. In this regime, the geometry is expected to
transition between the ``large'' and ``small''
regimes. In the absence of an exact brane-black-hole solution describing
this intermediate configuration and its time dependence,
the detailed evaporation law and the approach to a
spherically symmetric five-dimensional configuration
remain uncertain. It is reasonable to expect,
however, that an approximate interpolation between the
``large'' and ``small''
black hole phases should yield suitable results.
Non-exact numerical solutions (e.g.~\cite{Figueras_2011},~\cite{Tanahashi_2008}
and~\cite{Biggs_2022}) should also be good enough for a PBH lifetime modeller.

In a cosmological context, for ``large'' black holes to have evaporated sufficiently
to find themselves at this crossover today, they must have formed shortly after $t_c$ and the AdS radius must be very small ($l\lesssim 10^{-14}$m).
\section{Population abundance}
For a monochromatic mass function, the PBH population is described by
a single number density $n_{\scriptscriptstyle\mathrm{PBH}}$, which dilutes like matter,
\begin{equation}
    n_{\scriptscriptstyle\mathrm{PBH}} \propto a^{-3}\,.
\end{equation}
Its present value is related to the dark
matter fraction $f_{\scriptscriptstyle\mathrm{PBH}}$ by
\begin{equation}
    n_{\scriptscriptstyle\mathrm{PBH}} = \frac{f_{\scriptscriptstyle\mathrm{PBH}}\,\rho_{{\scriptscriptstyle\mathrm{DM}},0}}{M}\,,
    \label{eq:npbh-fpbh}
\end{equation}
where $\rho_{{_{\scriptscriptstyle\mathrm{DM}},0}}$ is the dark matter energy density today.

As we learnt in Chapter~\ref{chap6:hawking-evap}, the emission spectrum of a five-dimensional black hole depends solely on its temperature. Constraints derived from evaporation signals therefore limit the combination $(n_{\scriptscriptstyle\mathrm{PBH}}, T_{\scriptscriptstyle\mathrm{BH}})$: at fixed
temperature, they bound the number density, irrespective of the mass that realises that temperature. This has a useful consequence.
Equations~\eqref{st-temp} and~\eqref{horizonradius} show that, at
fixed $T_{\scriptscriptstyle\mathrm{BH}}$ and $M_4$, the black hole mass scales as
$M \propto \ell^{-1/2}$: the mass required to achieve a given
temperature decreases as the AdS radius grows. Through
Eq.~\eqref{eq:npbh-fpbh}, a bound on $n_{\scriptscriptstyle\mathrm{PBH}}$ at fixed $T_{\scriptscriptstyle\mathrm{BH}}$ then translates into a bound on $f_{\scriptscriptstyle\mathrm{PBH}}$ proportional to
$M$. Hence, comparing two AdS radii $\ell_1$ and $\ell_2$ at equal
temperature --- that is, at masses $M(T_{\scriptscriptstyle\mathrm{BH}},\ell_1)$ and $M(T_{\scriptscriptstyle\mathrm{BH}},\ell_2)$ ---
the corresponding bounds are related by
\begin{equation}
    f_{\scriptscriptstyle\mathrm{PBH}}(\ell_1)
    = \frac{M(T_{\scriptscriptstyle\mathrm{BH}},\ell_1)}{M(_{\scriptscriptstyle\mathrm{BH}},\ell_2)}\, f_{\scriptscriptstyle\mathrm{PBH}}(\ell_2)
    = \left(\frac{\ell_2}{\ell_1}\right)^{1/2} f_{\scriptscriptstyle\mathrm{PBH}}(\ell_2)\,,
\end{equation}
so evaporation constraints on $f_{\scriptscriptstyle\mathrm{PBH}}$ become \emph{stronger} at
larger AdS radii. This scaling, while not readily apparent in the
usual $f_{\scriptscriptstyle\mathrm{PBH}}$--$M$ plots, provides a direct translation of the
bounds between different values of $\ell$.
\section{Electromagnetic radiation}\label{section-EM-radiation}
\subsection{Production mechanisms}
\label{sec:em-production}
The total photon flux stemming from Hawking evaporation has
contributions from several production channels.
Direct emission is always present.
Secondary emission can occur
as a result of in-vacuo QED processes, involving
emitted quanta,
that produce final state photons.
Alternatively, it may occur due to
interactions between emitted quanta
and a medium.
Examples include
(non-)relativistic positron annihilation,
positronium formation and decay,
in-medium bremsstrahlung
and inverse Compton scattering.

For each process we require the instantaneous rate
for
Hawking emission
of the particles involved, obtained
from the emission rates~\eqref{emission-rate}.
In the following sections,
we discuss the channels that contribute most significantly
to the total photon flux.
We do not include processes with negligible impact.
For instance, secondary emission via hadron and gauge boson decay
is taken to be negligible. This is because primordial black holes
in the parameter space under consideration have temperatures
significantly lower than the QCD scale.
Therefore, hadron and gauge boson emission are expected to be
highly suppressed.%
    \footnote{In $(3+1)$-dimensional spacetime,
    the black hole temperature at which secondary photons
    start to dominate is 50 MeV~\cite{Carr_2021},
    which is around $\Lambda_{\scriptscriptstyle\mathrm{QCD}}/6$.
    We cannot know exactly when this transition occurs
    in RS-II without comparing the primary and secondary spectra
    to find the temperature at which the flux is comparable.
    However, we expect
    it should not differ by orders of magnitude.}
    
In Ref.~\cite{longair1992high,Strong_1998,lahav2004cosmologicalparameters} it was shown that,
for positrons with $E_{e^+}\lesssim 100\,\text{MeV}$,
in-medium bremsstrahlung and inverse Compton energy emission
can be neglected. Except for the final bursts in their evaporation,
braneworld black holes are generally too cold for these
production channels to be significant.

The total flux can be approximated
\begin{equation}\label{sum-prods}
    \frac{\dd\dot{N}_\gamma^{\scriptscriptstyle\mathrm{tot}}}{\dd E_{\gamma}}\approx  \frac{\dd \dot{N}^{\scriptscriptstyle\mathrm{P}}_\gamma}{\dd E_\gamma}+\frac{\dd \dot{N}^{\scriptscriptstyle\mathrm{FSR}}_\gamma}{\dd E_\gamma}+\frac{\dd\dot{N}^{\scriptscriptstyle\mathrm{IA}}_\gamma}{\dd E_\gamma }+\frac{\dd\dot{N}^{\scriptscriptstyle\mathrm{NR}}_\gamma}{\dd E_\gamma}\,,
\end{equation}
where ``FSR''
includes final state radiation from all relevant charged
particles emitted by the black hole;
``IA'' stands for inflight annihilation of relativistic positrons;
and ``NR'' includes all non-relativistic processes that take place
after positrons thermalise with the interstellar medium (ISM).

\begin{figure}
    \centering
    \includegraphics[width=0.58\linewidth]{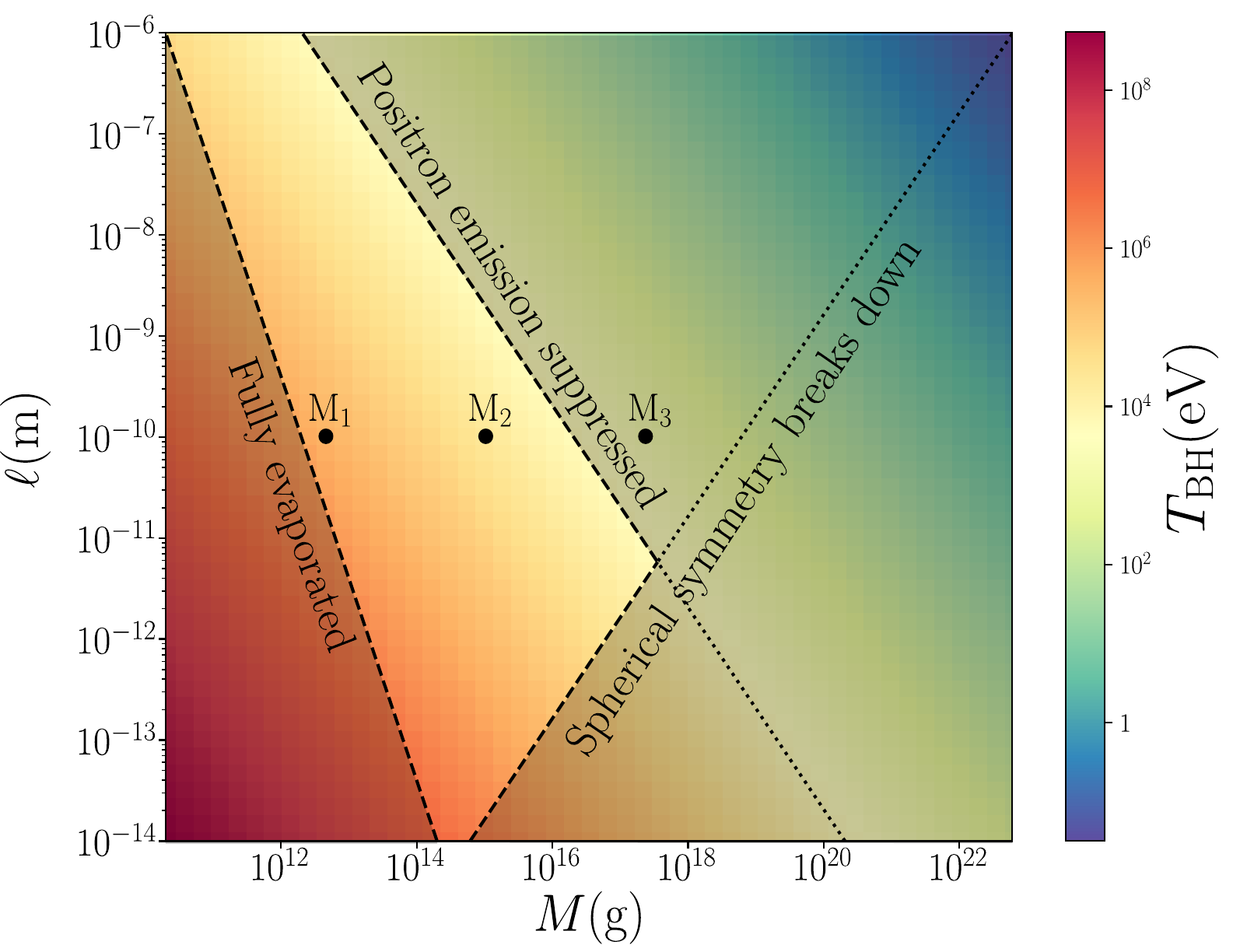}
    \caption[Positron emission vs.\ black hole temperature]{Dependence of positron emission on the black hole temperature. In the unshaded region of the parameter space $(M,\ell)$ positron/electron (or 511~keV photon) emission is favoured. The masses $M_1$, $M_2$ and $M_3$ for AdS radius $10^{-10}$m are used in Section~\ref{photon-distribution} for the study of the extragalactic photon flux, see Figure~\ref{EGB-plots}.}
    \label{emission-allowed}
\end{figure}

Since some of the channels listed above require the emission of
massive particles, their contribution to the total flux depends
strongly on the temperature of the black hole. The range of masses
where $e^\pm$ emission is not suppressed is shown in
Figure~\ref{emission-allowed}.
\subsubsection*{Direct Hawking emission}
The reduced temperature of higher-dimensional black holes, in contrast to the four-dimensional case, leads directly to a displacement of the photon emission spectrum toward lower energies. As discussed in earlier studies~\cite{Friedlander:2022ttk,Johnson_2020}, this implies that $\gamma$-ray constraints derived for standard PBHs are translated to substantially lower masses in higher-dimensional geometries.

\begin{figure}
    \centering
    \includegraphics[width=0.8\textwidth]{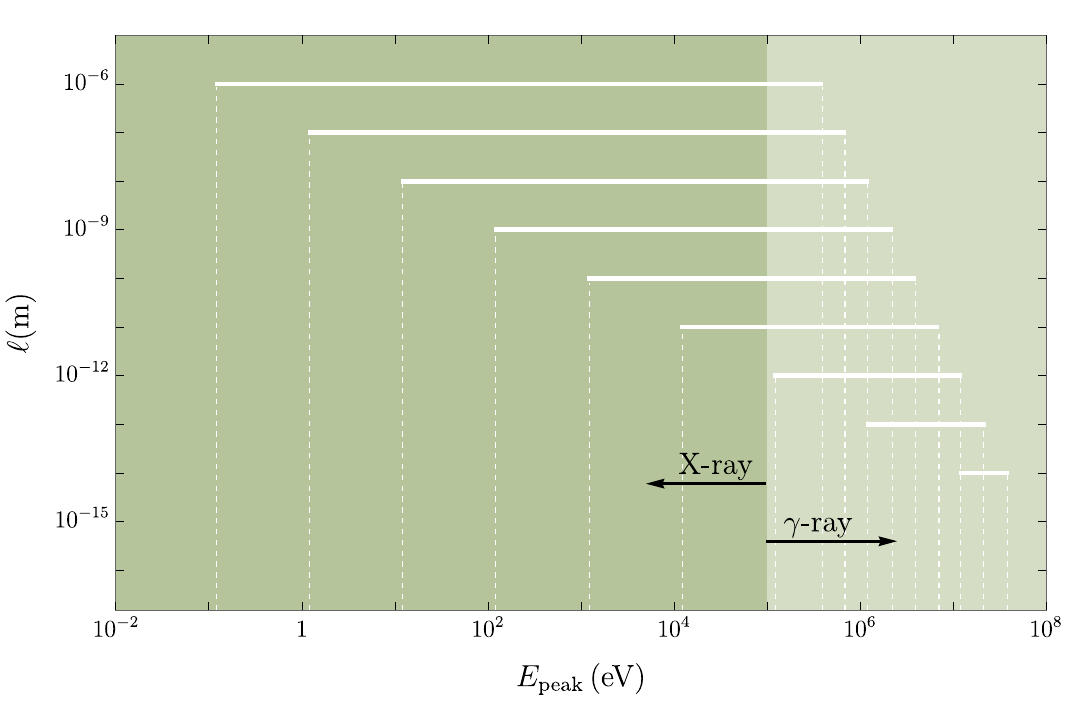}
    \caption[Peak emission energy of photons of 5D PBH dark matter candidates]{Each line covers the range of $E_{\scriptscriptstyle\mathrm{peak}}$ of five-dimensional PBHs that are DM candidates in RS-II, i.e.~ranging from the lightest mass ($M\sim3M_*$) to largest possible Schwarzschild--Tangherlini radius ($r_0(M)<\ell$) for different AdS radius values.}
    \label{fig-Epeak}
\end{figure}

As can be seen in Figure~\ref{fig-Epeak},
for larger values of the AdS radius $\ell$,
there is typically a wide range of masses where the spectral
peak lies in the X-ray band.
This suggests that to
distinguish signatures of PBHs
from direct photon production, it may be more profitable
to focus more on X-ray rather than $\gamma$-ray
observations.
\subsubsection*{Final state radiation}
High-energy charged particles emitted as Hawking quanta
are expected to radiate on-shell photons via
in-vacuo QED effects.
Radiative corrections predict photon emission shortly
after charged particles are produced. In the literature,
this is described
as final state radiation (FSR).
The process is described by the
Dokshitzer--Gribov--Lipatov--Altarelli--Parisi (DGLAP)
splitting functions~\cite{Dokshitzer:1977sg,Gribov:1972ri,Altarelli:1977zs}.
These represent
the differential probability densities for a charged particle
with energy $E_i$ to emit a photon with energy $E_\gamma$.
To leading order~\cite{Chen_2017},
\begin{equation}
    P(i\rightarrow i\gamma)=
    \left\{
    \begin{array}{ll}
        \displaystyle
        \frac{2(1-x)}{x} & \text{(bosons)} \\[2ex]
        \displaystyle
        \frac{1+(1-x)^2}{x} & \text{(fermions)}
    \end{array}
    \right.
\end{equation}
with $x\equiv 2E_\gamma/Q_i$, where $Q_i=2E_i$ is the energy scale. The FSR photon spectra per particle $i$ is given by
\begin{equation}
    f_{\scriptscriptstyle\mathrm{FSR}}=\frac{\alpha_{\scriptscriptstyle\mathrm{EW}}}{\pi Q_i}\left(\ln\frac{1-x}{\mu_i^2}-1\right)P(i\rightarrow i\gamma)\,,
\end{equation}
with $\mu_i\equiv m_i/Q_i$. 
This probabilistic photon emission process has to be applied to each emitted particle,
\begin{equation}\label{FSR-em}
\frac{\dd \dot{N}_\gamma ^{\scriptscriptstyle\mathrm{FSR}}}{\dd E_\gamma}=\sum_i\int_{m_i+E_\gamma}^\infty \dd E_i\frac{\dd \dot{N}_i}{\dd E_i}f_{\scriptscriptstyle\mathrm{FSR}}(E_i,m_i,E_\gamma)\,,
\end{equation}
which results in a photon flux with a
broad spectrum of frequencies $E_\gamma$.
This flux is
particularly significant at lower energies.
In our analysis, we include $e^\pm$ and $\mu^\pm$ but neglect other fermions and all bosons due to their large masses, which suppress emission. 

For charged particles,
final state radiation is expected to yield
a modest shift to lower energies in the spectrum.
Since it is produced promptly after emission of the Hawking quanta,
it alters the initial conditions for all in-medium interactions.
However, the energy loss in the positron flux is expected to be only a
few percent (see for example Ref.~\cite{Plestid_2025}),
even for highly energetic initial states.
Therefore, although we accurately track photons
produced by this mechanism,
for simplicity
we continue to use the undepleted,
initial spectrum
for the positrons themselves.
\subsubsection*{$\boldsymbol{e^-e^+}$ inflight annihilation}
Charged particles created via Hawking evaporation are expected to lose energy via in-medium interactions after emission. For positrons with $E_{e^+}\leq 1\,\text{GeV}$ travelling through the ISM, ionisation and excitation are the dominant energy-loss mechanisms. This energy loss is quantified via the relativistic Bethe--Bloch formula~\cite{Bethe:1930ku,Bethe:1932cbh,segre1953experimental}. To leading order, this can be approximated~\cite{Beacom_2006}
\begin{equation}\label{Bethe-Bloch}
    \bigg|\frac{\dd E}{\dd x}\bigg|\approx \frac{7.6\times 10^{-26}}{v^2}\left(\frac{n_H}{0.1\,\text{cm}^{-3}}\right)\left(\ln\gamma+6.6\right)\,\frac{\text{MeV}}{\text{cm}}\,,
\end{equation}
where $v$ is the velocity of the incident positron, $\gamma=1/(1-v^2)^{1/2}$ is the Lorentz factor and $n_H$ is the number density of neutral hydrogen in the medium.%
    \footnote{Reducing the complex ISM to pure neutral hydrogen
    is a reasonable approximation as long as the positrons have
    energies $E_{e^+}<1\,\text{GeV}$,
    since ionisation and excitation of neutral particles dominates over all other interactions~\cite{Jean_2009}.}

For sufficiently small initial kinetic energy, most positrons reach rest
via Bethe--Bloch emission~\eqref{Bethe-Bloch},
and then annihilate or form states with bound or free electrons of the ISM. However, for energies larger than 10 keV, a significant fraction of positrons annihilate inflight before reaching rest.
We calculate the probability of survival for a positron travelling through the ISM with initial energy $E_0$ and final energy $E$ using
\begin{equation}\label{prob-survival}
    P_{E_0\rightarrow E}=\exp \left(-\,n_{e^-}\int^{E_0}_E \frac{\sigma_\text{ann}(E')\dd E'}{\left|\dd E'/\dd x\right|}\right)\,,
\end{equation}
where $\sigma_\text{ann}(E)$ is the annihilation cross-section for a positron of energy $E$ with an electron approximately at rest,%
    \footnote{Even among the hottest ISM environments
    (e.g., supernova remnants), thermal electrons rarely
    reach speeds exceeding a few percent of MeV-scale
    positrons~\cite{Hughes_2000},
    so this approximation is valid.}
and $n_{e^-}=n_H$ for neutral hydrogen.
We take the tree-level cross-section~\cite{jauch1976theory}
\begin{equation}
    \sigma_{\scriptscriptstyle\mathrm{ann}}(E_{\scriptscriptstyle\mathrm{IA}})=\frac{\pi\, r_e^2}{\gamma_{\scriptscriptstyle\mathrm{IA}}+1}\left[\frac{\gamma_{\scriptscriptstyle\mathrm{IA}} ^2+4\,\gamma_{\scriptscriptstyle\mathrm{IA}}+1}{\gamma^2-1}\ln\left(\gamma_{\scriptscriptstyle\mathrm{IA}}+\sqrt{\gamma_{\scriptscriptstyle\mathrm{IA}}^2-1}\right)-\frac{\gamma_{\scriptscriptstyle\mathrm{IA}}+3}{\sqrt{\gamma_{\scriptscriptstyle\mathrm{IA}}^2-1}}
    \right]
\end{equation}
where $\gamma_{\scriptscriptstyle\mathrm{IA}}=E_{\scriptscriptstyle\mathrm{IA}}/m_e$, with $E_{\scriptscriptstyle\mathrm{IA}}$ being the energy the positron has at the time of annihilation, and $r_e$ is the classical electron radius. 
\\In order to calculate the photon spectrum from the annihilation process, we require the differential cross-section of $e^+e^-\rightarrow 2\gamma$~\cite{jauch1976theory},
\begin{equation}\label{diff-cross}
    \frac{\dd \sigma_{e^+e^-\rightarrow 2\gamma}}{\dd E_\gamma}=\frac{\pi\, r_e^2}{m_e\,\gamma_{\scriptscriptstyle\mathrm{IA}}\,v_{\scriptscriptstyle\mathrm{IA}}}\left[\frac{\left(\gamma_{\scriptscriptstyle\mathrm{IA}}+1\right)^2+\left(\gamma_{\scriptscriptstyle\mathrm{IA}}-1\right)^2-2\left(\gamma_{\scriptscriptstyle\mathrm{IA}}^2-1\right)E_\gamma/m_e}{\left(E_\gamma/m_e\right)^2\left(\gamma_{\scriptscriptstyle\mathrm{IA}}+1-E_\gamma/m_e\right)^2}
    \right]\,,
\end{equation}
where $E_\gamma$ is the photon energy in the lab frame ($e^-$ at rest) and lies in the range
\begin{equation}\label{energy-bounds}
    \frac{m_e}{2}\left(\gamma_{\scriptscriptstyle\mathrm{IA}}-1\right)\leq E_\gamma<\frac{m_e}{2}\left(\gamma_{\scriptscriptstyle\mathrm{IA}}+1\right)\,.
\end{equation}
With eq.~\eqref{diff-cross} one can calculate the 
distribution function of the emitted photons~\cite{Beacom_2006},
\begin{equation}\label{eq-fIA}
    f_{\scriptscriptstyle\mathrm{IA}}(E_\gamma,E_0)=n_e\int^{E_0}_{E_{\scriptscriptstyle\mathrm{min}}}\frac{\dd \sigma}{\dd E_{\gamma}}(E_\gamma,E_{\scriptscriptstyle\mathrm{IA}})\frac{P_{E_0 \rightarrow E_{\scriptscriptstyle\mathrm{IA}}}}{|\dd E/\dd x|(E_{\scriptscriptstyle\mathrm{IA}})}\dd E_{\scriptscriptstyle\mathrm{IA}}\,,
\end{equation}
where $E_{\scriptscriptstyle\mathrm{min}}$ is determined by~\eqref{energy-bounds}. Equation~\eqref{eq-fIA} describes a continuous spectrum of photons strongly shifted via the Doppler effect. If we use the notation $P_{E_0\rightarrow m_e} \equiv P(E_0)$, the percentage of the total positron flux that engages in inflight annihilation is $1-P(E_0)$. Therefore,
\begin{equation}\label{IA-em}
    \frac{\dd \dot{N}_{\scriptscriptstyle\mathrm{IA}}}{\dd E_{\gamma}}=\int_{m_e}^\infty\dd E_0 \frac{\dd \dot{N}_{e^+}(E_0)}{\dd E_{e^+}}(1-P(E_0))\,f_{\scriptscriptstyle\mathrm{IA}}(E_\gamma,E_0)\,.
\end{equation}
\subsubsection*{Positron thermalisation}
From eq.~\eqref{prob-survival} we learn that,
after Hawking emission,
most positrons radiate down to low kinetic energy.
When the kinetic energy of the positron becomes comparable
to those in the ambient medium, positrons start to thermalize
(i.e.~their energy distribution becomes Maxwellian~\cite{Prantzos_2011}).
They are expected to mostly annihilate with ISM electrons or to form
a positronium (Ps) bound state that rapidly decays.%
    \footnote{See~\cite{Prantzos_2011} for a complete list of possible interactions after thermalisation and likelihood in different Galactic regions.}

Utilizing data from INTEGRAL/SPI, the fraction of positrons that form positronium before annihilating (positronium fraction, $f_{\scriptscriptstyle\mathrm{Ps}}$) in the ISM of the Milky Way was estimated in~\cite{Siegert_2016}. In the Galactic bulge it is measured to be approximately $1.08\pm 0.03$, indicating that effectively all positrons form positronium before annihilation. In the Galactic disk, the positronium fraction is slightly lower but still consistent with almost complete positronium formation, measured at $0.90\pm 0.19$. These findings suggest that, throughout the Milky Way, positrons predominantly disappear via positronium formation and subsequent decay. 

Positronium is formed in a 3:1 ratio\footnote{This ratio is due to statistical spin considerations. Ortho-positronium is a triplet state with spin one (i.e.~three possible spin alignments) and para-positronium is a singlet state with spin zero and has only one alignment.} between ortho-positronium (oPs) and para-positronium (pPs). The former decays into three photons with $E_\gamma<511\,\text{keV}$ and the latter into two photons at exactly 511 keV\footnote{Annihilation into a larger number of photons in both cases is allowed but with negligible corresponding branching ratios~\cite{Prantzos_2011}.}. 
\\With this in mind, one quickly sees that the number of 511 keV photons produced per positron at rest is $2\times[(1-f_{\scriptscriptstyle\mathrm{Ps}})+f_{\scriptscriptstyle\mathrm{Ps}}/4]$,
where the first term represents direct annihilation of positrons and the second term photon emission from para-positronium decay. The number of $E_\gamma<511\,\text{keV}$ photons in turn is $3\times (3f_{\scriptscriptstyle\mathrm{Ps}}/4)$.

The positron flux that reaches thermalisation is $P\times\dd \dot{N}_{e^+}/\dd E_{e^+}$,
and the flux of 511 keV photons from this mechanism is
\begin{equation}\label{flux-511}
    \frac{\dd \dot{N}_{511}}{\dd E_\gamma}=\left[2(1-f_{\scriptscriptstyle\mathrm{Ps}})+f_{\scriptscriptstyle\mathrm{Ps}}/2\right]\int_{m_e}^{\infty}\dd E_{e^+}\,P(E_{e^+})\frac{\dd \dot{N}_{e^+}}{\dd E_{e^+}}\,f_{511}\,,
\end{equation}
where clearly the probability density distribution in this case is $f_{511}=\delta(E_\gamma-m_e)$. 

On the other hand, ortho-positronium decay results in a continuous spectrum of photon energies first derived by Ore and Powell~\cite{Ore:1949te}, where the following expression for the normalised probability distribution function was obtained\footnote{More recent calculations were carried out in~\cite{Manohar_2004} using non-relativistic quantum field theory. These corrections are important for photons in the UV range. Since the ISM strongly absorbs UV, we cannot employ this range of frequencies for constraints.}:
\begin{equation}
\frac{1}{\Gamma}\frac{\dd\Gamma}{\dd \xi}= \frac{2}{(\pi^2-9)}\left[\frac{\xi(1-\xi)}{(2-\xi)^2}+\frac{2-\xi}{\xi}+\frac{2(1-\xi)^2\ln(1-\xi)}{(2-\xi^3)}+\frac{2(1-\xi)\ln(1-\xi)}{\xi^2}\right]\,,
\end{equation}
with $\xi\equiv E_\gamma/m_e$. The probability density per unit energy is thus
\begin{equation}
   f_{\scriptscriptstyle\mathrm{oPs}}(E_\gamma)=\frac{1}{\Gamma_0}\frac{\dd\Gamma}{\dd E}=\frac{1}{m_e} \frac{1}{\Gamma}\frac{\dd\Gamma}{\dd \xi}
\end{equation}
and the integrated flux sourced via oPs decay,
\begin{equation}\label{OPS-em}
    \frac{\dd \dot{N}_\gamma^{\scriptscriptstyle\mathrm{oPs}}}{\dd E_\gamma}= \frac{9f_{\scriptscriptstyle\mathrm{Ps}}}{4}\int_{m_e}^{\infty}\dd E_{e^+}P(E_{e^+})\frac{\dd \dot{N}_{e^+}}{\dd E_{e^+}}\,f_{\scriptscriptstyle\mathrm{oPs}}(E_\gamma)\,.
\end{equation}
\subsection{Photon distribution}\label{photon-distribution}
\subsubsection*{Anisotropic component}
If primordial black holes account for the dark matter in the Universe, they are expected to have a specific energy density profile within our galaxy, just as any other dark matter candidate. The emitted electromagnetic flux is expected to have an anisotropic component separable from an isotropic background, with the ratio of the anisotropic and isotropic components depending on the Galactic coordinates. One expects intensity variations due to the local density enhancement in the Galaxy.

Due to the proximity of this PBH population we neglect redshifting
between emission and detection. We assume a spherically symmetric
PBH distribution about the centre of the Galaxy and adopt
the Navarro--Frenk--White (NFW)
dark matter density profile~\cite{Navarro_1996,Navarro_1997}
\begin{equation}\label{nfw-profile}
    \rho_{\scriptscriptstyle\mathrm{PBH}}(R,f_{\scriptscriptstyle\mathrm{PBH}})
    =
    \frac{
        f_{\scriptscriptstyle\mathrm{PBH}} \, \rho_\odot
    }{
        \frac{R}{R_\odot}
        \left(
            1
            +
            \frac{R}{R_\odot}
        \right)
    }
\end{equation}
where $R$ is the galactocentric distance,
\begin{equation}
R(s,b,l)=\sqrt{s^2-2s\,R_\odot \cos{b}\cos{l}+R_\odot^2 }\,,
\end{equation}
with $s$ the line-of-sight distance, $(b,l)$ the Galactic coordinates and $R_\odot$ the distance of the Sun from the Galactic centre.
To assign numerical values to the
parameters in~\eqref{nfw-profile},
we use the best-fit values obtained by Cautun et al.~\cite{Cautun_2020}, viz.,
$R_\odot=8.2\,\text{kpc}$
and
$\rho_\odot=0.33\,\text{GeV}/\text{cm}^3$.

In order to obtain the total flux over a region of interest ($s,\Delta\Omega$) we integrate the PBH number density along the line of sight and the sky region we are interested in, i.e.
\begin{equation}
    \label{phi-gal}
    \Phi_{\scriptscriptstyle\mathrm{gal}}(E_\gamma,\Omega)
    =
    \left(
        \frac{1}{4\pi\Delta\Omega}
    \right)
    \frac{
        \dd\dot{N}^{\scriptscriptstyle\mathrm{tot}}_\gamma
    }{
        \dd E_\gamma
    }
    \int_{\rm l.o.s}
    \int_{\Delta\Omega}
    n_{\scriptscriptstyle\mathrm{PBH}}(s,\Omega)
    \, \dd s
    \, \dd \Omega
    \qquad \text{[units:
    $\text{eV}^{-1} \, \text{cm}^{-2} \, \text{s}^{-1} \, \text{sr}^{-1}$]}
\end{equation}
where we have divided by the solid angle $\Delta\Omega$ in order to obtain the flux per steradian. The brightness associated with this flux follows the standard definition,
\begin{equation}
I(E_\gamma,\Omega)=E_\gamma\,\Phi_{\scriptscriptstyle\mathrm{gal}}\,.
\end{equation}
\subsubsection*{Isotropic component}
In the PBH-DM scenario,
black holes are expected to be distributed throughout the Universe.
As explained above, this will
source a nearly isotropic extragalactic background light.

\begin{figure}[htbp]
    \centering
       \begin{subfigure}[t]{0.49\textwidth}
        \includegraphics[width=\textwidth]{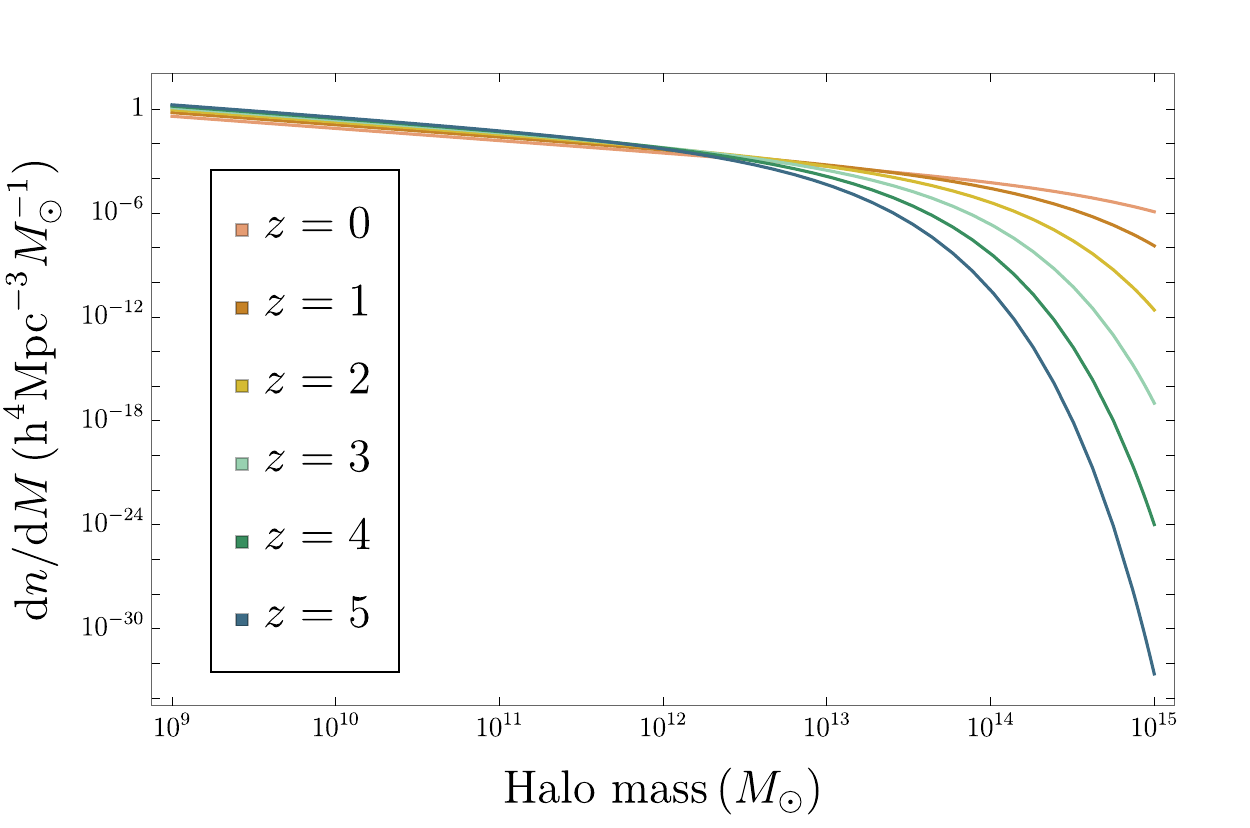}
    \end{subfigure}
    \begin{subfigure}[t]{0.49\textwidth}
        \includegraphics[width=\textwidth]{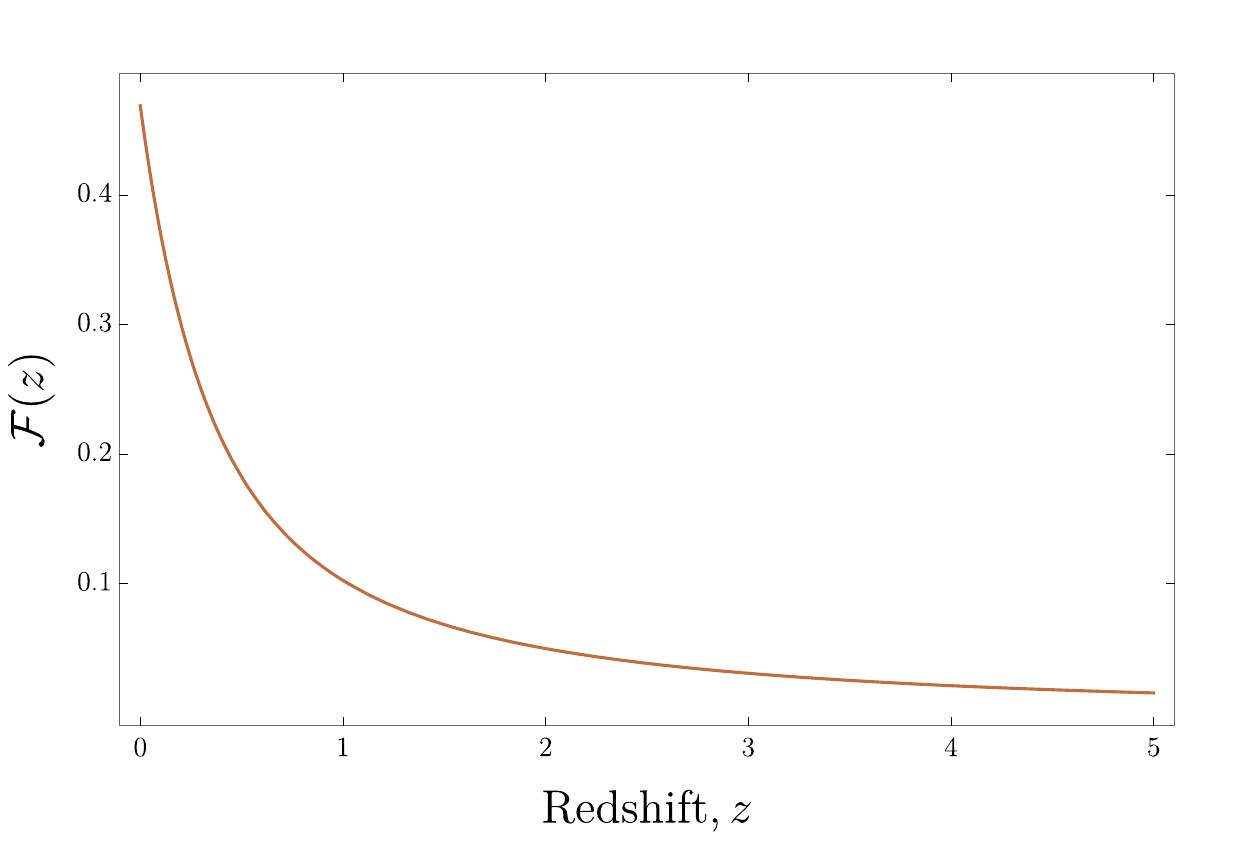}
    \end{subfigure}
    \caption[Halo number density and PBH fraction in haloes]{Comoving number density of haloes per mass interval for different redshift values in the Press-Schechter formalism (left) and the fraction of PBHs located in collapsed haloes for different redshifts (right).}
    \label{HMF-fig}
\end{figure}

Note that the location of the PBHs makes an important difference
when considering the
production mechanisms described in Section~\ref{sec:em-production}.
In our Galaxy,
the interstellar medium has
sufficiently large $n_{H}$ for efficient $e^-e^+$ interactions.
This is not necessarily the case for all extragalactic PBHs.
Positrons must be emitted in environments with a significant particle density, such that $n_{e^-}\geq0.01-0.1\,\rm cm^{-3}$.
This happens mostly in sufficiently massive haloes.
To quantify this, we use the Press--Schechter formalism%
    \footnote{The HMF provided by Tinker et al.~in 2008
   ~\cite{Tinker_2008} is a better fit to observations but the difference
    is less than an order of magnitude in the mass range we consider~\cite{cui2024knowhalomassfunction}.
    We use the Press--Schechter prescription for simplicity.}
\cite{Press:1973iz} to define the halo mass function (HMF),
i.e.~the comoving number density of haloes
per mass interval $\dd n/\dd M(M,z)$.
This is shown in Figure~\ref{HMF-fig}
as a function of the halo mass at different redshifts.
We see that at high redshift,
the number density of high-mass galaxies drop significantly, as expected. 

We compute the fraction of matter in the Universe contained in
sufficiently massive haloes
by integrating the HMF over a suitable mass range,
\begin{equation}
    \rho_{\scriptscriptstyle\mathrm{collapsed}}(z)=\int_{M_{\scriptscriptstyle\mathrm{min}}}^{M_{\scriptscriptstyle\mathrm{max}}}M\frac{\dd n}{\dd M}(M,z)\,\dd M\,.
\end{equation}
$M_{\scriptscriptstyle\mathrm{min}}$
should represent the lowest mass halo that can retain baryons
and cool them sufficiently to collapse into efficient annihilation zones.
We choose the conservative value
$M_{\scriptscriptstyle\mathrm{min}}\sim 10^{9}M_\odot$~\cite{Zheng_2024}.
At the upper end, we introduce a cutoff
$M_{\scriptscriptstyle\mathrm{max}}\sim 10^{16}M_\odot$.
It follows that
the fraction of PBHs in efficient annihilation zones at redshift $z$ is
\begin{equation}\label{Fofz}
    \mathcal{F}(z)=\frac{(1-f_{\rm b})\,\rho_{\scriptscriptstyle\mathrm{collapsed}}(z)}{(1-\bar{f}_{\rm b})\,\rho_{\rm m}(z)}\approx \frac{\rho_{\scriptscriptstyle\mathrm{collapsed}}(z)}{\rho_{\rm m}(z)} \,,
\end{equation}
where $f_{\rm b}$ is the fraction of matter in haloes that is of baryonic nature, $\bar{f}_{\rm b}$ the cosmic mean and $\rho_{\rm m}=\Omega_{\rm m}(z)\rho_{\scriptscriptstyle\mathrm{cr}}(z)$, where $\rho_{\scriptscriptstyle\mathrm{cr}}$ is the cosmological critical energy density. For this mass range,
the ratio $f_{\rm b}/\bar{f}_{\rm b}$ is expected to be in the range $0.2$--$1$~\cite{wright2024baryoncyclemoderncosmological}. 

We conclude that, when discussing extragalactic PBHs, only
a fraction $\mathcal{F}(z)$
of the total extragalactic positron flux contributes to the photon spectrum via in-medium interactions. In Figure~\ref{HMF-fig} we see that $\mathcal{F}(z)$ is non-negligible for small redshifts. Therefore, the photon flux produced via in-medium interactions in nearby galaxies is particularly relevant when estimating the total extragalactic flux. 

The total flux received from extragalactic sources is then
\begin{equation}\label{EGB-phi}
   \Phi^{\scriptscriptstyle\mathrm{EGB}}(E_\gamma)=\frac{1}{4\pi}\,n_{\scriptscriptstyle\mathrm{PBH},0}\int_0^{z_{\scriptscriptstyle\mathrm{max}}}\frac{\dd z}{H(z)}\left(\frac{\dd \dot{N}_\gamma^{\scriptscriptstyle\mathrm{P,\,FSR}}}{\dd E_\gamma}+\mathcal{F}(z)\frac{\dd \dot{N}_\gamma^{\scriptscriptstyle\mathrm{IA},\,\mathrm{NR}}}{\dd E_\gamma}\right)(E_\gamma(1+z)) \,,
\end{equation}
where
\begin{equation}
H(z)=H_0\sqrt{\Omega_\Lambda+\Omega_{\rm m}(1+z)^3}
\end{equation}
is the Hubble rate of expansion as a function of redshift
with $H_0=67.97\pm0.38\, \rm km/(s\,Mpc)$, $\Omega_\Lambda=1-\Omega_{\rm m}$, $\Omega_{\rm m}=0.307\pm0.005$~\cite{Adame_2025}.

Note, however, that extragalactic PBHs need not be
the only source of isotropic electromagnetic flux. It was noticed by Blanco and Hooper~\cite{Blanco_2019}
that the line-of-sight integral in Equation~\eqref{phi-gal}
deviates less than $10\%$ from its average value
over the high Galactic latitude region $|b|\geq20^{\circ}$.
This is often the area studied in analyses
of the isotropic diffuse background
in order to avoid Galactic bulge contamination.
The conclusion is that,
over this region,
a contribution from the Galactic halo
would be inseparable from the isotropic extragalactic component.
Its precise contribution depends on the parameter choice,
although in our analysis we see that the enhancement is not more than $\mathcal{O}(1)$.

\begin{figure}
    \centering
    \includegraphics[width=1\linewidth]{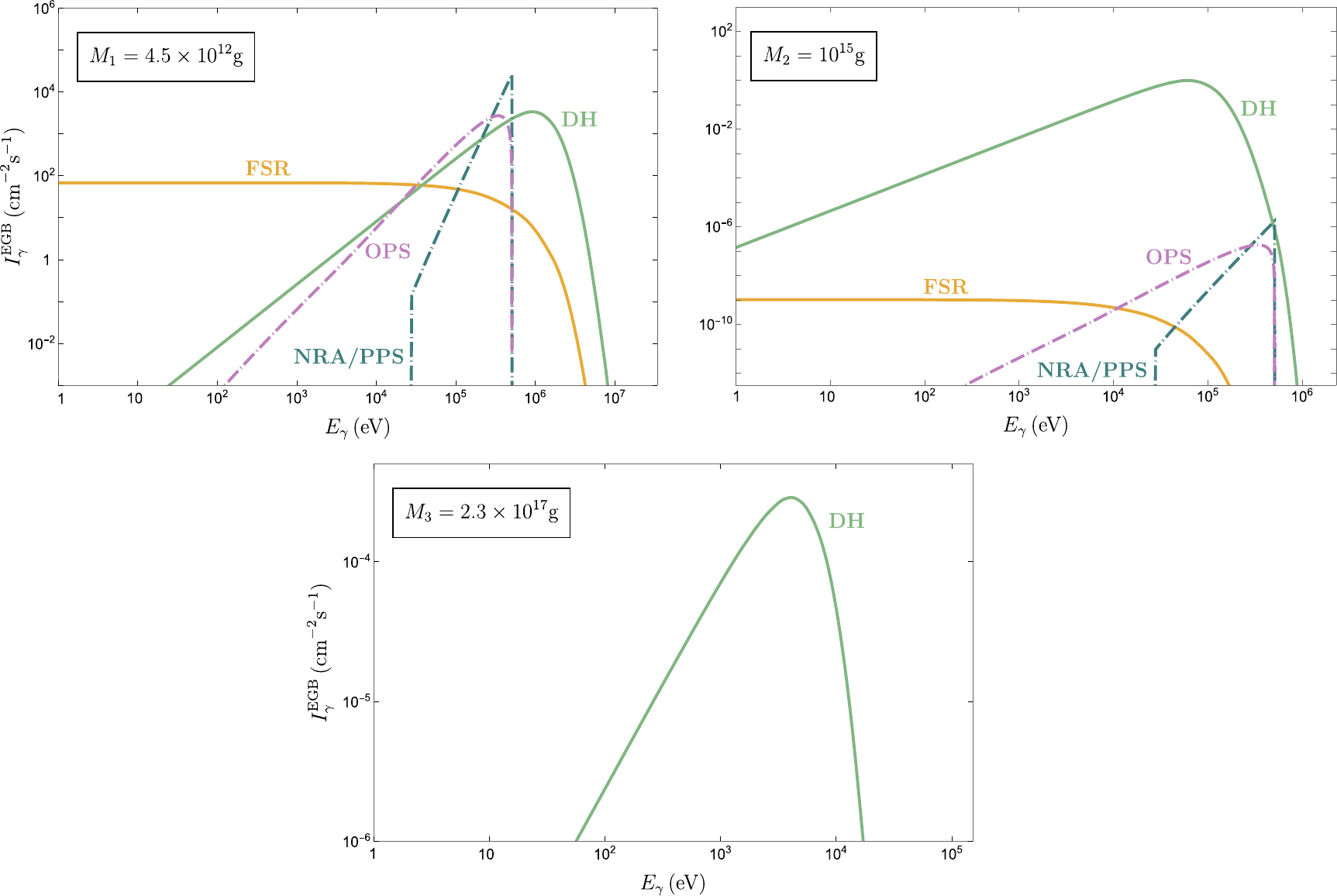}
    \caption[Extragalactic brightness by PBH Hawking emission]{Extragalactic brightness from different photon production mechanisms for three black hole masses (see Figure~\ref{emission-allowed} below) and $\ell=10^{-10}$m. Vacuum processes (FSR and DH) are depicted using solid lines, in-medium emission (NRA/PPS and OPS) using dashed lines. IA is omitted due to its relatively small impact.}
    \label{EGB-plots}
\end{figure}

The total isotropic brightness is then
\begin{equation}
    I^{\scriptscriptstyle\mathrm{iso}}_\gamma(E_\gamma)=E_\gamma (\Phi^{\scriptscriptstyle\mathrm{EGB}}+\Phi^{\scriptscriptstyle\mathrm{gal}}_{|b|\geq20^{\circ}}).
\end{equation}
In Figure~\ref{EGB-plots} we can see the flux of different photon emission channels sourced by extragalactic primordial black holes, following eq.~\eqref{EGB-phi}. We do not include the contribution from inflight-annihilation in the graphs since it is subdominant in all three cases. The masses in Figure~\ref{EGB-plots} are those shown in Figure~\ref{emission-allowed}.

\vspace{0.8cm}
\begin{mdframed}[style = mybbox]
Clearly, contributions from lepton emission cannot be neglected for hotter black holes. Since $(3+1)$-dimensional PBHs are hotter, this should also have a substantial effect in the conventional scenario. 
\end{mdframed}
\vspace{0.5cm}

%
\subsection{Unresolved electromagnetic signals in the Universe}
\label{sec:unresolved-electromagnetic}
There is unresolved background radiation
permeating the Universe at multiple wavelengths.
This includes both an isotropic component,
and anisotropic fluctuations that presumably
reflect the spatial clustering of unknown sources.
These signals may encode critical information about
exotic phenomena. In this section, we consider the possibility that Hawking evaporation from primordial black holes is the primary source of this
radiation.

\par
\begin{figure}[H]
    \centering
    \includegraphics[width=0.65\linewidth]{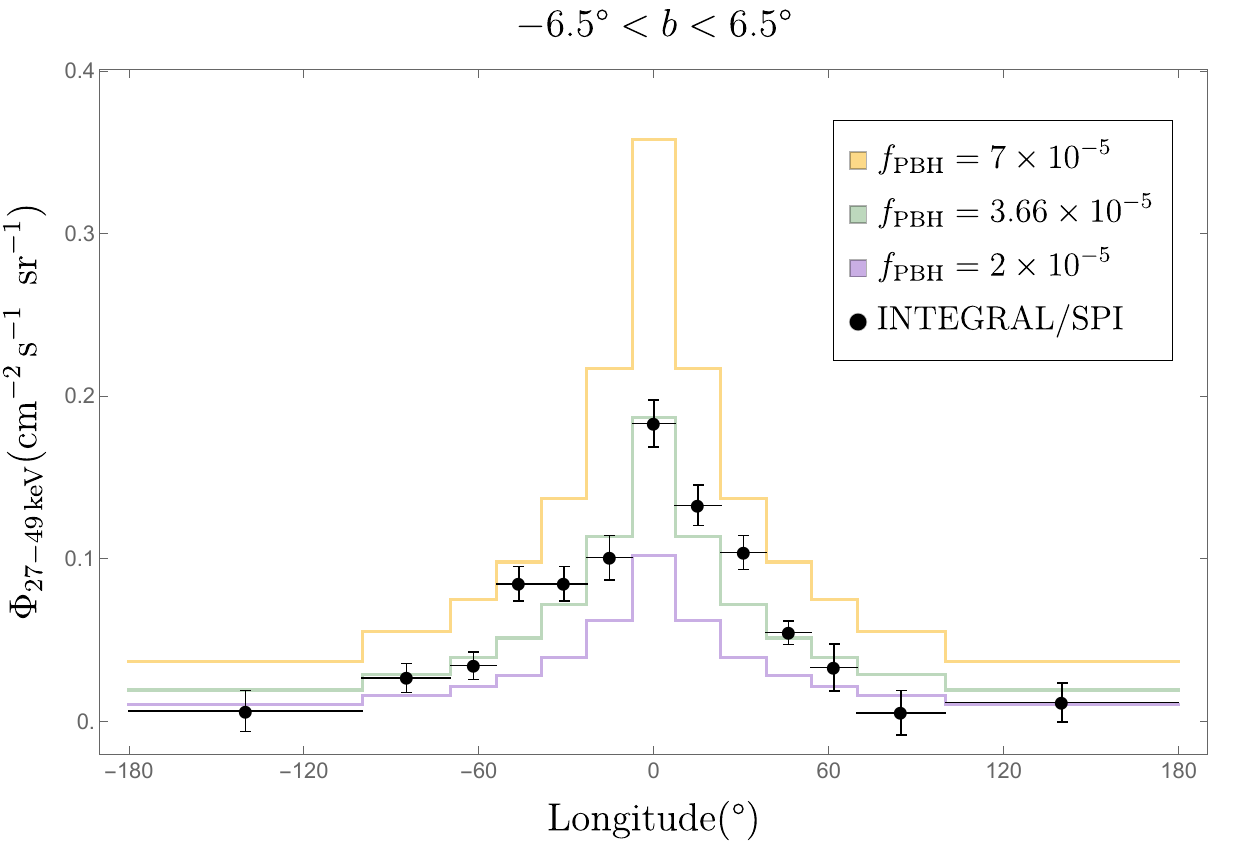}
    \caption[Integrated 27--49 keV flux vs.\ observations]{Integrated flux over several sky subregions and energy range 27--49 keV for $M=10^{11}$g and $\ell=10^{-6}$m for different PBH abundances (solid coloured lines), compared with empirical data (black dots). For this parameter choice, the abundance that fulfils $\chi^2_{\scriptscriptstyle\mathrm{eff}}=4$ is $f_{\scriptscriptstyle\mathrm{PBH}}=3.66\times 10^{-5}$.}
    \label{fig-GRE-example}
\end{figure}

\noindent\textbf{Datasets.}
We constrain the abundance of PBHs
using a simultaneous fit to
a number of observations.
We use the following measurements of the anisotropic
component.
\begin{itemize}
    \item \textsl{Diffuse emission from the Galactic ridge.}
    Background-subtracted measurements
    in this region were made the SPI spectrometer
    on board the INTEGRAL observatory~\cite{Bouchet_2011}.

    We fit a PBH population following an NFW distribution as
    the main source.%
        \footnote{This analysis was carried out for $(3+1)$-dimensional PBHs by Laha et al.~in~\cite{Laha_2020}.}
    To match SPI measurements of both the spatial morphology
    and spectral shape associated with the emission,
    we integrate over four energy bins
    ($27$--$49$~keV,
    $49$--$90$~keV,
    $100$--$200$~keV, and
    $200$--$600$~keV),
    and 13 $\Delta L$ bins covering the
    region $-6.5\degree < b < 6.5\degree$. See
    Figure~\ref{fig-GRE-example}.
    In this figure, one can clearly see the strong
    sensitivity to the abundance $f_{\scriptscriptstyle\mathrm{PBH}}$.

    \item \textsl{Chandra}. We use the background-subtracted
    spectrum reported in Ref.~\cite{Sicilian_2020}
    covering the range $3$--$5.5$~keV at latitudes $|b| \geq 10\degree$.

    \item \textsl{e-ROSITA and XMM-Newton}. We use data
    reported by e-ROSITA
    in the
    $0.2$--$10$~keV range,
    and by XMM-Newton
    in the
    $0.15$--$12$~keV range.
    The strictest constraints come from
    measurements in the sky area
    delimited by the third ring contour
    (see Ref.~\cite{Foster_2021}).
    For comparison
    to the PBH abundance, we employ the spectra expressed in experiment-independent units calculated in Ref.~\cite{Balaji_2025}.

    Unfortunately,
    bounds on the PBH abundance obtained from the eROSITA/XMM-Newton are not competitive with those arising from Chandra or the Galactic ridge signals anywhere in our parameter range.
\end{itemize}
We also use the following measurements of the isotropic component:
a Chandra
measurement at $0.5$--$3$~keV~\cite{Cappelluti_2017};
a HEAO 1 measurement at $3$--$500$~keV~\cite{Gruber_1999};
a COMPTEL measurement at $0.8$--$30$~MeV~\cite{Weidenspointner_2001};
an EGRET measurement at $30$--$200$~MeV~\cite{Strong_2004};
and a Fermi-LAT measurement at $200$--$10^4$~MeV~\cite{Abdo_2010}.

\par\noindent\textbf{Model fitting.}
Given a value of $f_{\scriptscriptstyle\mathrm{PBH}}$
and the PBH population models described above,
we obtain a theoretical prediction for the flux
observed in each of these measurements.
We report an upper limit
on $f_{\scriptscriptstyle\mathrm{PBH}}$
corresponding to the
$2\sigma$
($95.45\%$) confidence limit
for the one-sided goodness-of-fit
statistic
\begin{equation}
    \label{eq:unresolved-chisq}
    \chi^2_{\scriptscriptstyle\mathrm{eff}}=\sum_i\left(\frac{\text{Max}[\mathsf{X}^{\scriptscriptstyle\mathrm{th}}_i(f_{\scriptscriptstyle\mathrm{PBH}})-\mathsf{X}^{\scriptscriptstyle\mathrm{obs}}_i,0]}{2\,\sigma_i}\right) ,
\end{equation}
based on a global fit.
Here, $\mathsf{X}_i$
represent the $i^{\text{th}}$
observable,
$\sigma_i$ is the corresponding uncertainty,
``th'' denotes the prediction of our population model,
and ``obs'' denotes the measured value.
For the purposes of obtaining an estimate
we ignore covariances between measurements.

Because of the one-sided
fit,
we
drop negative bins where the model underpredicts
the measured flux.
This is because emission from PBHs
does not need to explain the measured data in
all wavebands.
Although the PBH prediction should not significantly
exceed the measured
value in any bin,
any deficit can be attributed to other unresolved sources.

\par
\begin{figure}
    \centering
    \includegraphics[width=0.7\linewidth]{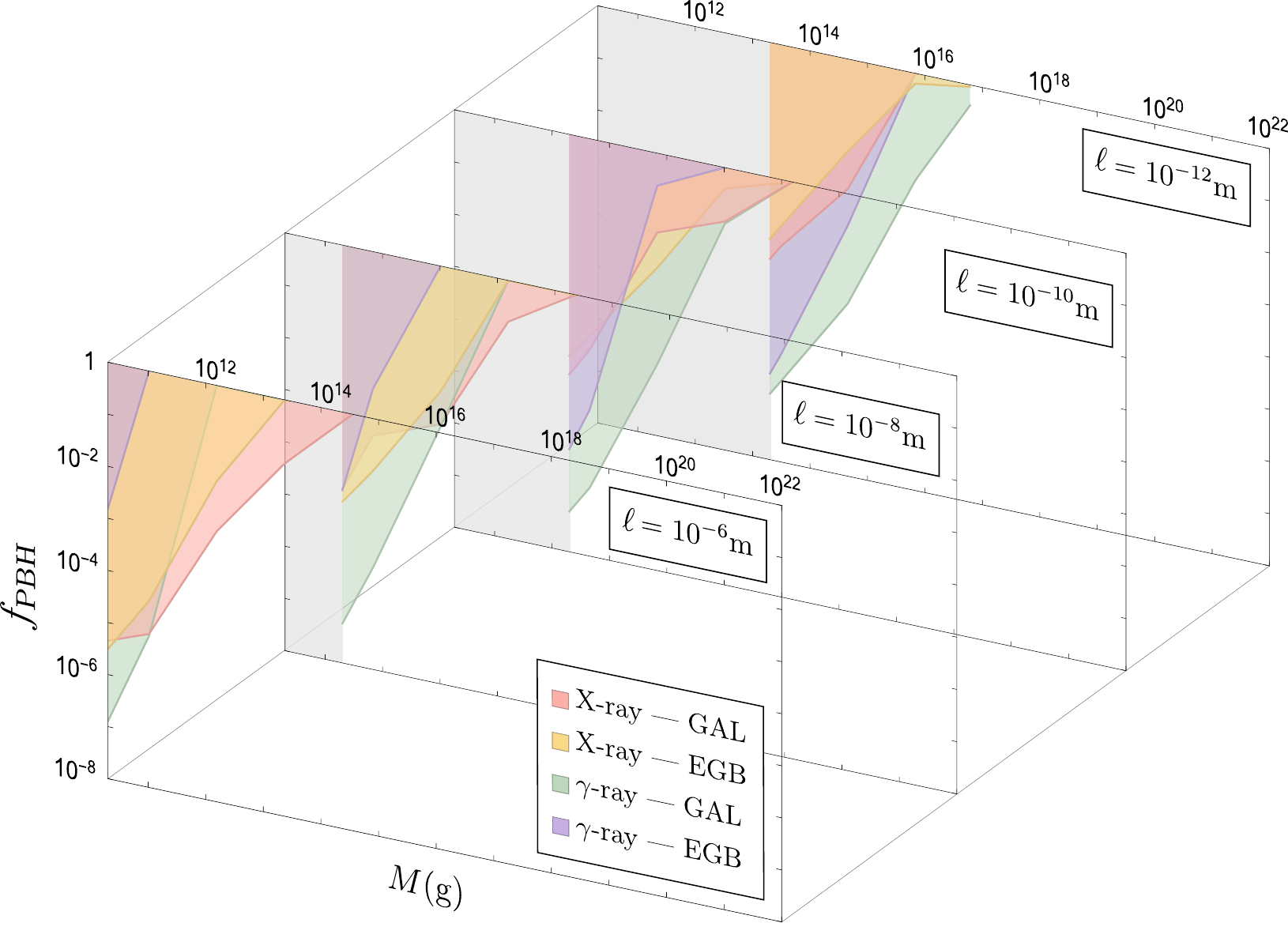}
    \caption[PBH abundance bounds from different forms of electromagnetic radiation]{Observational bounds on PBH abundance for a monochromatic population for different AdS radii (from front to back: $\ell=10^{-6},10^{-8},10^{-10}$ and $10^{-12}$m). Shaded regions are excluded by observations: purple by $\gamma$-ray detection of extragalactic sources, green by $\gamma$-radiation of Galactic signals, 
   yellow corresponds to X-ray radiation from extragalactic sources and pink to X-ray radiation from Galactic  sources. The grey-shaded regions cover the range of PBHs that have fully evaporated today or that would be rapidly evaporating today, $M<3M_*$.}
    \label{fig-3d-rad}
\end{figure}

\noindent\textbf{Results.}
Our results are shown in
Figure~\ref{fig-3d-rad}, where we have plotted
bounds on $f_{\scriptscriptstyle\mathrm{PBH}}$
for $\ell \in \{ 10^{-6} \, \text{m}, 10^{-8} \, \text{m},
10^{-10} \, \text{m}, 10^{-12} \, \text{m} \}$,
here plotted from front to back.%
    \footnote{For the case $\ell=10^{-12}\,\text{m}$,
    the envelopes end abruptly at $M=6\times 10^{16}$g.
    At this point
    the small-black-hole limit approximation is no longer valid.}
\newpage
    
\vspace{0.5cm}
\begin{mdframed}[style = mybbox]
The strong dependence on the AdS radius is clearly
visible.
In particular, the dark matter window $f_{\scriptscriptstyle\mathrm{PBH}}=1$ is wider
for larger AdS radii $\ell$ (i.e.~colder black holes).
Also, the relative constraining power of the different observational
data notably shifts. For a fixed mass, a lower temperature results in
emission of radiation at lower frequencies. At
large AdS radius, in contrast with the standard
$(3+1)$-dimensional scenario and small AdS radius,
radiation in the $\gamma$-ray band
has a limited constraining power.
Note that X-ray bounds, even if subdominant,
still usefully constrain the abundance of hotter black holes.
This is because particle production is favoured,
which results in FSR, IA and NRA photon emission
in the X-ray range, as shown in Figure~\ref{EGB-plots}. 
\end{mdframed}
\vspace{0.5cm}

\par\noindent\textbf{511 keV excess.}
Since the 1970s,
balloon and satellite experiments have consistently confirmed an excess of $511\,\text{keV}$
photons in the Milky Way~\cite{Harnden1972}. This $511\,\text{keV}$
emission features both disk and bulge components, with the bulge being
exceptionally narrow and bright. The most detailed observations have been
provided by INTEGRAL/SPI~\cite{Kn_dlseder_2005,Weidenspointner_2006,kierans2019positronannihilationgalaxy}. One of the biggest puzzles
is why the majority of the $511\,\text{keV}$ emission is concentrated in
the Galactic bulge, rather than being evenly spread throughout the disk,
as would be expected if supernovae were the dominant source. A variety of
astrophysical sources have been studied~\cite{Kalemci_2006,Cass_2004,Bertone_2006,Guessoum_2006,Bartels_2018,takhistov2019novelsignalsneutronstar,Fuller_2019},
but all candidates face significant challenges
in explaining the observed distribution and intensity of the
signal~\cite{Prantzos_2011}.

Dark matter models may naturally explain the dominance
of emission from the bulge,
because they typically predict a cusp in the dark matter density
near the Galactic centre.
The most studied
models are those involving
light particles, such as MeV-scale
WIMPs or axion-like particles~\cite{Ascasibar_2006,Rasera_2006}.
These may decay into positrons,
followed by annihilation in the Galactic medium
and decay into photons.
Primordial black holes have
also been considered as a possible origin~\cite{Okele:1980kwj}.
These results have been adapted to newer developments in the
field~\cite{luque2026refininggalacticprimordialblack,Keith_2021}.

To explore the possibility that braneworld black holes may explain the
$511\,\text{keV}$
excess,
we apply
the following standard procedure.
We integrate the total flux over Galactic latitudes in the range $-10.75\degree<b<10.25\degree$,
in an energy bin with width corresponding
to the uncertainty of the INTEGRAL/SPI detectors
at $E_\gamma=511$ keV.
We consider bins of angular size $\Delta L=12\degree$ and compare
our theoretical prediction
to the observed flux in each bin~\cite{Siegert_2019}.

Let us highlight that by introducing the survival
probability~\eqref{prob-survival} in our calculation
(including the IA photon spectrum and integrating over the
contributions from all channels),
we do not have to impose a cutoff on
the initial positron injection,
as discussed in Refs.~\cite{Friedlander:2022ttk,DeRocco_2019}.
We obtain results comparable to those 
for the UV-complete spectra
in~\cite{Friedlander:2022ttk,DeRocco_2019}, which suggests
that the condition $E_{e^+}<1$ MeV
used by these authors underestimates the total $511\,\text{keV}$ flux. The survival probability is still relatively high for such energies.
The probability of reaching thermalisation decreases as the energy
increases, but even for energies $E_{e^+}\sim 1$ GeV,
$P_{E_{e^+}\rightarrow m_e}\approx 0.64$. This still leads to a significant contribution to the integral in eq.~\eqref{flux-511}.  
%
%
%
%
\section{Low-energy electrons and positrons}\label{section-voyager}
Direct detection
of electrons or positrons
by spacecraft such as
Voyager also places constraints on the abundance
of evaporating primordial black holes.
In fact, this produces one of the most stringent limits on
$f_{\scriptscriptstyle\mathrm{PBH}}$ in the $(3+1)$-dimensions~\cite{Boudaud_2019}.

The Voyager spacecraft were designed to detect sub-GeV $e^\pm$,
among other charged cosmic rays.
This is ideal for studying PBHs,
since dark matter candidates are expected to produce
leptons in this energy range.
However, direct detection is severely hindered by the very low
probability of survival of positrons,
and the expected energy loss and
subsequent thermalisation of electrons.
As a result, we expect $e^\pm$ emitted only in nearby regions,
perhaps up to a few kiloparsecs away from the spacecraft,
to reach the detectors.

Measurements within the solar system are complicated
by the effect of the solar wind on charged cosmic rays.
However,
both Voyager
spacecraft have long crossed the heliopause;%
    \footnote{The \emph{heliopause} is the outermost boundary
    of the heliosphere, the region dominated by solar wind and the
    Sun’s magnetic field.}
Voyager 1 did so in 2012, and Voyager 2 in 2018.
They
have since been gathering more reliable data~\cite{burlaga2013,Cummings:2016pdr,Stone_2019}.
We calculate the local flux at the
position of the solar system,
and estimate limits on the PBH abundance.
We use
the Parker transport equation to model
steady injection and energy loss.
The total local flux is approximately given by
\begin{equation}\label{voy-flux}
   \Phi^{\odot}_{e^\pm}(E)\approx\int_E^\infty \frac{P_{E'\rightarrow E}}{ |\dd E/\dd x|(E')}Q(E',R_{\odot})\dd E'=\left(\frac{1}{4\pi}\right)\,n_{\scriptscriptstyle\mathrm{PBH}}(R_\odot)\int_E^\infty\frac{P_{E'\rightarrow E}}{ |\dd E/\dd x|(E')}\frac{\dd \dot{N}_{e^\pm}}{\dd E'}\dd E'\,,
\end{equation}
The label `$\odot$' denotes evaluation at the position of the
solar system, not the solar value,
and (as in eq.~\eqref{nfw-profile})
$R_{\odot}$ is the distance of the solar system
from the Galactic centre.
$Q(E,R_{\odot})$ is the local source term,
i.e.~the number of particles per unit energy,
per unit time, per unit volume.
In~\eqref{voy-flux} we have used again the
survival probability~\eqref{prob-survival}
and the Bethe--Bloch formula~\eqref{Bethe-Bloch}.
In both cases, the estimated number density of neutral hydrogen
at the position of the solar system
is taken to be $n_H\approx 0.12\,\text{cm}^{-3}$~\cite{Dialynas_2019}. 
%
%
\subsection{Detection at the heliopause}
We report a
$2\sigma$ upper limit
on $f_{\scriptscriptstyle\mathrm{PBH}}$
($>$95.45\% confidence level)
using the same technique described
in Section~\ref{sec:unresolved-electromagnetic};
see eq.~\eqref{eq:unresolved-chisq}.

\begin{figure}
    \centering
\includegraphics[width=0.83\linewidth]{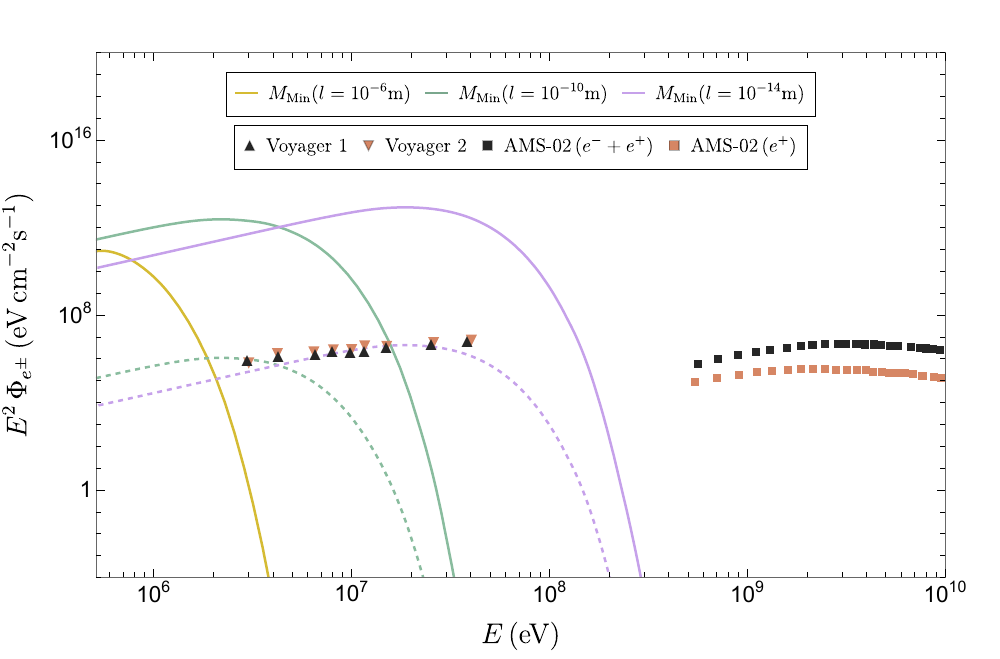}
    \caption[Voyager $e^\pm$ spectra of the lightest PBH candidates]{Voyager data and the spectra of $e^\pm$ of the lightest PBH-DM candidates for different values of AdS radius  (after propagation through the Galactic medium with no reacceleration). Solid lines are for a PBH population with $f_{\scriptscriptstyle\mathrm{PBH}}=1$ and the dashed lines for $f_{\scriptscriptstyle\mathrm{PBH}}$ values such that the flux agrees with observations, with ($\ell(\text{m}),f_{\scriptscriptstyle\mathrm{PBH}}$) approximately being $(10^{-10}, 1.7\times 10^{-6})$ and $(10^{-14}, 5.3\times 10^{-7})$.}
    \label{fig-voy}
\end{figure}

The result is plotted in
Figure~\ref{fig-voy}. For large AdS radii,
the emitted $e^\pm$ are too cold to be constrained by Voyager.
This differs significantly
from the $(3+1)$-dimensional case~\cite{Boudaud_2019},
where the behaviour is similar to the case
$\ell=10^{-14}\,\text{m}$,
with $f_{\scriptscriptstyle\mathrm{PBH}}\approx  6.6\times 10^{-7}$
for the lightest steadily evaporating PBH.

Note that diffuse reacceleration of leptons by the
astrophysical Galactic background can have a significant
impact on the calculation of upper bounds
for $f_{\scriptscriptstyle\mathrm{PBH}}$.
This effect is particularly important
for the emission of high-energy leptons in the range $0.1$--$1$ GeV.%
    \footnote{See Figure~1 in Ref.~\cite{Boudaud_2019},
    or Figure~2 in Ref.~\cite{luque2026refininggalacticprimordialblack}.}
In our case, this corresponds to the tail of the spectrum
$E>E_{\scriptscriptstyle\mathrm{peak}}$
for $M\sim M_{\scriptscriptstyle\mathrm{min}}$ at small AdS radii.
It is not relevant
for colder black holes.

For the former, we expect the emission of energies in the tail to be
enhanced (resulting in a broadened peak),
potentially reaching energies constrained by
AMS-02 detections~\cite{Aguilar_2021}.
For our parameter space, these bounds are expected to be
comparable or weaker to those from the Voyager spacecraft.
Meanwhile, for the case of colder black holes,
reacceleration of emitted leptons is subdominant,
as energy-loss mechanisms dominate the particle's evolution.
The influence of reacceleration is negligible unless the
turbulence is unusually strong (i.e.~high Alfv\'en speed environments),
which is not the case at the heliopause.
For this reason, braneworld primordial black holes are
expected to avoid useful bounds from AMS-02 positron data,
unlike the $(3+1)$-dimensional
scenario~\cite{Boudaud_2019,huang2025constraintsevaporatingprimordialblack}.
\begin{figure}
    \centering
    \includegraphics[width=0.6\linewidth]{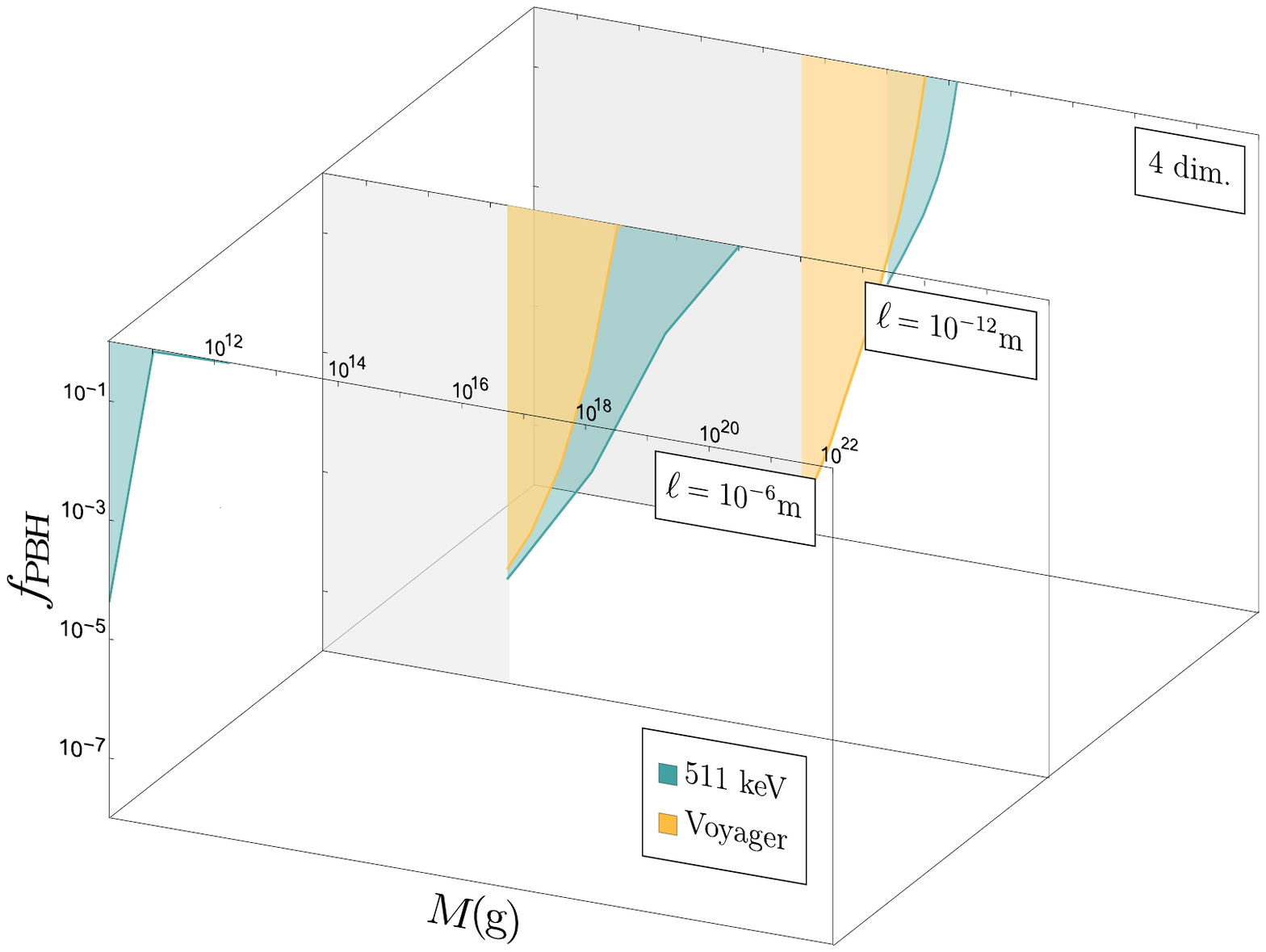}
    \caption[PBH bounds from $e^+e^-$ injection vs.\ conventional scenario]{Observational bounds on PBH abundance arising from $e^+e^-$ injection into the medium. We compare the cases $\ell=10^{-6}$m,
    $10^{-12}$m
    with the bounds obtained for the conventional scenario (511 keV line by DeRocco and Graham~\cite{DeRocco_2019} and Voyager by Boudaud and Cirelli~\cite{Boudaud_2019}) As already shown in Figure~\ref{fig-voy}, for $\ell=10^{-6}$m, PBHs  are unconstrained by direct $e^+e^-$ detection.}
    \label{fig:enter-label}
\end{figure}
%
%
\section{Impact on the cosmic microwave background}\label{section-CMB}
Predictions for energy injection into their environment
from
evaporating primordial black holes
can yield useful bounds on their abundance.
Electromagnetic energy injection
notably affects the evolution of the
free electron fraction $x_e\equiv n_e/n_H$,
the temperature of the intergalactic medium,
or even the CMB power spectrum. The rate at which energy is injected is
\begin{equation}\label{dEdVdt}
    \frac{\dd^2E}{\dd V\dd t}\Bigg|_{\rm inj}^i=n_{\scriptscriptstyle\mathrm{PBH},0}\,(1+z)^3\int E_i\frac{\dd\dot{N}}{\dd E_i}\dd E_i .
\end{equation}
Here $i$ labels the particle
species (photons, electrons, positrons),
and we integrate over the entire spectrum. 

The energy~\eqref{dEdVdt}
is deposited in its environment in three different ways.%
    \footnote{See~\cite{Poulin_2017} for a detailed description of these channels.}
These are: (a) ionisation (channel $i$),
(b) heating of the medium (channel $h$),
and (c) excitation of the Lyman-$\alpha$ transition (channel $\alpha$).
The relative importance of these channels
is accounted for by
dimensionless energy deposition
functions $f_c(z)$, 
\begin{equation}\label{dep-energy}
    \frac{\dd^2 E}{\dd V\dd t}\Bigg|_{\rm dep,c}(z)=f_c(z)\frac{\dd^2E}{\dd V\dd t}\Bigg|_{\rm inj}(z)\,,
\end{equation}
where $c=i,h,\alpha$. The $f_c(z)$
quantify how the energy deposition is
split between the channels.
They can be computed by~\cite{St_cker_2018}
\begin{equation}\label{def-fc}
    f_c(z)
    =
    \frac{
        \displaystyle
        \int_z^\infty \dd \ln(1+z')
        \,
        H^{-1}(z')
        \sum_i
            \int\dd E
            \,
            T_c^{i}(z',z,E)
            \,
            E
            \,
            \left.
                \frac{
                    \dd \dot{N}(E,t(z))
                }{
                    \dd E
                }
            \right|_{\rm inj}^{i}
    }{
        \displaystyle
        H^{-1}(z)
        \int\dd E
        \,
        E
        \,
        \left.
            \frac{
                \dd\dot{N}(E,t(z))
            }{
                \dd E
            }
        \right|^{\rm tot}_{\rm inj}
    }
    \,
    ,
\end{equation}
where $T_c^i$ are transfer functions for each particle $i$ and channel $c$. We employ the transfer functions computed in~\cite{Slatyer_2016,Slatyer_2016-b}. 
%
%
\subsection{Statistical analysis}
To calculate an upper bound on
$f_{\scriptscriptstyle\mathrm{PBH}}$
from the CMB
we follow the procedure of Ref.~\cite{Poulin_2017},
later refined in Ref.~\cite{St_cker_2018}.
In the context of higher dimensions, a similar analysis was carried
out by Friedlander et al.~\cite{Friedlander:2022ttk} in the LED scenario.
In each case, modifications to the
\textsc{Class} Boltzmann code
(Cosmic Linear Anisotropy Solving System~\cite{Diego_Blas_2011})
were implemented to model energy injection by PBHs.

For this analysis, we adapt the
\textsc{ExoClass}\_\textsc{led}
and \textsc{DarkAges}\_\textsc{led}
modules of \textsc{CosmoLed}~\cite{Friedlander:2022ttk}
to the Randall--Sundrum framework.
We consider a range of values for the black hole mass $M$
and AdS radius $\ell$,
and find the maximum abundance that keeps the CMB power
spectrum within 95\% C.L.~relative to the Planck 2018 data.

\begin{figure}
    \centering
\includegraphics[width=0.8\linewidth]{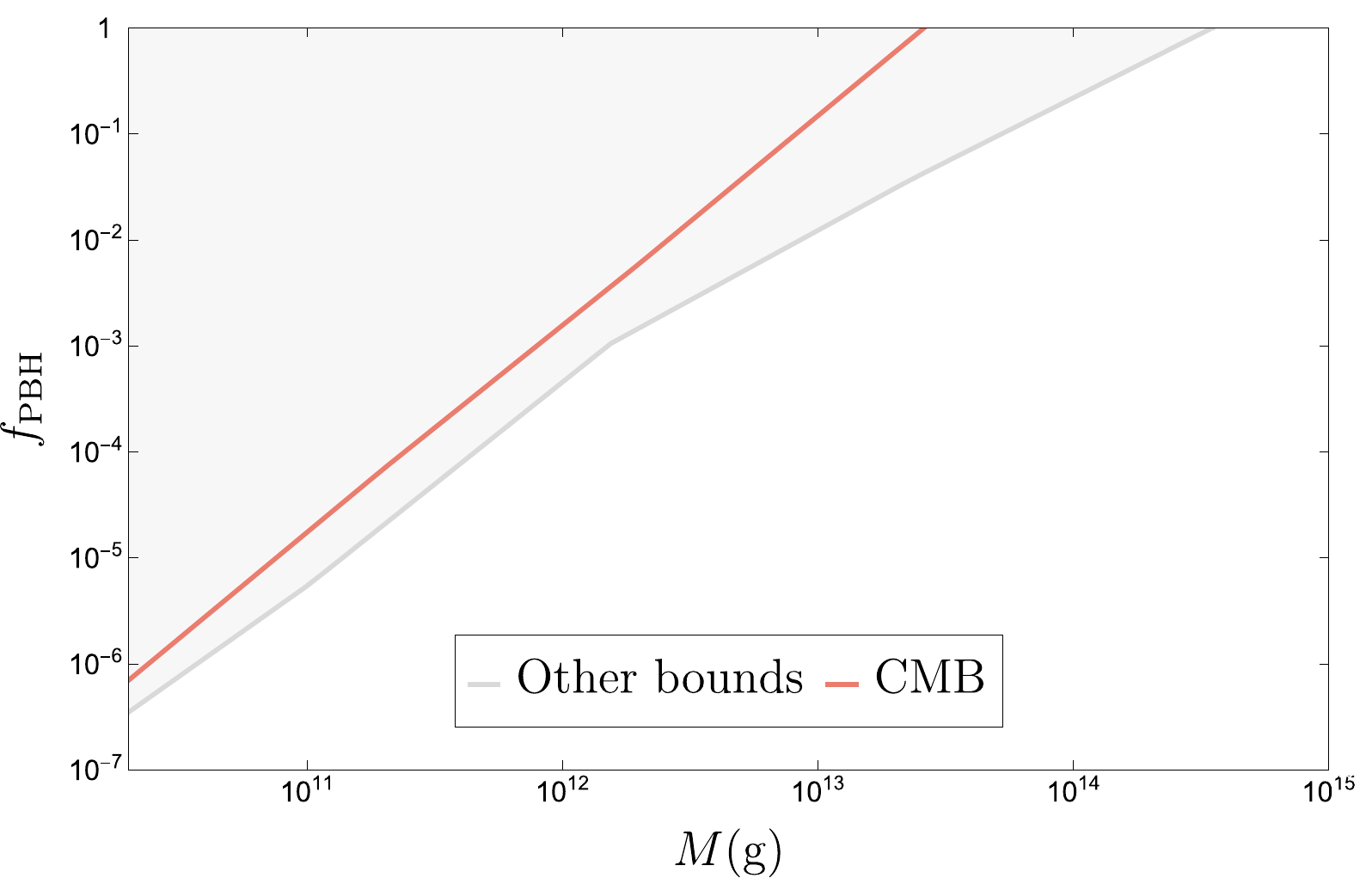}
    \caption[Evaporation and CMB bounds for $\ell = 10^{-6}$\,m]{PBH abundance bounds for AdS radius $\ell=10^{-6}$m. In grey,
    the region excluded by evaporation constraints. In red, conservative bound imposed by our statistical analysis of the CMB power spectra.}
    \label{cmb-line-l6}
\end{figure}

We keep the parameters of the background
$\Lambda$CDM cosmology fixed.
In principle,
one should carry out a Markov Chain Monte
Carlo analysis, as done in Refs.~\cite{St_cker_2018,Friedlander:2022ttk},
which would allow changes in the background cosmology to be
tensioned against the PBH energy deposition.
However, we will find that the CMB
bound does not dominate anywhere in our
$(M, \ell)$ parameter space
(see Figure~\ref{cmb-line-l6}).
Our analysis can therefore be regarded
as a consistency check
rather than a primary constraint,
for which purpose the simpler approach
is adequate.
It is conservative in the sense that
larger $f_{\scriptscriptstyle\mathrm{PBH}}$ may be allowed,
possibly at the expense of tensions with other
cosmological datasets.
For example,
if we compare the bounds
on $f_{\scriptscriptstyle\mathrm{PBH}}$
with and without a Monte Carlo analysis in
Ref.~\cite{St_cker_2018} and
Ref.~\cite{Friedlander:2022ttk}, the upper limit
relaxes by an order of magnitude.
We expect similar behaviour in our framework.

\begin{figure}
    \centering
    \begin{subfigure}[t]{0.48\textwidth}
        \includegraphics[width=\textwidth]{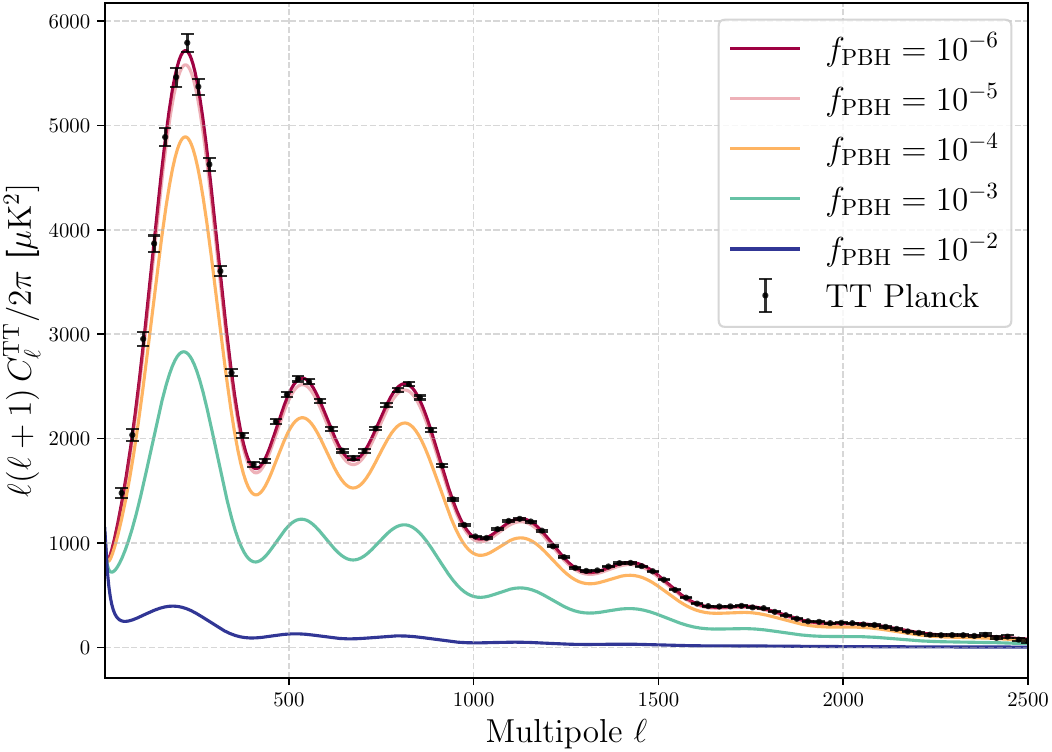}
    \end{subfigure}
    \hfill
    \begin{subfigure}[t]{0.495\textwidth}
        \includegraphics[width=\textwidth]{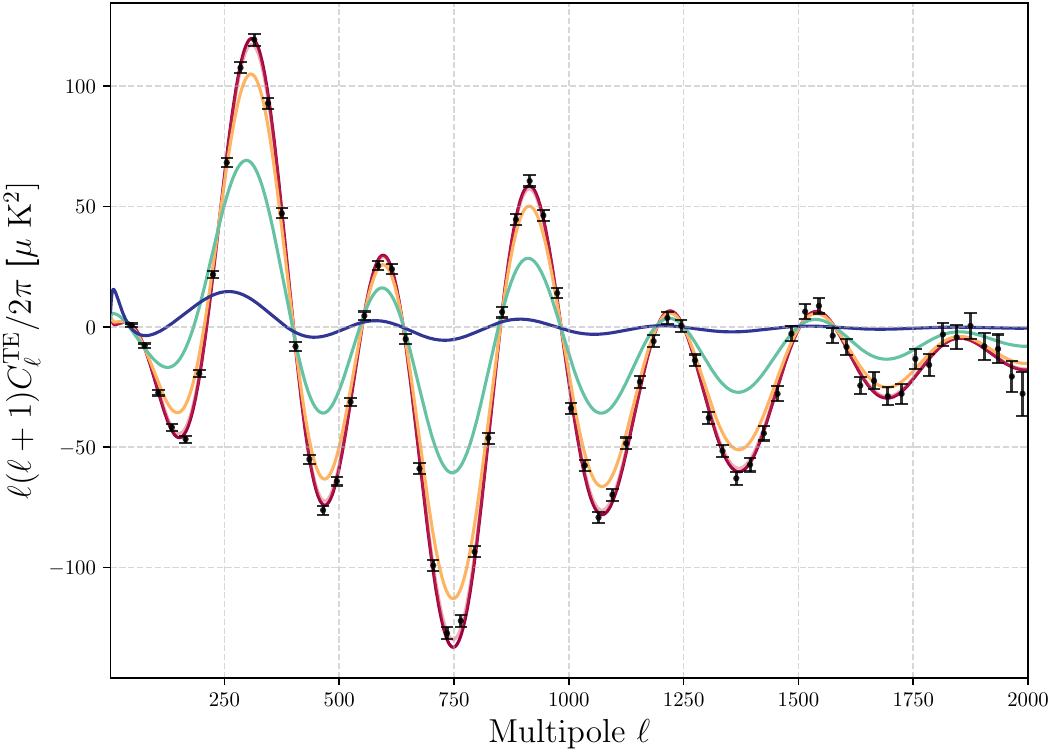}
    \end{subfigure}
    \\
    \vspace{0.4cm}
    \begin{subfigure}
     {0.495\textwidth}
        \includegraphics[width=\textwidth]{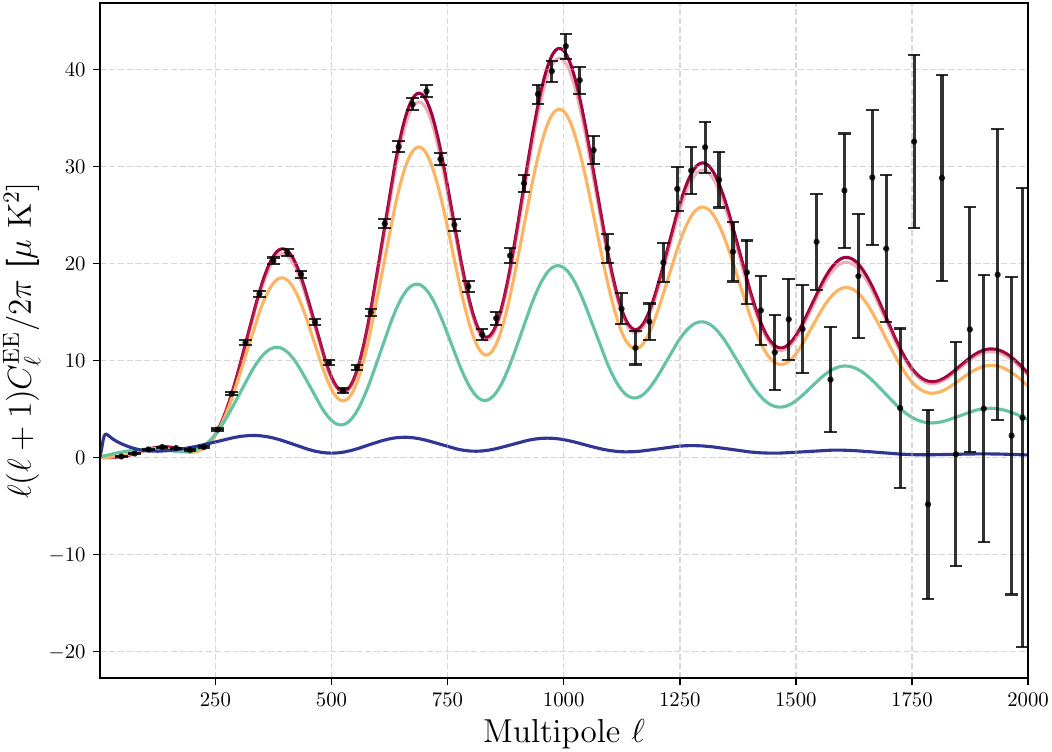}
    \end{subfigure}
    \caption[Planck spectra with PBH energy injection]{Example of spectra employed for the $\chi^2$ statistical analysis with Planck 2018 data. The coloured curves are the predicted TT-/TE-/EE-spectra if $\Lambda$CDM is combined with energy injection from a monochromatic PBH population with $(M_{\scriptscriptstyle\mathrm{min}}(\ell),\ell)=(2\times 10^{10}\,\text{g},10^{-6}\,\text{m})$ for different abundances. In the axes, $\ell$ is not the AdS radius, but the multipole of the spectra. We maintain the notation used in CMB literature for consistency.}
    \label{CMB-fig-1}
\end{figure}

Figure~\ref{CMB-fig-1}
plots
the modified temperature,
polarisation and cross-correlation spectra
for a range of values of
$f_{\scriptscriptstyle\mathrm{PBH}}$.
In these plots,
damping of anisotropies as the PBH abundance increases
is clearly visible.

%
%
\section{KM3-230213A event}
Recently there has been renewed interest in primordial black holes
within higher-dimensional frameworks~\cite{Anchordoqui_2025,anchordoqui2025primordialblackholes5d},
because these are a possible explanation for the
KM3NeT event~\cite{KM3NeT:2025npi}.
This is a detection of the most energetic neutrino to date.
The absence of multimessenger counterparts at the time of
detection~\cite{Klipfel_2025,airoldi2026primordialblackholeexplosion},
combined with the lack of similar high-energy events in
IceCube~\cite{li2025clashtitansultrahighenergy,km3netcollaboration2025ultrahighenergyeventkm3230213aglobal},
poses significant challenges for conventional interpretations.
Proposed exotic scenarios include higher-dimensional black holes
with sterile neutrinos~\cite{Anchordoqui_2025},
quasi-extremal PBHs~\cite{Baker_2026}
or extended PBH lifetimes via the
``memory burden effect''~\cite{dvali2018microscopicmodelholographysurvival,Dvali_2019,dvali2024memoryburdeneffectblack,boccia2025strikeluckkm3230213aevent,dondarini2025fastslowmergingprobes}.

In the higher-dimensional case,
and in the absence of further beyond-the-Standard-Model (BSM) physics,
Hawking emission rates for
photons and neutrinos
are expected to be comparable. Ref.~\cite{Anchordoqui_2025}
considered a LED framework in which, motivated by Swampland principles,
the extra-dimensional states form an explicit tower
of sterile neutrinos~\cite{Montero_2023}.
This yields an enhanced neutrino flux with a photon-neutrino
branching ratio of 1:6. As explained in Ref.~\cite{Anchordoqui_2025},
supersymmetric extensions could increase this ratio.

Note that an increased neutrino flux is not
an intrinsic feature of higher-dimensional black holes,
but requires BSM physics.
For standard evaporation in higher-dimensional frameworks,
we expect a comparable photon flux and hence $\gamma$-ray signal.

As discussed in Section~\ref{section-CMB},
constraints from energy injection into the medium during
Hawking evaporation can place stringent limits on the abundance of
evaporating black holes. Even if the photon-neutrino branching ratio
is altered by the introduction of additional neutrino degrees of freedom,
electron production can significantly impact the CMB.
Note that $T^{e^-}_c$ in~\eqref{def-fc} is generally
larger and peaks at higher redshift than $T^{\gamma}_c$.
Consequently, CMB constraints on candidate sources
prior to the final burst producing the KM3NeT neutrino
must be evaluated carefully for each PBH mass,
even if the multimessenger bounds are circumvented.
The impact on the CMB could be studied by introducing
BSM branching ratios in the \textsc{DarkAges} package
described in Ref.~\cite{St_cker_2018}
(alternatively, Ref.~\cite{Friedlander:2022ttk}
for higher-dimensional scenarios).
However, this has not yet been done.

Furthermore, emission into the brane is not suppressed,
contrary to early claims in Ref.~\cite{Anchordoqui_2025}.
As shown by Emparan et al.,
this misconception arose from the interpretation of brane
fields as bulk fields confined to a limited phase space,
rather than intrinsically four-dimensional~\cite{Emparan_2000}.
Proposed
explanations for the KM3NeT event should not rely on this suppression.

Observationally,
higher-dimensional PBHs differ from conventional
ones primarily in their lighter masses at fixed temperature.
This requires higher number densities for a given PBH fraction
$f_{\scriptscriptstyle\mathrm{PBH}}$.
Moreover, their longer lifetimes allow PBHs that would have
otherwise evaporated long ago to burst today.
Both features can be of use when the likelihood and energy range
of the event are highly constrained, as for the KM3NeT event.

The KM3NeT detection is very intriguing and has attracted
significant interest from the community.
Future observations will be crucial in determining whether
it has a
cosmogenic origin. If similar events are recorded,
multimessenger physics will play an essential role in
filtering source candidates.

     \chapter[Microlensing]{Microlensing}\label{chap12:microlensing}
     \thispagestyle{fancy}
\pagestyle{fancy}

    Gravitational lensing stands as one of the most profound and far-reaching predictions of general relativity. The deflection of light by massive objects acting as gravitational lenses has become a powerful probe of the Universe, enabling the detection and characterisation of compact objects across a vast range of scales.

While a number of black holes have been detected through gravitational observations, their measured masses generally do not allow one to determine whether they are of astrophysical or primordial origin. A definitive identification of a PBH would require a mass below the threshold expected from conventional stellar-collapse channels, ($M \lesssim M_\odot$)~\cite{LIGOScientific:2019kan}. Although no such object has yet been observed, gravitational lensing remains one of the most promising methods for probing this low-mass region of parameter space.

The lensing regime relevant to a given compact object depends sensitively on its mass. For the low-mass PBHs of interest as dark-matter candidates, the curvature radius is extremely small, and a detectable magnification signal requires near-perfect alignment between the source, lens, and observer. Lensing events satisfying this close-alignment condition are known as microlensing events. Consequently, the remainder of this chapter focuses on the microlensing signatures of braneworld PBHs and the constraints that can be derived from their non-observation.

We begin by reviewing the formalism of gravitational microlensing in four dimensions before discussing its extension to higher-dimensional braneworld geometries. Such extensions have been developed for the various black-hole solutions introduced in \href{part:higher-d}{\textbf{\textcolor{thesislilac}{Part 2}}}~\cite{Kar:2003gh,Frolov_2003,Majumdar:2004mz,Whisker_2005,Keeton_2005,Keeton_2006,Keeton:2006di,Gergely:2006qt,Pal_2008,Bin_Nun_2010}. A central issue in this context is determining which metric provides the physically relevant description of the lensing geometry at observationally accessible scales and how this affects the expected signatures of large and small braneworld black holes.

The answer depends first on the distance of closest approach of the light ray to the black hole. When photons pass close to the horizon, strong-deflection effects become important, and the lensing signal depends sensitively on the near-horizon geometry. Such effects have been studied in several braneworld scenarios \cite{Eiroa:2005vd,Whisker_2005}, but remain extremely challenging to detect observationally. In contrast, 
the lensing configurations relevant to current microlensing surveys are well described by weak-field gravity. In this limit, the observable signal is determined by the asymptotic gravitational field rather than the detailed structure of the metric near the horizon. 

Independently of this distinction, one must also consider the relationship between the wavelength of the observed radiation and the characteristic lensing scale. When the wavelength becomes comparable to this scale, diffraction effects become important, and the geometric-optics approximation breaks down. This is particularly relevant for low-mass PBHs. Among existing microlensing surveys, the Hyper Suprime-Cam (HSC) observations of M31 provide sensitivity to the smallest PBH masses. The HSC ($r$)-band filter probes wavelengths $\lambda \simeq 0.6,\mu\mathrm{m}$, placing the entire braneworld small-black-hole population in the wave-optics regime, where $\lambda \sim \ell \gg r_0$. For sufficiently massive black holes, however, the geometric-optics approximation remains valid.

These considerations establish that the microlensing configurations of interest lie entirely within the weak-field regime. In this limit, the lensing signal is determined by the asymptotic gravitational field rather than the near-horizon geometry, and the relevant description is therefore that of linearised gravity. In braneworld models, the unique solution to the linearised field equations for a localised mass source on the brane is the Garriga--Tanaka metric~\cite{Garriga_2000}, irrespective of the near-horizon black-hole solution. The analysis presented in this chapter is therefore based on this weak-field braneworld solution and explores the resulting microlensing signatures in both the wave-optics and geometric-optics regimes.
\section{Lensing geometry and key parameters}
\begin{figure}
    \centering
    \includegraphics[width=0.7\linewidth]{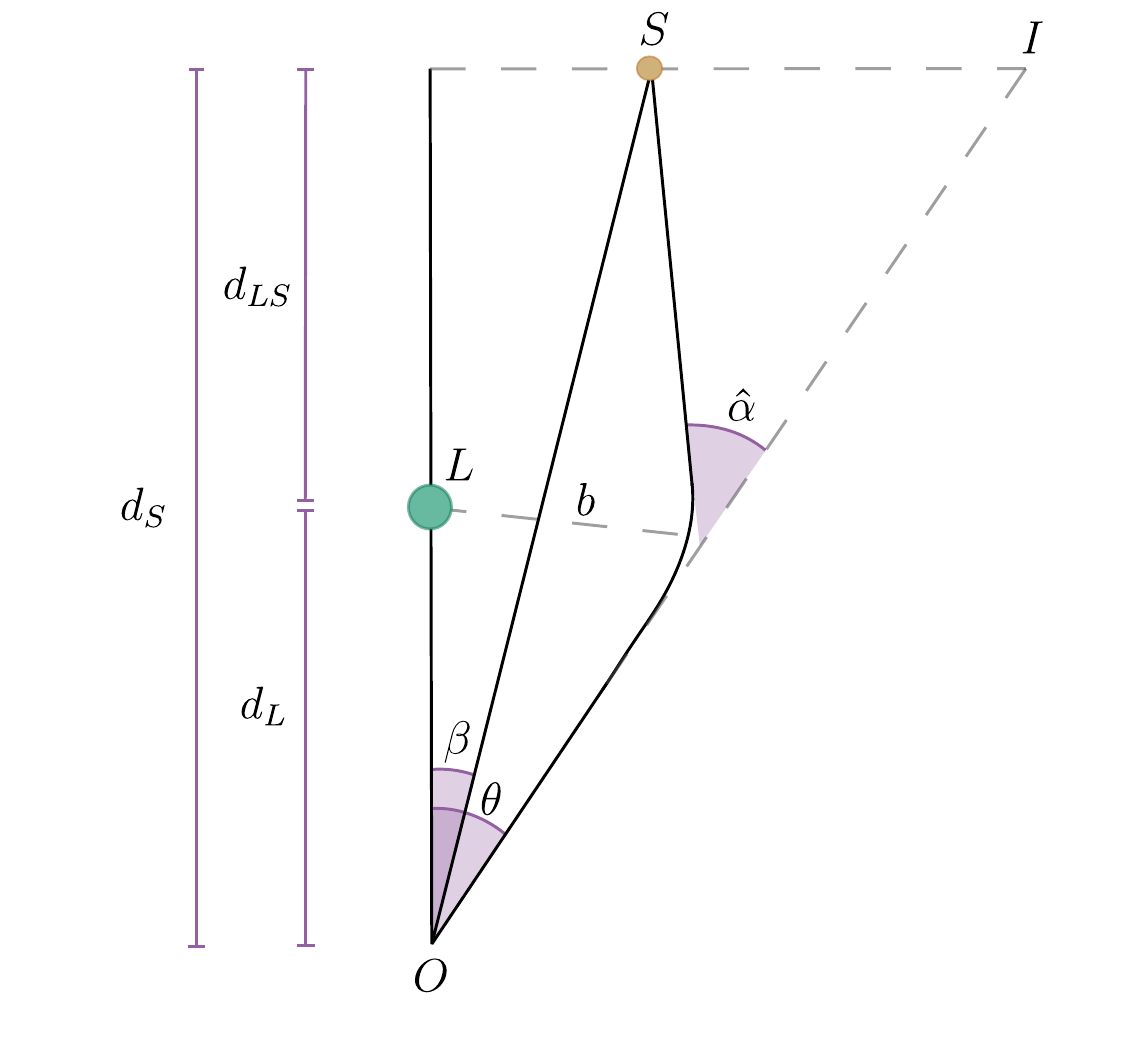}
    \caption{Microlensing diagram.}
    \label{fig:microlensing}
\end{figure}

The modern era of microlensing was inaugurated by Paczyński in 1986~\cite{Paczynski:1986}, who introduced the term and laid out the theoretical framework for the lensing of background stars in the Galactic bulge and the Magellanic Clouds by foreground stars or dark compact objects in the Milky Way. 
We begin by considering the simplest lensing configuration: a point source lensed by a compact object of mass $M$. The observer, lens and source are separated by the angular diameter distances $d_{\scriptscriptstyle\mathrm{L}}$, $d_{\scriptscriptstyle\mathrm{S}}$ and $d_{\scriptscriptstyle\mathrm{LS}}$, respectively. In the absence of the lens, light propagates along a straight path and the source appears at an angular position $\beta$ on the observer's sky. The gravitational field of the lens deflects the light rays, causing the source to be observed instead at an angular position $\theta$ (see Figure~\ref{fig:microlensing}).
In the weak-deflection regime, the geometry may be described using the thin-lens approximation, in which the deflection is assumed to occur instantaneously at the lens plane. The lens equation then relates the true angular position of the source to the observed angular position of an image according to~\cite{Virbhadra:2000vv}
\begin{equation}\label{lens-equation}
    \tan \beta=\tan \theta -\frac{d_{\scriptscriptstyle\mathrm{LS}}}{d_{\scriptscriptstyle\mathrm{S}}}\left(\tan\theta+\tan(\hat{\alpha}-\theta)\right)\,,
\end{equation}
where $\hat{\alpha}=\hat{\alpha}(\theta)$ is the deflection angle produced by the lens. For essentially all astrophysical microlensing configurations the small-angle approximation $\tan x\approx x$ is well justified and eq.~\eqref{lens-equation} simplifies to~\cite{10.1093/mnras/128.4.295}
\begin{equation}
\beta = \theta - \frac{d_{\scriptscriptstyle\mathrm{LS}}}{d_{\scriptscriptstyle\mathrm{S}}} \hat{\alpha}(\theta)\,.
\label{eq:lens-general}
\end{equation}
 For a point-mass lens in a $(3+1)$-dimensional Schwarzschild spacetime, the deflection angle takes the form
\begin{equation}\label{eq:deflection}
    \hat{\alpha}(\theta)=\frac{4G_4M}{ b}\,,
\end{equation}
where $b$ is the physical impact parameter --- the perpendicular distance between the light ray and the lens in the lens plane, with dimensions of length --- and it takes the form $b\approx \theta d_{\scriptscriptstyle\mathrm{L}}$ in the small-angle approximation. For convenience, we define the parameter
\begin{equation}
\theta_{\scriptscriptstyle\mathrm{E}} \equiv \sqrt{4G_4M \frac{d_{\scriptscriptstyle\mathrm{LS}}}{d_{\scriptscriptstyle\mathrm{L}} d_{\scriptscriptstyle\mathrm{S}}}}\,.
\label{eq:einstein-angle}
\end{equation}
Henceforth we refer to $\theta_{\scriptscriptstyle\mathrm{E}}$ as the Einstein angle. When substituting the deflection angle~\eqref{eq:deflection} and the Einstein angle~\eqref{eq:einstein-angle}, the lens equation~\eqref{eq:lens-general} takes the form
\begin{equation}
\beta = \theta - \frac{\theta_{\scriptscriptstyle\mathrm{E}}^2}{\theta}\,,
\label{eq:lens-point}
\end{equation}
which is quadratic in $\theta$
 and therefore admits exactly two solutions for any source position $\beta \neq 0$:
\begin{equation}
\theta_{\pm} = \frac{1}{2}\left(\beta \pm \sqrt{\beta^2 + 4\theta_{\scriptscriptstyle\mathrm{E}}^2}\right).
\label{eq:images}
\end{equation}
These correspond to two images of the source, located on opposite sides of the lens. The image $\theta_+$ lies outside the ring $\theta=\theta_{\scriptscriptstyle\mathrm{E}}$ on the same side as the source, $|\theta_+| > \theta_{\scriptscriptstyle\mathrm{E}}$, while $\theta_-
$ lies inside the ring ($|\theta_-| < \theta_{\scriptscriptstyle\mathrm{E}}$
) on the opposite side.

The Einstein angle thus sets the characteristic angular scale of the lensing problem, with its corresponding physical scale in the lens plane
\begin{equation}
R_E = \theta_{\scriptscriptstyle\mathrm{E}} d_{\scriptscriptstyle\mathrm{L}} = \sqrt{4G_4M \frac{d_{\scriptscriptstyle\mathrm{L}} d_{\scriptscriptstyle\mathrm{LS}}}{d_{\scriptscriptstyle\mathrm{S}}}}\,,
\label{eq:einstein-radius}
\end{equation}
which we refer to as the Einstein radius.  

\subsection{Einstein ring}
The special case of perfect alignment ($\beta = 0$) deserves particular attention. Setting $\beta = 0$
in eq.~\eqref{eq:images} gives $\theta_+ = \theta_{\scriptscriptstyle\mathrm{E}}$
 and $\theta_- = -\theta_{\scriptscriptstyle\mathrm{E}}$, i.e.~both images are located at the same angular distance from the lens, on exactly opposite sides. However, when the source, lens, and observer are perfectly aligned, the problem acquires full azimuthal symmetry about the line of sight. The lens equation fixes only the angular distance of the image from the lens and $\beta=0$ leaves the azimuthal direction unconstrained. Consequently, every point satisfying $\theta=\pm\theta_{\scriptscriptstyle\mathrm{E}}$ is an image of the source. The two-image solution therefore becomes degenerate and expands into a continuous ring of angular radius $\theta_{\scriptscriptstyle\mathrm{E}}$, known as the Einstein ring~\cite{Einstein:1936:LLA,Paczynski:1986,gould:2000}. The Einstein ring is thus not a separate solution of the lens equation, but rather the limiting form of the two-image solution in the presence of perfect axial symmetry.
\subsection{Magnification and the microlensing light curve}\label{micro-quants}
Since gravitational lensing preserves surface brightness but distorts solid angles, each image is magnified with respect to the unlensed source. The magnification of each image is given by the ratio of the image solid angle to the source solid angle:
\begin{equation}
\mu_{\pm} = \left|\frac{\theta_{\pm}}{\beta} \frac{\dd\theta_{\pm}}{\dd\beta}\right| = \frac{u^2 + 2}{2u\sqrt{u^2+4}} \pm \frac{1}{2}\,,
\label{eq:magnification-individual}
\end{equation}
where we have introduced the dimensionless source-lens separation
\begin{equation}
u \equiv \frac{\beta}{\theta_{\scriptscriptstyle\mathrm{E}}}\,,
\label{eq:u}
\end{equation}
which measures the angular separation between source and lens in units of the Einstein angle. However, for stellar-mass and substellar lenses the angular separation between these images is extremely small, typically of order microarcseconds to milliarcseconds. Current surveys are therefore unable to resolve the individual images directly and instead observe only the combined flux. The observable quantity is then the total magnification, obtained by summing the absolute magnifications of both images:
\begin{equation}
\mu_{\scriptscriptstyle\mathrm{tot}}= |\mu_+| + |\mu_-| = \frac{u^2 + 2}{u\sqrt{u^2 + 4}}\,.
\label{eq:magnification-total}
\end{equation}
This is the Paczyński magnification formula \cite{Paczynski:1986}.

In a typical microlensing event, the lens moves across the sky at a relative transverse velocity $v_\perp$ with respect to the source. The projected separation $u(t)$ therefore varies with time, tracing a trajectory that, for uniform rectilinear motion, follows
\begin{equation}
u(t) = \sqrt{u_0^2 + \left(\frac{t - t_0}{t_E}\right)^2},
\label{eq:u-t}
\end{equation}
where $u_0$ is the minimum projected angular separation between source and lens during the event, normalised to the Einstein angle, $u_0 = \beta_{\scriptscriptstyle\mathrm{min}}/\theta_{\scriptscriptstyle\mathrm{E}}$, reached at time of closest approach $t_0$ --- and $t_E$ is the Einstein crossing time, defined as the time for the lens to traverse one Einstein radius,
\begin{equation}
t_E = \frac{\theta_{\scriptscriptstyle\mathrm{E}} d_{\scriptscriptstyle\mathrm{L}}}{v_\perp} = \frac{R_E}{v_\perp}.
\end{equation}
Substituting equation~\eqref{eq:u-t} into the magnification formula~\eqref{eq:magnification-total} yields the characteristic symmetric, achromatic light curve of a point-source point-lens microlensing event. The shape of the light curve is determined entirely by $u_0$: a smaller impact parameter produces a higher peak magnification and a narrower, more pronounced event. For u$_0 \gg 1$ the source passes far outside the Einstein ring and the magnification remains close to unity throughout, producing no detectable signal. This is the fundamental reason why microlensing events require close alignment: only source trajectories with $u_0 \lesssim 1$, i.e.~passing within approximately one Einstein radius of the lens, produce magnifications sufficiently large to be detected above the photometric noise floor.
The three parameters $(u_0, t_0, t_E)$ fully describe a point-source point-lens light curve and are the primary observables extracted from a microlensing survey.
\section{Deflection angle in braneworlds}
The deflection angle $\hat{\alpha}$ for a static, spherically symmetric spacetime is obtained from the null geodesic equations by exploiting the two conserved quantities associated with time-translation and azimuthal symmetry --- the photon energy $E$ and angular momentum $L$, and defining the impact parameter $b\equiv E/L$. Eliminating the affine parameter in favour of the orbital angle and introducing the coordinate $u = 1/r$, one arrives at an orbit equation whose perturbative solution, expanded at leading order in $G_4M/b$, yields the deflection as the difference between the asymptotic incoming and outgoing directions of the ray~\cite{wald1984general}. For a general spherically symmetric solution, the deflection angle receives contributions from both the ($tt$) component and the ($rr$) component of the metric. The extension of this procedure to higher-dimensional and braneworld metrics follows the same steps with the appropriate metric functions, and for the Garriga--Tanaka metric is carried out in detail in Refs.~\cite{Kar:2003gh,Keeton_2005,Keeton:2006di}. Up to first-order correction, the deviation angle is
\begin{equation}\label{defl-angle}
    \hat{\alpha}=\frac{4M}{M_4^2r}\left(1+\frac{l^2}{r^2}\right)\,.
\end{equation}
In their work, Keeton and Petters~\cite{Keeton_2005,Keeton_2006,Keeton:2006di} emphasise that the deflection angle is inherently coordinate-dependent and develop an observationally motivated framework to ensure that the computed deflection corresponds to a genuine physical observable. This is particularly important in alternative theories of gravity, where the corrections arising from the choice of coordinates can be of the same order as the corrections predicted by the theory itself, and the two must therefore be carefully disentangled.
%
%
%
\section{Detectability restrictions}
There are three main factors that restrict the detectability of microlensing events:
\begin{itemize}
    \item \textcolor{thesislilac}{Finite source star size.} If the apparent size of the star in the lensing plane is larger than the Einstein radius of the black hole, the point-like source approximation breaks down and only a fraction of the star is lensed. This is the case when black holes are very light or close to the source star, and results in the net magnification being suppressed \cite{1994ApJ...430..505W}, often below our detection threshold. We introduce the parameter
\begin{equation}\label{finitesource}
    U_{E}=\frac{R_s/d_{\scriptscriptstyle\mathrm{S}}}{R_E/d_{\scriptscriptstyle\mathrm{L}}}
\end{equation}
with $R_s$ being the radius of the source star. Finite size effects dominate when $U_E\gg 1$. In the first analysis of the HSC survey \cite{Niikura_2019} M31 stars were assumed to have a radius of the order of $R_\odot$, whereas in Ref.~\cite{Smyth_2020} it is argued that stars with such a radius are too dim to be detected by the instrumentation employed for the HSC survey. The most abundant star population in such a galaxy with a brightness magnitude resolvable by the HSC survey have a radius of around $4R_\odot$. The brightest stars in M31 are argued to contribute the most to the constraints due to them being easier to detect and their radii are estimated to be of order $10R_\odot$. As argued in~\cite{Smyth_2020}, realistic star radii imply that PBHs with masses below $10^{23}$g do not produce detectable signals. Since the heaviest possible five-dimensional PBH has a mass of order $M\sim 10^{22}$g, the abundance of five-dimensional black holes cannot be constrained via the HSC survey. 
     \item \textcolor{thesislilac}{Effect of wave optics.} When the size of the curvature radius becomes comparable to the wavelength of the light being detected, the wave nature of light becomes important and the geometric interpretation is no longer valid. The HSC $r$-band filter detects light of wavelength $\lambda=6\times 10^{-7}$ m. Therefore, since $\ell\lesssim 10^{-6}$ m, lensing by small black holes in this particular setup is dominated by wave optic effects, see~\cite{Keeton:2006di} for a discussion.
      \item \textcolor{thesislilac}{Time scales.} These are roughly estimated by $t_E$, where $v_\perp=200-220$ km/s is the standard fiducial value~\cite{Niikura_2019}, irrespective of their mass. The timescale range of small black holes is $t_E<0.06$ hours, whereas methods employed for the HSC survey can recover about 60-70\% of microlensing events if the timescale is in the range 0.07-3 hours~\cite{Niikura_2019}.
\end{itemize}  
%
%
\section{Large black holes}
Following the same steps as in Section~\ref{micro-quants}, we evaluate eq.~\eqref{defl-angle} at $b$ and use the approximation $b\approx\theta d_{\scriptscriptstyle\mathrm{L}}$,
\begin{equation}\label{defl-angle-1}
 \hat{\alpha}=\frac{4M}{M_4^2\theta d_{\scriptscriptstyle\mathrm{L}}}\left(1+\frac{l^2}{\theta^2d_{\scriptscriptstyle\mathrm{L}}^2}\right)\,.
\end{equation}
Since $ \frac{\ell}{d_{\scriptscriptstyle\mathrm{L}}}=\tan \theta_\ell\approx \theta_\ell $, \eqref{defl-angle-1} becomes
\begin{equation}\label{defl-angle-2}
     \hat{\alpha}\approx\frac{4M}{M_4^2\theta d_{\scriptscriptstyle\mathrm{L}}}\left(1+\frac{\theta_\ell^2}{\theta^2}\right)\,.
\end{equation}
We see that the far-field condition of the Garriga--Tanaka metric implies
\begin{equation}
    \frac{\ell^2}{r^2}\ll 1 \qquad \rightarrow \qquad  \frac{\ell^2}{b^2}\ll 1  \qquad \rightarrow \qquad \frac{\theta_\ell^2}{\theta^2}\ll 1\,.
\end{equation}
In other words, the braneworld correction in eq.~\eqref{defl-angle-2} must be small.

We can solve for $\hat{\alpha}$ in the lens equation~\eqref{eq:lens-general} and set it equal to eq.~\eqref{defl-angle-2},
\begin{equation}\label{defl-angle-3}
    \frac{4M}{M_4^2\theta d_{\scriptscriptstyle\mathrm{L}}}\left(1+\frac{\theta_\ell^2}{\theta^2}\right)=\frac{d_{\scriptscriptstyle\mathrm{S}}}{d_{\scriptscriptstyle\mathrm{LS}}}\left(\theta-\beta\right)\,.
\end{equation}
After rearranging and imposing $\beta=0$, \eqref{defl-angle-3} becomes
\begin{equation}
    \theta^4-\theta_{\scriptscriptstyle\mathrm{E}}^2\theta^2-\theta^2_{\scriptscriptstyle\mathrm{E}}\theta_\ell^2=0\,.
\end{equation}
We can introduce an \textit{effective} Einstein angle,
\begin{equation}\label{thetaE-GT}
    \tilde{\theta}_{\scriptscriptstyle\mathrm{E}}^2=\frac{\theta_{\scriptscriptstyle\mathrm{E}}^2}{2}\left(1+\sqrt{1+\frac{4\theta_\ell^2}{\theta_{\scriptscriptstyle\mathrm{E}}^2}}\right)\approx\theta_{\scriptscriptstyle\mathrm{E}}^2\left(1+\frac{\varepsilon_{\theta}}{4}\right)\,,
\end{equation}
such that the images are located at 
\begin{equation}
    \theta=\pm \tilde{\theta}_{\scriptscriptstyle\mathrm{E}}\,.
\end{equation}
The braneworld correction to the conventional angle is encoded in $\varepsilon_{\theta}$,
\begin{equation}
\varepsilon_\theta=\frac{4\theta_\ell^2}{\theta_{\scriptscriptstyle\mathrm{E}}^2}=\frac{M_4^2\ell^2}{M}\left(\frac{d_{\scriptscriptstyle\mathrm{S}}}{d_{\scriptscriptstyle\mathrm{LS}}d_{\scriptscriptstyle\mathrm{L}}}\right)\,.
\end{equation}
The correction becomes larger with larger $\ell$ and lighter black hole masses. 

In what refers to magnification, we follow the full analytical description of weak-deflection
lensing in the Garriga--Tanaka framework developed in
Ref.~\cite{Keeton:2006di}. The total magnification in the Garriga--Tanaka regime is
\begin{equation}\label{tot-mag}
  \mu_{\scriptscriptstyle\mathrm{tot}}=\left(1-f(u)\,\varepsilon_\ell^2\right)\mu_{\scriptscriptstyle\mathrm{tot}}^{\scriptscriptstyle\mathrm{(3+1)}}+\mathcal{O}\left(\varepsilon_\ell^3\right)\,.
\end{equation}
Here, $\mu_{\scriptscriptstyle\mathrm{tot}}^{\scriptscriptstyle\mathrm{(3+1)}}$
is the expected result for $(3+1)$-dimensions,
and $\varepsilon_\mu$
is the parameter
\begin{equation}
    \label{eq:varepsilon-l}
    \varepsilon_{\mu}=\frac{\tan^{-1}(\ell/d_{\scriptscriptstyle\mathrm{L}})}{\theta_{\scriptscriptstyle\mathrm{E}}}
\end{equation}
that quantifies the deviation from the prediction from the $(3+1)$-dimensional framework. In eq.~\eqref{eq:varepsilon-l} the function $f(u)$ satisfies
\begin{equation}
    f(u)=\frac{4}{(2+u^2)(4+u^2)}\,.
\end{equation}
Because this factor is less than unity,
braneworld effects
\emph{decrease} the total magnification.

Equation~\eqref{tot-mag}
is obtained by making a Taylor expansion of the image positions
in powers of $\varepsilon_\mu$.
The zeroth-order term yields the usual expression for
the weak-deflection limit of the conventional framework. Braneworld corrections enter at order $\mathcal{O}(\varepsilon_\mu^2)$, and for the geometrical configuration of the HSC survey,
we have
\begin{equation}
\varepsilon_{\mu}\approx 10^{8}\left(\frac{10^{-8}M_\odot}{M}\right)^{1/2}\tan^{-1}\left(\frac{\ell}{10^{21}\rm m}\right)\,.
\end{equation}
Note that the prefactor of the $\varepsilon_\mu^2$ term
in~\eqref{tot-mag}
lies in the range (0, 0.5). We see that the correction depends heavily 
on the ratio of the AdS radius and the distance of the lens.

\vspace{0.5cm}
\begin{mdframed}[style = mybbox]
It follows that,
in any survey covering astrophysical scales,
the correction to the standard result is always negligible,
regardless the black hole mass (within the geometric optics approximation).
We conclude that black holes with masses that exceed the small black hole
limit are expected to be constrained in the same way as in a
$(3+1)$-dimensional scenario.
\end{mdframed}
\vspace{0.5cm}
Therefore, we conclude that for black holes in the mass range probed by the HSC survey, the braneworld corrections to the lensing signal are negligible and the four-dimensional results apply to excellent approximation. We adopt the results of Smyth et al.~\cite{Smyth_2020} for our figures.
%
%
%
%
\section{Small black holes}
For the small-black-hole limit, we must study wave optics effects
in the braneworld framework. These were presented in
Ref.~\cite{Keeton:2006di},
where the phenomenon of \textit{attolensing} was introduced for the
first time. Like femtolensing in the case of
standard gravity~\cite{Gould1992},
one expects to see interference fringes in the frequency spectrum
once wave optics becomes non-negligible.

In the standard scenario,
femtolensing was discarded as a constraint for primordial black holes
due to arguments presented in~\cite{Katz_2018,Sugiyama_2020}, mostly based on the finite source star size problem. These apply equally to attolensing.

\vspace{0.5cm}
\begin{mdframed}[style = mybbox]
Interference patterns are observable only if the star source size
projected onto the lens plane is not much larger than the Einstein
radius of the black hole. In the HSC setup, this is not the case
for any lens in the small black hole limit.
Not only are the interference patterns undetectable,
but so is any sort of signal magnification.
Small black holes are thus undetectable in this astrophysical setup.
\end{mdframed}
\vspace{0.5cm}

If a different detection setup were considered where the
finite source size effect is subdominant,
these interference fringes would be hints of braneworld cosmology,
since the patterns predicted in Refs.~\cite{Keeton:2006di} and \cite{Gould1992}
are notably different.
The setup proposed in Ref.~\cite{Sugiyama_2020},
where a denser cadence ($10$s)
and g-band monitoring for a sample of white dwarfs over a
year timescale is considered,
would most likely discriminate between models
if interference fringes were detected.

    \chapter{Findings and open horizons}\label{sec:conclusions}
    \section*{\textcolor{thesislilac}{Intricacies of the covariant approach}}
Much of the appeal of braneworld physics lies in how little it allows
one to take for granted: constructions that are conceptually trivial
in four dimensions --- a top-hat density profile, the propagation of
perturbations in an ordinary fluid, an equation of state, a test fluid falling onto a fixed
background --- import assumptions that must be revisited one by one
once a bulk is present. The covariant approach is an invaluable tool
for this purpose, but it does not contain the whole picture: the
quantities it deals in are projections of five-dimensional objects
onto the brane --- shadows of the bulk dynamics rather than the
dynamics itself. Throughout this thesis, we have found that keeping the bulk
perspective in view is not optional: it is what decides which
four-dimensional intuitions survive.

Gravitational collapse offers a known but instructive example, which
we revisited via the junction conditions and the Misner--Sharp
formalism. The symmetries of the Oppenheimer--Snyder configuration
suggest that nothing essential should change on the brane, yet the
exterior of a collapsing region cannot remain
static~\cite{Bruni_2001}. Braneworlds thus lose a shortcut that the
conventional scenario enjoys: there, idealised top-hat profiles
suffice for pressureless matter, and only radiation demands the full
machinery of numerical critical collapse, on which the quantitative
picture of formation thresholds and mass functions
rests~\cite{Musco_2019,Escriv__2025}. On the brane, that machinery
becomes necessary regardless of the nature of the fluid. To our knowledge, no such simulations exist for braneworlds in the context of primordial black hole formation; extending them to the RS-II scenario, in both the radiation and dust regimes, would link the initial conditions that calculations like ours are currently forced to assume, with the formation threshold feeding directly into the predicted primordial black hole abundance today."

The core of our contribution concerns accretion. We analysed spherical
accretion onto five-dimensional Schwarzschild--Tangherlini black holes
in the RS-II framework, adopting the fully relativistic treatment
introduced by Michel~\cite{Michel1972}, ensuring consistency with the
higher-dimensional setting at each step and quantifying the impact of
braneworld corrections.

By performing a complete analysis of scalar perturbations, we
determined the propagation speed of pressure-supported disturbances in
the accreting fluid, which is essential for assessing the sonic
properties and relevant time scales of the system. We find that this
speed can be approximated by the thermodynamic sound speed $c_s$ in
the quadratic regime, but only locally ($k\eta \to 0$) and for a
barotropic equation of state. We then computed the critical radius of
Schwarzschild--Tangherlini black holes, obtaining $r_{\rm cr} =
\sqrt{2}\,r_0$ for $w = 1/3$, in agreement with
Ref.~\cite{Tikhomirov:2005bt}. This allowed us to determine the velocity and energy density profiles.

Using these results, we derived the black hole mass evolution. While
its functional dependence matches that found in previous
studies~\cite{Guedens_2002-2,Majumdar_2003,Tikhomirov:2005bt}, we
obtain a different value and interpretation for the parameter $F$. In
particular, we find significantly stronger accretion than previously
reported, while maintaining consistency with causality. By imposing
observational constraints on the tensor power spectrum, we established
a lower bound on the end of inflation in the RS-II scenario, thereby
fixing the earliest possible time for primordial black hole formation
in this particular framework. Combining this with the mass evolution,
we find that earlier formation leads to a prolonged strong-accretion
phase and enhanced growth: strong AdS warping implies an earlier end
of inflation and a delayed transition out of the quadratic regime,
extending the epoch of strong accretion, whereas weaker warping
results in a later end of inflation and an earlier transition,
reducing its duration. We show that, if formed sufficiently early,
black holes with initial masses below the effective four-dimensional
Planck scale may survive until today, evolving as discussed in
\hyperref[part:pbhs]{\textbf{\textcolor{thesislilac}{Part~III}}}, and
thus constitute viable dark matter candidates.

Finally, we assessed the validity of the assumptions underlying our
accretion framework. Despite the enhanced radial accretion scale
relative to the four-dimensional case, and thus the longer accretion
time scales, the quasi-static approximation is robust in
braneworlds across a broader parameter space than in the $(3+1)$-dimensional scenario. Moreover, the test-fluid approximation
remains well satisfied throughout the quadratic regime, as the energy
density of the matter fields decreases in the RS-II scenario and the
corresponding contribution to the enclosed gravitational energy
remains subdominant compared to that of the Weyl cloud surrounding the
black hole; in $(3+1)$ dimensions, by contrast, it is only valid in a
restricted parameter space.

We conclude that strong accretion significantly alters the mapping
between the conditions required for primordial black hole formation
and the resulting mass function at later times, compared to the
standard scenario. In particular, cosmological conditions associated
with specific formation mechanisms --- such as peaks in the primordial
power spectrum in overdensity collapse --- correspond to markedly
different scales for a given present-day mass. The formation time and
the size of the black hole relative to the cosmological horizon
depend sensitively on the formation mechanism and the details of the
collapse; a more precise characterisation therefore requires a
dedicated treatment of these processes --- the numerical collapse
programme noted above --- and their inclusion in the calculation.
\section*{\textcolor{thesislilac}{Phenomenology of PBHs in braneworlds}}
 In this thesis we have also investigated the phenomenology of
primordial black holes in Randall--Sundrum Type-II braneworld scenarios,
with particular emphasis on their role as dark matter candidates.
Building on earlier analytic estimates that established the
main qualitative features of primordial black holes in warped geometries,
we have revisited the problem with a more detailed treatment of evaporation
and cosmological evolution.
We have also incorporated
improved astrophysical and cosmological constraints that are
now available. 

At first sight, the restriction of the AdS curvature radius to sub-micrometer values from table-top experiments might seem like a very strong constraint when expressed in everyday or astrophysical units, but in the context of the primordial black hole dark matter window such scales are still comparatively large. This allows a large parameter space with cosmological and astrophysical signatures that can differ substantially from those expected in the conventional flat (3+1)-dimensional scenario. These differences depend greatly on the mass of the black hole. 

A particularly interesting feature of these models is that for
sufficiently large AdS curvature radii, very light black holes --- with masses as low as $M\sim 10^9\,\text{g}$ for $\ell\sim 10^{-6}\,\text{m}$ --- 
could survive until today.
This stands in contrast to the conventional four-dimensional case,
where such black holes would have evaporated long ago unless additional
effects (e.g.~memory burden) are considered. Moreover, their
emission spectra are qualitatively distinct, and could serve as a
discriminant
between the conventional scenario and higher-dimensional
frameworks if a small black hole were to be detected. 

\begin{figure}
\centering
\includegraphics[width=0.7\linewidth]{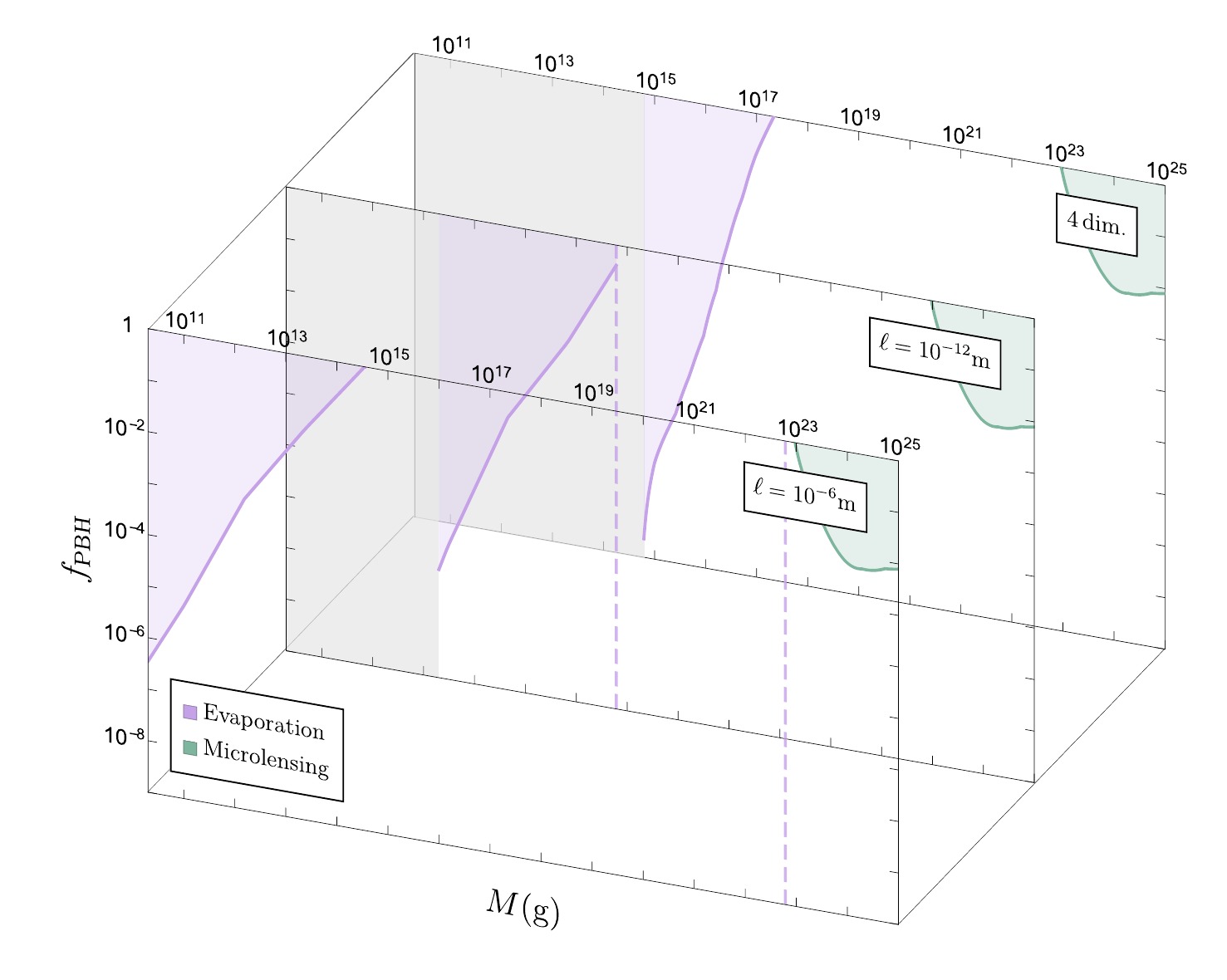}
    \caption[Dark matter window: 5D vs.\ conventional scenario]{Dark matter window for $\ell=10^{-6}\,,\,10^{-12}$m and the conventional scenario \cite{Carr_2021} for a monochromatic PBH distribution. Grey-shaded regions indicate black holes that have already evaporated or are currently rapidly evaporating;
    purple envelopes are ($f_{\scriptscriptstyle\mathrm{PBH}},M$) regions
    excluded by observations related to Hawking evaporation;
    and the green region by the HSC microlensing survey \cite{Smyth_2020}.
    The dashed purple line indicates the limit where $r_0\sim \ell$ for each particular $\ell$ choice, i.e.~only black holes lighter than this threshold can be described by the Schwarzschild--Tangherlini solution.}
    \label{final-plot}
\end{figure}

Due to their lower effective temperatures, higher-dimensional
black holes can either evade some observational bounds
altogether, or significantly reduce their severity,
compared to four-dimensional models.
This is particularly true for bounds associated with particle
emission.
Figure~\ref{final-plot} shows that this broadens the allowed
dark matter window by up to about three orders of magnitude. For context, the width is about six orders of magnitude in the conventional scenario.
Similar observations that a non-conventional geometry
can
significantly change the PBH dark matter parameter space
have been made before; see Refs.~\cite{Johnson_2020, Friedlander:2022ttk}.
In the particular case of RS-II, the enlargement of the DM window
is more pronounced for larger AdS radii. However, bounds obtained
from X-ray detections remain especially significant in this regime
(see Figure~\ref{fig-3d-rad}). Consequently,
the true parameter space is narrower than one might na\"{\i}vely
infer by simply relaxing the most stringent bounds
from the conventional four-dimensional scenario. 

For AdS radii in the range
$10^{-6}\,\text{m} \lesssim \ell \lesssim 10^{-11}\,\text{m}$,
we find that dark matter could (in principle) be composed
entirely of higher-dimensional black holes.
For smaller curvature scales,
evaporation constraints rule out this possibility.
Conversely, the case of $\ell\sim 10^{-6}\,\text{m}$ represents an
extreme situation in which the dark matter window is
almost exclusively higher-dimensional. 

What are the prospects for future PBH detection, and especially
for discriminating between the conventional $(3+1)$-dimensional
model and a braneworld scenario?
This picture here is complex.
Interference patterns produced in microlensing
would serve as a clear discriminant between models.
Although not discussed in this paper,
strong-field lensing is also expected to differ
significantly~\cite{Eiroa:2005vd}.
Unfortunately, as is often the case,
our current observational capabilities impose significant limitations.

We have emphasised the importance of accurately modelling the photon flux.
This is crucial if one aims to obtain a realistic estimate
for astrophysical observables---both in local environments,
and for extragalactic surveys. However,
it should be understood that much of our analysis is still
(semi)-analytic.
Therefore, our results are still a simplified
approximation in which many intricacies
are absent.
For example, we do not fully capture
the details of interactions in the interstellar medium,
or the effect of magnetic fields.
We have also assumed a monochromatic PBH mass spectrum.
Extending our results to a broader mass distribution
is straightforward in principle.
We have chosen the idealised monochromatic scenario for
the purpose of maintaining a manageable parameter
space, and to aid visualisation.
Similar remarks apply to black hole spin, which we have taken to be
zero.
In the standard formation scenario we do not expect
significant angular momentum.
However, in more exotic formation mechanisms,
such as particle collisions in the early universe,
the spin distribution may be more complicated.
We defer all these issues to future work.

    \thispagestyle{fancy}
\pagestyle{fancy}

    
    

\backmatter 

\pagestyle{plain}
\begingroup
\footnotesize
\addcontentsline{toc}{chapter}{Bibliography}
\setlength{\bibitemsep}{2pt}
\setstretch{0.95}
\printbibliography

@article{Riemann1854,
  author = {Riemann, Bernhard},
  title = {Über die Hypothesen, welche der Geometrie zu Grunde liegen},
  year = {1854},
  note = {Habilitationsvortrag; published in 1868}
}

@book{Kragh2012,
  author = {Kragh, Helge},
  title = {Higher Speculations: Grand Theories and Failed Revolutions in Physics and Cosmology},
  publisher = {Oxford University Press},
  year = {2012}
}

@article{Kaluza1921,
  author = {Kaluza, Theodor},
  title = {Zum Unitätsproblem der Physik},
  journal = {Sitzungsberichte der Preussischen Akademie der Wissenschaften},
  pages = {966--972},
  year = {1921}
}

@article{Klein1926,
  author = {Klein, Oskar},
  title = {Quantentheorie und fünfdimensionale Relativitätstheorie},
  journal = {Zeitschrift für Physik},
  volume = {37},
  pages = {895--906},
  year = {1926}
}

@article{OverduinWesson1997,
  author = {Overduin, James M. and Wesson, Paul S.},
  title = {Kaluza-Klein Gravity},
  journal = {Physics Reports},
  volume = {283},
  number = {5-6},
  pages = {303--378},
  year = {1997},
  doi = {10.1016/S0370-1573(96)00046-4}
}

@article{RubakovShaposhnikov1983,
  author = {Rubakov, V. A. and Shaposhnikov, M. E.},
  title = {Do We Live Inside a Domain Wall?},
  journal = {Physics Letters B},
  volume = {125},
  number = {2-3},
  pages = {136--138},
  year = {1983},
  doi = {10.1016/0370-2693(83)91253-4}
}

@article{Rubakov2001,
  author = {Rubakov, V. A.},
  title = {Large and Infinite Extra Dimensions},
  journal = {Physics-Uspekhi},
  volume = {44},
  number = {9},
  pages = {871--893},
  year = {2001},
  doi = {10.1070/PU2001v044n09ABEH001009},
  eprint = {hep-ph/0104152},
  archivePrefix = {arXiv}
}

@article{ArkaniHamedDimopoulosDvali1998,
  author = {Arkani-Hamed, Nima and Dimopoulos, Savas and Dvali, Gia},
  title = {The Hierarchy Problem and New Dimensions at a Millimeter},
  journal = {Physics Letters B},
  volume = {429},
  number = {3-4},
  pages = {263--272},
  year = {1998},
  doi = {10.1016/S0370-2693(98)00466-3},
  eprint = {hep-ph/9803315},
  archivePrefix = {arXiv}
}

@book{GreenSchwarzWitten1987,
  author = {Green, Michael B. and Schwarz, John H. and Witten, Edward},
  title = {Superstring Theory},
  volume = {1},
  publisher = {Cambridge University Press},
  year = {1987}
}

@book{Polchinski1998,
  author = {Polchinski, Joseph},
  title = {String Theory, Volume 1: An Introduction to the Bosonic String},
  publisher = {Cambridge University Press},
  year = {1998}
}

@article{RandallSundrum1999a,
  author = {Randall, Lisa and Sundrum, Raman},
  title = {A Large Mass Hierarchy from a Small Extra Dimension},
  journal = {Physical Review Letters},
  volume = {83},
  number = {17},
  pages = {3370--3373},
  year = {1999},
  doi = {10.1103/PhysRevLett.83.3370},
  eprint = {hep-ph/9905221},
  archivePrefix = {arXiv}
}

@article{RandallSundrum1999b,
  author = {Randall, Lisa and Sundrum, Raman},
  title = {An Alternative to Compactification},
  journal = {Physical Review Letters},
  volume = {83},
  number = {23},
  pages = {4690--4693},
  year = {1999},
  doi = {10.1103/PhysRevLett.83.4690},
  eprint = {hep-th/9906064},
  archivePrefix = {arXiv}
}

@article{tHooft1993,
  author = {'t Hooft, Gerard},
  title = {Dimensional Reduction in Quantum Gravity},
  eprint = {gr-qc/9310026},
  archivePrefix = {arXiv},
  year = {1993}
}

@article{Susskind1995,
  author = {Susskind, Leonard},
  title = {The World as a Hologram},
  journal = {Journal of Mathematical Physics},
  volume = {36},
  pages = {6377--6396},
  year = {1995},
  doi = {10.1063/1.531249},
  eprint = {hep-th/9409089},
  archivePrefix = {arXiv}
}

@article{Maldacena1998,
  author = {Maldacena, Juan Martin},
  title = {The Large N Limit of Superconformal Field Theories and Supergravity},
  journal = {Advances in Theoretical and Mathematical Physics},
  volume = {2},
  pages = {231--252},
  year = {1998},
  doi = {10.4310/ATMP.1998.v2.n2.a1},
  eprint = {hep-th/9711200},
  archivePrefix = {arXiv}
}

@article{Witten1998,
  author = {Witten, Edward},
  title = {Anti de Sitter Space and Holography},
  journal = {Advances in Theoretical and Mathematical Physics},
  volume = {2},
  number = {2},
  pages = {253--291},
  year = {1998},
  doi = {10.4310/ATMP.1998.v2.n2.a2},
  eprint = {hep-th/9802150},
  archivePrefix = {arXiv}
}

@article{Israel:1966rt,
    author = "Israel, W.",
    title = "{Singular hypersurfaces and thin shells in general relativity}",
    doi = "10.1007/BF02710419",
    journal = "Nuovo Cim. B",
    volume = "44S10",
    pages = "1",
    year = "1966",
    note = "[Erratum: Nuovo Cim.B 48, 463 (1967)]"
}

@article{Goldberger:1999uk,
    author = "Goldberger, Walter D. and Wise, Mark B.",
    title = "{Modulus stabilization with bulk fields}",
    eprint = "hep-ph/9907447",
    archivePrefix = "arXiv",
    reportNumber = "CALT-68-2232",
    doi = "10.1103/PhysRevLett.83.4922",
    journal = "Phys. Rev. Lett.",
    volume = "83",
    pages = "4922--4925",
    year = "1999"
}

@article{Binetruy:1999ut,
    author = "Binetruy, Pierre and Deffayet, Cedric and Langlois, David",
    title = "{Nonconventional cosmology from a brane universe}",
    eprint = "hep-th/9905012",
    archivePrefix = "arXiv",
    reportNumber = "LPT-ORSAY-99-25",
    doi = "10.1016/S0550-3213(99)00696-3",
    journal = "Nucl. Phys. B",
    volume = "565",
    pages = "269--287",
    year = "2000"
}

@article{Walker:1991ap,
    author = "Walker, Terry P. and Steigman, Gary and Schramm, David N. and Olive, Keith A. and Kang, Ho-Shik",
    title = "{Primordial Nucleosynthesis Redux}",
    reportNumber = "OSU-TA-10-90, UMN-TH-826-90, FERMILAB-PUB-91-036-A",
    doi = "10.1086/170255",
    journal = "Astrophys. J.",
    volume = "376",
    pages = "51--69",
    year = "1991"
}

@article{Cline_1999,
   title={Cosmological Expansion in the Presence of an Extra Dimension},
   volume={83},
   ISSN={1079-7114},
   url={http://dx.doi.org/10.1103/PhysRevLett.83.4245},
   DOI={10.1103/physrevlett.83.4245},
   number={21},
   journal={Physical Review Letters},
   publisher={American Physical Society (APS)},
   author={Cline, James M. and Grojean, Christophe and Servant, Géraldine},
   year={1999},
   month=Nov, pages={4245–4248} }

@article{Cs_ki_1999,
   title={Cosmology of one extra dimension with localized gravity},
   volume={462},
   ISSN={0370-2693},
   url={http://dx.doi.org/10.1016/S0370-2693(99)00896-5},
   DOI={10.1016/s0370-2693(99)00896-5},
   number={1-2},
   journal={Physics Letters B},
   publisher={Elsevier BV},
   author={Csáki, Csaba and Graesser, Michael and Kolda, Christopher and Terning, John},
   year={1999},
   month=sep, pages={34–40} }

@article{Bowcock_2000,
   title={General brane cosmologies and their global spacetime structure},
   volume={17},
   ISSN={1361-6382},
   url={http://dx.doi.org/10.1088/0264-9381/17/22/313},
   DOI={10.1088/0264-9381/17/22/313},
   number={22},
   journal={Classical and Quantum Gravity},
   publisher={IOP Publishing},
   author={Bowcock, Peter and Charmousis, Christos and Gregory, Ruth},
   year={2000},
   month=Oct, pages={4745–4763} }

@article{Kraus_1999,
   title={Dynamics of anti-de Sitter domain walls},
   volume={1999},
   ISSN={1029-8479},
   url={http://dx.doi.org/10.1088/1126-6708/1999/12/011},
   DOI={10.1088/1126-6708/1999/12/011},
   number={12},
   journal={Journal of High Energy Physics},
   publisher={Springer Science and Business Media LLC},
   author={Kraus, Per},
   year={1999},
   month=Dec, pages={011–011} }

@article{Ida_2000,
   title={Brane-world cosmology},
   volume={2000},
   ISSN={1029-8479},
   url={http://dx.doi.org/10.1088/1126-6708/2000/09/014},
   DOI={10.1088/1126-6708/2000/09/014},
   number={09},
   journal={Journal of High Energy Physics},
   publisher={Springer Science and Business Media LLC},
   author={Ida, Daisuke},
   year={2000},
   month=sep, pages={014–014} }

@article{Mukohyama:1999wi,
    author = "Mukohyama, Shinji and Shiromizu, Tetsuya and Maeda, Kei-ichi",
    title = "{Global structure of exact cosmological solutions in the brane world}",
    eprint = "hep-th/9912287",
    archivePrefix = "arXiv",
    reportNumber = "CITA-99-44, DAMTP-1999-177, RESCEU-50-99, UTAP-357",
    doi = "10.1103/PhysRevD.63.029901",
    journal = "Phys. Rev. D",
    volume = "62",
    pages = "024028",
    year = "2000",
    note = "[Erratum: Phys.Rev.D 63, 029901 (2001)]"
}

@article{Gregory_2002,
   title={Exact braneworld cosmology induced from bulk black holes},
   volume={19},
   ISSN={0264-9381},
   url={http://dx.doi.org/10.1088/0264-9381/19/15/313},
   DOI={10.1088/0264-9381/19/15/313},
   number={15},
   journal={Classical and Quantum Gravity},
   publisher={IOP Publishing},
   author={Gregory, James P and Padilla, Antonio},
   year={2002},
   month=jul, pages={4071–4083} }

@article{Shiromizu_2000,
   title={The Einstein equations on the 3-brane world},
   volume={62},
   ISSN={1089-4918},
   url={http://dx.doi.org/10.1103/PhysRevD.62.024012},
   DOI={10.1103/physrevd.62.024012},
   number={2},
   journal={Physical Review D},
   publisher={American Physical Society (APS)},
   author={Shiromizu, Tetsuya and Maeda, Kei-ichi and Sasaki, Misao},
   year={2000},
   month=jun }

@book{wald1984general,
  title     = {General Relativity},
  author    = {Wald, Robert M.},
  year      = {1984},
  publisher = {University of Chicago Press},
  address   = {Chicago, IL},
  isbn      = {978-0-226-87033-5}
}

@article{van_de_Bruck_2000,
   title={Cosmological perturbations in brane-world theories: Formalism},
   volume={62},
   ISSN={1089-4918},
   url={http://dx.doi.org/10.1103/PhysRevD.62.123515},
   DOI={10.1103/physrevd.62.123515},
   number={12},
   journal={Physical Review D},
   publisher={American Physical Society (APS)},
   author={van de Bruck, Carsten and Dorca, Miquel and Brandenberger, Robert H. and Lukas, André},
   year={2000},
   month=Nov }

@article{PhysRevD.62.124009,
  title = {Dilaton gravity on the brane},
  author = {Maeda, Kei-ichi and Wands, David},
  journal = {Phys. Rev. D},
  volume = {62},
  issue = {12},
  pages = {124009},
  numpages = {9},
  year = {2000},
  month = nov,
  publisher = {American Physical Society},
  doi = {10.1103/PhysRevD.62.124009},
  url = {https://link.aps.org/doi/10.1103/PhysRevD.62.124009}
}

@article{Sasankan_2017-1,
   title={New observational limits on dark radiation in braneworld cosmology},
   volume={95},
   ISSN={2470-0029},
   url={http://dx.doi.org/10.1103/PhysRevD.95.083516},
   DOI={10.1103/physrevd.95.083516},
   number={8},
   journal={Physical Review D},
   publisher={American Physical Society (APS)},
   author={Sasankan, Nishanth and Gangopadhyay, Mayukh Raj and Mathews, Grant J. and Kusakabe, Motohiko},
   year={2017},
   month=apr }

@article{Sasankan_2017-2,
   title={Limits on brane-world and particle dark radiation from big bang nucleosynthesis and the CMB},
   volume={26},
   ISSN={1793-6608},
   url={http://dx.doi.org/10.1142/S0218301317410075},
   DOI={10.1142/s0218301317410075},
   number={08},
   journal={International Journal of Modern Physics E},
   publisher={World Scientific Pub Co Pte Lt},
   author={Sasankan, N. and Gangopadhyay, Mayukh R. and Mathews, G. J. and Kusakabe, M.},
   year={2017},
   month=aug, pages={1741007} }

@book{Brillouin1960,
  author    = {Léon Brillouin},
  title     = {Wave Propagation and Group Velocity},
  series    = {Pure and Applied Physics},
  volume    = {8},
  publisher = {Academic Press},
  address   = {New York},
  year      = {1960},
  isbn      = {0121349683}
}

@article{Maartens_2010,
   title={Brane-World Gravity},
   volume={13},
   ISSN={1433-8351},
   url={http://dx.doi.org/10.12942/lrr-2010-5},
   DOI={10.12942/lrr-2010-5},
   number={1},
   journal={Living Reviews in Relativity},
   publisher={Springer Science and Business Media LLC},
   author={Maartens, Roy and Koyama, Kazuya},
   year={2010},
   month=sep }

@article{Langlois_2001,
   title={Large-scale cosmological perturbations on the brane},
   volume={63},
   ISSN={1089-4918},
   url={http://dx.doi.org/10.1103/PhysRevD.63.084009},
   DOI={10.1103/physrevd.63.084009},
   number={8},
   journal={Physical Review D},
   publisher={American Physical Society (APS)},
   author={Langlois, David and Maartens, Roy and Sasaki, Misao and Wands, David},
   year={2001},
   month=mar }

@article{Kodama:1984ziu,
    author = "Kodama, Hideo and Sasaki, Misao",
    title = "{Cosmological Perturbation Theory}",
    doi = "10.1143/PTPS.78.1",
    journal = "Prog. Theor. Phys. Suppl.",
    volume = "78",
    pages = "1--166",
    year = "1984"
}

@article{Kodama_2000,
   title={Brane world cosmology: Gauge-invariant formalism for perturbation},
   volume={62},
   ISSN={1089-4918},
   url={http://dx.doi.org/10.1103/PhysRevD.62.064022},
   DOI={10.1103/physrevd.62.064022},
   number={6},
   journal={Physical Review D},
   publisher={American Physical Society (APS)},
   author={Kodama, Hideo and Ishibashi, Akihiro and Seto, Osamu},
   year={2000},
   month=Aug }

@article{Cardoso_2007,
   title={Scalar perturbations in braneworld cosmology},
   volume={2007},
   ISSN={1475-7516},
   url={http://dx.doi.org/10.1088/1475-7516/2007/07/008},
   DOI={10.1088/1475-7516/2007/07/008},
   number={07},
   journal={Journal of Cosmology and Astroparticle Physics},
   publisher={IOP Publishing},
   author={Cardoso, Antonio and Hiramatsu, Takashi and Koyama, Kazuya and Seahra, Sanjeev S},
   year={2007},
   month=jul, pages={008–008} }

@ARTICLE{2002PhRvD,
       author = {{Deffayet}, C{\'e}dric},
        title = "{On brane world cosmological perturbations}",
      journal = {\prd},
         year = 2002,
        month = nov,
       volume = {66},
       number = {10},
          eid = {103504},
        pages = {103504},
          doi = {10.1103/PhysRevD.66.103504},
archivePrefix = {arXiv},
       eprint = {hep-th/0205084},
 primaryClass = {hep-th},
       adsurl = {https://ui.adsabs.harvard.edu/abs/2002PhRvD..66j3504D}
}

@article{Mukohyama_2000,
   title={Gauge-invariant gravitational perturbations of maximally symmetric spacetimes},
   volume={62},
   ISSN={1089-4918},
   url={http://dx.doi.org/10.1103/PhysRevD.62.084015},
   DOI={10.1103/physrevd.62.084015},
   number={8},
   journal={Physical Review D},
   publisher={American Physical Society (APS)},
   author={Mukohyama, Shinji},
   year={2000},
   month=sep }

@article{Tangherlini:1963bw,
    author = "Tangherlini, F. R.",
    title = "{Schwarzschild field in n dimensions and the dimensionality of space problem}",
    doi = "10.1007/BF02784569",
    journal = "Nuovo Cim.",
    volume = "27",
    pages = "636--651",
    year = "1963"
}

@article{Myers:1986un,
    author = "Myers, Robert C. and Perry, M. J.",
    title = "{Black Holes in Higher Dimensional Space-Times}",
    reportNumber = "PRINT-86-0067 (PRINCETON)",
    doi = "10.1016/0003-4916(86)90186-7",
    journal = "Annals Phys.",
    volume = "172",
    pages = "304",
    year = "1986"
}

@article{Giddings_2002,
   title={High energy colliders as black hole factories: The end of short distance physics},
   volume={65},
   ISSN={1089-4918},
   url={http://dx.doi.org/10.1103/PhysRevD.65.056010},
   DOI={10.1103/physrevd.65.056010},
   number={5},
   journal={Physical Review D},
   publisher={American Physical Society (APS)},
   author={Giddings, Steven B. and Thomas, Scott},
   year={2002},
   month=Feb }

@article{Dimopoulos_2001,
   title={Black Holes at the Large Hadron Collider},
   volume={87},
   ISSN={1079-7114},
   url={http://dx.doi.org/10.1103/PhysRevLett.87.161602},
   DOI={10.1103/physrevlett.87.161602},
   number={16},
   journal={Physical Review Letters},
   publisher={American Physical Society (APS)},
   author={Dimopoulos, Savas and Landsberg, Greg},
   year={2001},
   month=sep }

@article{Dimopoulos_2002,
   title={String balls at the LHC and beyond},
   volume={526},
   ISSN={0370-2693},
   url={http://dx.doi.org/10.1016/S0370-2693(01)01525-8},
   DOI={10.1016/s0370-2693(01)01525-8},
   number={3-4},
   journal={Physics Letters B},
   publisher={Elsevier BV},
   author={Dimopoulos, Savas and Emparan, Roberto},
   year={2002},
   month=Feb, pages={393–398} }

@article{Cheung_2002-a,
   title={Black hole, string ball, and $p$-brane production at hadronic supercolliders},
   volume={66},
   ISSN={1089-4918},
   url={http://dx.doi.org/10.1103/PhysRevD.66.036007},
   DOI={10.1103/physrevd.66.036007},
   number={3},
   journal={Physical Review D},
   publisher={American Physical Society (APS)},
   author={Cheung, Kingman},
   year={2002},
   month=Aug }

@article{Cheung_2002,
   title={Black Hole Production and Large Extra Dimensions},
   volume={88},
   ISSN={1079-7114},
   url={http://dx.doi.org/10.1103/PhysRevLett.88.221602},
   DOI={10.1103/physrevlett.88.221602},
   number={22},
   journal={Physical Review Letters},
   publisher={American Physical Society (APS)},
   author={Cheung, Kingman},
   year={2002},
   month=May }

@article{Park:2001xc,
    author = "Park, Seong Chan and Song, H. S.",
    title = "{Production of spinning black holes at colliders}",
    eprint = "hep-ph/0111069",
    archivePrefix = "arXiv",
    reportNumber = "SNUTP-01-038",
    journal = "J. Korean Phys. Soc.",
    volume = "43",
    pages = "30--33",
    year = "2003"
}

@article{Giudice_2002,
   title={Transplanckian collisions at the LHC and beyond},
   volume={630},
   ISSN={0550-3213},
   url={http://dx.doi.org/10.1016/S0550-3213(02)00142-6},
   DOI={10.1016/s0550-3213(02)00142-6},
   number={1-2},
   journal={Nuclear Physics B},
   publisher={Elsevier BV},
   author={Giudice, Gian F. and Rattazzi, Riccardo and Wells, James D.},
   year={2002},
   month=May, pages={293–325} }

@article{CASADIO_2002,
   title={Can black holes and naked singularities be detected in accelerators?},
   volume={17},
   ISSN={1793-656X},
   url={http://dx.doi.org/10.1142/S0217751X02012259},
   DOI={10.1142/s0217751x02012259},
   number={31},
   journal={International Journal of Modern Physics A},
   publisher={World Scientific Pub Co Pte Lt},
   author={Casadio, Roberto and Harms, Benjamin},
   year={2002},
   month=Dec, pages={4635–4645} }

@article{Mocioiu:2003gi,
    author = "Mocioiu, Irina and Nara, Yasushi and Sarcevic, Ina",
    title = "{Hadrons as signature of black hole production at the LHC}",
    eprint = "hep-ph/0301073",
    archivePrefix = "arXiv",
    doi = "10.1016/S0370-2693(03)00181-3",
    journal = "Phys. Lett. B",
    volume = "557",
    pages = "87--93",
    year = "2003"
}

@article{Kanti:2004nr,
    author = "Kanti, Panagiota",
    title = "{Black holes in theories with large extra dimensions: A Review}",
    eprint = "hep-ph/0402168",
    archivePrefix = "arXiv",
    doi = "10.1142/S0217751X04018324",
    journal = "Int. J. Mod. Phys. A",
    volume = "19",
    pages = "4899--4951",
    year = "2004"
}

@article{Emparan_2008,
   title={Black Holes in Higher Dimensions},
   volume={11},
   ISSN={1433-8351},
   url={http://dx.doi.org/10.12942/lrr-2008-6},
   DOI={10.12942/lrr-2008-6},
   number={1},
   journal={Living Reviews in Relativity},
   publisher={Springer Science and Business Media LLC},
   author={Emparan, Roberto and Reall, Harvey S.},
   year={2008},
   month=sep }

@article{2016,
   title={Search for strong gravity in multijet final states produced in pp collisions at $ \sqrt{s}=13 $
               TeV using the ATLAS detector at the LHC},
   volume={2016},
   ISSN={1029-8479},
   url={http://dx.doi.org/10.1007/JHEP03(2016)026},
   DOI={10.1007/jhep03(2016)026},
   number={3},
   journal={Journal of High Energy Physics},
   publisher={Springer Science and Business Media LLC},
   author={ and Aad, G. and Abbott, B. and Abdallah, J. and Abdinov, O. and Abeloos, B. and Aben, R. and Abolins, M. and AbouZeid, O. S. and Abramowicz, H. and Abreu, H. and Abreu, R. and Abulaiti, Y. and Acharya, B. S. and Adamczyk, L. and Adams, D. L. and Adelman, J. and Adomeit, S. and Adye, T. and Affolder, A. A. and Agatonovic-Jovin, T. and Agricola, J. and Aguilar-Saavedra, J. A. and Ahlen, S. P. and Ahmadov, F. and Aielli, G. and Akerstedt, H. and Åkesson, T. P. A. and Akimov, A. V. and Alberghi, G. L. and Albert, J. and Albrand, S. and Alconada Verzini, M. J. and Aleksa, M. and Aleksandrov, I. N. and Alexa, C. and Alexander, G. and Alexopoulos, T. and Alhroob, M. and Alimonti, G. and Alio, L. and Alison, J. and Alkire, S. P. and Allbrooke, B. M. M. and Allen, B. W. and Allport, P. P. and Aloisio, A. and Alonso, A. and Alonso, F. and Alpigiani, C. and Alvarez Gonzalez, B. and Álvarez Piqueras, D. and Alviggi, M. G. and Amadio, B. T. and Amako, K. and Amaral Coutinho, Y. and Amelung, C. and Amidei, D. and Amor Dos Santos, S. P. and Amorim, A. and Amoroso, S. and Amram, N. and Amundsen, G. and Anastopoulos, C. and Ancu, L. S. and Andari, N. and Andeen, T. and Anders, C. F. and Anders, G. and Anders, J. K. and Anderson, K. J. and Andreazza, A. and Andrei, V. and Angelidakis, S. and Angelozzi, I. and Anger, P. and Angerami, A. and Anghinolfi, F. and Anisenkov, A. V. and Anjos, N. and Annovi, A. and Antonelli, M. and Antonov, A. and Antos, J. and Anulli, F. and Aoki, M. and Aperio Bella, L. and Arabidze, G. and Arai, Y. and Araque, J. P. and Arce, A. T. H. and Arduh, F. A. and Arguin, J-F. and Argyropoulos, S. and Arik, M. and Armbruster, A. J. and Armitage, L. J. and Arnaez, O. and Arnold, H. and Arratia, M. and Arslan, O. and Artamonov, A. and Artoni, G. and Artz, S. and Asai, S. and Asbah, N. and Ashkenazi, A. and Åsman, B. and Asquith, L. and Assamagan, K. and Astalos, R. and Atkinson, M. and Atlay, N. B. and Augsten, K. and Avolio, G. and Axen, B. and Ayoub, M. K. and Azuelos, G. and Baak, M. A. and Baas, A. E. and Baca, M. J. and Bachacou, H. and Bachas, K. and Backes, M. and Backhaus, M. and Bagiacchi, P. and Bagnaia, P. and Bai, Y. and Baines, J. T. and Baker, O. K. and Baldin, E. M. and Balek, P. and Balestri, T. and Balli, F. and Balunas, W. K. and Banas, E. and Banerjee, Sw. and Bannoura, A. A. E. and Barak, L. and Barberio, E. L. and Barberis, D. and Barbero, M. and Barillari, T. and Barisonzi, M. and Barklow, T. and Barlow, N. and Barnes, S. L. and Barnett, B. M. and Barnett, R. M. and Barnovska, Z. and Baroncelli, A. and Barone, G. and Barr, A. J. and Barranco Navarro, L. and Barreiro, F. and Barreiro Guimarães da Costa, J. and Bartoldus, R. and Barton, A. E. and Bartos, P. and Basalaev, A. and Bassalat, A. and Basye, A. and Bates, R. L. and Batista, S. J. and Batley, J. R. and Battaglia, M. and Bauce, M. and Bauer, F. and Bawa, H. S. and Beacham, J. B. and Beattie, M. D. and Beau, T. and Beauchemin, P. H. and Beccherle, R. and Bechtle, P. and Beck, H. P. and Becker, K. and Becker, M. and Beckingham, M. and Becot, C. and Beddall, A. J. and Beddall, A. and Bednyakov, V. A. and Bedognetti, M. and Bee, C. P. and Beemster, L. J. and Beermann, T. A. and Begel, M. and Behr, J. K. and Belanger-Champagne, C. and Bell, A. S. and Bell, W. H. and Bella, G. and Bellagamba, L. and Bellerive, A. and Bellomo, M. and Belotskiy, K. and Beltramello, O. and Benary, O. and Benchekroun, D. and Bender, M. and Bendtz, K. and Benekos, N. and Benhammou, Y. and Benhar Noccioli, E. and Benitez, J. and Benitez Garcia, J. A. and Benjamin, D. P. and Bensinger, J. R. and Bentvelsen, S. and Beresford, L. and Beretta, M. and Berge, D. and Bergeaas Kuutmann, E. and Berger, N. and Berghaus, F. and Beringer, J. and Bernard, C. and Bernard, N. R. and Bernius, C. and Bernlochner, F. U. and Berry, T. and Berta, P. and Bertella, C. and Bertoli, G. and Bertolucci, F. and Bertsche, C. and Bertsche, D. and Besjes, G. J. and Bessidskaia Bylund, O. and Bessner, M. and Besson, N. and Betancourt, C. and Bethke, S. and Bevan, A. J. and Bhimji, W. and Bianchi, R. M. and Bianchini, L. and Bianco, M. and Biebel, O. and Biedermann, D. and Bielski, R. and Biesuz, N. V. and Biglietti, M. and Bilbao De Mendizabal, J. and Bilokon, H. and Bindi, M. and Binet, S. and Bingul, A. and Bini, C. and Biondi, S. and Bjergaard, D. M. and Black, C. W. and Black, J. E. and Black, K. M. and Blackburn, D. and Blair, R. E. and Blanchard, J.-B. and Blanco, J. E. and Blazek, T. and Bloch, I. and Blocker, C. and Blum, W. and Blumenschein, U. and Blunier, S. and Bobbink, G. J. and Bobrovnikov, V. S. and Bocchetta, S. S. and Bocci, A. and Bock, C. and Boehler, M. and Boerner, D. and Bogaerts, J. A. and Bogavac, D. and Bogdanchikov, A. G. and Bohm, C. and Boisvert, V. and Bold, T. and Boldea, V. and Boldyrev, A. S. and Bomben, M. and Bona, M. and Boonekamp, M. and Borisov, A. and Borissov, G. and Bortfeldt, J. and Bortoletto, D. and Bortolotto, V. and Bos, K. and Boscherini, D. and Bosman, M. and Bossio Sola, J. D. and Boudreau, J. and Bouffard, J. and Bouhova-Thacker, E. V. and Boumediene, D. and Bourdarios, C. and Bousson, N. and Boutle, S. K. and Boveia, A. and Boyd, J. and Boyko, I. R. and Bracinik, J. and Brandt, A. and Brandt, G. and Brandt, O. and Bratzler, U. and Brau, B. and Brau, J. E. and Braun, H. M. and Breaden Madden, W. D. and Brendlinger, K. and Brennan, A. J. and Brenner, L. and Brenner, R. and Bressler, S. and Bristow, T. M. and Britton, D. and Britzger, D. and Brochu, F. M. and Brock, I. and Brock, R. and Brooijmans, G. and Brooks, T. and Brooks, W. K. and Brosamer, J. and Brost, E. and Bruckman de Renstrom, P. A. and Bruncko, D. and Bruneliere, R. and Bruni, A. and Bruni, G. and Brunt, BH and Bruschi, M. and Bruscino, N. and Bryant, P. and Bryngemark, L. and Buanes, T. and Buat, Q. and Buchholz, P. and Buckley, A. G. and Budagov, I. A. and Buehrer, F. and Bugge, M. K. and Bulekov, O. and Bullock, D. and Burckhart, H. and Burdin, S. and Burgard, C. D. and Burghgrave, B. and Burka, K. and Burke, S. and Burmeister, I. and Busato, E. and Büscher, D. and Büscher, V. and Bussey, P. and Butler, J. M. and Butt, A. I. and Buttar, C. M. and Butterworth, J. M. and Butti, P. and Buttinger, W. and Buzatu, A. and Buzykaev, A. R. and Cabrera Urbán, S. and Caforio, D. and Cairo, V. M. and Cakir, O. and Calace, N. and Calafiura, P. and Calandri, A. and Calderini, G. and Calfayan, P. and Caloba, L. P. and Calvet, D. and Calvet, S. and Calvet, T. P. and Camacho Toro, R. and Camarda, S. and Camarri, P. and Cameron, D. and Caminal Armadans, R. and Camincher, C. and Campana, S. and Campanelli, M. and Campoverde, A. and Canale, V. and Canepa, A. and Cano Bret, M. and Cantero, J. and Cantrill, R. and Cao, T. and Capeans Garrido, M. D. M. and Caprini, I. and Caprini, M. and Capua, M. and Caputo, R. and Carbone, R. M. and Cardarelli, R. and Cardillo, F. and Carli, T. and Carlino, G. and Carminati, L. and Caron, S. and Carquin, E. and Carrillo-Montoya, G. D. and Carter, J. R. and Carvalho, J. and Casadei, D. and Casado, M. P. and Casolino, M. and Casper, D. W. and Castaneda-Miranda, E. and Castelli, A. and Castillo Gimenez, V. and Castro, N. F. and Catinaccio, A. and Catmore, J. R. and Cattai, A. and Caudron, J. and Cavaliere, V. and Cavalli, D. and Cavalli-Sforza, M. and Cavasinni, V. and Ceradini, F. and Cerda Alberich, L. and Cerio, B. C. and Cerqueira, A. S. and Cerri, A. and Cerrito, L. and Cerutti, F. and Cerv, M. and Cervelli, A. and Cetin, S. A. and Chafaq, A. and Chakraborty, D. and Chalupkova, I. and Chan, Y. L. and Chang, P. and Chapman, J. D. and Charlton, D. G. and Chau, C. C. and Chavez Barajas, C. A. and Che, S. and Cheatham, S. and Chegwidden, A. and Chekanov, S. and Chekulaev, S. V. and Chelkov, G. A. and Chelstowska, M. A. and Chen, C. and Chen, H. and Chen, K. and Chen, S. and Chen, S. and Chen, X. and Chen, Y. and Cheng, H. C. and Cheng, Y. and Cheplakov, A. and Cheremushkina, E. and Cherkaoui El Moursli, R. and Chernyatin, V. and Cheu, E. and Chevalier, L. and Chiarella, V. and Chiarelli, G. and Chiodini, G. and Chisholm, A. S. and Chislett, R. T. and Chitan, A. and Chizhov, M. V. and Choi, K. and Chouridou, S. and Chow, B. K. B. and Christodoulou, V. and Chromek-Burckhart, D. and Chudoba, J. and Chuinard, A. J. and Chwastowski, J. J. and Chytka, L. and Ciapetti, G. and Ciftci, A. K. and Cinca, D. and Cindro, V. and Cioara, I. A. and Ciocio, A. and Cirotto, F. and Citron, Z. H. and Ciubancan, M. and Clark, A. and Clark, B. L. and Clark, P. J. and Clarke, R. N. and Clement, C. and Coadou, Y. and Cobal, M. and Coccaro, A. and Cochran, J. and Coffey, L. and Colasurdo, L. and Cole, B. and Cole, S. and Colijn, A. P. and Collot, J. and Colombo, T. and Compostella, G. and Conde Muiño, P. and Coniavitis, E. and Connell, S. H. and Connelly, I. A. and Consorti, V. and Constantinescu, S. and Conta, C. and Conti, G. and Conventi, F. and Cooke, M. and Cooper, B. D. and Cooper-Sarkar, A. M. and Cornelissen, T. and Corradi, M. and Corriveau, F. and Corso-Radu, A. and Cortes-Gonzalez, A. and Cortiana, G. and Costa, G. and Costa, M. J. and Costanzo, D. and Cottin, G. and Cowan, G. and Cox, B. E. and Cranmer, K. and Crawley, S. J. and Cree, G. and Crépé-Renaudin, S. and Crescioli, F. and Cribbs, W. A. and Crispin Ortuzar, M. and Cristinziani, M. and Croft, V. and Crosetti, G. and Cuhadar Donszelmann, T. and Cummings, J. and Curatolo, M. and Cúth, J. and Cuthbert, C. and Czirr, H. and Czodrowski, P. and D’Auria, S. and D’Onofrio, M. and Da Cunha Sargedas De Sousa, M. J. and Da Via, C. and Dabrowski, W. and Dai, T. and Dale, O. and Dallaire, F. and Dallapiccola, C. and Dam, M. and Dandoy, J. R. and Dang, N. P. and Daniells, A. C. and Danninger, M. and Dano Hoffmann, M. and Dao, V. and Darbo, G. and Darmora, S. and Dassoulas, J. and Dattagupta, A. and Davey, W. and David, C. and Davidek, T. and Davies, M. and Davison, P. and Davygora, Y. and Dawe, E. and Dawson, I. and Daya-Ishmukhametova, R. K. and De, K. and de Asmundis, R. and De Benedetti, A. and De Castro, S. and De Cecco, S. and De Groot, N. and de Jong, P. and De la Torre, H. and De Lorenzi, F. and De Pedis, D. and De Salvo, A. and De Sanctis, U. and De Santo, A. and De Vivie De Regie, J. B. and Dearnaley, W. J. and Debbe, R. and Debenedetti, C. and Dedovich, D. V. and Deigaard, I. and Del Peso, J. and Del Prete, T. and Delgove, D. and Deliot, F. and Delitzsch, C. M. and Deliyergiyev, M. and Dell’Acqua, A. and Dell’Asta, L. and Dell’Orso, M. and Della Pietra, M. and della Volpe, D. and Delmastro, M. and Delsart, P. A. and Deluca, C. and DeMarco, D. A. and Demers, S. and Demichev, M. and Demilly, A. and Denisov, S. P. and Denysiuk, D. and Derendarz, D. and Derkaoui, J. E. and Derue, F. and Dervan, P. and Desch, K. and Deterre, C. and Dette, K. and Deviveiros, P. O. and Dewhurst, A. and Dhaliwal, S. and Di Ciaccio, A. and Di Ciaccio, L. and Di Clemente, W. K. and Di Domenico, A. and Di Donato, C. and Di Girolamo, A. and Di Girolamo, B. and Di Mattia, A. and Di Micco, B. and Di Nardo, R. and Di Simone, A. and Di Sipio, R. and Di Valentino, D. and Diaconu, C. and Diamond, M. and Dias, F. A. and Diaz, M. A. and Dickinson, J. and Diehl, E. B. and Dietrich, J. and Diglio, S. and Dimitrievska, A. and Dingfelder, J. and Dita, P. and Dita, S. and Dittus, F. and Djama, F. and Djobava, T. and Djuvsland, J. I. and do Vale, M. A. B. and Dobos, D. and Dobre, M. and Doglioni, C. and Dohmae, T. and Dolejsi, J. and Dolezal, Z. and Dolgoshein, B. A. and Donadelli, M. and Donati, S. and Dondero, P. and Donini, J. and Dopke, J. and Doria, A. and Dova, M. T. and Doyle, A. T. and Drechsler, E. and Dris, M. and Du, Y. and Duarte-Campderros, J. and Duchovni, E. and Duckeck, G. and Ducu, O. A. and Duda, D. and Dudarev, A. and Duflot, L. and Duguid, L. and Dührssen, M. and Dunford, M. and Duran Yildiz, H. and Düren, M. and Durglishvili, A. and Duschinger, D. and Dutta, B. and Dyndal, M. and Eckardt, C. and Ecker, K. M. and Edgar, R. C. and Edson, W. and Edwards, N. C. and Eifert, T. and Eigen, G. and Einsweiler, K. and Ekelof, T. and El Kacimi, M. and Ellajosyula, V. and Ellert, M. and Elles, S. and Ellinghaus, F. and Elliot, A. A. and Ellis, N. and Elmsheuser, J. and Elsing, M. and Emeliyanov, D. and Enari, Y. and Endner, O. C. and Endo, M. and Ennis, J. S. and Erdmann, J. and Ereditato, A. and Ernis, G. and Ernst, J. and Ernst, M. and Errede, S. and Ertel, E. and Escalier, M. and Esch, H. and Escobar, C. and Esposito, B. and Etienvre, A. I. and Etzion, E. and Evans, H. and Ezhilov, A. and Fabbri, L. and Facini, G. and Fakhrutdinov, R. M. and Falciano, S. and Falla, R. J. and Faltova, J. and Fang, Y. and Fanti, M. and Farbin, A. and Farilla, A. and Farina, C. and Farooque, T. and Farrell, S. and Farrington, S. M. and Farthouat, P. and Fassi, F. and Fassnacht, P. and Fassouliotis, D. and Faucci Giannelli, M. and Favareto, A. and Fayard, L. and Fedin, O. L. and Fedorko, W. and Feigl, S. and Feligioni, L. and Feng, C. and Feng, E. J. and Feng, H. and Fenyuk, A. B. and Feremenga, L. and Fernandez Martinez, P. and Fernandez Perez, S. and Ferrando, J. and Ferrari, A. and Ferrari, P. and Ferrari, R. and Ferreira de Lima, D. E. and Ferrer, A. and Ferrere, D. and Ferretti, C. and Ferretto Parodi, A. and Fiedler, F. and Filipčič, A. and Filipuzzi, M. and Filthaut, F. and Fincke-Keeler, M. and Finelli, K. D. and Fiolhais, M. C. N. and Fiorini, L. and Firan, A. and Fischer, A. and Fischer, C. and Fischer, J. and Fisher, W. C. and Flaschel, N. and Fleck, I. and Fleischmann, P. and Fletcher, G. T. and Fletcher, G. and Fletcher, R. R. M. and Flick, T. and Floderus, A. and Flores Castillo, L. R. and Flowerdew, M. J. and Forcolin, G. T. and Formica, A. and Forti, A. and Fournier, D. and Fox, H. and Fracchia, S. and Francavilla, P. and Franchini, M. and Francis, D. and Franconi, L. and Franklin, M. and Frate, M. and Fraternali, M. and Freeborn, D. and Fressard-Batraneanu, S. M. and Friedrich, F. and Froidevaux, D. and Frost, J. A. and Fukunaga, C. and Fullana Torregrosa, E. and Fusayasu, T. and Fuster, J. and Gabaldon, C. and Gabizon, O. and Gabrielli, A. and Gabrielli, A. and Gach, G. P. and Gadatsch, S. and Gadomski, S. and Gagliardi, G. and Gagnon, P. and Galea, C. and Galhardo, B. and Gallas, E. J. and Gallop, B. J. and Gallus, P. and Galster, G. and Gan, K. K. and Gao, J. and Gao, Y. and Gao, Y. S. and Garay Walls, F. M. and García, C. and García Navarro, J. E. and Garcia-Sciveres, M. and Gardner, R. W. and Garelli, N. and Garonne, V. and Gascon Bravo, A. and Gatti, C. and Gaudiello, A. and Gaudio, G. and Gaur, B. and Gauthier, L. and Gavrilenko, I. L. and Gay, C. and Gaycken, G. and Gazis, E. N. and Gecse, Z. and Gee, C. N. P. and Geich-Gimbel, Ch. and Geisler, M. P. and Gemme, C. and Genest, M. H. and Geng, C. and Gentile, S. and George, S. and Gerbaudo, D. and Gershon, A. and Ghasemi, S. and Ghazlane, H. and Giacobbe, B. and Giagu, S. and Giannetti, P. and Gibbard, B. and Gibson, S. M. and Gignac, M. and Gilchriese, M. and Gillam, T. P. S. and Gillberg, D. and Gilles, G. and Gingrich, D. M. and Giokaris, N. and Giordani, M. P. and Giorgi, F. M. and Giorgi, F. M. and Giraud, P. F. and Giromini, P. and Giugni, D. and Giuliani, C. and Giulini, M. and Gjelsten, B. K. and Gkaitatzis, S. and Gkialas, I. and Gkougkousis, E. L. and Gladilin, L. K. and Glasman, C. and Glatzer, J. and Glaysher, P. C. F. and Glazov, A. and Goblirsch-Kolb, M. and Godlewski, J. and Goldfarb, S. and Golling, T. and Golubkov, D. and Gomes, A. and Gonçalo, R. and Goncalves Pinto Firmino Da Costa, J. and Gonella, L. and Gongadze, A. and González de la Hoz, S. and Gonzalez Parra, G. and Gonzalez-Sevilla, S. and Goossens, L. and Gorbounov, P. A. and Gordon, H. A. and Gorelov, I. and Gorini, B. and Gorini, E. and Gorišek, A. and Gornicki, E. and Goshaw, A. T. and Gössling, C. and Gostkin, M. I. and Goudet, C. R. and Goujdami, D. and Goussiou, A. G. and Govender, N. and Gozani, E. and Graber, L. and Grabowska-Bold, I. and Gradin, P. O. J. and Grafström, P. and Gramling, J. and Gramstad, E. and Grancagnolo, S. and Gratchev, V. and Gray, H. M. and Graziani, E. and Greenwood, Z. D. and Grefe, C. and Gregersen, K. and Gregor, I. M. and Grenier, P. and Grevtsov, K. and Griffiths, J. and Grillo, A. A. and Grimm, K. and Grinstein, S. and Gris, Ph. and Grivaz, J.-F. and Groh, S. and Grohs, J. P. and Gross, E. and Grosse-Knetter, J. and Grossi, G. C. and Grout, Z. J. and Guan, L. and Guenther, J. and Guescini, F. and Guest, D. and Gueta, O. and Guido, E. and Guillemin, T. and Guindon, S. and Gul, U. and Gumpert, C. and Guo, J. and Guo, Y. and Gupta, S. and Gustavino, G. and Gutierrez, P. and Gutierrez Ortiz, N. G. and Gutschow, C. and Guyot, C. and Gwenlan, C. and Gwilliam, C. B. and Haas, A. and Haber, C. and Hadavand, H. K. and Haddad, N. and Hadef, A. and Haefner, P. and Hageböck, S. and Hajduk, Z. and Hakobyan, H. and Haleem, M. and Haley, J. and Hall, D. and Halladjian, G. and Hallewell, G. D. and Hamacher, K. and Hamal, P. and Hamano, K. and Hamilton, A. and Hamity, G. N. and Hamnett, P. G. and Han, L. and Hanagaki, K. and Hanawa, K. and Hance, M. and Haney, B. and Hanke, P. and Hanna, R. and Hansen, J. B. and Hansen, J. D. and Hansen, M. C. and Hansen, P. H. and Hara, K. and Hard, A. S. and Harenberg, T. and Hariri, F. and Harkusha, S. and Harrington, R. D. and Harrison, P. F. and Hartjes, F. and Hasegawa, M. and Hasegawa, Y. and Hasib, A. and Hassani, S. and Haug, S. and Hauser, R. and Hauswald, L. and Havranek, M. and Hawkes, C. M. and Hawkings, R. J. and Hawkins, A. D. and Hayashi, T. and Hayden, D. and Hays, C. P. and Hays, J. M. and Hayward, H. S. and Haywood, S. J. and Head, S. J. and Heck, T. and Hedberg, V. and Heelan, L. and Heim, S. and Heim, T. and Heinemann, B. and Heinrich, J. J. and Heinrich, L. and Heinz, C. and Hejbal, J. and Helary, L. and Hellman, S. and Helsens, C. and Henderson, J. and Henderson, R. C. W. and Heng, Y. and Henkelmann, S. and Henriques Correia, A. M. and Henrot-Versille, S. and Herbert, G. H. and Hernández Jiménez, Y. and Herten, G. and Hertenberger, R. and Hervas, L. and Hesketh, G. G. and Hessey, N. P. and Hetherly, J. W. and Hickling, R. and Higón-Rodriguez, E. and Hill, E. and Hill, J. C. and Hiller, K. H. and Hillier, S. J. and Hinchliffe, I. and Hines, E. and Hinman, R. R. and Hirose, M. and Hirschbuehl, D. and Hobbs, J. and Hod, N. and Hodgkinson, M. C. and Hodgson, P. and Hoecker, A. and Hoeferkamp, M. R. and Hoenig, F. and Hohlfeld, M. and Hohn, D. and Holmes, T. R. and Homann, M. and Hong, T. M. and Hooberman, B. H. and Hopkins, W. H. and Horii, Y. and Horton, A. J. and Hostachy, J-Y. and Hou, S. and Hoummada, A. and Howard, J. and Howarth, J. and Hrabovsky, M. and Hristova, I. and Hrivnac, J. and Hryn’ova, T. and Hrynevich, A. and Hsu, C. and Hsu, P. J. and Hsu, S.-C. and Hu, D. and Hu, Q. and Huang, Y. and Hubacek, Z. and Hubaut, F. and Huegging, F. and Huffman, T. B. and Hughes, E. W. and Hughes, G. and Huhtinen, M. and Hülsing, T. A. and Huseynov, N. and Huston, J. and Huth, J. and Iacobucci, G. and Iakovidis, G. and Ibragimov, I. and Iconomidou-Fayard, L. and Ideal, E. and Idrissi, Z. and Iengo, P. and Igonkina, O. and Iizawa, T. and Ikegami, Y. and Ikeno, M. and Ilchenko, Y. and Iliadis, D. and Ilic, N. and Ince, T. and Introzzi, G. and Ioannou, P. and Iodice, M. and Iordanidou, K. and Ippolito, V. and Irles Quiles, A. and Isaksson, C. and Ishino, M. and Ishitsuka, M. and Ishmukhametov, R. and Issever, C. and Istin, S. and Iturbe Ponce, J. M. and Iuppa, R. and Ivarsson, J. and Iwanski, W. and Iwasaki, H. and Izen, J. M. and Izzo, V. and Jabbar, S. and Jackson, B. and Jackson, M. and Jackson, P. and Jain, V. and Jakobi, K. B. and Jakobs, K. and Jakobsen, S. and Jakoubek, T. and Jamin, D. O. and Jana, D. K. and Jansen, E. and Jansky, R. and Janssen, J. and Janus, M. and Jarlskog, G. and Javadov, N. and Javůrek, T. and Jeanneau, F. and Jeanty, L. and Jejelava, J. and Jeng, G.-Y. and Jennens, D. and Jenni, P. and Jentzsch, J. and Jeske, C. and Jézéquel, S. and Ji, H. and Jia, J. and Jiang, H. and Jiang, Y. and Jiggins, S. and Jimenez Pena, J. and Jin, S. and Jinaru, A. and Jinnouchi, O. and Johansson, P. and Johns, K. A. and Johnson, W. J. and Jon-And, K. and Jones, G. and Jones, R. W. L. and Jones, S. and Jones, T. J. and Jongmanns, J. and Jorge, P. M. and Jovicevic, J. and Ju, X. and Juste Rozas, A. and Köhler, M. K. and Kaci, M. and Kaczmarska, A. and Kado, M. and Kagan, H. and Kagan, M. and Kahn, S. J. and Kajomovitz, E. and Kalderon, C. W. and Kaluza, A. and Kama, S. and Kamenshchikov, A. and Kanaya, N. and Kaneti, S. and Kantserov, V. A. and Kanzaki, J. and Kaplan, B. and Kaplan, L. S. and Kapliy, A. and Kar, D. and Karakostas, K. and Karamaoun, A. and Karastathis, N. and Kareem, M. J. and Karentzos, E. and Karnevskiy, M. and Karpov, S. N. and Karpova, Z. M. and Karthik, K. and Kartvelishvili, V. and Karyukhin, A. N. and Kasahara, K. and Kashif, L. and Kass, R. D. and Kastanas, A. and Kataoka, Y. and Kato, C. and Katre, A. and Katzy, J. and Kawade, K. and Kawagoe, K. and Kawamoto, T. and Kawamura, G. and Kazama, S. and Kazanin, V. F. and Keeler, R. and Kehoe, R. and Keller, J. S. and Kempster, J. J. and Keoshkerian, H. and Kepka, O. and Kerševan, B. P. and Kersten, S. and Keyes, R. A. and Khalil-zada, F. and Khandanyan, H. and Khanov, A. and Kharlamov, A. G. and Khoo, T. J. and Khovanskiy, V. and Khramov, E. and Khubua, J. and Kido, S. and Kim, H. Y. and Kim, S. H. and Kim, Y. K. and Kimura, N. and Kind, O. M. and King, B. T. and King, M. and King, S. B. and Kirk, J. and Kiryunin, A. E. and Kishimoto, T. and Kisielewska, D. and Kiss, F. and Kiuchi, K. and Kivernyk, O. and Kladiva, E. and Klein, M. H. and Klein, M. and Klein, U. and Kleinknecht, K. and Klimek, P. and Klimentov, A. and Klingenberg, R. and Klinger, J. A. and Klioutchnikova, T. and Kluge, E.-E. and Kluit, P. and Kluth, S. and Knapik, J. and Kneringer, E. and Knoops, E. B. F. G. and Knue, A. and Kobayashi, A. and Kobayashi, D. and Kobayashi, T. and Kobel, M. and Kocian, M. and Kodys, P. and Koffas, T. and Koffeman, E. and Kogan, L. A. and Kohlmann, S. and Kohriki, T. and Koi, T. and Kolanoski, H. and Kolb, M. and Koletsou, I. and Komar, A. A. and Komori, Y. and Kondo, T. and Kondrashova, N. and Köneke, K. and König, A. C. and Kono, T. and Konoplich, R. and Konstantinidis, N. and Kopeliansky, R. and Koperny, S. and Köpke, L. and Kopp, A. K. and Korcyl, K. and Kordas, K. and Korn, A. and Korol, A. A. and Korolkov, I. and Korolkova, E. V. and Kortner, O. and Kortner, S. and Kosek, T. and Kostyukhin, V. V. and Kotov, V. M. and Kotwal, A. and Kourkoumeli-Charalampidi, A. and Kourkoumelis, C. and Kouskoura, V. and Koutsman, A. and Kowalewski, R. and Kowalski, T. Z. and Kozanecki, W. and Kozhin, A. S. and Kramarenko, V. A. and Kramberger, G. and Krasnopevtsev, D. and Krasny, M. W. and Krasznahorkay, A. and Kraus, J. K. and Kravchenko, A. and Kretz, M. and Kretzschmar, J. and Kreutzfeldt, K. and Krieger, P. and Krizka, K. and Kroeninger, K. and Kroha, H. and Kroll, J. and Kroseberg, J. and Krstic, J. and Kruchonak, U. and Krüger, H. and Krumnack, N. and Kruse, A. and Kruse, M. C. and Kruskal, M. and Kubota, T. and Kucuk, H. and Kuday, S. and Kuechler, J. T. and Kuehn, S. and Kugel, A. and Kuger, F. and Kuhl, A. and Kuhl, T. and Kukhtin, V. and Kukla, R. and Kulchitsky, Y. and Kuleshov, S. and Kuna, M. and Kunigo, T. and Kupco, A. and Kurashige, H. and Kurochkin, Y. A. and Kus, V. and Kuwertz, E. S. and Kuze, M. and Kvita, J. and Kwan, T. and Kyriazopoulos, D. and La Rosa, A. and La Rosa Navarro, J. L. and La Rotonda, L. and Lacasta, C. and Lacava, F. and Lacey, J. and Lacker, H. and Lacour, D. and Lacuesta, V. R. and Ladygin, E. and Lafaye, R. and Laforge, B. and Lagouri, T. and Lai, S. and Lammers, S. and Lampen, C. L. and Lampl, W. and Lançon, E. and Landgraf, U. and Landon, M. P. J. and Lang, V. S. and Lange, J. C. and Lankford, A. J. and Lanni, F. and Lantzsch, K. and Lanza, A. and Laplace, S. and Lapoire, C. and Laporte, J. F. and Lari, T. and Lasagni Manghi, F. and Lassnig, M. and Laurelli, P. and Lavrijsen, W. and Law, A. T. and Laycock, P. and Lazovich, T. and Le Dortz, O. and Le Guirriec, E. and Le Menedeu, E. and LeBlanc, M. and LeCompte, T. and Ledroit-Guillon, F. and Lee, C. A. and Lee, S. C. and Lee, L. and Lefebvre, G. and Lefebvre, M. and Legger, F. and Leggett, C. and Lehan, A. and Lehmann Miotto, G. and Lei, X. and Leight, W. A. and Leisos, A. and Leister, A. G. and Leite, M. A. L. and Leitner, R. and Lellouch, D. and Lemmer, B. and Leney, K. J. C. and Lenz, T. and Lenzi, B. and Leone, R. and Leone, S. and Leonidopoulos, C. and Leontsinis, S. and Leroy, C. and Lester, C. G. and Levchenko, M. and Levêque, J. and Levin, D. and Levinson, L. J. and Levy, M. and Leyko, A. M. and Leyton, M. and Li, B. and Li, H. and Li, H. L. and Li, L. and Li, L. and Li, Q. and Li, S. and Li, X. and Li, Y. and Liang, Z. and Liao, H. and Liberti, B. and Liblong, A. and Lichard, P. and Lie, K. and Liebal, J. and Liebig, W. and Limbach, C. and Limosani, A. and Lin, S. C. and Lin, T. H. and Lindquist, B. E. and Lipeles, E. and Lipniacka, A. and Lisovyi, M. and Liss, T. M. and Lissauer, D. and Lister, A. and Litke, A. M. and Liu, B. and Liu, D. and Liu, H. and Liu, H. and Liu, J. and Liu, J. B. and Liu, K. and Liu, L. and Liu, M. and Liu, M. and Liu, Y. L. and Liu, Y. and Livan, M. and Lleres, A. and Llorente Merino, J. and Lloyd, S. L. and Lo Sterzo, F. and Lobodzinska, E. and Loch, P. and Lockman, W. S. and Loebinger, F. K. and Loevschall-Jensen, A. E. and Loew, K. M. and Loginov, A. and Lohse, T. and Lohwasser, K. and Lokajicek, M. and Long, B. A. and Long, J. D. and Long, R. E. and Longo, L. and Looper, K. A. and Lopes, L. and Lopez Mateos, D. and Lopez Paredes, B. and Lopez Paz, I. and Lopez Solis, A. and Lorenz, J. and Lorenzo Martinez, N. and Losada, M. and Lösel, P. J. and Lou, X. and Lounis, A. and Love, J. and Love, P. A. and Lu, H. and Lu, N. and Lubatti, H. J. and Luci, C. and Lucotte, A. and Luedtke, C. and Luehring, F. and Lukas, W. and Luminari, L. and Lundberg, O. and Lund-Jensen, B. and Lynn, D. and Lysak, R. and Lytken, E. and Ma, H. and Ma, L. L. and Maccarrone, G. and Macchiolo, A. and Macdonald, C. M. and Maček, B. and Machado Miguens, J. and Madaffari, D. and Madar, R. and Maddocks, H. J. and Mader, W. F. and Madsen, A. and Maeda, J. and Maeland, S. and Maeno, T. and Maevskiy, A. and Magradze, E. and Mahlstedt, J. and Maiani, C. and Maidantchik, C. and Maier, A. A. and Maier, T. and Maio, A. and Majewski, S. and Makida, Y. and Makovec, N. and Malaescu, B. and Malecki, Pa. and Maleev, V. P. and Malek, F. and Mallik, U. and Malon, D. and Malone, C. and Maltezos, S. and Malyshev, V. M. and Malyukov, S. and Mamuzic, J. and Mancini, G. and Mandelli, B. and Mandelli, L. and Mandić, I. and Maneira, J. and Manhaes de Andrade Filho, L. and Manjarres Ramos, J. and Mann, A. and Mansoulie, B. and Mantifel, R. and Mantoani, M. and Manzoni, S. and Mapelli, L. and March, L. and Marchiori, G. and Marcisovsky, M. and Marjanovic, M. and Marley, D. E. and Marroquim, F. and Marsden, S. P. and Marshall, Z. and Marti, L. F. and Marti-Garcia, S. and Martin, B. and Martin, T. A. and Martin, V. J. and Martin dit Latour, B. and Martinez, M. and Martin-Haugh, S. and Martoiu, V. S. and Martyniuk, A. C. and Marx, M. and Marzano, F. and Marzin, A. and Masetti, L. and Mashimo, T. and Mashinistov, R. and Masik, J. and Maslennikov, A. L. and Massa, I. and Massa, L. and Mastrandrea, P. and Mastroberardino, A. and Masubuchi, T. and Mättig, P. and Mattmann, J. and Maurer, J. and Maxfield, S. J. and Maximov, D. A. and Mazini, R. and Mazza, S. M. and Mc Fadden, N. C. and Mc Goldrick, G. and Mc Kee, S. P. and McCarn, A. and McCarthy, R. L. and McCarthy, T. G. and McFarlane, K. W. and Mcfayden, J. A. and Mchedlidze, G. and McMahon, S. J. and McPherson, R. A. and Medinnis, M. and Meehan, S. and Mehlhase, S. and Mehta, A. and Meier, K. and Meineck, C. and Meirose, B. and Mellado Garcia, B. R. and Meloni, F. and Mengarelli, A. and Menke, S. and Meoni, E. and Mercurio, K. M. and Mergelmeyer, S. and Mermod, P. and Merola, L. and Meroni, C. and Merritt, F. S. and Messina, A. and Metcalfe, J. and Mete, A. S. and Meyer, C. and Meyer, C. and Meyer, J-P. and Meyer, J. and Meyer Zu Theenhausen, H. and Middleton, R. P. and Miglioranzi, S. and Mijović, L. and Mikenberg, G. and Mikestikova, M. and Mikuž, M. and Milesi, M. and Milic, A. and Miller, D. W. and Mills, C. and Milov, A. and Milstead, D. A. and Minaenko, A. A. and Minami, Y. and Minashvili, I. A. and Mincer, A. I. and Mindur, B. and Mineev, M. and Ming, Y. and Mir, L. M. and Mistry, K. P. and Mitani, T. and Mitrevski, J. and Mitsou, V. A. and Miucci, A. and Miyagawa, P. S. and Mjörnmark, J. U. and Moa, T. and Mochizuki, K. and Mohapatra, S. and Mohr, W. and Molander, S. and Moles-Valls, R. and Monden, R. and Mondragon, M. C. and Mönig, K. and Monk, J. and Monnier, E. and Montalbano, A. and Montejo Berlingen, J. and Monticelli, F. and Monzani, S. and Moore, R. W. and Morange, N. and Moreno, D. and Moreno Llácer, M. and Morettini, P. and Mori, D. and Mori, T. and Morii, M. and Morinaga, M. and Morisbak, V. and Moritz, S. and Morley, A. K. and Mornacchi, G. and Morris, J. D. and Mortensen, S. S. and Morvaj, L. and Mosidze, M. and Moss, J. and Motohashi, K. and Mount, R. and Mountricha, E. and Mouraviev, S. V. and Moyse, E. J. W. and Muanza, S. and Mudd, R. D. and Mueller, F. and Mueller, J. and Mueller, R. S. P. and Mueller, T. and Muenstermann, D. and Mullen, P. and Mullier, G. A. and Munoz Sanchez, F. J. and Murillo Quijada, J. A. and Murray, W. J. and Musheghyan, H. and Myagkov, A. G. and Myska, M. and Nachman, B. P. and Nackenhorst, O. and Nadal, J. and Nagai, K. and Nagai, R. and Nagai, Y. and Nagano, K. and Nagasaka, Y. and Nagata, K. and Nagel, M. and Nagy, E. and Nairz, A. M. and Nakahama, Y. and Nakamura, K. and Nakamura, T. and Nakano, I. and Namasivayam, H. and Naranjo Garcia, R. F. and Narayan, R. and Narrias Villar, D. I. and Naryshkin, I. and Naumann, T. and Navarro, G. and Nayyar, R. and Neal, H. A. and Nechaeva, P. Yu. and Neep, T. J. and Nef, P. D. and Negri, A. and Negrini, M. and Nektarijevic, S. and Nellist, C. and Nelson, A. and Nemecek, S. and Nemethy, P. and Nepomuceno, A. A. and Nessi, M. and Neubauer, M. S. and Neumann, M. and Neves, R. M. and Nevski, P. and Newman, P. R. and Nguyen, D. H. and Nickerson, R. B. and Nicolaidou, R. and Nicquevert, B. and Nielsen, J. and Nikiforov, A. and Nikolaenko, V. and Nikolic-Audit, I. and Nikolopoulos, K. and Nilsen, J. K. and Nilsson, P. and Ninomiya, Y. and Nisati, A. and Nisius, R. and Nobe, T. and Nodulman, L. and Nomachi, M. and Nomidis, I. and Nooney, T. and Norberg, S. and Nordberg, M. and Novgorodova, O. and Nowak, S. and Nozaki, M. and Nozka, L. and Ntekas, K. and Nurse, E. and Nuti, F. and O’grady, F. and O’Neil, D. C. and O’Shea, V. and Oakham, F. G. and Oberlack, H. and Obermann, T. and Ocariz, J. and Ochi, A. and Ochoa, I. and Ochoa-Ricoux, J. P. and Oda, S. and Odaka, S. and Ogren, H. and Oh, A. and Oh, S. H. and Ohm, C. C. and Ohman, H. and Oide, H. and Okawa, H. and Okumura, Y. and Okuyama, T. and Olariu, A. and Oleiro Seabra, L. F. and Olivares Pino, S. A. and Oliveira Damazio, D. and Olszewski, A. and Olszowska, J. and Onofre, A. and Onogi, K. and Onyisi, P. U. E. and Oram, C. J. and Oreglia, M. J. and Oren, Y. and Orestano, D. and Orlando, N. and Orr, R. S. and Osculati, B. and Ospanov, R. and Otero y Garzon, G. and Otono, H. and Ouchrif, M. and Ould-Saada, F. and Ouraou, A. and Oussoren, K. P. and Ouyang, Q. and Ovcharova, A. and Owen, M. and Owen, R. E. and Ozcan, V. E. and Ozturk, N. and Pachal, K. and Pacheco Pages, A. and Padilla Aranda, C. and Pagáčová, M. and Pagan Griso, S. and Paige, F. and Pais, P. and Pajchel, K. and Palacino, G. and Palestini, S. and Palka, M. and Pallin, D. and Palma, A. and Panagiotopoulou, E. St. and Pandini, C. E. and Panduro Vazquez, J. G. and Pani, P. and Panitkin, S. and Pantea, D. and Paolozzi, L. and Papadopoulou, Th. D. and Papageorgiou, K. and Paramonov, A. and Paredes Hernandez, D. and Parker, M. A. and Parker, K. A. and Parodi, F. and Parsons, J. A. and Parzefall, U. and Pascuzzi, V. and Pasqualucci, E. and Passaggio, S. and Pastore, F. and Pastore, Fr. and Pásztor, G. and Pataraia, S. and Patel, N. D. and Pater, J. R. and Pauly, T. and Pearce, J. and Pearson, B. and Pedersen, L. E. and Pedersen, M. and Pedraza Lopez, S. and Pedro, R. and Peleganchuk, S. V. and Pelikan, D. and Penc, O. and Peng, C. and Peng, H. and Penwell, J. and Peralva, B. S. and Perepelitsa, D. V. and Perez Codina, E. and Perini, L. and Pernegger, H. and Perrella, S. and Peschke, R. and Peshekhonov, V. D. and Peters, K. and Peters, R. F. Y. and Petersen, B. A. and Petersen, T. C. and Petit, E. and Petridis, A. and Petridou, C. and Petroff, P. and Petrolo, E. and Petrucci, F. and Pettersson, N. E. and Peyaud, A. and Pezoa, R. and Phillips, P. W. and Piacquadio, G. and Pianori, E. and Picazio, A. and Piccaro, E. and Piccinini, M. and Pickering, M. A. and Piegaia, R. and Pilcher, J. E. and Pilkington, A. D. and Pin, A. W. J. and Pina, J. and Pinamonti, M. and Pinfold, J. L. and Pingel, A. and Pires, S. and Pirumov, H. and Pitt, M. and Plazak, L. and Pleier, M.-A. and Pleskot, V. and Plotnikova, E. and Plucinski, P. and Pluth, D. and Poettgen, R. and Poggioli, L. and Pohl, D. and Polesello, G. and Poley, A. and Policicchio, A. and Polifka, R. and Polini, A. and Pollard, C. S. and Polychronakos, V. and Pommès, K. and Pontecorvo, L. and Pope, B. G. and Popeneciu, G. A. and Popovic, D. S. and Poppleton, A. and Pospisil, S. and Potamianos, K. and Potrap, I. N. and Potter, C. J. and Potter, C. T. and Poulard, G. and Poveda, J. and Pozdnyakov, V. and Pozo Astigarraga, M. E. and Pralavorio, P. and Pranko, A. and Prell, S. and Price, D. and Price, L. E. and Primavera, M. and Prince, S. and Proissl, M. and Prokofiev, K. and Prokoshin, F. and Protopopescu, S. and Proudfoot, J. and Przybycien, M. and Puddu, D. and Puldon, D. and Purohit, M. and Puzo, P. and Qian, J. and Qin, G. and Qin, Y. and Quadt, A. and Quarrie, D. R. and Quayle, W. B. and Queitsch-Maitland, M. and Quilty, D. and Raddum, S. and Radeka, V. and Radescu, V. and Radhakrishnan, S. K. and Radloff, P. and Rados, P. and Ragusa, F. and Rahal, G. and Rajagopalan, S. and Rammensee, M. and Rangel-Smith, C. and Rauscher, F. and Rave, S. and Ravenscroft, T. and Raymond, M. and Read, A. L. and Readioff, N. P. and Rebuzzi, D. M. and Redelbach, A. and Redlinger, G. and Reece, R. and Reeves, K. and Rehnisch, L. and Reichert, J. and Reisin, H. and Rembser, C. and Ren, H. and Rescigno, M. and Resconi, S. and Rezanova, O. L. and Reznicek, P. and Rezvani, R. and Richter, R. and Richter, S. and Richter-Was, E. and Ricken, O. and Ridel, M. and Rieck, P. and Riegel, C. J. and Rieger, J. and Rifki, O. and Rijssenbeek, M. and Rimoldi, A. and Rinaldi, L. and Ristić, B. and Ritsch, E. and Riu, I. and Rizatdinova, F. and Rizvi, E. and Robertson, S. H. and Robichaud-Veronneau, A. and Robinson, D. and Robinson, J. E. M. and Robson, A. and Roda, C. and Rodina, Y. and Rodriguez Perez, A. and Roe, S. and Rogan, C. S. and Røhne, O. and Romaniouk, A. and Romano, M. and Romano Saez, S. M. and Romero Adam, E. and Rompotis, N. and Ronzani, M. and Roos, L. and Ros, E. and Rosati, S. and Rosbach, K. and Rose, P. and Rosenthal, O. and Rossetti, V. and Rossi, E. and Rossi, L. P. and Rosten, J. H. N. and Rosten, R. and Rotaru, M. and Roth, I. and Rothberg, J. and Rousseau, D. and Royon, C. R. and Rozanov, A. and Rozen, Y. and Ruan, X. and Rubbo, F. and Rubinskiy, I. and Rud, V. I. and Rudolph, M. S. and Rühr, F. and Ruiz-Martinez, A. and Rurikova, Z. and Rusakovich, N. A. and Ruschke, A. and Russell, H. L. and Rutherfoord, J. P. and Ruthmann, N. and Ryabov, Y. F. and Rybar, M. and Rybkin, G. and Ryder, N. C. and Ryu, S. and Ryzhov, A. and Saavedra, A. F. and Sabato, G. and Sacerdoti, S. and Sadrozinski, H. F-W. and Sadykov, R. and Safai Tehrani, F. and Saha, P. and Sahinsoy, M. and Saimpert, M. and Saito, T. and Sakamoto, H. and Sakurai, Y. and Salamanna, G. and Salamon, A. and Salazar Loyola, J. E. and Salek, D. and Sales De Bruin, P. H. and Salihagic, D. and Salnikov, A. and Salt, J. and Salvatore, D. and Salvatore, F. and Salvucci, A. and Salzburger, A. and Sammel, D. and Sampsonidis, D. and Sanchez, A. and Sánchez, J. and Sanchez Martinez, V. and Sandaker, H. and Sandbach, R. L. and Sander, H. G. and Sanders, M. P. and Sandhoff, M. and Sandoval, C. and Sandstroem, R. and Sankey, D. P. C. and Sannino, M. and Sansoni, A. and Santoni, C. and Santonico, R. and Santos, H. and Santoyo Castillo, I. and Sapp, K. and Sapronov, A. and Saraiva, J. G. and Sarrazin, B. and Sasaki, O. and Sasaki, Y. and Sato, K. and Sauvage, G. and Sauvan, E. and Savage, G. and Savard, P. and Sawyer, C. and Sawyer, L. and Saxon, J. and Sbarra, C. and Sbrizzi, A. and Scanlon, T. and Scannicchio, D. A. and Scarcella, M. and Scarfone, V. and Schaarschmidt, J. and Schacht, P. and Schaefer, D. and Schaefer, R. and Schaeffer, J. and Schaepe, S. and Schaetzel, S. and Schäfer, U. and Schaffer, A. C. and Schaile, D. and Schamberger, R. D. and Scharf, V. and Schegelsky, V. A. and Scheirich, D. and Schernau, M. and Schiavi, C. and Schillo, C. and Schioppa, M. and Schlenker, S. and Schmieden, K. and Schmitt, C. and Schmitt, S. and Schmitt, S. and Schmitz, S. and Schneider, B. and Schnellbach, Y. J. and Schnoor, U. and Schoeffel, L. and Schoening, A. and Schoenrock, B. D. and Schopf, E. and Schorlemmer, A. L. S. and Schott, M. and Schouten, D. and Schovancova, J. and Schramm, S. and Schreyer, M. and Schuh, N. and Schultens, M. J. and Schultz-Coulon, H.-C. and Schulz, H. and Schumacher, M. and Schumm, B. A. and Schune, Ph. and Schwanenberger, C. and Schwartzman, A. and Schwarz, T. A. and Schwegler, Ph. and Schweiger, H. and Schwemling, Ph. and Schwienhorst, R. and Schwindling, J. and Schwindt, T. and Sciolla, G. and Scuri, F. and Scutti, F. and Searcy, J. and Seema, P. and Seidel, S. C. and Seiden, A. and Seifert, F. and Seixas, J. M. and Sekhniaidze, G. and Sekhon, K. and Sekula, S. J. and Seliverstov, D. M. and Semprini-Cesari, N. and Serfon, C. and Serin, L. and Serkin, L. and Sessa, M. and Seuster, R. and Severini, H. and Sfiligoj, T. and Sforza, F. and Sfyrla, A. and Shabalina, E. and Shaikh, N. W. and Shan, L. Y. and Shang, R. and Shank, J. T. and Shapiro, M. and Shatalov, P. B. and Shaw, K. and Shaw, S. M. and Shcherbakova, A. and Shehu, C. Y. and Sherwood, P. and Shi, L. and Shimizu, S. and Shimmin, C. O. and Shimojima, M. and Shiyakova, M. and Shmeleva, A. and Shoaleh Saadi, D. and Shochet, M. J. and Shojaii, S. and Shrestha, S. and Shulga, E. and Shupe, M. A. and Sicho, P. and Sidebo, P. E. and Sidiropoulou, O. and Sidorov, D. and Sidoti, A. and Siegert, F. and Sijacki, Dj. and Silva, J. and Silva, M. and Silverstein, S. B. and Simak, V. and Simard, O. and Simic, Lj. and Simion, S. and Simioni, E. and Simmons, B. and Simon, D. and Simon, M. and Sinervo, P. and Sinev, N. B. and Sioli, M. and Siragusa, G. and Sivoklokov, S. Yu. and Sjölin, J. and Sjursen, T. B. and Skinner, M. B. and Skottowe, H. P. and Skubic, P. and Slater, M. and Slavicek, T. and Slawinska, M. and Sliwa, K. and Smakhtin, V. and Smart, B. H. and Smestad, L. and Smirnov, S. Yu. and Smirnov, Y. and Smirnova, L. N. and Smirnova, O. and Smith, M. N. K. and Smith, R. W. and Smizanska, M. and Smolek, K. and Snesarev, A. A. and Snidero, G. and Snyder, S. and Sobie, R. and Socher, F. and Soffer, A. and Soh, D. A. and Sokhrannyi, G. and Solans Sanchez, C. A. and Solar, M. and Soldatov, E. Yu. and Soldevila, U. and Solodkov, A. A. and Soloshenko, A. and Solovyanov, O. V. and Solovyev, V. and Sommer, P. and Song, H. Y. and Soni, N. and Sood, A. and Sopczak, A. and Sopko, V. and Sorin, V. and Sosa, D. and Sotiropoulou, C. L. and Soualah, R. and Soukharev, A. M. and South, D. and Sowden, B. C. and Spagnolo, S. and Spalla, M. and Spangenberg, M. and Spanò, F. and Sperlich, D. and Spettel, F. and Spighi, R. and Spigo, G. and Spiller, L. A. and Spousta, M. and St. Denis, R. D. and Stabile, A. and Staerz, S. and Stahlman, J. and Stamen, R. and Stamm, S. and Stanecka, E. and Stanek, R. W. and Stanescu, C. and Stanescu-Bellu, M. and Stanitzki, M. M. and Stapnes, S. and Starchenko, E. A. and Stark, G. H. and Stark, J. and Staroba, P. and Starovoitov, P. and Staszewski, R. and Steinberg, P. and Stelzer, B. and Stelzer, H. J. and Stelzer-Chilton, O. and Stenzel, H. and Stewart, G. A. and Stillings, J. A. and Stockton, M. C. and Stoebe, M. and Stoicea, G. and Stolte, P. and Stonjek, S. and Stradling, A. R. and Straessner, A. and Stramaglia, M. E. and Strandberg, J. and Strandberg, S. and Strandlie, A. and Strauss, M. and Strizenec, P. and Ströhmer, R. and Strom, D. M. and Stroynowski, R. and Strubig, A. and Stucci, S. A. and Stugu, B. and Styles, N. A. and Su, D. and Su, J. and Subramaniam, R. and Suchek, S. and Sugaya, Y. and Suk, M. and Sulin, V. V. and Sultansoy, S. and Sumida, T. and Sun, S. and Sun, X. and Sundermann, J. E. and Suruliz, K. and Susinno, G. and Sutton, M. R. and Suzuki, S. and Svatos, M. and Swiatlowski, M. and Sykora, I. and Sykora, T. and Ta, D. and Taccini, C. and Tackmann, K. and Taenzer, J. and Taffard, A. and Tafirout, R. and Taiblum, N. and Takai, H. and Takashima, R. and Takeda, H. and Takeshita, T. and Takubo, Y. and Talby, M. and Talyshev, A. A. and Tam, J. Y. C. and Tan, K. G. and Tanaka, J. and Tanaka, R. and Tanaka, S. and Tannenwald, B. B. and Tapia Araya, S. and Tapprogge, S. and Tarem, S. and Tartarelli, G. F. and Tas, P. and Tasevsky, M. and Tashiro, T. and Tassi, E. and Tavares Delgado, A. and Tayalati, Y. and Taylor, A. C. and Taylor, G. N. and Taylor, P. T. E. and Taylor, W. and Teischinger, F. A. and Teixeira-Dias, P. and Temming, K. K. and Temple, D. and Ten Kate, H. and Teng, P. K. and Teoh, J. J. and Tepel, F. and Terada, S. and Terashi, K. and Terron, J. and Terzo, S. and Testa, M. and Teuscher, R. J. and Theveneaux-Pelzer, T. and Thomas, J. P. and Thomas-Wilsker, J. and Thompson, E. N. and Thompson, P. D. and Thompson, R. J. and Thompson, A. S. and Thomsen, L. A. and Thomson, E. and Thomson, M. and Tibbetts, M. J. and Ticse Torres, R. E. and Tikhomirov, V. O. and Tikhonov, Yu. A. and Timoshenko, S. and Tiouchichine, E. and Tipton, P. and Tisserant, S. and Todome, K. and Todorov, T. and Todorova-Nova, S. and Tojo, J. and Tokár, S. and Tokushuku, K. and Tolley, E. and Tomlinson, L. and Tomoto, M. and Tompkins, L. and Toms, K. and Tong, B. and Torrence, E. and Torres, H. and Torró Pastor, E. and Toth, J. and Touchard, F. and Tovey, D. R. and Trefzger, T. and Tremblet, L. and Tricoli, A. and Trigger, I. M. and Trincaz-Duvoid, S. and Tripiana, M. F. and Trischuk, W. and Trocmé, B. and Trofymov, A. and Troncon, C. and Trottier-McDonald, M. and Trovatelli, M. and Truong, L. and Trzebinski, M. and Trzupek, A. and Tseng, J. C-L. and Tsiareshka, P. V. and Tsipolitis, G. and Tsirintanis, N. and Tsiskaridze, S. and Tsiskaridze, V. and Tskhadadze, E. G. and Tsui, K. M. and Tsukerman, I. I. and Tsulaia, V. and Tsuno, S. and Tsybychev, D. and Tudorache, A. and Tudorache, V. and Tuna, A. N. and Tupputi, S. A. and Turchikhin, S. and Turecek, D. and Turgeman, D. and Turra, R. and Turvey, A. J. and Tuts, P. M. and Tylmad, M. and Tyndel, M. and Ueda, I. and Ueno, R. and Ughetto, M. and Ukegawa, F. and Unal, G. and Undrus, A. and Unel, G. and Ungaro, F. C. and Unno, Y. and Unverdorben, C. and Urban, J. and Urquijo, P. and Urrejola, P. and Usai, G. and Usanova, A. and Vacavant, L. and Vacek, V. and Vachon, B. and Valderanis, C. and Valencic, N. and Valentinetti, S. and Valero, A. and Valery, L. and Valkar, S. and Vallecorsa, S. and Valls Ferrer, J. A. and Van Den Wollenberg, W. and Van Der Deijl, P. C. and van der Geer, R. and van der Graaf, H. and van Eldik, N. and van Gemmeren, P. and Van Nieuwkoop, J. and van Vulpen, I. and van Woerden, M. C. and Vanadia, M. and Vandelli, W. and Vanguri, R. and Vaniachine, A. and Vardanyan, G. and Vari, R. and Varnes, E. W. and Varol, T. and Varouchas, D. and Vartapetian, A. and Varvell, K. E. and Vazeille, F. and Vazquez Schroeder, T. and Veatch, J. and Veloce, L. M. and Veloso, F. and Veneziano, S. and Ventura, A. and Venturi, M. and Venturi, N. and Venturini, A. and Vercesi, V. and Verducci, M. and Verkerke, W. and Vermeulen, J. C. and Vest, A. and Vetterli, M. C. and Viazlo, O. and Vichou, I. and Vickey, T. and Vickey Boeriu, O. E. and Viehhauser, G. H. A. and Viel, S. and Vigne, R. and Villa, M. and Villaplana Perez, M. and Vilucchi, E. and Vincter, M. G. and Vinogradov, V. B. and Vivarelli, I. and Vlachos, S. and Vlasak, M. and Vogel, M. and Vokac, P. and Volpi, G. and Volpi, M. and von der Schmitt, H. and von Toerne, E. and Vorobel, V. and Vorobev, K. and Vos, M. and Voss, R. and Vossebeld, J. H. and Vranjes, N. and Vranjes Milosavljevic, M. and Vrba, V. and Vreeswijk, M. and Vuillermet, R. and Vukotic, I. and Vykydal, Z. and Wagner, P. and Wagner, W. and Wahlberg, H. and Wahrmund, S. and Wakabayashi, J. and Walder, J. and Walker, R. and Walkowiak, W. and Wallangen, V. and Wang, C. and Wang, C. and Wang, F. and Wang, H. and Wang, H. and Wang, J. and Wang, J. and Wang, K. and Wang, R. and Wang, S. M. and Wang, T. and Wang, T. and Wang, X. and Wanotayaroj, C. and Warburton, A. and Ward, C. P. and Wardrope, D. R. and Washbrook, A. and Watkins, P. M. and Watson, A. T. and Watson, I. J. and Watson, M. F. and Watts, G. and Watts, S. and Waugh, B. M. and Webb, S. and Weber, M. S. and Weber, S. W. and Webster, J. S. and Weidberg, A. R. and Weinert, B. and Weingarten, J. and Weiser, C. and Weits, H. and Wells, P. S. and Wenaus, T. and Wengler, T. and Wenig, S. and Wermes, N. and Werner, M. and Werner, P. and Wessels, M. and Wetter, J. and Whalen, K. and Wharton, A. M. and White, A. and White, M. J. and White, R. and White, S. and Whiteson, D. and Wickens, F. J. and Wiedenmann, W. and Wielers, M. and Wienemann, P. and Wiglesworth, C. and Wiik-Fuchs, L. A. M. and Wildauer, A. and Wilkens, H. G. and Williams, H. H. and Williams, S. and Willis, C. and Willocq, S. and Wilson, J. A. and Wingerter-Seez, I. and Winklmeier, F. and Winter, B. T. and Wittgen, M. and Wittkowski, J. and Wollstadt, S. J. and Wolter, M. W. and Wolters, H. and Wosiek, B. K. and Wotschack, J. and Woudstra, M. J. and Wozniak, K. W. and Wu, M. and Wu, M. and Wu, S. L. and Wu, X. and Wu, Y. and Wyatt, T. R. and Wynne, B. M. and Xella, S. and Xu, D. and Xu, L. and Yabsley, B. and Yacoob, S. and Yakabe, R. and Yamaguchi, D. and Yamaguchi, Y. and Yamamoto, A. and Yamamoto, S. and Yamanaka, T. and Yamauchi, K. and Yamazaki, Y. and Yan, Z. and Yang, H. and Yang, H. and Yang, Y. and Yang, Z. and Yao, W-M. and Yap, Y. C. and Yasu, Y. and Yatsenko, E. and Yau Wong, K. H. and Ye, J. and Ye, S. and Yeletskikh, I. and Yen, A. L. and Yildirim, E. and Yorita, K. and Yoshida, R. and Yoshihara, K. and Young, C. and Young, C. J. S. and Youssef, S. and Yu, D. R. and Yu, J. and Yu, J. M. and Yu, J. and Yuan, L. and Yuen, S. P. Y. and Yusuff, I. and Zabinski, B. and Zaidan, R. and Zaitsev, A. M. and Zakharchuk, N. and Zalieckas, J. and Zaman, A. and Zambito, S. and Zanello, L. and Zanzi, D. and Zeitnitz, C. and Zeman, M. and Zemla, A. and Zeng, J. C. and Zeng, Q. and Zengel, K. and Zenin, O. and Ženiš, T. and Zerwas, D. and Zhang, D. and Zhang, F. and Zhang, G. and Zhang, H. and Zhang, J. and Zhang, L. and Zhang, R. and Zhang, R. and Zhang, X. and Zhang, Z. and Zhao, X. and Zhao, Y. and Zhao, Z. and Zhemchugov, A. and Zhong, J. and Zhou, B. and Zhou, C. and Zhou, L. and Zhou, L. and Zhou, M. and Zhou, N. and Zhu, C. G. and Zhu, H. and Zhu, J. and Zhu, Y. and Zhuang, X. and Zhukov, K. and Zibell, A. and Zieminska, D. and Zimine, N. I. and Zimmermann, C. and Zimmermann, S. and Zinonos, Z. and Zinser, M. and Ziolkowski, M. and Živković, L. and Zobernig, G. and Zoccoli, A. and zur Nedden, M. and Zurzolo, G. and Zwalinski, L.},
   year={2016},
   month=Mar }

@article{2018,
   title={Search for black holes and sphalerons in high-multiplicity final states in proton-proton collisions at $ \sqrt{s}=13 $ TeV},
   volume={2018},
   ISSN={1029-8479},
   url={http://dx.doi.org/10.1007/JHEP11(2018)042},
   DOI={10.1007/jhep11(2018)042},
   number={11},
   journal={Journal of High Energy Physics},
   publisher={Springer Science and Business Media LLC},
   author={ and Sirunyan, A. M. and Tumasyan, A. and Adam, W. and Ambrogi, F. and Asilar, E. and Bergauer, T. and Brandstetter, J. and Dragicevic, M. and Erö, J. and Escalante Del Valle, A. and Flechl, M. and Frühwirth, R. and Ghete, V. M. and Hrubec, J. and Jeitler, M. and Krammer, N. and Krätschmer, I. and Liko, D. and Madlener, T. and Mikulec, I. and Rad, N. and Rohringer, H. and Schieck, J. and Schöfbeck, R. and Spanring, M. and Spitzbart, D. and Taurok, A. and Waltenberger, W. and Wittmann, J. and Wulz, C.-E. and Zarucki, M. and Chekhovsky, V. and Mossolov, V. and Suarez Gonzalez, J. and De Wolf, E. A. and Di Croce, D. and Janssen, X. and Lauwers, J. and Pieters, M. and Van De Klundert, M. and Van Haevermaet, H. and Van Mechelen, P. and Van Remortel, N. and Abu Zeid, S. and Blekman, F. and D’Hondt, J. and De Bruyn, I. and De Clercq, J. and Deroover, K. and Flouris, G. and Lontkovskyi, D. and Lowette, S. and Marchesini, I. and Moortgat, S. and Moreels, L. and Python, Q. and Skovpen, K. and Tavernier, S. and Van Doninck, W. and Van Mulders, P. and Van Parijs, I. and Beghin, D. and Bilin, B. and Brun, H. and Clerbaux, B. and De Lentdecker, G. and Delannoy, H. and Dorney, B. and Fasanella, G. and Favart, L. and Goldouzian, R. and Grebenyuk, A. and Kalsi, A. K. and Lenzi, T. and Luetic, J. and Postiau, N. and Starling, E. and Thomas, L. and Vander Velde, C. and Vanlaer, P. and Vannerom, D. and Wang, Q. and Cornelis, T. and Dobur, D. and Fagot, A. and Gul, M. and Khvastunov, I. and Poyraz, D. and Roskas, C. and Trocino, D. and Tytgat, M. and Verbeke, W. and Vermassen, B. and Vit, M. and Zaganidis, N. and Bakhshiansohi, H. and Bondu, O. and Brochet, S. and Bruno, G. and Caputo, C. and David, P. and Delaere, C. and Delcourt, M. and Francois, B. and Giammanco, A. and Krintiras, G. and Lemaitre, V. and Magitteri, A. and Mertens, A. and Musich, M. and Piotrzkowski, K. and Saggio, A. and Vidal Marono, M. and Wertz, S. and Zobec, J. and Alves, F. L. and Alves, G. A. and Brito, L. and Correa Martins Junior, M. and Correia Silva, G. and Hensel, C. and Moraes, A. and Pol, M. E. and Rebello Teles, P. and Belchior Batista Das Chagas, E. and Carvalho, W. and Chinellato, J. and Coelho, E. and Da Costa, E. M. and Da Silveira, G. G. and De Jesus Damiao, D. and De Oliveira Martins, C. and Fonseca De Souza, S. and Malbouisson, H. and Matos Figueiredo, D. and Melo De Almeida, M. and Mora Herrera, C. and Mundim, L. and Nogima, H. and Prado Da Silva, W. L. and Sanchez Rosas, L. J. and Santoro, A. and Sznajder, A. and Thiel, M. and Tonelli Manganote, E. J. and Torres Da Silva De Araujo, F. and Vilela Pereira, A. and Ahuja, S. and Bernardes, C. A. and Calligaris, L. and Fernandez Perez Tomei, T. R. and Gregores, E. M. and Mercadante, P. G. and Novaes, S. F. and Padula, SandraS. and Romero Abad, D. and Aleksandrov, A. and Hadjiiska, R. and Iaydjiev, P. and Marinov, A. and Misheva, M. and Rodozov, M. and Shopova, M. and Sultanov, G. and Dimitrov, A. and Litov, L. and Pavlov, B. and Petkov, P. and Fang, W. and Gao, X. and Yuan, L. and Ahmad, M. and Bian, J. G. and Chen, G. M. and Chen, H. S. and Chen, M. and Chen, Y. and Jiang, C. H. and Leggat, D. and Liao, H. and Liu, Z. and Romeo, F. and Shaheen, S. M. and Spiezia, A. and Tao, J. and Wang, C. and Wang, Z. and Yazgan, E. and Zhang, H. and Zhao, J. and Ban, Y. and Chen, G. and Levin, A. and Li, J. and Li, L. and Li, Q. and Mao, Y. and Qian, S. J. and Wang, D. and Xu, Z. and Wang, Y. and Avila, C. and Cabrera, A. and Carrillo Montoya, C. A. and Chaparro Sierra, L. F. and Florez, C. and González Hernández, C. F. and Segura Delgado, M. A. and Courbon, B. and Godinovic, N. and Lelas, D. and Puljak, I. and Sculac, T. and Antunovic, Z. and Kovac, M. and Brigljevic, V. and Ferencek, D. and Kadija, K. and Mesic, B. and Starodumov, A. and Susa, T. and Ather, M. W. and Attikis, A. and Kolosova, M. and Mavromanolakis, G. and Mousa, J. and Nicolaou, C. and Ptochos, F. and Razis, P. A. and Rykaczewski, H. and Finger, M. and Finger, M. and Ayala, E. and Carrera Jarrin, E. and Assran, Y. and Elgammal, S. and Khalil, S. and Bhowmik, S. and Carvalho Antunes De Oliveira, A. and Dewanjee, R. K. and Ehataht, K. and Kadastik, M. and Raidal, M. and Veelken, C. and Eerola, P. and Kirschenmann, H. and Pekkanen, J. and Voutilainen, M. and Havukainen, J. and Heikkilä, J. K. and Järvinen, T. and Karimäki, V. and Kinnunen, R. and Lampén, T. and Lassila-Perini, K. and Laurila, S. and Lehti, S. and Lindén, T. and Luukka, P. and Mäenpää, T. and Siikonen, H. and Tuominen, E. and Tuominiemi, J. and Tuuva, T. and Besancon, M. and Couderc, F. and Dejardin, M. and Denegri, D. and Faure, J. L. and Ferri, F. and Ganjour, S. and Givernaud, A. and Gras, P. and Hamel de Monchenault, G. and Jarry, P. and Leloup, C. and Locci, E. and Malcles, J. and Negro, G. and Rander, J. and Rosowsky, A. and Sahin, M. Ö. and Titov, M. and Abdulsalam, A. and Amendola, C. and Antropov, I. and Beaudette, F. and Busson, P. and Charlot, C. and Granier de Cassagnac, R. and Kucher, I. and Lisniak, S. and Lobanov, A. and Martin Blanco, J. and Nguyen, M. and Ochando, C. and Ortona, G. and Pigard, P. and Salerno, R. and Sauvan, J. B. and Sirois, Y. and Stahl Leiton, A. G. and Zabi, A. and Zghiche, A. and Agram, J.-L. and Andrea, J. and Bloch, D. and Brom, J.-M. and Chabert, E. C. and Cherepanov, V. and Collard, C. and Conte, E. and Fontaine, J.-C. and Gelé, D. and Goerlach, U. and Jansová, M. and Le Bihan, A.-C. and Tonon, N. and Van Hove, P. and Gadrat, S. and Beauceron, S. and Bernet, C. and Boudoul, G. and Chanon, N. and Chierici, R. and Contardo, D. and Depasse, P. and El Mamouni, H. and Fay, J. and Finco, L. and Gascon, S. and Gouzevitch, M. and Grenier, G. and Ille, B. and Lagarde, F. and Laktineh, I. B. and Lattaud, H. and Lethuillier, M. and Mirabito, L. and Pequegnot, A. L. and Perries, S. and Popov, A. and Sordini, V. and Vander Donckt, M. and Viret, S. and Zhang, S. and Khvedelidze, A. and Tsamalaidze, Z. and Autermann, C. and Feld, L. and Kiesel, M. K. and Klein, K. and Lipinski, M. and Preuten, M. and Rauch, M. P. and Schomakers, C. and Schulz, J. and Teroerde, M. and Wittmer, B. and Zhukov, V. and Albert, A. and Duchardt, D. and Endres, M. and Erdmann, M. and Esch, T. and Fischer, R. and Ghosh, S. and Güth, A. and Hebbeker, T. and Heidemann, C. and Hoepfner, K. and Keller, H. and Knutzen, S. and Mastrolorenzo, L. and Merschmeyer, M. and Meyer, A. and Millet, P. and Mukherjee, S. and Pook, T. and Radziej, M. and Reithler, H. and Rieger, M. and Scheuch, F. and Schmidt, A. and Teyssier, D. and Flügge, G. and Hlushchenko, O. and Kargoll, B. and Kress, T. and Künsken, A. and Müller, T. and Nehrkorn, A. and Nowack, A. and Pistone, C. and Pooth, O. and Sert, H. and Stahl, A. and Aldaya Martin, M. and Arndt, T. and Asawatangtrakuldee, C. and Babounikau, I. and Beernaert, K. and Behnke, O. and Behrens, U. and Bermúdez Martínez, A. and Bertsche, D. and Bin Anuar, A. A. and Borras, K. and Botta, V. and Campbell, A. and Connor, P. and Contreras-Campana, C. and Costanza, F. and Danilov, V. and De Wit, A. and Defranchis, M. M. and Diez Pardos, C. and Domínguez Damiani, D. and Eckerlin, G. and Eichhorn, T. and Elwood, A. and Eren, E. and Gallo, E. and Geiser, A. and Grados Luyando, J. M. and Grohsjean, A. and Gunnellini, P. and Guthoff, M. and Haranko, M. and Harb, A. and Hauk, J. and Jung, H. and Kasemann, M. and Keaveney, J. and Kleinwort, C. and Knolle, J. and Krücker, D. and Lange, W. and Lelek, A. and Lenz, T. and Lipka, K. and Lohmann, W. and Mankel, R. and Melzer-Pellmann, I.-A. and Meyer, A. B. and Meyer, M. and Missiroli, M. and Mittag, G. and Mnich, J. and Myronenko, V. and Pflitsch, S. K. and Pitzl, D. and Raspereza, A. and Savitskyi, M. and Saxena, P. and Schütze, P. and Schwanenberger, C. and Shevchenko, R. and Singh, A. and Stefaniuk, N. and Tholen, H. and Turkot, O. and Vagnerini, A. and Van Onsem, G. P. and Walsh, R. and Wen, Y. and Wichmann, K. and Wissing, C. and Zenaiev, O. and Aggleton, R. and Bein, S. and Benato, L. and Benecke, A. and Blobel, V. and Centis Vignali, M. and Dreyer, T. and Garutti, E. and Gonzalez, D. and Haller, J. and Hinzmann, A. and Karavdina, A. and Kasieczka, G. and Klanner, R. and Kogler, R. and Kovalchuk, N. and Kurz, S. and Kutzner, V. and Lange, J. and Marconi, D. and Multhaup, J. and Niedziela, M. and Nowatschin, D. and Perieanu, A. and Reimers, A. and Rieger, O. and Scharf, C. and Schleper, P. and Schumann, S. and Schwandt, J. and Sonneveld, J. and Stadie, H. and Steinbrück, G. and Stober, F. M. and Stöver, M. and Troendle, D. and Vanhoefer, A. and Vormwald, B. and Akbiyik, M. and Barth, C. and Baselga, M. and Baur, S. and Butz, E. and Caspart, R. and Chwalek, T. and Colombo, F. and De Boer, W. and Dierlamm, A. and Faltermann, N. and Freund, B. and Giffels, M. and Harrendorf, M. A. and Hartmann, F. and Heindl, S. M. and Husemann, U. and Kassel, F. and Katkov, I. and Kudella, S. and Mildner, H. and Mitra, S. and Mozer, M. U. and Müller, Th. and Plagge, M. and Quast, G. and Rabbertz, K. and Schröder, M. and Shvetsov, I. and Sieber, G. and Simonis, H. J. and Ulrich, R. and Wayand, S. and Weber, M. and Weiler, T. and Williamson, S. and Wöhrmann, C. and Wolf, R. and Anagnostou, G. and Daskalakis, G. and Geralis, T. and Kyriakis, A. and Loukas, D. and Paspalaki, G. and Topsis-Giotis, I. and Karathanasis, G. and Kesisoglou, S. and Kontaxakis, P. and Panagiotou, A. and Saoulidou, N. and Tziaferi, E. and Vellidis, K. and Kousouris, K. and Papakrivopoulos, I. and Tsipolitis, G. and Evangelou, I. and Foudas, C. and Gianneios, P. and Katsoulis, P. and Kokkas, P. and Mallios, S. and Manthos, N. and Papadopoulos, I. and Paradas, E. and Strologas, J. and Triantis, F. A. and Tsitsonis, D. and Bartók, M. and Csanad, M. and Filipovic, N. and Major, P. and Nagy, M. I. and Pasztor, G. and Surányi, O. and Veres, G. I. and Bencze, G. and Hajdu, C. and Horvath, D. and Hunyadi, Á. and Sikler, F. and Vámi, T. Á. and Veszpremi, V. and Vesztergombi, G. and Beni, N. and Czellar, S. and Karancsi, J. and Makovec, A. and Molnar, J. and Szillasi, Z. and Raics, P. and Trocsanyi, Z. L. and Ujvari, B. and Choudhury, S. and Komaragiri, J. R. and Tiwari, P. C. and Bahinipati, S. and Kar, C. and Mal, P. and Mandal, K. and Nayak, A. and Sahoo, D. K. and Swain, S. K. and Bansal, S. and Beri, S. B. and Bhatnagar, V. and Chauhan, S. and Chawla, R. and Dhingra, N. and Gupta, R. and Kaur, A. and Kaur, A. and Kaur, M. and Kaur, S. and Kumar, R. and Kumari, P. and Lohan, M. and Mehta, A. and Sandeep, K. and Sharma, S. and Singh, J. B. and Walia, G. and Bhardwaj, A. and Choudhary, B. C. and Garg, R. B. and Gola, M. and Keshri, S. and Kumar, Ashok and Malhotra, S. and Naimuddin, M. and Priyanka, P. and Ranjan, K. and Shah, Aashaq and Sharma, R. and Bhardwaj, R. and Bharti, M. and Bhattacharya, R. and Bhattacharya, S. and Bhawandeep, U. and Bhowmik, D. and Dey, S. and Dutt, S. and Dutta, S. and Ghosh, S. and Mondal, K. and Nandan, S. and Purohit, A. and Rout, P. K. and Roy, A. and Roy Chowdhury, S. and Sarkar, S. and Sharan, M. and Singh, B. and Thakur, S. and Behera, P. K. and Chudasama, R. and Dutta, D. and Jha, V. and Kumar, V. and Netrakanti, P. K. and Pant, L. M. and Shukla, P. and Aziz, T. and Bhat, M. A. and Dugad, S. and Mohanty, G. B. and Sur, N. and Sutar, B. and Verma, RavindraKumar and Banerjee, S. and Bhattacharya, S. and Chatterjee, S. and Das, P. and Guchait, M. and Jain, Sa. and Karmakar, S. and Kumar, S. and Maity, M. and Majumder, G. and Mazumdar, K. and Sahoo, N. and Sarkar, T. and Chauhan, S. and Dube, S. and Hegde, V. and Kapoor, A. and Kothekar, K. and Pandey, S. and Rane, A. and Sharma, S. and Chenarani, S. and Eskandari Tadavani, E. and Etesami, S. M. and Khakzad, M. and Mohammadi Najafabadi, M. and Naseri, M. and Rezaei Hosseinabadi, F. and Safarzadeh, B. and Zeinali, M. and Felcini, M. and Grunewald, M. and Abbrescia, M. and Calabria, C. and Colaleo, A. and Creanza, D. and Cristella, L. and De Filippis, N. and De Palma, M. and Di Florio, A. and Errico, F. and Fiore, L. and Gelmi, A. and Iaselli, G. and Ince, M. and Lezki, S. and Maggi, G. and Maggi, M. and Miniello, G. and My, S. and Nuzzo, S. and Pompili, A. and Pugliese, G. and Radogna, R. and Ranieri, A. and Selvaggi, G. and Sharma, A. and Silvestris, L. and Venditti, R. and Verwilligen, P. and Zito, G. and Abbiendi, G. and Battilana, C. and Bonacorsi, D. and Borgonovi, L. and Braibant-Giacomelli, S. and Campanini, R. and Capiluppi, P. and Castro, A. and Cavallo, F. R. and Chhibra, S. S. and Ciocca, C. and Codispoti, G. and Cuffiani, M. and Dallavalle, G. M. and Fabbri, F. and Fanfani, A. and Giacomelli, P. and Grandi, C. and Guiducci, L. and Iemmi, F. and Marcellini, S. and Masetti, G. and Montanari, A. and Navarria, F. L. and Perrotta, A. and Primavera, F. and Rossi, A. M. and Rovelli, T. and Siroli, G. P. and Tosi, N. and Albergo, S. and Di Mattia, A. and Potenza, R. and Tricomi, A. and Tuve, C. and Barbagli, G. and Chatterjee, K. and Ciulli, V. and Civinini, C. and D’Alessandro, R. and Focardi, E. and Latino, G. and Lenzi, P. and Meschini, M. and Paoletti, S. and Russo, L. and Sguazzoni, G. and Strom, D. and Viliani, L. and Benussi, L. and Bianco, S. and Fabbri, F. and Piccolo, D. and Ferro, F. and Ravera, F. and Robutti, E. and Tosi, S. and Benaglia, A. and Beschi, A. and Brianza, L. and Brivio, F. and Ciriolo, V. and Di Guida, S. and Dinardo, M. E. and Fiorendi, S. and Gennai, S. and Ghezzi, A. and Govoni, P. and Malberti, M. and Malvezzi, S. and Massironi, A. and Menasce, D. and Moroni, L. and Paganoni, M. and Pedrini, D. and Ragazzi, S. and Tabarelli de Fatis, T. and Buontempo, S. and Cavallo, N. and Di Crescenzo, A. and Fabozzi, F. and Fienga, F. and Galati, G. and Iorio, A. O. M. and Khan, W. A. and Lista, L. and Meola, S. and Paolucci, P. and Sciacca, C. and Voevodina, E. and Azzi, P. and Bacchetta, N. and Bisello, D. and Boletti, A. and Bragagnolo, A. and Dall’Osso, M. and De Castro Manzano, P. and Dorigo, T. and Dosselli, U. and Gasparini, F. and Gasparini, U. and Gozzelino, A. and Lacaprara, S. and Lujan, P. and Margoni, M. and Meneguzzo, A. T. and Montecassiano, F. and Pazzini, J. and Pozzobon, N. and Ronchese, P. and Rossin, R. and Simonetto, F. and Tiko, A. and Torassa, E. and Zanetti, M. and Zotto, P. and Zumerle, G. and Braghieri, A. and Magnani, A. and Montagna, P. and Ratti, S. P. and Re, V. and Ressegotti, M. and Riccardi, C. and Salvini, P. and Vai, I. and Vitulo, P. and Alunni Solestizi, L. and Biasini, M. and Bilei, G. M. and Cecchi, C. and Ciangottini, D. and Fanò, L. and Lariccia, P. and Leonardi, R. and Manoni, E. and Mantovani, G. and Mariani, V. and Menichelli, M. and Rossi, A. and Santocchia, A. and Spiga, D. and Androsov, K. and Azzurri, P. and Bagliesi, G. and Bianchini, L. and Boccali, T. and Borrello, L. and Castaldi, R. and Ciocci, M. A. and Dell’Orso, R. and Fedi, G. and Fiori, F. and Giannini, L. and Giassi, A. and Grippo, M. T. and Ligabue, F. and Manca, E. and Mandorli, G. and Messineo, A. and Palla, F. and Rizzi, A. and Spagnolo, P. and Tenchini, R. and Tonelli, G. and Venturi, A. and Verdini, P. G. and Barone, L. and Cavallari, F. and Cipriani, M. and Daci, N. and Del Re, D. and Di Marco, E. and Diemoz, M. and Gelli, S. and Longo, E. and Marzocchi, B. and Meridiani, P. and Organtini, G. and Pandolfi, F. and Paramatti, R. and Preiato, F. and Rahatlou, S. and Rovelli, C. and Santanastasio, F. and Amapane, N. and Arcidiacono, R. and Argiro, S. and Arneodo, M. and Bartosik, N. and Bellan, R. and Biino, C. and Cartiglia, N. and Cenna, F. and Cometti, S. and Costa, M. and Covarelli, R. and Demaria, N. and Kiani, B. and Mariotti, C. and Maselli, S. and Migliore, E. and Monaco, V. and Monteil, E. and Monteno, M. and Obertino, M. M. and Pacher, L. and Pastrone, N. and Pelliccioni, M. and Pinna Angioni, G. L. and Romero, A. and Ruspa, M. and Sacchi, R. and Shchelina, K. and Sola, V. and Solano, A. and Soldi, D. and Staiano, A. and Belforte, S. and Candelise, V. and Casarsa, M. and Cossutti, F. and Della Ricca, G. and Vazzoler, F. and Zanetti, A. and Kim, D. H. and Kim, G. N. and Kim, M. S. and Lee, J. and Lee, S. and Lee, S. W. and Moon, C. S. and Oh, Y. D. and Sekmen, S. and Son, D. C. and Yang, Y. C. and Kim, H. and Moon, D. H. and Oh, G. and Goh, J. and Kim, T. J. and Cho, S. and Choi, S. and Go, Y. and Gyun, D. and Ha, S. and Hong, B. and Jo, Y. and Lee, K. and Lee, K. S. and Lee, S. and Lim, J. and Park, S. K. and Roh, Y. and Kim, H. S. and Almond, J. and Kim, J. and Kim, J. S. and Lee, H. and Lee, K. and Nam, K. and Oh, S. B. and Radburn-Smith, B. C. and Seo, S. h. and Yang, U. K. and Yoo, H. D. and Yu, G. B. and Jeon, D. and Kim, H. and Kim, J. H. and Lee, J. S. H. and Park, I. C. and Choi, Y. and Hwang, C. and Lee, J. and Yu, I. and Dudenas, V. and Juodagalvis, A. and Vaitkus, J. and Ahmed, I. and Ibrahim, Z. A. and Md Ali, M. A. B. and Mohamad Idris, F. and Wan Abdullah, W. A. T. and Yusli, M. N. and Zolkapli, Z. and Castaneda Hernandez, A. and Murillo Quijada, J. A. and Duran-Osuna, M. C. and Castilla-Valdez, H. and De La Cruz-Burelo, E. and Ramirez-Sanchez, G. and Heredia-De La Cruz, I. and Rabadan-Trejo, R. I. and Lopez-Fernandez, R. and Mejia Guisao, J. and Reyes-Almanza, R and Sanchez-Hernandez, A. and Carrillo Moreno, S. and Oropeza Barrera, C. and Vazquez Valencia, F. and Eysermans, J. and Pedraza, I. and Salazar Ibarguen, H. A. and Uribe Estrada, C. and Morelos Pineda, A. and Krofcheck, D. and Bheesette, S. and Butler, P. H. and Ahmad, A. and Ahmad, M. and Asghar, M. I. and Hassan, Q. and Hoorani, H. R. and Saddique, A. and Shah, M. A. and Shoaib, M. and Waqas, M. and Bialkowska, H. and Bluj, M. and Boimska, B. and Frueboes, T. and Górski, M. and Kazana, M. and Nawrocki, K. and Szleper, M. and Traczyk, P. and Zalewski, P. and Bunkowski, K. and Byszuk, A. and Doroba, K. and Kalinowski, A. and Konecki, M. and Krolikowski, J. and Misiura, M. and Olszewski, M. and Pyskir, A. and Walczak, M. and Bargassa, P. and Beirão Da Cruz E Silva, C. and Di Francesco, A. and Faccioli, P. and Galinhas, B. and Gallinaro, M. and Hollar, J. and Leonardo, N. and Lloret Iglesias, L. and Nemallapudi, M. V. and Seixas, J. and Strong, G. and Toldaiev, O. and Vadruccio, D. and Varela, J. and Afanasiev, S. and Alexakhin, V. and Bunin, P. and Gavrilenko, M. and Golunov, A. and Golutvin, I. and Gorbounov, N. and Karjavin, V. and Lanev, A. and Malakhov, A. and Matveev, V. and Moisenz, P. and Palichik, V. and Perelygin, V. and Savina, M. and Shmatov, S. and Smirnov, V. and Voytishin, N. and Zarubin, A. and Golovtsov, V. and Ivanov, Y. and Kim, V. and Kuznetsova, E. and Levchenko, P. and Murzin, V. and Oreshkin, V. and Smirnov, I. and Sosnov, D. and Sulimov, V. and Uvarov, L. and Vavilov, S. and Vorobyev, A. and Andreev, Yu. and Dermenev, A. and Gninenko, S. and Golubev, N. and Karneyeu, A. and Kirsanov, M. and Krasnikov, N. and Pashenkov, A. and Tlisov, D. and Toropin, A. and Epshteyn, V. and Gavrilov, V. and Lychkovskaya, N. and Popov, V. and Pozdnyakov, I. and Safronov, G. and Spiridonov, A. and Stepennov, A. and Stolin, V. and Toms, M. and Vlasov, E. and Zhokin, A. and Aushev, T. and Chistov, R. and Danilov, M. and Parygin, P. and Philippov, D. and Polikarpov, S. and Tarkovskii, E. and Andreev, V. and Azarkin, M. and Dremin, I. and Kirakosyan, M. and Rusakov, S. V. and Terkulov, A. and Baskakov, A. and Belyaev, A. and Boos, E. and Dubinin, M. and Dudko, L. and Ershov, A. and Gribushin, A. and Klyukhin, V. and Kodolova, O. and Lokhtin, I. and Miagkov, I. and Obraztsov, S. and Petrushanko, S. and Savrin, V. and Snigirev, A. and Blinov, V. and Dimova, T. and Kardapoltsev, L. and Shtol, D. and Skovpen, Y. and Azhgirey, I. and Bayshev, I. and Bitioukov, S. and Elumakhov, D. and Godizov, A. and Kachanov, V. and Kalinin, A. and Konstantinov, D. and Mandrik, P. and Petrov, V. and Ryutin, R. and Slabospitskii, S. and Sobol, A. and Troshin, S. and Tyurin, N. and Uzunian, A. and Volkov, A. and Babaev, A. and Baidali, S. and Adzic, P. and Cirkovic, P. and Devetak, D. and Dordevic, M. and Milosevic, J. and Alcaraz Maestre, J. and Álvarez Fernández, A. and Bachiller, I. and Barrio Luna, M. and Brochero Cifuentes, J. A. and Cerrada, M. and Colino, N. and De La Cruz, B. and Delgado Peris, A. and Fernandez Bedoya, C. and Fernández Ramos, J. P. and Flix, J. and Fouz, M. C. and Gonzalez Lopez, O. and Goy Lopez, S. and Hernandez, J. M. and Josa, M. I. and Moran, D. and Pérez-Calero Yzquierdo, A. and Puerta Pelayo, J. and Redondo, I. and Romero, L. and Soares, M. S. and Triossi, A. and Albajar, C. and de Trocóniz, J. F. and Cuevas, J. and Erice, C. and Fernandez Menendez, J. and Folgueras, S. and Gonzalez Caballero, I. and González Fernández, J. R. and Palencia Cortezon, E. and Rodríguez Bouza, V. and Sanchez Cruz, S. and Vischia, P. and Vizan Garcia, J. M. and Cabrillo, I. J. and Calderon, A. and Chazin Quero, B. and Duarte Campderros, J. and Fernandez, M. and Fernández Manteca, P. J. and García Alonso, A. and Garcia-Ferrero, J. and Gomez, G. and Lopez Virto, A. and Marco, J. and Martinez Rivero, C. and Martinez Ruiz del Arbol, P. and Matorras, F. and Piedra Gomez, J. and Prieels, C. and Rodrigo, T. and Ruiz-Jimeno, A. and Scodellaro, L. and Trevisani, N. and Vila, I. and Vilar Cortabitarte, R. and Abbaneo, D. and Akgun, B. and Auffray, E. and Baillon, P. and Ball, A. H. and Barney, D. and Bendavid, J. and Bianco, M. and Bocci, A. and Botta, C. and Brondolin, E. and Camporesi, T. and Cepeda, M. and Cerminara, G. and Chapon, E. and Chen, Y. and Cucciati, G. and d’Enterria, D. and Dabrowski, A. and Daponte, V. and David, A. and De Roeck, A. and Deelen, N. and Dobson, M. and du Pree, T. and Dünser, M. and Dupont, N. and Peisert, A. Elliott- and Everaerts, P. and Fallavollita, F. and Fasanella, D. and Franzoni, G. and Fulcher, J. and Funk, W. and Gigi, D. and Gilbert, A. and Gill, K. and Glege, F. and Guilbaud, M. and Gulhan, D. and Hegeman, J. and Innocente, V. and Jafari, A. and Janot, P. and Karacheban, O. and Kieseler, J. and Kornmayer, A. and Krammer, M. and Lange, C. and Lecoq, P. and Lourenço, C. and Malgeri, L. and Mannelli, M. and Meijers, F. and Merlin, J. A. and Mersi, S. and Meschi, E. and Milenovic, P. and Moortgat, F. and Mulders, M. and Ngadiuba, J. and Orfanelli, S. and Orsini, L. and Pantaleo, F. and Pape, L. and Perez, E. and Peruzzi, M. and Petrilli, A. and Petrucciani, G. and Pfeiffer, A. and Pierini, M. and Pitters, F. M. and Rabady, D. and Racz, A. and Reis, T. and Rolandi, G. and Rovere, M. and Sakulin, H. and Schäfer, C. and Schwick, C. and Seidel, M. and Selvaggi, M. and Sharma, A. and Silva, P. and Sphicas, P. and Stakia, A. and Steggemann, J. and Tosi, M. and Treille, D. and Tsirou, A. and Veckalns, V. and Zeuner, W. D. and Caminada, L. and Deiters, K. and Erdmann, W. and Horisberger, R. and Ingram, Q. and Kaestli, H. C. and Kotlinski, D. and Langenegger, U. and Rohe, T. and Wiederkehr, S. A. and Backhaus, M. and Bäni, L. and Berger, P. and Chernyavskaya, N. and Dissertori, G. and Dittmar, M. and Donegà, M. and Dorfer, C. and Grab, C. and Heidegger, C. and Hits, D. and Hoss, J. and Klijnsma, T. and Lustermann, W. and Manzoni, R. A. and Marionneau, M. and Meinhard, M. T. and Micheli, F. and Musella, P. and Nessi-Tedaldi, F. and Pata, J. and Pauss, F. and Perrin, G. and Perrozzi, L. and Pigazzini, S. and Quittnat, M. and Ruini, D. and Sanz Becerra, D. A. and Schönenberger, M. and Shchutska, L. and Tavolaro, V. R. and Theofilatos, K. and Vesterbacka Olsson, M. L. and Wallny, R. and Zhu, D. H. and Aarrestad, T. K. and Amsler, C. and Brzhechko, D. and Canelli, M. F. and De Cosa, A. and Del Burgo, R. and Donato, S. and Galloni, C. and Hreus, T. and Kilminster, B. and Neutelings, I. and Pinna, D. and Rauco, G. and Robmann, P. and Salerno, D. and Schweiger, K. and Seitz, C. and Takahashi, Y. and Zucchetta, A. and Chang, Y. H. and Cheng, K. y. and Doan, T. H. and Jain, Sh. and Khurana, R. and Kuo, C. M. and Lin, W. and Pozdnyakov, A. and Yu, S. S. and Chang, P. and Chao, Y. and Chen, K. F. and Chen, P. H. and Hou, W.-S. and Kumar, Arun and Li, Y. y. and Liu, Y. F. and Lu, R.-S. and Paganis, E. and Psallidas, A. and Steen, A. and Tsai, J. f. and Asavapibhop, B. and Srimanobhas, N. and Suwonjandee, N. and Bat, A. and Boran, F. and Cerci, S. and Damarseckin, S. and Demiroglu, Z. S. and Dolek, F. and Dozen, C. and Dumanoglu, I. and Girgis, S. and Gokbulut, G. and Guler, Y. and Gurpinar, E. and Hos, I. and Isik, C. and Kangal, E. E. and Kara, O. and Kayis Topaksu, A. and Kiminsu, U. and Oglakci, M. and Onengut, G. and Ozdemir, K. and Ozturk, S. and Sunar Cerci, D. and Tali, B. and Tok, U. G. and Turkcapar, S. and Zorbakir, I. S. and Zorbilmez, C. and Isildak, B. and Karapinar, G. and Yalvac, M. and Zeyrek, M. and Atakisi, I. O. and Gülmez, E. and Kaya, M. and Kaya, O. and Tekten, S. and Yetkin, E. A. and Agaras, M. N. and Atay, S. and Cakir, A. and Cankocak, K. and Komurcu, Y. and Sen, S. and Grynyov, B. and Levchuk, L. and Ball, F. and Beck, L. and Brooke, J. J. and Burns, D. and Clement, E. and Cussans, D. and Davignon, O. and Flacher, H. and Goldstein, J. and Heath, G. P. and Heath, H. F. and Kreczko, L. and Newbold, D. M. and Paramesvaran, S. and Penning, B. and Sakuma, T. and Smith, D. and Smith, V. J. and Taylor, J. and Titterton, A. and Bell, K. W. and Belyaev, A. and Brew, C. and Brown, R. M. and Cieri, D. and Cockerill, D. J. A. and Coughlan, J. A. and Harder, K. and Harper, S. and Linacre, J. and Olaiya, E. and Petyt, D. and Shepherd-Themistocleous, C. H. and Thea, A. and Tomalin, I. R. and Williams, T. and Womersley, W. J. and Auzinger, G. and Bainbridge, R. and Bloch, P. and Borg, J. and Breeze, S. and Buchmuller, O. and Bundock, A. and Casasso, S. and Colling, D. and Corpe, L. and Dauncey, P. and Davies, G. and Della Negra, M. and Di Maria, R. and Haddad, Y. and Hall, G. and Iles, G. and James, T. and Komm, M. and Laner, C. and Lyons, L. and Magnan, A.-M. and Malik, S. and Martelli, A. and Nash, J. and Nikitenko, A. and Palladino, V. and Pesaresi, M. and Richards, A. and Rose, A. and Scott, E. and Seez, C. and Shtipliyski, A. and Singh, G. and Stoye, M. and Strebler, T. and Summers, S. and Tapper, A. and Uchida, K. and Virdee, T. and Wardle, N. and Winterbottom, D. and Wright, J. and Zenz, S. C. and Cole, J. E. and Hobson, P. R. and Khan, A. and Kyberd, P. and Mackay, C. K. and Morton, A. and Reid, I. D. and Teodorescu, L. and Zahid, S. and Call, K. and Dittmann, J. and Hatakeyama, K. and Liu, H. and Madrid, C. and Mcmaster, B. and Pastika, N. and Smith, C. and Bartek, R. and Dominguez, A. and Buccilli, A. and Cooper, S. I. and Henderson, C. and Rumerio, P. and West, C. and Arcaro, D. and Bose, T. and Gastler, D. and Rankin, D. and Richardson, C. and Rohlf, J. and Sulak, L. and Zou, D. and Benelli, G. and Coubez, X. and Cutts, D. and Hadley, M. and Hakala, J. and Heintz, U. and Hogan, J. M. and Kwok, K. H. M. and Laird, E. and Landsberg, G. and Lee, J. and Mao, Z. and Narain, M. and Piperov, S. and Sagir, S. and Syarif, R. and Usai, E. and Yu, D. and Band, R. and Brainerd, C. and Breedon, R. and Burns, D. and Calderon De La Barca Sanchez, M. and Chertok, M. and Conway, J. and Conway, R. and Cox, P. T. and Erbacher, R. and Flores, C. and Funk, G. and Ko, W. and Kukral, O. and Lander, R. and Mclean, C. and Mulhearn, M. and Pellett, D. and Pilot, J. and Shalhout, S. and Shi, M. and Stolp, D. and Taylor, D. and Tos, K. and Tripathi, M. and Wang, Z. and Zhang, F. and Bachtis, M. and Bravo, C. and Cousins, R. and Dasgupta, A. and Florent, A. and Hauser, J. and Ignatenko, M. and Mccoll, N. and Regnard, S. and Saltzberg, D. and Schnaible, C. and Valuev, V. and Bouvier, E. and Burt, K. and Clare, R. and Gary, J. W. and Ghiasi Shirazi, S. M. A. and Hanson, G. and Karapostoli, G. and Kennedy, E. and Lacroix, F. and Long, O. R. and Olmedo Negrete, M. and Paneva, M. I. and Si, W. and Wang, L. and Wei, H. and Wimpenny, S. and Yates, B. R. and Branson, J. G. and Cittolin, S. and Derdzinski, M. and Gerosa, R. and Gilbert, D. and Hashemi, B. and Holzner, A. and Klein, D. and Kole, G. and Krutelyov, V. and Letts, J. and Masciovecchio, M. and Olivito, D. and Padhi, S. and Pieri, M. and Sani, M. and Sharma, V. and Simon, S. and Tadel, M. and Vartak, A. and Wasserbaech, S. and Wood, J. and Würthwein, F. and Yagil, A. and Zevi Della Porta, G. and Amin, N. and Bhandari, R. and Bradmiller-Feld, J. and Campagnari, C. and Citron, M. and Dishaw, A. and Dutta, V. and Franco Sevilla, M. and Gouskos, L. and Heller, R. and Incandela, J. and Ovcharova, A. and Qu, H. and Richman, J. and Stuart, D. and Suarez, I. and Wang, S. and Yoo, J. and Anderson, D. and Bornheim, A. and Lawhorn, J. M. and Newman, H. B. and Nguyen, T. Q. and Spiropulu, M. and Vlimant, J. R. and Wilkinson, R. and Xie, S. and Zhang, Z. and Zhu, R. Y. and Andrews, M. B. and Ferguson, T. and Mudholkar, T. and Paulini, M. and Sun, M. and Vorobiev, I. and Weinberg, M. and Cumalat, J. P. and Ford, W. T. and Jensen, F. and Johnson, A. and Krohn, M. and Leontsinis, S. and MacDonald, E. and Mulholland, T. and Stenson, K. and Ulmer, K. A. and Wagner, S. R. and Alexander, J. and Chaves, J. and Cheng, Y. and Chu, J. and Datta, A. and Mcdermott, K. and Mirman, N. and Patterson, J. R. and Quach, D. and Rinkevicius, A. and Ryd, A. and Skinnari, L. and Soffi, L. and Tan, S. M. and Tao, Z. and Thom, J. and Tucker, J. and Wittich, P. and Zientek, M. and Abdullin, S. and Albrow, M. and Alyari, M. and Apollinari, G. and Apresyan, A. and Apyan, A. and Banerjee, S. and Bauerdick, L. A. T. and Beretvas, A. and Berryhill, J. and Bhat, P. C. and Bolla, G. and Burkett, K. and Butler, J. N. and Canepa, A. and Cerati, G. B. and Cheung, H. W. K. and Chlebana, F. and Cremonesi, M. and Duarte, J. and Elvira, V. D. and Freeman, J. and Gecse, Z. and Gottschalk, E. and Gray, L. and Green, D. and Grünendahl, S. and Gutsche, O. and Hanlon, J. and Harris, R. M. and Hasegawa, S. and Hirschauer, J. and Hu, Z. and Jayatilaka, B. and Jindariani, S. and Johnson, M. and Joshi, U. and Klima, B. and Kortelainen, M. J. and Kreis, B. and Lammel, S. and Lincoln, D. and Lipton, R. and Liu, M. and Liu, T. and Lykken, J. and Maeshima, K. and Marraffino, J. M. and Mason, D. and McBride, P. and Merkel, P. and Mrenna, S. and Nahn, S. and O’Dell, V. and Pedro, K. and Pena, C. and Prokofyev, O. and Rakness, G. and Ristori, L. and Savoy-Navarro, A. and Schneider, B. and Sexton-Kennedy, E. and Soha, A. and Spalding, W. J. and Spiegel, L. and Stoynev, S. and Strait, J. and Strobbe, N. and Taylor, L. and Tkaczyk, S. and Tran, N. V. and Uplegger, L. and Vaandering, E. W. and Vernieri, C. and Verzocchi, M. and Vidal, R. and Wang, M. and Weber, H. A. and Whitbeck, A. and Acosta, D. and Avery, P. and Bortignon, P. and Bourilkov, D. and Brinkerhoff, A. and Cadamuro, L. and Carnes, A. and Carver, M. and Curry, D. and Field, R. D. and Gleyzer, S. V. and Joshi, B. M. and Konigsberg, J. and Korytov, A. and Ma, P. and Matchev, K. and Mei, H. and Mitselmakher, G. and Shi, K. and Sperka, D. and Wang, J. and Wang, S. and Joshi, Y. R. and Linn, S. and Ackert, A. and Adams, T. and Askew, A. and Hagopian, S. and Hagopian, V. and Johnson, K. F. and Kolberg, T. and Martinez, G. and Perry, T. and Prosper, H. and Saha, A. and Sharma, V. and Yohay, R. and Baarmand, M. M. and Bhopatkar, V. and Colafranceschi, S. and Hohlmann, M. and Noonan, D. and Rahmani, M. and Roy, T. and Yumiceva, F. and Adams, M. R. and Apanasevich, L. and Berry, D. and Betts, R. R. and Cavanaugh, R. and Chen, X. and Dittmer, S. and Evdokimov, O. and Gerber, C. E. and Hangal, D. A. and Hofman, D. J. and Jung, K. and Kamin, J. and Mills, C. and Sandoval Gonzalez, I. D. and Tonjes, M. B. and Varelas, N. and Wang, H. and Wang, X. and Wu, Z. and Zhang, J. and Alhusseini, M. and Bilki, B. and Clarida, W. and Dilsiz, K. and Durgut, S. and Gandrajula, R. P. and Haytmyradov, M. and Khristenko, V. and Merlo, J.-P. and Mestvirishvili, A. and Moeller, A. and Nachtman, J. and Ogul, H. and Onel, Y. and Ozok, F. and Penzo, A. and Snyder, C. and Tiras, E. and Wetzel, J. and Blumenfeld, B. and Cocoros, A. and Eminizer, N. and Fehling, D. and Feng, L. and Gritsan, A. V. and Hung, W. T. and Maksimovic, P. and Roskes, J. and Sarica, U. and Swartz, M. and Xiao, M. and You, C. and Al-bataineh, A. and Baringer, P. and Bean, A. and Boren, S. and Bowen, J. and Bylinkin, A. and Castle, J. and Khalil, S. and Kropivnitskaya, A. and Majumder, D. and Mcbrayer, W. and Murray, M. and Rogan, C. and Sanders, S. and Schmitz, E. and Tapia ∋Takaki, J. D. and Wang, Q. and Duric, S. and Ivanov, A. and Kaadze, K. and Kim, D. and Maravin, Y. and Mendis, D. R. and Mitchell, T. and Modak, A. and Mohammadi, A. and Saini, L. K. and Skhirtladze, N. and Rebassoo, F. and Wright, D. and Baden, A. and Baron, O. and Belloni, A. and Eno, S. C. and Feng, Y. and Ferraioli, C. and Hadley, N. J. and Jabeen, S. and Jeng, G. Y. and Kellogg, R. G. and Kunkle, J. and Mignerey, A. C. and Ricci-Tam, F. and Shin, Y. H. and Skuja, A. and Tonwar, S. C. and Wong, K. and Abercrombie, D. and Allen, B. and Azzolini, V. and Baty, A. and Bauer, G. and Bi, R. and Brandt, S. and Busza, W. and Cali, I. A. and D’Alfonso, M. and Demiragli, Z. and Gomez Ceballos, G. and Goncharov, M. and Harris, P. and Hsu, D. and Hu, M. and Iiyama, Y. and Innocenti, G. M. and Klute, M. and Kovalskyi, D. and Lee, Y.-J. and Luckey, P. D. and Maier, B. and Marini, A. C. and Mcginn, C. and Mironov, C. and Narayanan, S. and Niu, X. and Paus, C. and Roland, C. and Roland, G. and Stephans, G. S. F. and Sumorok, K. and Tatar, K. and Velicanu, D. and Wang, J. and Wang, T. W. and Wyslouch, B. and Zhaozhong, S. and Benvenuti, A. C. and Chatterjee, R. M. and Evans, A. and Hansen, P. and Kalafut, S. and Kubota, Y. and Lesko, Z. and Mans, J. and Nourbakhsh, S. and Ruckstuhl, N. and Rusack, R. and Turkewitz, J. and Wadud, M. A. and Acosta, J. G. and Oliveros, S. and Avdeeva, E. and Bloom, K. and Claes, D. R. and Fangmeier, C. and Golf, F. and Gonzalez Suarez, R. and Kamalieddin, R. and Kravchenko, I. and Monroy, J. and Siado, J. E. and Snow, G. R. and Stieger, B. and Godshalk, A. and Harrington, C. and Iashvili, I. and Kharchilava, A. and Nguyen, D. and Parker, A. and Rappoccio, S. and Roozbahani, B. and Barberis, E. and Freer, C. and Hortiangtham, A. and Morse, D. M. and Orimoto, T. and Teixeira De Lima, R. and Wamorkar, T. and Wang, B. and Wisecarver, A. and Wood, D. and Bhattacharya, S. and Charaf, O. and Hahn, K. A. and Mucia, N. and Odell, N. and Schmitt, M. H. and Sung, K. and Trovato, M. and Velasco, M. and Bucci, R. and Dev, N. and Hildreth, M. and Hurtado Anampa, K. and Jessop, C. and Karmgard, D. J. and Kellams, N. and Lannon, K. and Li, W. and Loukas, N. and Marinelli, N. and Meng, F. and Mueller, C. and Musienko, Y. and Planer, M. and Reinsvold, A. and Ruchti, R. and Siddireddy, P. and Smith, G. and Taroni, S. and Wayne, M. and Wightman, A. and Wolf, M. and Woodard, A. and Alimena, J. and Antonelli, L. and Bylsma, B. and Durkin, L. S. and Flowers, S. and Francis, B. and Hart, A. and Hill, C. and Ji, W. and Ling, T. Y. and Luo, W. and Winer, B. L. and Wulsin, H. W. and Cooperstein, S. and Elmer, P. and Hardenbrook, J. and Hebda, P. and Higginbotham, S. and Kalogeropoulos, A. and Lange, D. and Lucchini, M. T. and Luo, J. and Marlow, D. and Mei, K. and Ojalvo, I. and Olsen, J. and Palmer, C. and Piroué, P. and Salfeld-Nebgen, J. and Stickland, D. and Tully, C. and Malik, S. and Norberg, S. and Barker, A. and Barnes, V. E. and Das, S. and Gutay, L. and Jones, M. and Jung, A. W. and Khatiwada, A. and Mahakud, B. and Miller, D. H. and Neumeister, N. and Peng, C. C. and Qiu, H. and Schulte, J. F. and Sun, J. and Wang, F. and Xiao, R. and Xie, W. and Cheng, T. and Dolen, J. and Parashar, N. and Chen, Z. and Ecklund, K. M. and Freed, S. and Geurts, F. J. M. and Kilpatrick, M. and Li, W. and Michlin, B. and Padley, B. P. and Roberts, J. and Rorie, J. and Shi, W. and Tu, Z. and Zabel, J. and Zhang, A. and Bodek, A. and de Barbaro, P. and Demina, R. and Duh, Y. t. and Dulemba, J. L. and Fallon, C. and Ferbel, T. and Galanti, M. and Garcia-Bellido, A. and Han, J. and Hindrichs, O. and Khukhunaishvili, A. and Lo, K. H. and Tan, P. and Taus, R. and Verzetti, M. and Agapitos, A. and Chou, J. P. and Gershtein, Y. and Gómez Espinosa, T. A. and Halkiadakis, E. and Heindl, M. and Hughes, E. and Kaplan, S. and Kunnawalkam Elayavalli, R. and Kyriacou, S. and Lath, A. and Montalvo, R. and Nash, K. and Osherson, M. and Saka, H. and Salur, S. and Schnetzer, S. and Sheffield, D. and Somalwar, S. and Stone, R. and Thomas, S. and Thomassen, P. and Walker, M. and Delannoy, A. G. and Heideman, J. and Riley, G. and Rose, K. and Spanier, S. and Thapa, K. and Bouhali, O. and Celik, A. and Dalchenko, M. and De Mattia, M. and Delgado, A. and Dildick, S. and Eusebi, R. and Gilmore, J. and Huang, T. and Kamon, T. and Luo, S. and Mueller, R. and Pakhotin, Y. and Patel, R. and Perloff, A. and Perniè, L. and Rathjens, D. and Safonov, A. and Tatarinov, A. and Akchurin, N. and Damgov, J. and De Guio, F. and Dudero, P. R. and Kunori, S. and Lamichhane, K. and Lee, S. W. and Mengke, T. and Muthumuni, S. and Peltola, T. and Undleeb, S. and Volobouev, I. and Wang, Z. and Greene, S. and Gurrola, A. and Janjam, R. and Johns, W. and Maguire, C. and Melo, A. and Ni, H. and Padeken, K. and Ruiz Alvarez, J. D. and Sheldon, P. and Tuo, S. and Velkovska, J. and Verweij, M. and Xu, Q. and Arenton, M. W. and Barria, P. and Cox, B. and Dezoort, G. and Hirosky, R. and Jiwon, H. and Joyce, M. and Ledovskoy, A. and Li, H. and Neu, C. and Sinthuprasith, T. and Wang, Y. and Wolfe, E. and Xia, F. and Harr, R. and Karchin, P. E. and Poudyal, N. and Sturdy, J. and Thapa, P. and Zaleski, S. and Brodski, M. and Buchanan, J. and Caillol, C. and Carlsmith, D. and Dasu, S. and Dodd, L. and Gomber, B. and Grothe, M. and Herndon, M. and Hervé, A. and Hussain, U. and Klabbers, P. and Lanaro, A. and Levine, A. and Long, K. and Loveless, R. and Ruggles, T. and Savin, A. and Smith, N. and Smith, W. H. and Woods, N.},
   year={2018},
   month=Nov }

@article{Emparan_2002,
   title={Quantum Black Holes as Holograms in AdS Braneworlds},
   volume={2002},
   ISSN={1029-8479},
   url={http://dx.doi.org/10.1088/1126-6708/2002/08/043},
   DOI={10.1088/1126-6708/2002/08/043},
   number={08},
   journal={Journal of High Energy Physics},
   publisher={Springer Science and Business Media LLC},
   author={Emparan, Roberto and Fabbri, Alessandro and Kaloper, Nemanja},
   year={2002},
   month=Aug, pages={043–043} }

@article{Frost_2009,
   title={Phenomenology of production and decay of spinning extra-dimensional black holes at hadron colliders},
   volume={2009},
   ISSN={1029-8479},
   url={http://dx.doi.org/10.1088/1126-6708/2009/10/014},
   DOI={10.1088/1126-6708/2009/10/014},
   number={10},
   journal={Journal of High Energy Physics},
   publisher={Springer Science and Business Media LLC},
   author={Frost, James A and Gaunt, Jonathan R and Sampaio, Marco O.P and Casals, Marc and Dolan, Sam R and Parker, M. Andrew and Webber, Bryan R},
   year={2009},
   month=Oct, pages={014–014} }

@article{Gregory_1993,
   title={Black strings and $p$-branes are unstable},
   volume={70},
   ISSN={0031-9007},
   url={http://dx.doi.org/10.1103/PhysRevLett.70.2837},
   DOI={10.1103/physrevlett.70.2837},
   number={19},
   journal={Physical Review Letters},
   publisher={American Physical Society (APS)},
   author={Gregory, Ruth and Laflamme, Raymond},
   year={1993},
   month=May, pages={2837–2840} }

@inbook{Gregory,
   title={Braneworld Black Holes},
   ISBN={9783540884606},
   ISSN={1616-6361},
   url={http://dx.doi.org/10.1007/978-3-540-88460-6_7},
   DOI={10.1007/978-3-540-88460-6_7},
   booktitle={Physics of Black Holes},
   publisher={Springer Berlin Heidelberg},
   author={Gregory, R.},
   pages={259–298} }

@article{Olasagasti_2000,
   title={Gravity of higher-dimensional global defects},
   volume={62},
   ISSN={1089-4918},
   url={http://dx.doi.org/10.1103/PhysRevD.62.044014},
   DOI={10.1103/physrevd.62.044014},
   number={4},
   journal={Physical Review D},
   publisher={American Physical Society (APS)},
   author={Olasagasti, Itsaso and Vilenkin, Alexander},
   year={2000},
   month=jul }

@article{Emparan_2000-b,
   title={Exact description of black holes on branes},
   volume={2000},
   ISSN={1029-8479},
   url={http://dx.doi.org/10.1088/1126-6708/2000/01/007},
   DOI={10.1088/1126-6708/2000/01/007},
   number={01},
   journal={Journal of High Energy Physics},
   publisher={Springer Science and Business Media LLC},
   author={Emparan, Roberto and Horowitz, Gary T and Myers, Robert C},
   year={2000},
   month=Jan, pages={007–007} }

@article{Fraser_2015,
   title={Stability and binding energy of small asymptotically Randall-Sundrum black holes},
   volume={92},
   ISSN={1550-2368},
   url={http://dx.doi.org/10.1103/PhysRevD.92.024032},
   DOI={10.1103/physrevd.92.024032},
   number={2},
   journal={Physical Review D},
   publisher={American Physical Society (APS)},
   author={Fraser, Scott and Eardley, Douglas M.},
   year={2015},
   month=jul }

@article{Frolov_2002,
   title={Black Hole as a Point Radiator and Recoil Effect on the Brane World},
   volume={89},
   ISSN={1079-7114},
   url={http://dx.doi.org/10.1103/PhysRevLett.89.151302},
   DOI={10.1103/physrevlett.89.151302},
   number={15},
   journal={Physical Review Letters},
   publisher={American Physical Society (APS)},
   author={Frolov, Valeri and Stojković, Dejan},
   year={2002},
   month=sep }

@article{Frolov_2002-b,
   title={Black hole radiation in the brane world and the recoil effect},
   volume={66},
   ISSN={1089-4918},
   url={http://dx.doi.org/10.1103/PhysRevD.66.084002},
   DOI={10.1103/physrevd.66.084002},
   number={8},
   journal={Physical Review D},
   publisher={American Physical Society (APS)},
   author={Frolov, Valeri and Stojković, Dejan},
   year={2002},
   month=Oct }

@article{Flachi_2005,
   title={Escape of Black Holes from the Brane},
   volume={95},
   ISSN={1079-7114},
   url={http://dx.doi.org/10.1103/PhysRevLett.95.161302},
   DOI={10.1103/physrevlett.95.161302},
   number={16},
   journal={Physical Review Letters},
   publisher={American Physical Society (APS)},
   author={Flachi, Antonino and Tanaka, Takahiro},
   year={2005},
   month=Oct }

@article{Gregory_2004,
   title={Observing braneworld black holes},
   volume={2004},
   ISSN={1475-7516},
   url={http://dx.doi.org/10.1088/1475-7516/2004/10/013},
   DOI={10.1088/1475-7516/2004/10/013},
   number={10},
   journal={Journal of Cosmology and Astroparticle Physics},
   publisher={IOP Publishing},
   author={Gregory, Ruth and Whisker, Richard and Beckwith, Kris and Done, Chris},
   year={2004},
   month=Oct, pages={013–013} }

@article{Dadhich_2000,
   title={Black holes on the brane},
   volume={487},
   ISSN={0370-2693},
   url={http://dx.doi.org/10.1016/S0370-2693(00)00798-X},
   DOI={10.1016/s0370-2693(00)00798-x},
   number={1-2},
   journal={Physics Letters B},
   publisher={Elsevier BV},
   author={Dadhich, Naresh and Maartens, Roy and Papadopoulos, Philippos and Rezania, Vahid},
   year={2000},
   month=Aug, pages={1–6} }

@article{Garriga_2000,
   title={Gravity in the Randall-Sundrum Brane World},
   volume={84},
   ISSN={1079-7114},
   url={http://dx.doi.org/10.1103/PhysRevLett.84.2778},
   DOI={10.1103/physrevlett.84.2778},
   number={13},
   journal={Physical Review Letters},
   publisher={American Physical Society (APS)},
   author={Garriga, Jaume and Tanaka, Takahiro},
   year={2000},
   month=Mar, pages={2778–2781} }

@misc{whisker2008braneworldblackholes,
      title={Braneworld Black Holes}, 
      author={Richard Whisker},
      year={2008},
      eprint={0810.1534},
      archivePrefix={arXiv},
      primaryClass={gr-qc},
      url={https://arxiv.org/abs/0810.1534}, 
}

@article{Kapner:2006si,
    author = "Kapner, D. J. and Cook, T. S. and Adelberger, E. G. and Gundlach, J. H. and Heckel, Blayne R. and Hoyle, C. D. and Swanson, H. E.",
    title = "{Tests of the gravitational inverse-square law below the dark-energy length scale}",
    eprint = "hep-ph/0611184",
    archivePrefix = "arXiv",
    doi = "10.1103/PhysRevLett.98.021101",
    journal = "Phys. Rev. Lett.",
    volume = "98",
    pages = "021101",
    year = "2007"
}

@article{Hawking:1974rv,
    author = "Hawking, S. W.",
    title = "{Black hole explosions}",
    doi = "10.1038/248030a0",
    journal = "Nature",
    volume = "248",
    pages = "30--31",
    year = "1974"
}

@article{Hawking:1975vcx,
    author = "Hawking, S. W.",
    editor = "Gibbons, G. W. and Hawking, S. W.",
    title = "{Particle Creation by Black Holes}",
    doi = "10.1007/BF02345020",
    journal = "Commun. Math. Phys.",
    volume = "43",
    pages = "199--220",
    year = "1975",
    note = "[Erratum: Commun.Math.Phys. 46, 206 (1976)]"
}

@article{Page:1976df,
    author = "Page, Don N.",
    title = "{Particle Emission Rates from a Black Hole: Massless Particles from an Uncharged, Nonrotating Hole}",
    doi = "10.1103/PhysRevD.13.198",
    journal = "Phys. Rev. D",
    volume = "13",
    pages = "198--206",
    year = "1976"
}

@article{Harris:2003eg,
    author = "Harris, Chris M. and Kanti, Panagiota",
    title = "{Hawking radiation from a (4+n)-dimensional black hole: Exact results for the Schwarzschild phase}",
    eprint = "hep-ph/0309054",
    archivePrefix = "arXiv",
    reportNumber = "CAVENDISH-HEP-03-17, CERN-TH-2003-205",
    doi = "10.1088/1126-6708/2003/10/014",
    journal = "JHEP",
    volume = "10",
    pages = "014",
    year = "2003"
}

@article{Cardoso:2005mh,
    author = "Cardoso, Vitor and Cavaglia, Marco and Gualtieri, Leonardo",
    title = "{Hawking emission of gravitons in higher dimensions: Non-rotating black holes}",
    eprint = "hep-th/0512116",
    archivePrefix = "arXiv",
    doi = "10.1088/1126-6708/2006/02/021",
    journal = "JHEP",
    volume = "02",
    pages = "021",
    year = "2006"
}

@article{Kanti:2002nr,
    author = "Kanti, Panagiota and March-Russell, John",
    title = "{Calculable corrections to brane black hole decay. 1. The scalar case}",
    eprint = "hep-ph/0203223",
    archivePrefix = "arXiv",
    reportNumber = "CERN-TH-2002-014",
    doi = "10.1103/PhysRevD.66.024023",
    journal = "Phys. Rev. D",
    volume = "66",
    pages = "024023",
    year = "2002"
}

@article{Creek_2006,
   title={Graviton emission in the bulk from a higher-dimensional Schwarzschild black hole},
   volume={635},
   ISSN={0370-2693},
   url={http://dx.doi.org/10.1016/j.physletb.2006.02.030},
   DOI={10.1016/j.physletb.2006.02.030},
   number={1},
   journal={Physics Letters B},
   publisher={Elsevier BV},
   author={Creek, S. and Efthimiou, O. and Kanti, P. and Tamvakis, K.},
   year={2006},
   month=Mar, pages={39–49} }

@article{Cardoso_2006,
   title={Hawking emission of gravitons in higher dimensions: non-rotating black holes},
   volume={2006},
   ISSN={1029-8479},
   url={http://dx.doi.org/10.1088/1126-6708/2006/02/021},
   DOI={10.1088/1126-6708/2006/02/021},
   number={02},
   journal={Journal of High Energy Physics},
   publisher={Springer Science and Business Media LLC},
   author={Cardoso, Vitor and Cavaglià, Marco and Gualtieri, Leonardo},
   year={2006},
   month=Feb, pages={021–021} }

@article{Cornell_2006,
   title={Graviton emission from a higher-dimensional black hole},
   volume={2006},
   ISSN={1029-8479},
   url={http://dx.doi.org/10.1088/1126-6708/2006/02/012},
   DOI={10.1088/1126-6708/2006/02/012},
   number={02},
   journal={Journal of High Energy Physics},
   publisher={Springer Science and Business Media LLC},
   author={Cornell, Alan S and Naylor, Wade and Sasaki, Misao},
   year={2006},
   month=Feb, pages={012–012} }

@article{Emparan_2000,
   title={Black Holes Radiate Mainly on the Brane},
   volume={85},
   ISSN={1079-7114},
   url={http://dx.doi.org/10.1103/PhysRevLett.85.499},
   DOI={10.1103/physrevlett.85.499},
   number={3},
   journal={Physical Review Letters},
   publisher={American Physical Society (APS)},
   author={Emparan, Roberto and Horowitz, Gary T. and Myers, Robert C.},
   year={2000},
   month=jul, pages={499–502} }

@article{Carr_2016,
   title={Constraints on primordial black holes from the Galactic gamma-ray background},
   volume={94},
   ISSN={2470-0029},
   url={http://dx.doi.org/10.1103/PhysRevD.94.044029},
   DOI={10.1103/physrevd.94.044029},
   number={4},
   journal={Physical Review D},
   publisher={American Physical Society (APS)},
   author={Carr, B. J. and Kohri, Kazunori and Sendouda, Yuuiti and Yokoyama, Jun’ichi},
   year={2016},
   month=Aug }

@article{Fitzpatrick_2006,
   title={On the existence and dynamics of braneworld black holes},
   volume={2006},
   ISSN={1029-8479},
   url={http://dx.doi.org/10.1088/1126-6708/2006/11/033},
   DOI={10.1088/1126-6708/2006/11/033},
   number={11},
   journal={Journal of High Energy Physics},
   publisher={Springer Science and Business Media LLC},
   author={Fitzpatrick, Andrew Liam and Randall, Lisa and Wiseman, Toby},
   year={2006},
   month=Nov, pages={033–033} }

@article{Figueras_2011,
   title={Gravity and Large Black Holes in Randall-Sundrum II Braneworlds},
   volume={107},
   ISSN={1079-7114},
   url={http://dx.doi.org/10.1103/PhysRevLett.107.081101},
   DOI={10.1103/physrevlett.107.081101},
   number={8},
   journal={Physical Review Letters},
   publisher={American Physical Society (APS)},
   author={Figueras, Pau and Wiseman, Toby},
   year={2011},
   month=Aug }

@article{Emparan_2023,
   title={Holographic duals of evaporating black holes},
   volume={2023},
   ISSN={1029-8479},
   url={http://dx.doi.org/10.1007/JHEP05(2023)182},
   DOI={10.1007/jhep05(2023)182},
   number={5},
   journal={Journal of High Energy Physics},
   publisher={Springer Science and Business Media LLC},
   author={Emparan, Roberto and Luna, Raimon and Suzuki, Ryotaku and Tomašević, Marija and Way, Benson},
   year={2023},
   month=May }

@article{Casals_2006,
   title={Brane decay of a (4+n)-dimensional rotating black hole II: spin-1 particles},
   volume={2006},
   ISSN={1029-8479},
   url={http://dx.doi.org/10.1088/1126-6708/2006/02/051},
   DOI={10.1088/1126-6708/2006/02/051},
   number={02},
   journal={Journal of High Energy Physics},
   publisher={Springer Science and Business Media LLC},
   author={Casals, Marc and Kanti, Panagiota and Winstanley, Elizabeth},
   year={2006},
   month=Feb, pages={051–051} }

@article{Ida_2006,
   title={Rotating black holes at future colliders. III. Determination of black hole evolution},
   volume={73},
   ISSN={1550-2368},
   url={http://dx.doi.org/10.1103/PhysRevD.73.124022},
   DOI={10.1103/physrevd.73.124022},
   number={12},
   journal={Physical Review D},
   publisher={American Physical Society (APS)},
   author={Ida, Daisuke and Oda, Kin-ya and Park, Seong Chan},
   year={2006},
   month=jun }

@article{Gingrich:2009hj,
    author = "Gingrich, Douglas M.",
    title = "{Quantum black holes with charge, colour, and spin at the LHC}",
    eprint = "0912.0826",
    archivePrefix = "arXiv",
    primaryClass = "hep-ph",
    doi = "10.1088/0954-3899/37/10/105008",
    journal = "J. Phys. G",
    volume = "37",
    pages = "105008",
    year = "2010"
}

@article{CDF:2002nno,
    author = "Acosta, D. and others",
    collaboration = "CDF",
    title = "{Momentum Distribution of Charged Particles in Jets in Dijet Events in $p \bar{p}$ Collisions at $\sqrt{s} = 1.8$ TeV and Comparisons to Perturbative QCD Predictions}",
    reportNumber = "FERMILAB-PUB-02-096-E",
    doi = "10.1103/PhysRevD.68.012003",
    journal = "Phys. Rev. D",
    volume = "68",
    pages = "012003",
    year = "2003"
}

@article{Cavaglia:2006uk,
    author = "Cavaglia, M. and Godang, R. and Cremaldi, L. and Summers, D.",
    title = "{Catfish: A Monte Carlo simulator for black holes at the LHC}",
    eprint = "hep-ph/0609001",
    archivePrefix = "arXiv",
    doi = "10.1016/j.cpc.2007.05.011",
    journal = "Comput. Phys. Commun.",
    volume = "177",
    pages = "506--517",
    year = "2007"
}

@article{Dai:2007ki,
    author = "Dai, De-Chang and Starkman, Glenn and Stojkovic, Dejan and Issever, Cigdem and Rizvi, Eram and Tseng, Jeff",
    title = "{BlackMax: A black-hole event generator with rotation, recoil, split branes, and brane tension}",
    eprint = "0711.3012",
    archivePrefix = "arXiv",
    primaryClass = "hep-ph",
    doi = "10.1103/PhysRevD.77.076007",
    journal = "Phys. Rev. D",
    volume = "77",
    pages = "076007",
    year = "2008"
}

@article{Arbey:2019mbc,
    author = "Arbey, Alexandre and Auffinger, J{\'e}r{\'e}my",
    title = "{BlackHawk: A public code for calculating the Hawking evaporation spectra of any black hole distribution}",
    eprint = "1905.04268",
    archivePrefix = "arXiv",
    primaryClass = "gr-qc",
    reportNumber = "CERN-TH-2019-067",
    doi = "10.1140/epjc/s10052-019-7161-1",
    journal = "Eur. Phys. J. C",
    volume = "79",
    number = "8",
    pages = "693",
    year = "2019"
}

@article{Arbey:2021mbl,
    author = "Arbey, Alexandre and Auffinger, J{\'e}r{\'e}my",
    title = "{Physics Beyond the Standard Model with BlackHawk v2.0}",
    eprint = "2108.02737",
    archivePrefix = "arXiv",
    primaryClass = "gr-qc",
    reportNumber = "CERN-TH-2021-117",
    doi = "10.1140/epjc/s10052-021-09702-8",
    journal = "Eur. Phys. J. C",
    volume = "81",
    pages = "910",
    year = "2021"
}

@misc{aldecoatamayo2026cosmologicalaccretionbraneworldblack,
      title={Cosmological accretion onto braneworld black holes: a relativistic treatment}, 
      author={Itzi Aldecoa-Tamayo and Christian T. Byrnes and David Seery},
      year={2026},
      eprint={2605.30446},
      archivePrefix={arXiv},
      primaryClass={gr-qc},
      url={https://arxiv.org/abs/2605.30446}, 
}

@article{Green_2021,
   title={Primordial black holes as a dark matter candidate},
   volume={48},
   ISSN={1361-6471},
   url={http://dx.doi.org/10.1088/1361-6471/abc534},
   DOI={10.1088/1361-6471/abc534},
   number={4},
   journal={Journal of Physics G: Nuclear and Particle Physics},
   publisher={IOP Publishing},
   author={Green, Anne M and Kavanagh, Bradley J},
   year={2021},
   month=feb, pages={043001} }

@inbook{Escriv__2024,
   title={Primordial black holes},
   ISBN={9780323956369},
   url={http://dx.doi.org/10.1016/B978-0-32-395636-9.00012-8},
   DOI={10.1016/b978-0-32-395636-9.00012-8},
   booktitle={Black Holes in the Era of Gravitational-Wave Astronomy},
   publisher={Elsevier},
   author={Escrivà, Albert and Kühnel, Florian and Tada, Yuichiro},
   year={2024},
   pages={261–377} }

@article{hawking1971gravitationally,
  title={Gravitationally collapsed objects of very low mass},
  author={Hawking, Stephen},
  journal={Monthly Notices of the Royal Astronomical Society},
  volume={152},
  number={1},
  pages={75--78},
  year={1971},
  publisher={Oxford University Press},
  doi={10.1093/mnras/152.1.75},
  url={https://academic.oup.com/mnras/article/152/1/75/2604549}
}

@article{Aldecoa-Tamayo_2026,
doi = {10.1088/1475-7516/2026/02/002},
url = {https://doi.org/10.1088/1475-7516/2026/02/002},
year = {2026},
month = feb,
publisher = {IOP Publishing},
volume = {2026},
number = {02},
pages = {002},
author = {Aldecoa-Tamayo, Itzi and Byrnes, Christian T. and Seery, David},
title = {Primordial black holes in Randall-Sundrum: cosmological signatures},
journal = {Journal of Cosmology and Astroparticle Physics}
}

@article{Guedens_2002-1,
   title={Primordial black holes in braneworld cosmologies: Formation, cosmological evolution, and evaporation},
   volume={66},
   ISSN={1089-4918},
   url={http://dx.doi.org/10.1103/PhysRevD.66.043513},
   DOI={10.1103/physrevd.66.043513},
   number={4},
   journal={Physical Review D},
   publisher={American Physical Society (APS)},
   author={Guedens, Raf and Clancy, Dominic and Liddle, Andrew R.},
   year={2002},
   month=aug }

@article{Guedens_2002-2,
   title={Primordial black holes in braneworld cosmologies: Accretion after formation},
   volume={66},
   ISSN={1089-4918},
   url={http://dx.doi.org/10.1103/PhysRevD.66.083509},
   DOI={10.1103/physrevd.66.083509},
   number={8},
   journal={Physical Review D},
   publisher={American Physical Society (APS)},
   author={Guedens, Raf and Clancy, Dominic and Liddle, Andrew R.},
   year={2002},
   month=oct }

@article{Clancy:2003zd,
    author = "Clancy, Dominic and Guedens, Raf and Liddle, Andrew R.",
    title = "{Primordial black holes in brane world cosmologies: Astrophysical constraints}",
    eprint = "astro-ph/0301568",
    archivePrefix = "arXiv",
    doi = "10.1103/PhysRevD.68.023507",
    journal = "Phys. Rev. D",
    volume = "68",
    pages = "023507",
    year = "2003"
}

@article{Majumdar_2003,
   title={Domination of Black Hole Accretion in Brane Cosmology},
   volume={90},
   ISSN={1079-7114},
   url={http://dx.doi.org/10.1103/PhysRevLett.90.031303},
   DOI={10.1103/physrevlett.90.031303},
   number={3},
   journal={Physical Review Letters},
   publisher={American Physical Society (APS)},
   author={Majumdar, A. S.},
   year={2003},
   month=jan }

@article{Majumdar:2003sf,
    author = "Majumdar, Archan S. and Mehta, Anita and Luck, J. M.",
    title = "{Interacting black holes on the brane: The Seeding of binaries}",
    eprint = "astro-ph/0311148",
    archivePrefix = "arXiv",
    doi = "10.1016/j.physletb.2004.12.059",
    journal = "Phys. Lett. B",
    volume = "607",
    pages = "219--224",
    year = "2005"
}

@article{Majumdar:2004mz,
    author = "Majumdar, Archan S. and Mukherjee, Nupur",
    title = "{Gravitational lensing in the weak field limit by a braneworld black hole}",
    eprint = "astro-ph/0403405",
    archivePrefix = "arXiv",
    doi = "10.1142/S0217732305017676",
    journal = "Mod. Phys. Lett. A",
    volume = "20",
    pages = "2487--2496",
    year = "2005"
}

@article{Majumdar:2005ba,
    author = "Majumdar, Archan S. and Mukherjee, N.",
    title = "{Braneworld black holes in cosmology and astrophysics}",
    eprint = "astro-ph/0503473",
    archivePrefix = "arXiv",
    doi = "10.1142/S0218271805006948",
    journal = "Int. J. Mod. Phys. D",
    volume = "14",
    pages = "1095",
    year = "2005"
}

@article{Inoue:2003di,
    author = "Inoue, Kaiki Taro and Tanaka, Takahiro",
    title = "{Gravitational waves from sublunar mass primordial black hole binaries: A New probe of extradimensions}",
    eprint = "gr-qc/0303058",
    archivePrefix = "arXiv",
    doi = "10.1103/PhysRevLett.91.021101",
    journal = "Phys. Rev. Lett.",
    volume = "91",
    pages = "021101",
    year = "2003"
}

@article{Sendouda:2003dc,
    author = "Sendouda, Yuuiti and Nagataki, Shigehiro and Sato, Katsuhiko",
    title = "{Constraints on the mass spectrum of primordial black holes and braneworld parameters from the high-energy diffuse photon background}",
    eprint = "astro-ph/0309170",
    archivePrefix = "arXiv",
    reportNumber = "UTAP-449, RESCEU-20-03",
    doi = "10.1103/PhysRevD.68.103510",
    journal = "Phys. Rev. D",
    volume = "68",
    pages = "103510",
    year = "2003"
}

@article{Sendouda:2004hz,
    author = "Sendouda, Yuuiti and Kohri, Kazunori and Nagataki, Shigehiro and Sato, Katsuhiko",
    title = "{Sub-GeV galactic cosmic-ray antiprotons from PBHs in the Randall-Sundrum braneworld}",
    eprint = "astro-ph/0408369",
    archivePrefix = "arXiv",
    reportNumber = "UTAP-493, OU-TAP-233, YITP-04-43, RESCEU-30-04",
    doi = "10.1103/PhysRevD.71.063512",
    journal = "Phys. Rev. D",
    volume = "71",
    pages = "063512",
    year = "2005"
}

@article{Tikhomirov:2005bt,
    author = "Tikhomirov, Victor V. and Tsalkou, Y. A.",
    title = "{How particle collisions increase the rate of accretion from cosmological background onto primordial black holes in braneworld cosmology}",
    eprint = "astro-ph/0510212",
    archivePrefix = "arXiv",
    doi = "10.1103/PhysRevD.72.121301",
    journal = "Phys. Rev. D",
    volume = "72",
    pages = "121301",
    year = "2005"
}

@article{Eiroa:2005vd,
    author = "Eiroa, Ernesto F.",
    title = "{Braneworld black holes as gravitational lenses}",
    eprint = "gr-qc/0511004",
    archivePrefix = "arXiv",
    doi = "10.1590/S0103-97332005000700026",
    journal = "Braz. J. Phys.",
    volume = "35",
    pages = "1113--1116",
    year = "2005"
}

@article{Keeton_2005,
   title={Formalism for testing theories of gravity using lensing by compact objects: Static, spherically symmetric case},
   volume={72},
   ISSN={1550-2368},
   url={http://dx.doi.org/10.1103/PhysRevD.72.104006},
   DOI={10.1103/physrevd.72.104006},
   number={10},
   journal={Physical Review D},
   publisher={American Physical Society (APS)},
   author={Keeton, Charles R. and Petters, A. O.},
   year={2005},
   month=nov }

@article{Keeton_2006,
   title={Formalism for testing theories of gravity using lensing by compact objects. II. Probing post-post-Newtonian metrics},
   volume={73},
   ISSN={1550-2368},
   url={http://dx.doi.org/10.1103/PhysRevD.73.044024},
   DOI={10.1103/physrevd.73.044024},
   number={4},
   journal={Physical Review D},
   publisher={American Physical Society (APS)},
   author={Keeton, Charles R. and Petters, A. O.},
   year={2006},
   month=feb }

@article{Keeton:2006di,
    author = "Keeton, Charles R. and Petters, A. O.",
    title = "{Formalism for testing theories of gravity using lensing by compact objects. III. Braneworld gravity}",
    eprint = "gr-qc/0603061",
    archivePrefix = "arXiv",
    doi = "10.1103/PhysRevD.73.104032",
    journal = "Phys. Rev. D",
    volume = "73",
    pages = "104032",
    year = "2006"
}

@article{Sendouda:2006nu,
    author = "Sendouda, Yuuiti and Nagataki, Shigehiro and Sato, Katsuhiko",
    title = "{Mass spectrum of primordial black holes from inflationary perturbation in the randall-sundrum braneworld: a limit on blue spectra}",
    eprint = "astro-ph/0603509",
    archivePrefix = "arXiv",
    reportNumber = "UTAP-552, RESCEU-1-06",
    doi = "10.1088/1475-7516/2006/06/003",
    journal = "JCAP",
    volume = "06",
    pages = "003",
    year = "2006"
}

@article{Sendouda:2006yc,
    author = "Sendouda, Y.",
    editor = "Alimi, J. M. and Fuzfa, A.",
    title = "{Cosmic rays from primordial black holes in the Randall-Sundrum braneworld}",
    doi = "10.1063/1.2399694",
    journal = "AIP Conf. Proc.",
    volume = "861",
    number = "1",
    pages = "1023--1030",
    year = "2006"
}

@article{Friedlander:2022ttk,
    author = "Friedlander, Avi and Mack, Katherine J. and Schon, Sarah and Song, Ningqiang and Vincent, Aaron C.",
    title = "{Primordial black hole dark matter in the context of extra dimensions}",
    eprint = "2201.11761",
    archivePrefix = "arXiv",
    primaryClass = "hep-ph",
    doi = "10.1103/PhysRevD.105.103508",
    journal = "Phys. Rev. D",
    volume = "105",
    number = "10",
    pages = "103508",
    year = "2022"
}

@article{Friedlander:2023qmc,
    author = "Friedlander, Avi and Song, Ningqiang and Vincent, Aaron C.",
    title = "{Dark matter from higher-dimensional primordial black holes}",
    eprint = "2306.01520",
    archivePrefix = "arXiv",
    primaryClass = "hep-ph",
    doi = "10.1103/PhysRevD.108.043523",
    journal = "Phys. Rev. D",
    volume = "108",
    number = "4",
    pages = "043523",
    year = "2023"
}

@article{ghoshal2025complementaryprobeswarpedextra,
      title={Complementary Probes of Warped Extra Dimension: Colliders, Gravitational Waves and Primordial Black Holes from Phase Transitions}, 
      author={Anish Ghoshal and Eugenio Megias and Germano Nardini and Mariano Quiros},
      year={2025},
      eprint={2502.03588},
      archivePrefix={arXiv},
      primaryClass={hep-ph},
      url={https://arxiv.org/abs/2502.03588}, 
}

@article{Bondi1952,
  title={On spherically symmetric accretion},
  author={Bondi, Hermann},
  journal={Monthly Notices of the Royal Astronomical Society},
  volume={112},
  number={2},
  pages={195--204},
  year={1952},
  publisher={Oxford University Press}
}

@article{Michel1972,
  author = {Michel, F. C.},
  title = {Accretion of matter by condensed objects},
  journal = {Astrophysics and Space Science},
  year = {1972},
  volume = {15},
  number = {1},
  pages = {153--160},
  doi = {10.1007/BF00649949},
  url = {https://doi.org/10.1007/BF00649949}
}

@article{Ricotti_2007,
   title={Bondi Accretion in the Early Universe},
   volume={662},
   ISSN={1538-4357},
   url={http://dx.doi.org/10.1086/516562},
   DOI={10.1086/516562},
   number={1},
   journal={The Astrophysical Journal},
   publisher={American Astronomical Society},
   author={Ricotti, Massimo},
   year={2007},
   month=jun, pages={53–61} }

@article{Maartens_2004,
   title={Brane-World Gravity},
   volume={7},
   ISSN={1433-8351},
   url={http://dx.doi.org/10.12942/lrr-2004-7},
   DOI={10.12942/lrr-2004-7},
   number={1},
   journal={Living Reviews in Relativity},
   publisher={Springer Science and Business Media LLC},
   author={Maartens, Roy},
   year={2004},
   month=jun }

@article{Harko_2009,
   title={Matter Accretion by Brane-World Black Holes},
   volume={54},
   ISSN={0374-4884},
   url={http://dx.doi.org/10.3938/jkps.54.2583},
   DOI={10.3938/jkps.54.2583},
   number={6(1)},
   journal={Journal of the Korean Physical Society},
   publisher={Korean Physical Society},
   author={Harko, Tiberiu},
   year={2009},
   month=jun, pages={2583–2594} }

@article{Casadio_2005,
   title={Gravitational Collapse and Black Hole Evolution: -- Do Holographic Black Holes Eventually “Anti-Evaporate”? --},
   volume={114},
   ISSN={1347-4081},
   url={http://dx.doi.org/10.1143/PTP.114.23},
   DOI={10.1143/ptp.114.23},
   number={1},
   journal={Progress of Theoretical Physics},
   publisher={Oxford University Press (OUP)},
   author={Casadio, R. and Germani, C.},
   year={2005},
   month=jul, pages={23–56} }

@inbook{shapiro-book,
author = {Shapiro, Stuart L. and Teukolsky, Saul A.},
publisher = {John Wiley \& Sons, Ltd},
isbn = {9783527617661},
title = {Appendix G: Spherical Accretion onto a Black Hole: The Relativistic Equations},
booktitle = {Black Holes, White Dwarfs, and Neutron Stars},
pages = {568-575},
doi = {10.1002/9783527617661.app7},
year = {1983}
}

@article{John_2013,
   title={Accretion onto a higher dimensional black hole},
   volume={88},
   ISSN={1550-2368},
   url={http://dx.doi.org/10.1103/PhysRevD.88.104005},
   DOI={10.1103/physrevd.88.104005},
   number={10},
   journal={Physical Review D},
   publisher={American Physical Society (APS)},
   author={John, Anslyn J. and Ghosh, Sushant G. and Maharaj, Sunil D.},
   year={2013},
   month=nov }

@article{Seery_2005,
   title={Consistency relation in general braneworld inflation},
   volume={71},
   ISSN={1550-2368},
   url={http://dx.doi.org/10.1103/PhysRevD.71.063508},
   DOI={10.1103/physrevd.71.063508},
   number={6},
   journal={Physical Review D},
   publisher={American Physical Society (APS)},
   author={Seery, David and Taylor, Andy},
   year={2005},
   month=mar }

@article{Langlois_2000,
   title={Gravitational waves from inflation on the brane},
   volume={489},
   ISSN={0370-2693},
   url={http://dx.doi.org/10.1016/S0370-2693(00)00957-6},
   DOI={10.1016/s0370-2693(00)00957-6},
   number={3–4},
   journal={Physics Letters B},
   publisher={Elsevier BV},
   author={Langlois, David and Maartens, Roy and Wands, David},
   year={2000},
   month=sep, pages={259–267} }

@article{Gorbunov_2001,
   title={Gravity waves from inflating brane or Mirrors moving in adS5},
   volume={2001},
   ISSN={1029-8479},
   url={http://dx.doi.org/10.1088/1126-6708/2001/10/015},
   DOI={10.1088/1126-6708/2001/10/015},
   number={10},
   journal={Journal of High Energy Physics},
   publisher={Springer Science and Business Media LLC},
   author={Gorbunov, Dmitry S and Rubakov, Valery A and Sibiryakov, Sergei M},
   year={2001},
   month=oct, pages={015–015} }

@article{Tristram_2022,
   title={Improved limits on the tensor-to-scalar ratio using BICEP and 
Planck},
   volume={105},
   ISSN={2470-0029},
   url={http://dx.doi.org/10.1103/PhysRevD.105.083524},
   DOI={10.1103/physrevd.105.083524},
   number={8},
   journal={Physical Review D},
   publisher={American Physical Society (APS)},
   author={Tristram, M. and Banday, A.J. and Górski, K.M. and Keskitalo, R. and Lawrence, C.R. and Andersen, K.J. and Barreiro, R.B. and Borrill, J. and Colombo, L.P.L. and Eriksen, H.K. and Fernandez-Cobos, R. and Kisner, T.S. and Martínez-González, E. and Partridge, B. and Scott, D. and Svalheim, T.L. and Wehus, I.K.},
   year={2022},
   month=apr }

@article{2020,
   title={Planck
                    2018 results: VI. Cosmological parameters},
   volume={641},
   ISSN={1432-0746},
   url={http://dx.doi.org/10.1051/0004-6361/201833910},
   DOI={10.1051/0004-6361/201833910},
   journal={Astronomy \& Astrophysics},
   publisher={EDP Sciences},
   author={Aghanim, N. and Akrami, Y. and Ashdown, M. and Aumont, J. and Baccigalupi, C. and Ballardini, M. and Banday, A. J. and Barreiro, R. B. and Bartolo, N. and Basak, S. and Battye, R. and Benabed, K. and Bernard, J.-P. and Bersanelli, M. and Bielewicz, P. and Bock, J. J. and Bond, J. R. and Borrill, J. and Bouchet, F. R. and Boulanger, F. and Bucher, M. and Burigana, C. and Butler, R. C. and Calabrese, E. and Cardoso, J.-F. and Carron, J. and Challinor, A. and Chiang, H. C. and Chluba, J. and Colombo, L. P. L. and Combet, C. and Contreras, D. and Crill, B. P. and Cuttaia, F. and de Bernardis, P. and de Zotti, G. and Delabrouille, J. and Delouis, J.-M. and Di Valentino, E. and Diego, J. M. and Doré, O. and Douspis, M. and Ducout, A. and Dupac, X. and Dusini, S. and Efstathiou, G. and Elsner, F. and Enßlin, T. A. and Eriksen, H. K. and Fantaye, Y. and Farhang, M. and Fergusson, J. and Fernandez-Cobos, R. and Finelli, F. and Forastieri, F. and Frailis, M. and Fraisse, A. A. and Franceschi, E. and Frolov, A. and Galeotta, S. and Galli, S. and Ganga, K. and Génova-Santos, R. T. and Gerbino, M. and Ghosh, T. and González-Nuevo, J. and Górski, K. M. and Gratton, S. and Gruppuso, A. and Gudmundsson, J. E. and Hamann, J. and Handley, W. and Hansen, F. K. and Herranz, D. and Hildebrandt, S. R. and Hivon, E. and Huang, Z. and Jaffe, A. H. and Jones, W. C. and Karakci, A. and Keihänen, E. and Keskitalo, R. and Kiiveri, K. and Kim, J. and Kisner, T. S. and Knox, L. and Krachmalnicoff, N. and Kunz, M. and Kurki-Suonio, H. and Lagache, G. and Lamarre, J.-M. and Lasenby, A. and Lattanzi, M. and Lawrence, C. R. and Le Jeune, M. and Lemos, P. and Lesgourgues, J. and Levrier, F. and Lewis, A. and Liguori, M. and Lilje, P. B. and Lilley, M. and Lindholm, V. and López-Caniego, M. and Lubin, P. M. and Ma, Y.-Z. and Macías-Pérez, J. F. and Maggio, G. and Maino, D. and Mandolesi, N. and Mangilli, A. and Marcos-Caballero, A. and Maris, M. and Martin, P. G. and Martinelli, M. and Martínez-González, E. and Matarrese, S. and Mauri, N. and McEwen, J. D. and Meinhold, P. R. and Melchiorri, A. and Mennella, A. and Migliaccio, M. and Millea, M. and Mitra, S. and Miville-Deschênes, M.-A. and Molinari, D. and Montier, L. and Morgante, G. and Moss, A. and Natoli, P. and Nørgaard-Nielsen, H. U. and Pagano, L. and Paoletti, D. and Partridge, B. and Patanchon, G. and Peiris, H. V. and Perrotta, F. and Pettorino, V. and Piacentini, F. and Polastri, L. and Polenta, G. and Puget, J.-L. and Rachen, J. P. and Reinecke, M. and Remazeilles, M. and Renzi, A. and Rocha, G. and Rosset, C. and Roudier, G. and Rubiño-Martín, J. A. and Ruiz-Granados, B. and Salvati, L. and Sandri, M. and Savelainen, M. and Scott, D. and Shellard, E. P. S. and Sirignano, C. and Sirri, G. and Spencer, L. D. and Sunyaev, R. and Suur-Uski, A.-S. and Tauber, J. A. and Tavagnacco, D. and Tenti, M. and Toffolatti, L. and Tomasi, M. and Trombetti, T. and Valenziano, L. and Valiviita, J. and Van Tent, B. and Vibert, L. and Vielva, P. and Villa, F. and Vittorio, N. and Wandelt, B. D. and Wehus, I. K. and White, M. and White, S. D. M. and Zacchei, A. and Zonca, A.},
   year={2020},
   month=sep, pages={A6} }

@article{vitale2026microscopicprimordialblackholes,
      title={Microscopic primordial black holes as macroscopic dark matter from large extra dimensions}, 
      author={Giuseppe Filiberto Vitale and Gaetano Lambiase and Tanmay Kumar Poddar and Luca Visinelli},
      year={2026},
      eprint={2604.14871},
      archivePrefix={arXiv},
      primaryClass={astro-ph.CO},
      url={https://arxiv.org/abs/2604.14871}, 
}

@book{hawking1996nature,
  title     = {The Nature of Space and Time},
  author    = {Hawking, Stephen and Penrose, Roger},
  year      = {1996},
  publisher = {Princeton University Press},
  address   = {Princeton, N.J.},
  isbn      = {9780691037912}
}

@article{Carr_2021,
   title={Constraints on primordial black holes},
   volume={84},
   ISSN={1361-6633},
   url={http://dx.doi.org/10.1088/1361-6633/ac1e31},
   DOI={10.1088/1361-6633/ac1e31},
   number={11},
   journal={Reports on Progress in Physics},
   publisher={IOP Publishing},
   author={Carr, Bernard and Kohri, Kazunori and Sendouda, Yuuiti and Yokoyama, Jun’ichi},
   year={2021},
   month=nov, pages={116902} }

@article{Strong_1998,
   title={Propagation of Cosmic‐Ray Nucleons in the Galaxy},
   volume={509},
   ISSN={1538-4357},
   url={http://dx.doi.org/10.1086/306470},
   DOI={10.1086/306470},
   number={1},
   journal={The Astrophysical Journal},
   publisher={American Astronomical Society},
   author={Strong, Andrew W. and Moskalenko, Igor V.},
   year={1998},
   month=dec, pages={212–228} }

@article{lahav2004cosmologicalparameters,
      title={The Cosmological Parameters}, 
      author={Ofer Lahav and Andrew R Liddle},
      year={2004},
      eprint={astro-ph/0406681},
      archivePrefix={arXiv},
      primaryClass={astro-ph},
      url={https://arxiv.org/abs/astro-ph/0406681}, 
}

@book{longair1992high,
  title     = {High Energy Astrophysics: Volume 1, Particles, Photons and their Detection},
  author    = {Longair, Malcolm S.},
  edition   = {2nd},
  year      = {1992},
  publisher = {Cambridge University Press},
  address   = {Cambridge, England},
  isbn      = {9780521387736}
}

@article{Johnson_2020,
   title={Primordial black hole constraints with large extra dimensions},
   volume={2020},
   ISSN={1475-7516},
   url={http://dx.doi.org/10.1088/1475-7516/2020/09/046},
   DOI={10.1088/1475-7516/2020/09/046},
   number={09},
   journal={Journal of Cosmology and Astroparticle Physics},
   publisher={IOP Publishing},
   author={Johnson, George},
   year={2020},
   month=sep, pages={046–046} }

@article{Dokshitzer:1977sg,
    author = "Dokshitzer, Yuri L.",
    title = "{Calculation of the Structure Functions for Deep Inelastic Scattering and e+ e- Annihilation by Perturbation Theory in Quantum Chromodynamics.}",
    journal = "Sov. Phys. JETP",
    volume = "46",
    pages = "641--653",
    year = "1977"
}

@article{Gribov:1972ri,
    author = "Gribov, V. N. and Lipatov, L. N.",
    title = "{Deep inelastic e p scattering in perturbation theory}",
    reportNumber = "IPTI-381-71",
    journal = "Sov. J. Nucl. Phys.",
    volume = "15",
    pages = "438--450",
    year = "1972"
}

@article{Altarelli:1977zs,
    author = "Altarelli, Guido and Parisi, G.",
    title = "{Asymptotic Freedom in Parton Language}",
    reportNumber = "LPTENS-77-6",
    doi = "10.1016/0550-3213(77)90384-4",
    journal = "Nucl. Phys. B",
    volume = "126",
    pages = "298--318",
    year = "1977"
}

@article{Chen_2017,
   title={Electroweak splitting functions and high energy showering},
   volume={2017},
   ISSN={1029-8479},
   url={http://dx.doi.org/10.1007/JHEP11(2017)093},
   DOI={10.1007/jhep11(2017)093},
   number={11},
   journal={Journal of High Energy Physics},
   publisher={Springer Science and Business Media LLC},
   author={Chen, Junmou and Han, Tao and Tweedie, Brock},
   year={2017},
   month=nov }

@article{Plestid_2025,
   title={Final state radiation from high and ultrahigh energy neutrino interactions},
   volume={111},
   ISSN={2470-0029},
   url={http://dx.doi.org/10.1103/PhysRevD.111.043007},
   DOI={10.1103/physrevd.111.043007},
   number={4},
   journal={Physical Review D},
   publisher={American Physical Society (APS)},
   author={Plestid, Ryan and Zhou, Bei},
   year={2025},
   month=feb }

@article{Bethe:1930ku,
    author = "Bethe, H.",
    title = "{Theory of the Passage of Fast Corpuscular Rays Through Matter}",
    doi = "10.1002/andp.19303970303",
    journal = "Annalen Phys.",
    volume = "5",
    pages = "325--400",
    year = "1930"
}

@article{Bethe:1932cbh,
    author = "Bethe, H.",
    title = {{Bremsformel f{\"u}r Elektronen relativistischer Geschwindigkeit}},
    doi = "10.1007/BF01342532",
    journal = "Z. Phys.",
    volume = "76",
    number = "5",
    pages = "293--299",
    year = "1932"
}

@book{segre1953experimental,
  title     = {Experimental Nuclear Physics},
  author    = {Segrè, Emilio},
  volume    = {1},
  year      = {1953},
  publisher = {John Wiley \& Sons},
  address   = {New York}
}

@article{Beacom_2006,
   title={Stringent Constraint on Galactic Positron Production},
   volume={97},
   ISSN={1079-7114},
   url={http://dx.doi.org/10.1103/PhysRevLett.97.071102},
   DOI={10.1103/physrevlett.97.071102},
   number={7},
   journal={Physical Review Letters},
   publisher={American Physical Society (APS)},
   author={Beacom, John F. and Yüksel, Hasan},
   year={2006},
   month=aug }

@article{Jean_2009,
   title={Positron transport in the interstellar medium},
   volume={508},
   ISSN={1432-0746},
   url={http://dx.doi.org/10.1051/0004-6361/200809830},
   DOI={10.1051/0004-6361/200809830},
   number={3},
   journal={Astronomy \& Astrophysics},
   publisher={EDP Sciences},
   author={Jean, P. and Gillard, W. and Marcowith, A. and Ferrière, K.},
   year={2009},
   month=oct, pages={1099–1116} }

@article{Hughes_2000,
   title={Electron Heating and Cosmic Rays at a Supernova Shock from [ITAL]Chandra[/ITAL] X-Ray Observations of 1E 0102.2−7219},
   volume={543},
   ISSN={0004-637X},
   url={http://dx.doi.org/10.1086/312945},
   DOI={10.1086/312945},
   number={1},
   journal={The Astrophysical Journal},
   publisher={American Astronomical Society},
   author={Hughes, John P. and Rakowski, Cara E. and Decourchelle, Anne},
   year={2000},
   month=nov, pages={L000–L000} }

@book{jauch1976theory,
  title     = {The Theory of Photons and Electrons: The Relativistic Quantum Field Theory of Charged Particles with Spin One-half},
  author    = {Jauch, Josef M. and Rohrlich, F.},
  year      = {1976},
  edition   = {2nd},
  publisher = {Springer-Verlag},
  address   = {New York},
  isbn      = {978-0-387-07295-1}
}

@article{Siegert_2016,
   title={Gamma-ray spectroscopy of positron annihilation in the Milky Way},
   volume={586},
   ISSN={1432-0746},
   url={http://dx.doi.org/10.1051/0004-6361/201527510},
   DOI={10.1051/0004-6361/201527510},
   journal={Astronomy \& Astrophysics},
   publisher={EDP Sciences},
   author={Siegert, Thomas and Diehl, Roland and Khachatryan, Gerasim and Krause, Martin G. H. and Guglielmetti, Fabrizia and Greiner, Jochen and Strong, Andrew W. and Zhang, Xiaoling},
   year={2016},
   month=jan, pages={A84} }

@article{Ore:1949te,
    author = "Ore, A. and Powell, J. L.",
    title = "{Three photon annihilation of an electron - positron pair}",
    doi = "10.1103/PhysRev.75.1696",
    journal = "Phys. Rev.",
    volume = "75",
    pages = "1696--1699",
    year = "1949"
}

@article{Manohar_2004,
   title={Orthopositronium decay spectrum using NRQED},
   volume={69},
   ISSN={1550-2368},
   url={http://dx.doi.org/10.1103/PhysRevD.69.053003},
   DOI={10.1103/physrevd.69.053003},
   number={5},
   journal={Physical Review D},
   publisher={American Physical Society (APS)},
   author={Manohar, Aneesh V. and Ruiz-Femenía, Pedro},
   year={2004},
   month=mar }

@article{Navarro_1996,
   title={The Structure of Cold Dark Matter Halos},
   volume={462},
   ISSN={1538-4357},
   url={http://dx.doi.org/10.1086/177173},
   DOI={10.1086/177173},
   journal={The Astrophysical Journal},
   publisher={American Astronomical Society},
   author={Navarro, Julio F. and Frenk, Carlos S. and White, Simon D. M.},
   year={1996},
   month=may, pages={563} }

@article{Navarro_1997,
   title={A Universal Density Profile from Hierarchical Clustering},
   volume={490},
   ISSN={1538-4357},
   url={http://dx.doi.org/10.1086/304888},
   DOI={10.1086/304888},
   number={2},
   journal={The Astrophysical Journal},
   publisher={American Astronomical Society},
   author={Navarro, Julio F. and Frenk, Carlos S. and White, Simon D. M.},
   year={1997},
   month=dec, pages={493–508} }

@article{Cautun_2020,
   title={The milky way total mass profile as inferred from Gaia DR2},
   volume={494},
   ISSN={1365-2966},
   url={http://dx.doi.org/10.1093/mnras/staa1017},
   DOI={10.1093/mnras/staa1017},
   number={3},
   journal={Monthly Notices of the Royal Astronomical Society},
   publisher={Oxford University Press (OUP)},
   author={Cautun, Marius and Benítez-Llambay, Alejandro and Deason, Alis J and Frenk, Carlos S and Fattahi, Azadeh and Gómez, Facundo A and Grand, Robert J J and Oman, Kyle A and Navarro, Julio F and Simpson, Christine M},
   year={2020},
   month=apr, pages={4291–4313} }

@article{Tinker_2008,
   title={Toward a Halo Mass Function for Precision Cosmology: The Limits of Universality},
   volume={688},
   ISSN={1538-4357},
   url={http://dx.doi.org/10.1086/591439},
   DOI={10.1086/591439},
   number={2},
   journal={The Astrophysical Journal},
   publisher={American Astronomical Society},
   author={Tinker, Jeremy and Kravtsov, Andrey V. and Klypin, Anatoly and Abazajian, Kevork and Warren, Michael and Yepes, Gustavo and Gottlöber, Stefan and Holz, Daniel E.},
   year={2008},
   month=dec, pages={709–728} }

@article{cui2024knowhalomassfunction,
      title={How much do we know the halo mass function? Predictions beyond resolution}, 
      author={Weiguang Cui},
      year={2024},
      eprint={2406.03829},
      archivePrefix={arXiv},
      primaryClass={astro-ph.CO},
      url={https://arxiv.org/abs/2406.03829}, 
}

@article{Press:1973iz,
    author = "Press, William H. and Schechter, Paul",
    title = "{Formation of galaxies and clusters of galaxies by selfsimilar gravitational condensation}",
    doi = "10.1086/152650",
    journal = "Astrophys. J.",
    volume = "187",
    pages = "425--438",
    year = "1974"
}

@article{Zheng_2024,
   title={The influence of baryons on low-mass haloes},
   volume={532},
   ISSN={1365-2966},
   url={http://dx.doi.org/10.1093/mnras/stae1666},
   DOI={10.1093/mnras/stae1666},
   number={3},
   journal={Monthly Notices of the Royal Astronomical Society},
   publisher={Oxford University Press (OUP)},
   author={Zheng, Haonan and Bose, Sownak and Frenk, Carlos S and Gao, Liang and Jenkins, Adrian and Liao, Shihong and Springel, Volker and Wang, Jie and White, Simon D M},
   year={2024},
   month=jul, pages={3151–3165} }

@article{wright2024baryoncyclemoderncosmological,
      title={The baryon cycle in modern cosmological hydrodynamical simulations}, 
      author={Ruby J. Wright and Rachel S. Somerville and Claudia del P. Lagos and Matthieu Schaller and Romeel Davé and Daniel Anglés-Alcázar and Shy Genel},
      year={2024},
      eprint={2402.08408},
      archivePrefix={arXiv},
      primaryClass={astro-ph.GA},
      url={https://arxiv.org/abs/2402.08408}, 
}

@article{Adame_2025,
   title={DESI 2024 VI:  cosmological constraints from the measurements of baryon acoustic oscillations},
   volume={2025},
   ISSN={1475-7516},
   url={http://dx.doi.org/10.1088/1475-7516/2025/02/021},
   DOI={10.1088/1475-7516/2025/02/021},
   number={02},
   journal={Journal of Cosmology and Astroparticle Physics},
   publisher={IOP Publishing},
   author={Adame, A.G. and Aguilar, J. and Ahlen, S. and Alam, S. and Alexander, D.M. and Alvarez, M. and Alves, O. and Anand, A. and Andrade, U. and Armengaud, E. and Avila, S. and Aviles, A. and Awan, H. and Bahr-Kalus, B. and Bailey, S. and Baltay, C. and Bault, A. and Behera, J. and BenZvi, S. and Bera, A. and Beutler, F. and Bianchi, D. and Blake, C. and Blum, R. and Brieden, S. and Brodzeller, A. and Brooks, D. and Buckley-Geer, E. and Burtin, E. and Calderon, R. and Canning, R. and Carnero Rosell, A. and Cereskaite, R. and Cervantes-Cota, J.L. and Chabanier, S. and Chaussidon, E. and Chaves-Montero, J. and Chen, S. and Chen, X. and Claybaugh, T. and Cole, S. and Cuceu, A. and Davis, T.M. and Dawson, K. and de la Macorra, A. and de Mattia, A. and Deiosso, N. and Dey, A. and Dey, B. and Ding, Z. and Doel, P. and Edelstein, J. and Eftekharzadeh, S. and Eisenstein, D.J. and Elliott, A. and Fagrelius, P. and Fanning, K. and Ferraro, S. and Ereza, J. and Findlay, N. and Flaugher, B. and Font-Ribera, A. and Forero-Sánchez, D. and Forero-Romero, J.E. and Frenk, C.S. and Garcia-Quintero, C. and Gaztañaga, E. and Gil-Marín, H. and Gontcho, S.Gontcho A. and Gonzalez-Morales, A.X. and Gonzalez-Perez, V. and Gordon, C. and Green, D. and Gruen, D. and Gsponer, R. and Gutierrez, G. and Guy, J. and Hadzhiyska, B. and Hahn, C. and Hanif, M.M.S. and Herrera-Alcantar, H.K. and Honscheid, K. and Howlett, C. and Huterer, D. and Iršič, V. and Ishak, M. and junau, S. and Karaçaylı, N.G. and Kehoe, R. and Kent, S. and Kirkby, D. and Kremin, A. and Krolewski, A. and Lai, Y. and Lan, T.-W. and Landriau, M. and Lang, D. and Lasker, J. and Le Goff, J.M. and Le Guillou, L. and Leauthaud, A. and Levi, M.E. and Li, T.S. and Linder, E. and Lodha, K. and Magneville, C. and Manera, M. and Margala, D. and Martini, P. and Maus, M. and McDonald, P. and Medina-Varela, L. and Meisner, A. and Mena-Fernández, J. and Miquel, R. and Moon, J. and Moore, S. and Moustakas, J. and Mueller, E. and Muñoz-Gutiérrez, A. and Myers, A.D. and Nadathur, S. and Napolitano, L. and Neveux, R. and Newman, J.A. and Nguyen, N.M. and Nie, J. and Niz, G. and Noriega, H.E. and Padmanabhan, N. and Paillas, E. and Palanque-Delabrouille, N. and Pan, J. and Penmetsa, S. and Percival, W.J. and Pieri, M.M. and Pinon, M. and Poppett, C. and Porredon, A. and Prada, F. and Pérez-Fernández, A. and Pérez-Ràfols, I. and Rabinowitz, D. and Raichoor, A. and Ramírez-Pérez, C. and Ramirez-Solano, S. and Rashkovetskyi, M. and Ravoux, C. and Rezaie, M. and Rich, J. and Rocher, A. and Rockosi, C. and Roe, N.A. and Rosado-Marin, A. and Ross, A.J. and Rossi, G. and Ruggeri, R. and Ruhlmann-Kleider, V. and Samushia, L. and Sanchez, E. and Saulder, C. and Schlafly, E.F. and Schlegel, D. and Schubnell, M. and Seo, H. and Shafieloo, A. and Sharples, R. and Silber, J. and Slosar, A. and Smith, A. and Sprayberry, D. and Tan, T. and Tarlé, G. and Taylor, P. and Trusov, S. and Ureña-López, L.A. and Vaisakh, R. and Valcin, D. and Valdes, F. and Vargas-Magaña, M. and Verde, L. and Walther, M. and Wang, B. and Wang, M.S. and Weaver, B.A. and Weaverdyck, N. and Wechsler, R.H. and Weinberg, D.H. and White, M. and Yu, J. and Yu, Y. and Yuan, S. and Yèche, C. and Zaborowski, E.A. and Zarrouk, P. and Zhang, H. and Zhao, C. and Zhao, R. and Zhou, R. and Zhuang, T. and Zou, H. and },
   year={2025},
   month=feb, pages={021} }

@article{Blanco_2019,
   title={Constraints on decaying dark matter from the isotropic gamma-ray background},
   volume={2019},
   ISSN={1475-7516},
   url={http://dx.doi.org/10.1088/1475-7516/2019/03/019},
   DOI={10.1088/1475-7516/2019/03/019},
   number={03},
   journal={Journal of Cosmology and Astroparticle Physics},
   publisher={IOP Publishing},
   author={Blanco, Carlos and Hooper, Dan},
   year={2019},
   month=mar, pages={019–019} }

@article{Bouchet_2011,
   title={DIFFUSE EMISSION MEASUREMENT WITH THE SPECTROMETER ON INTEGRAL AS AN INDIRECT PROBE OF COSMIC-RAY ELECTRONS AND POSITRONS},
   volume={739},
   ISSN={1538-4357},
   url={http://dx.doi.org/10.1088/0004-637X/739/1/29},
   DOI={10.1088/0004-637x/739/1/29},
   number={1},
   journal={The Astrophysical Journal},
   publisher={American Astronomical Society},
   author={Bouchet, Laurent and Strong, Andrew W. and Porter, Troy A. and Moskalenko, Igor V. and Jourdain, Elisabeth and Roques, Jean-Pierre},
   year={2011},
   month=sep, pages={29} }

@article{Laha_2020,
   title={INTEGRAL constraints on primordial black holes and particle dark matter},
   volume={101},
   ISSN={2470-0029},
   url={http://dx.doi.org/10.1103/PhysRevD.101.123514},
   DOI={10.1103/physrevd.101.123514},
   number={12},
   journal={Physical Review D},
   publisher={American Physical Society (APS)},
   author={Laha, Ranjan and Muñoz, Julian B. and Slatyer, Tracy R.},
   year={2020},
   month=jun }

@article{Sicilian_2020,
   title={Probing the Milky Way’s Dark Matter Halo for the 3.5 keV Line},
   volume={905},
   ISSN={1538-4357},
   url={http://dx.doi.org/10.3847/1538-4357/abbee9},
   DOI={10.3847/1538-4357/abbee9},
   number={2},
   journal={The Astrophysical Journal},
   publisher={American Astronomical Society},
   author={Sicilian, Dominic and Cappelluti, Nico and Bulbul, Esra and Civano, Francesca and Moscetti, Massimo and Reynolds, Christopher S.},
   year={2020},
   month=dec, pages={146} }

@article{Foster_2021,
   title={Deep Search for Decaying Dark Matter with XMM-Newton Blank-Sky Observations},
   volume={127},
   ISSN={1079-7114},
   url={http://dx.doi.org/10.1103/PhysRevLett.127.051101},
   DOI={10.1103/physrevlett.127.051101},
   number={5},
   journal={Physical Review Letters},
   publisher={American Physical Society (APS)},
   author={Foster, Joshua W. and Kongsore, Marius and Dessert, Christopher and Park, Yujin and Rodd, Nicholas L. and Cranmer, Kyle and Safdi, Benjamin R.},
   year={2021},
   month=jul }

@article{Balaji_2025,
   title={Dark matter in X-rays: revised XMM-Newton limits and new constraints from eROSITA},
   volume={2025},
   ISSN={1475-7516},
   url={http://dx.doi.org/10.1088/1475-7516/2025/11/053},
   DOI={10.1088/1475-7516/2025/11/053},
   number={11},
   journal={Journal of Cosmology and Astroparticle Physics},
   publisher={IOP Publishing},
   author={Balaji, Shyam and Cleaver, Damon and De la Torre Luque, Pedro and Michailidis, Miltiadis},
   year={2025},
   month=nov, pages={053} }

@article{Cappelluti_2017,
   title={The Chandra COSMOS Legacy Survey: Energy Spectrum of the Cosmic X-Ray Background and Constraints on Undetected Populations},
   volume={837},
   ISSN={1538-4357},
   url={http://dx.doi.org/10.3847/1538-4357/aa5ea4},
   DOI={10.3847/1538-4357/aa5ea4},
   number={1},
   journal={The Astrophysical Journal},
   publisher={American Astronomical Society},
   author={Cappelluti, Nico and Li, Yanxia and Ricarte, Angelo and Agarwal, Bhaskar and Allevato, Viola and Ananna, Tonima Tasnim and Ajello, Marco and Civano, Francesca and Comastri, Andrea and Elvis, Martin and Finoguenov, Alexis and Gilli, Roberto and Hasinger, Günther and Marchesi, Stefano and Natarajan, Priyamvada and Pacucci, Fabio and Treister, E. and Urry, C. Megan},
   year={2017},
   month=feb, pages={19} }

@article{Gruber_1999,
   title={The Spectrum of Diffuse Cosmic Hard X‐Rays Measured with HEAO 1},
   volume={520},
   ISSN={1538-4357},
   url={http://dx.doi.org/10.1086/307450},
   DOI={10.1086/307450},
   number={1},
   journal={The Astrophysical Journal},
   publisher={American Astronomical Society},
   author={Gruber, D. E. and Matteson, J. L. and Peterson, L. E. and Jung, G. V.},
   year={1999},
   month=jul, pages={124–129} }

@article{Weidenspointner_2001,
   title={The COMPTEL instrumental line background},
   volume={368},
   ISSN={1432-0746},
   url={http://dx.doi.org/10.1051/0004-6361:20000489},
   DOI={10.1051/0004-6361:20000489},
   number={1},
   journal={Astronomy \& Astrophysics},
   publisher={EDP Sciences},
   author={Weidenspointner, G. and Varendorff, M. and Oberlack, U. and Morris, D. and Plüschke, S. and Diehl, R. and Kappadath, S. C. and McConnell, M. and Ryan, J. and Schönfelder, V. and Steinle, H.},
   year={2001},
   month=mar, pages={347–368} }

@article{Strong_2004,
   title={A New Determination of the Extragalactic Diffuse Gamma‐Ray Background from EGRET Data},
   volume={613},
   ISSN={1538-4357},
   url={http://dx.doi.org/10.1086/423196},
   DOI={10.1086/423196},
   number={2},
   journal={The Astrophysical Journal},
   publisher={American Astronomical Society},
   author={Strong, Andrew W. and Moskalenko, Igor V. and Reimer, Olaf},
   year={2004},
   month=oct, pages={956–961} }

@article{Abdo_2010,
   title={Spectrum of the Isotropic Diffuse Gamma-Ray Emission Derived from First-Year Fermi Large Area Telescope Data},
   volume={104},
   ISSN={1079-7114},
   url={http://dx.doi.org/10.1103/PhysRevLett.104.101101},
   DOI={10.1103/physrevlett.104.101101},
   number={10},
   journal={Physical Review Letters},
   publisher={American Physical Society (APS)},
   author={Abdo, A. A. and Ackermann, M. and Ajello, M. and Atwood, W. B. and Baldini, L. and Ballet, J. and Barbiellini, G. and Bastieri, D. and Baughman, B. M. and Bechtol, K. and Bellazzini, R. and Berenji, B. and Blandford, R. D. and Bloom, E. D. and Bonamente, E. and Borgland, A. W. and Bregeon, J. and Brez, A. and Brigida, M. and Bruel, P. and Burnett, T. H. and Buson, S. and Caliandro, G. A. and Cameron, R. A. and Caraveo, P. A. and Casandjian, J. M. and Cavazzuti, E. and Cecchi, C. and Çelik, Ö. and Charles, E. and Chekhtman, A. and Cheung, C. C. and Chiang, J. and Ciprini, S. and Claus, R. and Cohen-Tanugi, J. and Cominsky, L. R. and Conrad, J. and Cutini, S. and Dermer, C. D. and de Angelis, A. and de Palma, F. and Digel, S. W. and Di Bernardo, G. and e Silva, E. do Couto and Drell, P. S. and Drlica-Wagner, A. and Dubois, R. and Dumora, D. and Farnier, C. and Favuzzi, C. and Fegan, S. J. and Focke, W. B. and Fortin, P. and Frailis, M. and Fukazawa, Y. and Funk, S. and Fusco, P. and Gaggero, D. and Gargano, F. and Gasparrini, D. and Gehrels, N. and Germani, S. and Giebels, B. and Giglietto, N. and Giommi, P. and Giordano, F. and Glanzman, T. and Godfrey, G. and Grenier, I. A. and Grondin, M.-H. and Grove, J. E. and Guillemot, L. and Guiriec, S. and Gustafsson, M. and Hanabata, Y. and Harding, A. K. and Hayashida, M. and Hughes, R. E. and Itoh, R. and Jackson, M. S. and Jóhannesson, G. and Johnson, A. S. and Johnson, R. P. and Johnson, T. J. and Johnson, W. N. and Kamae, T. and Katagiri, H. and Kataoka, J. and Kawai, N. and Kerr, M. and Knödlseder, J. and Kocian, M. L. and Kuehn, F. and Kuss, M. and Lande, J. and Latronico, L. and Lemoine-Goumard, M. and Longo, F. and Loparco, F. and Lott, B. and Lovellette, M. N. and Lubrano, P. and Madejski, G. M. and Makeev, A. and Mazziotta, M. N. and McConville, W. and McEnery, J. E. and Meurer, C. and Michelson, P. F. and Mitthumsiri, W. and Mizuno, T. and Moiseev, A. A. and Monte, C. and Monzani, M. E. and Morselli, A. and Moskalenko, I. V. and Murgia, S. and Nolan, P. L. and Norris, J. P. and Nuss, E. and Ohsugi, T. and Omodei, N. and Orlando, E. and Ormes, J. F. and Paneque, D. and Panetta, J. H. and Parent, D. and Pelassa, V. and Pepe, M. and Pesce-Rollins, M. and Piron, F. and Porter, T. A. and Rainò, S. and Rando, R. and Razzano, M. and Reimer, A. and Reimer, O. and Reposeur, T. and Ritz, S. and Rochester, L. S. and Rodriguez, A. Y. and Roth, M. and Ryde, F. and Sadrozinski, H. F.-W. and Sanchez, D. and Sander, A. and Parkinson, P. M. Saz and Scargle, J. D. and Sellerholm, A. and Sgrò, C. and Shaw, M. S. and Siskind, E. J. and Smith, D. A. and Smith, P. D. and Spandre, G. and Spinelli, P. and Starck, J.-L. and Strickman, M. S. and Strong, A. W. and Suson, D. J. and Tajima, H. and Takahashi, H. and Takahashi, T. and Tanaka, T. and Thayer, J. B. and Thayer, J. G. and Thompson, D. J. and Tibaldo, L. and Torres, D. F. and Tosti, G. and Tramacere, A. and Uchiyama, Y. and Usher, T. L. and Vasileiou, V. and Vilchez, N. and Vitale, V. and Waite, A. P. and Wang, P. and Winer, B. L. and Wood, K. S. and Ylinen, T. and Ziegler, M.},
   year={2010},
   month=mar }

@article{Harnden1972,title = {Evidence for Hard X-Ray Pulsations from the VELA Pulsar},author = {Harnden, F. R., Jr. and Johnson, W. N., III and Haymes, R. C.},journal = {The Astrophysical Journal},year = {1972},volume = {172},pages = {L91--L94},doi = {10.1086/180897},bibcode = {1972ApJ...172L..91H}}

@article{Kn_dlseder_2005,
   title={The all-sky distribution of 511 keV electron-positron annihilation emission},
   volume={441},
   ISSN={1432-0746},
   url={http://dx.doi.org/10.1051/0004-6361:20042063},
   DOI={10.1051/0004-6361:20042063},
   number={2},
   journal={Astronomy \& Astrophysics},
   publisher={EDP Sciences},
   author={Knödlseder, J. and Jean, P. and Lonjou, V. and Weidenspointner, G. and Guessoum, N. and Gillard, W. and Skinner, G. and von Ballmoos, P. and Vedrenne, G. and Roques, J.-P. and Schanne, S. and Teegarden, B. and Schönfelder, V. and Winkler, C.},
   year={2005},
   month=sep, pages={513–532} }

@article{Weidenspointner_2006,
   title={The sky distribution of positronium annihilation continuum emission
measured with SPI/INTEGRAL},
   volume={450},
   ISSN={1432-0746},
   url={http://dx.doi.org/10.1051/0004-6361:20054046},
   DOI={10.1051/0004-6361:20054046},
   number={3},
   journal={Astronomy \& Astrophysics},
   publisher={EDP Sciences},
   author={Weidenspointner, G. and Shrader, C. R. and Knödlseder, J. and Jean, P. and Lonjou, V. and Guessoum, N. and Diehl, R. and Gillard, W. and Harris, M. J. and Skinner, G. K. and von Ballmoos, P. and Vedrenne, G. and Roques, J.-P. and Schanne, S. and Sizun, P. and Teegarden, B. J. and Schönfelder, V. and Winkler, C.},
   year={2006},
   month=apr, pages={1013–1021} }

@article{kierans2019positronannihilationgalaxy,
      title={Positron Annihilation in the Galaxy}, 
      author={Carolyn A. Kierans and John F. Beacom and Steve Boggs and Matthew Buckley and Regina Caputo and Roland Crocker and Michael De Becker and Roland Diehl and Chris L. Fryer and Sean Griffin and Dieter Hartmann and Elizabeth Hays and Pierre Jean and Martin G. H. Krause and Tim Linden and Alexandre Marcowith and Pierrick Martin and Alexander Moiseev and Uwe Oberlack and Elena Orlando and Fiona Panther and Nikos Prantzos and Richard Rothschild and Ivo Seitenzahl and Chris Shrader and Thomas Siegert and Andy Strong and John Tomsick and W. Thomas Vestrand and Andreas Zoglauer},
      year={2019},
      eprint={1903.05569},
      archivePrefix={arXiv},
      primaryClass={astro-ph.HE},
      url={https://arxiv.org/abs/1903.05569}, 
}

@article{Kalemci_2006,
   title={Searching for Annihilation Radiation from SN 1006 with SPI on INTEGRAL},
   volume={640},
   ISSN={1538-4357},
   url={http://dx.doi.org/10.1086/503289},
   DOI={10.1086/503289},
   number={1},
   journal={The Astrophysical Journal},
   publisher={American Astronomical Society},
   author={Kalemci, E. and Boggs, S. E. and Milne, P. A. and Reynolds, S. P.},
   year={2006},
   month=mar, pages={L55–L57} }

@article{Cass_2004,
   title={Hypernovae/Gamma-Ray Bursts in the Galactic Center as Possible Sources of Galactic Positrons},
   volume={602},
   ISSN={1538-4357},
   url={http://dx.doi.org/10.1086/381884},
   DOI={10.1086/381884},
   number={1},
   journal={The Astrophysical Journal},
   publisher={American Astronomical Society},
   author={Cass, M. and Cordier, B. and Paul, J. and Schanne, S.},
   year={2004},
   month=jan, pages={L17–L20} }

@article{Bertone_2006,
   title={Gamma-ray bursts and the origin of galactic positrons},
   volume={636},
   ISSN={0370-2693},
   url={http://dx.doi.org/10.1016/j.physletb.2006.03.022},
   DOI={10.1016/j.physletb.2006.03.022},
   number={1},
   journal={Physics Letters B},
   publisher={Elsevier BV},
   author={Bertone, Gianfranco and Kusenko, Alexander and Palomares-Ruiz, Sergio and Pascoli, Silvia and Semikoz, Dmitry},
   year={2006},
   month=apr, pages={20–24} }

@article{Guessoum_2006,
   title={Microquasars as sources of positron annihilation radiation},
   volume={457},
   ISSN={1432-0746},
   url={http://dx.doi.org/10.1051/0004-6361:20065240},
   DOI={10.1051/0004-6361:20065240},
   number={3},
   journal={Astronomy \& Astrophysics},
   publisher={EDP Sciences},
   author={Guessoum, N. and Jean, P. and Prantzos, N.},
   year={2006},
   month=sep, pages={753–762} }

@article{Bartels_2018,
   title={Galactic binaries can explain the Fermi Galactic centre excess and 511 keV emission},
   volume={480},
   ISSN={1365-2966},
   url={http://dx.doi.org/10.1093/mnras/sty2135},
   DOI={10.1093/mnras/sty2135},
   number={3},
   journal={Monthly Notices of the Royal Astronomical Society},
   publisher={Oxford University Press (OUP)},
   author={Bartels, R and Calore, F and Storm, E and Weniger, C},
   year={2018},
   month=aug, pages={3826–3841} }

@article{takhistov2019novelsignalsneutronstar,
      title={Novel Signals from Neutron Star Mergers at 511 keV}, 
      author={Volodymyr Takhistov},
      year={2019},
      eprint={1908.01100},
      archivePrefix={arXiv},
      primaryClass={astro-ph.HE},
      url={https://arxiv.org/abs/1908.01100}, 
}

@article{Fuller_2019,
   title={Positrons and 511 keV Radiation as Tracers of Recent Binary Neutron Star Mergers},
   volume={122},
   ISSN={1079-7114},
   url={http://dx.doi.org/10.1103/PhysRevLett.122.121101},
   DOI={10.1103/physrevlett.122.121101},
   number={12},
   journal={Physical Review Letters},
   publisher={American Physical Society (APS)},
   author={Fuller, George M. and Kusenko, Alexander and Radice, David and Takhistov, Volodymyr},
   year={2019},
   month=mar }

@article{Prantzos_2011,
   title={The 511 keV emission from positron annihilation in the Galaxy},
   volume={83},
   ISSN={1539-0756},
   url={http://dx.doi.org/10.1103/RevModPhys.83.1001},
   DOI={10.1103/revmodphys.83.1001},
   number={3},
   journal={Reviews of Modern Physics},
   publisher={American Physical Society (APS)},
   author={Prantzos, N. and Boehm, C. and Bykov, A. M. and Diehl, R. and Ferrière, K. and Guessoum, N. and Jean, P. and Knoedlseder, J. and Marcowith, A. and Moskalenko, I. V. and Strong, A. and Weidenspointner, G.},
   year={2011},
   month=sep, pages={1001–1056} }

@article{Ascasibar_2006,
   title={Constraints on dark matter and the shape of the Milky Way dark halo from the 511-keV line},
   volume={368},
   ISSN={1365-2966},
   url={http://dx.doi.org/10.1111/j.1365-2966.2006.10226.x},
   DOI={10.1111/j.1365-2966.2006.10226.x},
   number={4},
   journal={Monthly Notices of the Royal Astronomical Society},
   publisher={Oxford University Press (OUP)},
   author={Ascasibar, Y. and Jean, P. and Boehm, C. and Knodlseder, J.},
   year={2006},
   month=jun, pages={1695–1705} }

@article{Rasera_2006,
   title={Soft gamma-ray background and light dark matter annihilation},
   volume={73},
   ISSN={1550-2368},
   url={http://dx.doi.org/10.1103/PhysRevD.73.103518},
   DOI={10.1103/physrevd.73.103518},
   number={10},
   journal={Physical Review D},
   publisher={American Physical Society (APS)},
   author={Rasera, Yann and Teyssier, Romain and Sizun, Patrick and Cassé, Michel and Fayet, Pierre and Cordier, Bertrand and Paul, Jacques},
   year={2006},
   month=may }

@article{Okele:1980kwj,
    author = "Okele, P. N. and Rees, M. J.",
    title = "{Observational consequences of positron production by evaporating black holes}",
    journal = "Astron. Astrophys.",
    volume = "81",
    pages = "263--264",
    year = "1980"
}

@article{luque2026refininggalacticprimordialblack,
      title={Refining Galactic primordial black hole evaporation constraints}, 
      author={Pedro De la Torre Luque and Jordan Koechler and Shyam Balaji},
      year={2026},
      eprint={2406.11949},
      archivePrefix={arXiv},
      primaryClass={astro-ph.HE},
      url={https://arxiv.org/abs/2406.11949}, 
}

@article{Keith_2021,
   title={511 keV excess and primordial black holes},
   volume={104},
   ISSN={2470-0029},
   url={http://dx.doi.org/10.1103/PhysRevD.104.063033},
   DOI={10.1103/physrevd.104.063033},
   number={6},
   journal={Physical Review D},
   publisher={American Physical Society (APS)},
   author={Keith, Celeste and Hooper, Dan},
   year={2021},
   month=sep }

@article{Siegert_2019,
   title={Constraints on positron annihilation kinematics in the inner Galaxy},
   volume={627},
   ISSN={1432-0746},
   url={http://dx.doi.org/10.1051/0004-6361/201833856},
   DOI={10.1051/0004-6361/201833856},
   journal={Astronomy \& Astrophysics},
   publisher={EDP Sciences},
   author={Siegert, Thomas and Crocker, Roland M. and Diehl, Roland and Krause, Martin G. H. and Panther, Fiona H. and Pleintinger, Moritz M. M. and Weinberger, Christoph},
   year={2019},
   month=jul, pages={A126} }

@article{DeRocco_2019,
   title={Constraining Primordial Black Hole Abundance with the Galactic 511 keV Line},
   volume={123},
   ISSN={1079-7114},
   url={http://dx.doi.org/10.1103/PhysRevLett.123.251102},
   DOI={10.1103/physrevlett.123.251102},
   number={25},
   journal={Physical Review Letters},
   publisher={American Physical Society (APS)},
   author={DeRocco, William and Graham, Peter W.},
   year={2019},
   month=dec }

@article{Boudaud_2019,
   title={Voyager 1 $e^{\pm}$
 Further Constrain Primordial Black Holes as Dark Matter},
   volume={122},
   ISSN={1079-7114},
   url={http://dx.doi.org/10.1103/PhysRevLett.122.041104},
   DOI={10.1103/physrevlett.122.041104},
   number={4},
   journal={Physical Review Letters},
   publisher={American Physical Society (APS)},
   author={Boudaud, Mathieu and Cirelli, Marco},
   year={2019},
   month=jan }

@article{burlaga2013,
  title = {Magnetic Field Observations as {{Voyager}} 1 Entered the Heliosheath Depletion Region},
  author = {Burlaga, L. F. and Ness, N. F. and Stone, E. C.},
  year = {2013},
  journal = {Science},
  volume = {341},
  number = {6141},
  pages = {147--150},
  doi = {10.1126/science.1235451},
  url = {https://www.science.org/doi/abs/10.1126/science.1235451}
}

@article{Cummings:2016pdr,
    author = "Cummings, A. C. and Stone, E. C. and Heikkila, B. C. and Lal, N. and Webber, W. R. and J{\'o}hannesson, G. and Moskalenko, I. V. and Orlando, E. and Porter, T. A.",
    title = "{Galactic Cosmic Rays in the Local Interstellar Medium: Voyager 1 Observations and Model Results}",
    doi = "10.3847/0004-637X/831/1/18",
    journal = "Astrophys. J.",
    volume = "831",
    number = "1",
    pages = "18",
    year = "2016"
}

@article{Stone_2019,
	title = {Cosmic ray measurements from Voyager 2 as it crossed into interstellar space},
	volume = {3},
	issn = {2397-3366},
	url = {https://doi.org},
	doi = {10.1038/s41550-019-0928-3},
	number = {11},
	journal = {Nature Astronomy},
	author = {Stone, E. C. and Cummings, A. C. and Heikkila, B. C. and Lal, N.},
	month = nov,
	year = {2019},
	pages = {1013--1018}
}

@article{Dialynas_2019,
   title={Plasma Pressures in the Heliosheath From Cassini ENA and Voyager 2 Measurements: Validation by the Voyager 2 Heliopause Crossing},
   volume={46},
   ISSN={1944-8007},
   url={http://dx.doi.org/10.1029/2019GL083924},
   DOI={10.1029/2019gl083924},
   number={14},
   journal={Geophysical Research Letters},
   publisher={American Geophysical Union (AGU)},
   author={Dialynas, Konstantinos and Krimigis, Stamatios M. and Decker, Robert B. and Mitchell, Donald G.},
   year={2019},
   month=jul, pages={7911–7919} }

@article{Aguilar_2021,
	title = {The Alpha Magnetic Spectrometer ({AMS}) on the international space station: {Part} {II} — {Results} from the first seven years},
	volume = {894},
	ISSN = {0370-1573},
	url = {https://doi.org/10.1016/j.physrep.2020.09.003},
	DOI = {10.1016/j.physrep.2020.09.003},
	journal = {Physics Reports},
	publisher = {Elsevier BV},
	author = {Aguilar, M. and Ali Cavasonza, L. and Allen, M.S. and Alpat, B. and Ambrosi, G. and Andeen, K. and Arruda, L. and Attig, N. and Azzarello, P. and et al.},
	year = {2021},
	pages = {1--116}
}

@article{huang2025constraintsevaporatingprimordialblack,
      title={Constraints on evaporating primordial black holes from the AMS-02 positron data}, 
      author={Jia-Zhi Huang and Yu-Feng Zhou},
      year={2025},
      eprint={2403.04987},
      archivePrefix={arXiv},
      primaryClass={hep-ph},
      url={https://arxiv.org/abs/2403.04987}, 
}

@article{Poulin_2017,
   title={Cosmological constraints on exotic injection of electromagnetic energy},
   volume={2017},
   ISSN={1475-7516},
   url={http://dx.doi.org/10.1088/1475-7516/2017/03/043},
   DOI={10.1088/1475-7516/2017/03/043},
   number={03},
   journal={Journal of Cosmology and Astroparticle Physics},
   publisher={IOP Publishing},
   author={Poulin, Vivian and Lesgourgues, Julien and Serpico, Pasquale D.},
   year={2017},
   month=mar, pages={043–043} }

@article{St_cker_2018,
   title={Exotic energy injection with ExoCLASS: application to the Higgs portal model and evaporating black holes},
   volume={2018},
   ISSN={1475-7516},
   url={http://dx.doi.org/10.1088/1475-7516/2018/03/018},
   DOI={10.1088/1475-7516/2018/03/018},
   number={03},
   journal={Journal of Cosmology and Astroparticle Physics},
   publisher={IOP Publishing},
   author={Stöcker, Patrick and Krämer, Michael and Lesgourgues, Julien and Poulin, Vivian},
   year={2018},
   month=mar, pages={018–018} }

@article{Slatyer_2016,
   title={Indirect dark matter signatures in the cosmic dark ages. I. Generalizing the bound on $s$-wave dark matter annihilation from Planck results},
   volume={93},
   ISSN={2470-0029},
   url={http://dx.doi.org/10.1103/PhysRevD.93.023527},
   DOI={10.1103/physrevd.93.023527},
   number={2},
   journal={Physical Review D},
   publisher={American Physical Society (APS)},
   author={Slatyer, Tracy R.},
   year={2016},
   month=jan }

@article{Slatyer_2016-b,
   title={Indirect dark matter signatures in the cosmic dark ages. II. Ionization, heating, and photon production from arbitrary energy injections},
   volume={93},
   ISSN={2470-0029},
   url={http://dx.doi.org/10.1103/PhysRevD.93.023521},
   DOI={10.1103/physrevd.93.023521},
   number={2},
   journal={Physical Review D},
   publisher={American Physical Society (APS)},
   author={Slatyer, Tracy R.},
   year={2016},
   month=jan }

@article{Diego_Blas_2011,
   title={The Cosmic Linear Anisotropy Solving System (CLASS).
 Part II: Approximation schemes},
   volume={2011},
   ISSN={1475-7516},
   url={http://dx.doi.org/10.1088/1475-7516/2011/07/034},
   DOI={10.1088/1475-7516/2011/07/034},
   number={07},
   journal={Journal of Cosmology and Astroparticle Physics},
   publisher={IOP Publishing},
   author={Diego Blas and Julien Lesgourgues and Thomas Tram},
   year={2011},
   month=jul, pages={034–034} }

@article{Anchordoqui_2025,
   title={Neutrinos from primordial black holes in theories with extra dimensions},
   volume={112},
   ISSN={2470-0029},
   url={http://dx.doi.org/10.1103/5kt2-5pvj},
   DOI={10.1103/5kt2-5pvj},
   number={8},
   journal={Physical Review D},
   publisher={American Physical Society (APS)},
   author={Anchordoqui, Luis A. and Halzen, Francis and Lüst, Dieter},
   year={2025},
   month=oct }

@article{anchordoqui2025primordialblackholes5d,
      title={Primordial Black Holes are 5D}, 
      author={Luis A. Anchordoqui and Alek Bedroya and Dieter Lüst},
      year={2025},
      eprint={2506.14874},
      archivePrefix={arXiv},
      primaryClass={hep-ph},
      url={https://arxiv.org/abs/2506.14874}, 
}

@article{KM3NeT:2025npi,
    author = "Aiello, S. and others",
    collaboration = "KM3NeT",
    title = "{Observation of an ultra-high-energy cosmic neutrino with KM3NeT}",
    doi = "10.1038/s41586-024-08543-1",
    journal = "Nature",
    volume = "638",
    number = "8050",
    pages = "376--382",
    year = "2025",
    note = "[Erratum: Nature 640, E3 (2025)]"
}

@article{Klipfel_2025,
   title={Ultrahigh-Energy Neutrinos from Primordial Black Holes},
   volume={135},
   ISSN={1079-7114},
   url={http://dx.doi.org/10.1103/vnm4-7wdc},
   DOI={10.1103/vnm4-7wdc},
   number={12},
   journal={Physical Review Letters},
   publisher={American Physical Society (APS)},
   author={Klipfel, Alexandra P. and Kaiser, David I.},
   year={2025},
   month=sep }

@article{airoldi2026primordialblackholeexplosion,
      title={Could a Primordial Black Hole Explosion Explain the extremely high-energy KM3NeT neutrino Event?}, 
      author={Lua F. T. Airoldi and Gustavo F. S. Alves and Yuber F. Perez-Gonzalez and Gabriel M. Salla and Renata Zukanovich Funchal},
      year={2026},
      eprint={2505.24666},
      archivePrefix={arXiv},
      primaryClass={hep-ph},
      url={https://arxiv.org/abs/2505.24666}, 
}

@article{km3netcollaboration2025ultrahighenergyeventkm3230213aglobal,
      title={The ultra-high-energy event KM3-230213A within the global neutrino landscape}, 
      author={KM3NeT Collaboration and O. Adriani and S. Aiello and A. Albert and A. R. Alhebsi and M. Alshamsi and S. Alves Garre and A. Ambrosone and F. Ameli and M. Andre and L. Aphecetche and M. Ardid and S. Ardid and C. Argüelles and J. Aublin and F. Badaracco and L. Bailly-Salins and Z. Bardačová and B. Baret and A. Bariego-Quintana and Y. Becherini and M. Bendahman and F. Benfenati Gualandi and M. Benhassi and M. Bennani and D. M. Benoit and E. Berbee and E. Berti and V. Bertin and P. Betti and S. Biagi and M. Boettcher and D. Bonanno and S. Bottai and A. B. Bouasla and J. Boumaaza and M. Bouta and M. Bouwhuis and C. Bozza and R. M. Bozza and H. Brânzaš and F. Bretaudeau and M. Breuhaus and R. Bruijn and J. Brunner and R. Bruno and E. Buis and R. Buompane and J. Busto and B. Caiffi and D. Calvo and A. Capone and F. Carenini and V. Carretero and T. Cartraud and P. Castaldi and V. Cecchini and S. Celli and L. Cerisy and M. Chabab and A. Chen and S. Cherubini and T. Chiarusi and M. Circella and R. Clark and R. Cocimano and J. A. B. Coelho and A. Coleiro and A. Condorelli and R. Coniglione and P. Coyle and A. Creusot and G. Cuttone and R. Dallier and A. De Benedittis and G. De Wasseige and V. Decoene and P. Deguire and I. Del Rosso and L. S. Di Mauro and I. Di Palma and A. F. Díaz and D. Diego-Tortosa and C. Distefano and A. Domi and C. Donzaud and D. Dornic and E. Drakopoulou and D. Drouhin and J. -G. Ducoin and P. Duverne and R. Dvornický and T. Eberl and E. Eckerová and A. Eddymaoui and T. van Eeden and M. Eff and D. van Eijk and I. El Bojaddaini and S. El Hedri and S. El Mentawi and V. Ellajosyula and A. Enzenhöfer and G. Ferrara and M. D. Filipović and F. Filippini and D. Franciotti and L. A. Fusco and T. Gal and J. García Méndez and A. Garcia Soto and C. Gatius Oliver and N. Geißelbrecht and E. Genton and H. Ghaddari and L. Gialanella and B. K. Gibson and E. Giorgio and I. Goos and P. Goswami and S. R. Gozzini and R. Gracia and C. Guidi and B. Guillon and M. Gutiérrez and C. Haack and H. van Haren and A. Heijboer and L. Hennig and J. J. Hernández-Rey and A. Idrissi and W. Idrissi Ibnsalih and G. Illuminati and O. Janik and D. Joly and M. de Jong and P. de Jong and B. J. Jung and P. Kalaczyński and N. Kamp and J. Keegans and V. Kikvadze and G. Kistauri and C. Kopper and A. Kouchner and Y. Y. Kovalev and L. Krupa and V. Kueviakoe and V. Kulikovskiy and R. Kvatadze and M. Labalme and R. Lahmann and M. Lamoureux and G. Larosa and C. Lastoria and J. Lazar and A. Lazo and S. Le Stum and G. Lehaut and V. Lemaître and E. Leonora and N. Lessing and G. Levi and M. Lindsey Clark and F. Longhitano and F. Magnani and J. Majumdar and L. Malerba and F. Mamedov and A. Manfreda and A. Manousakis and M. Marconi and A. Margiotta and A. Marinelli and C. Markou and L. Martin and M. Mastrodicasa and S. Mastroianni and J. Mauro and K. C. K. Mehta and A. Meskar and G. Miele and P. Migliozzi and E. Migneco and M. L. Mitsou and C. M. Mollo and L. Morales-Gallegos and N. Mori and A. Moussa and I. Mozun Mateo and R. Muller and M. R. Musone and M. Musumeci and S. Navas and A. Nayerhoda and C. A. Nicolau and B. Nkosi and B. Ó Fearraigh and V. Oliviero and A. Orlando and E. Oukacha and L. Pacini and D. Paesani and J. Palacios González and G. Papalashvili and P. Papini and V. Parisi and A. Parmar and E. J. Pastor Gomez and C. Pastore and A. M. Păun and G. E. Păvălaš and S. Peña Martínez and M. Perrin-Terrin and V. Pestel and R. Pestes and M. Petropavlova and P. Piattelli and A. Plavin and C. Poirè and V. Popa and T. Pradier and J. Prado and S. Pulvirenti and C. A. Quiroz-Rangel and N. Randazzo and A. Ratnani and S. Razzaque and I. C. Rea and D. Real and G. Riccobene and J. Robinson and A. Romanov and E. Ros and A. Šaina and F. Salesa Greus and D. F. E. Samtleben and A. Sánchez Losa and S. Sanfilippo and M. Sanguineti and D. Santonocito and P. Sapienza and M. Scaringella and M. Scarnera and J. Schnabel and J. Schumann and H. M. Schutte and J. Seneca and N. Sennan and P. A. Sevle Myhr and I. Sgura and R. Shanidze and A. Sharma and Y. Shitov and F. Šimkovic and A. Simonelli and A. Sinopoulou and B. Spisso and M. Spurio and O. Starodubtsev and D. Stavropoulos and I. Štekl and D. Stocco and M. Taiuti and G. Takadze and Y. Tayalati and H. Thiersen and S. Thoudam and I. Tosta e Melo and B. Trocmé and V. Tsourapis and E. Tzamariudaki and A. Ukleja and A. Vacheret and V. Valsecchi and V. Van Elewyck and G. Vannoye and E. Vannuccini and G. Vasileiadis and F. Vazquez de Sola and A. Veutro and S. Viola and D. Vivolo and A. van Vliet and A. Y. Wen and E. de Wolf and I. Lhenry-Yvon and S. Zavatarelli and A. Zegarelli and D. Zito and J. D. Zornoza and J. Zúñiga and N. Zywucka},
      year={2025},
      eprint={2502.08173},
      archivePrefix={arXiv},
      primaryClass={astro-ph.HE},
      url={https://arxiv.org/abs/2502.08173}, 
}

@article{li2025clashtitansultrahighenergy,
      title={Clash of the Titans: ultra-high energy KM3NeT event versus IceCube data}, 
      author={Shirley Weishi Li and Pedro Machado and Daniel Naredo-Tuero and Thomas Schwemberger},
      year={2025},
      eprint={2502.04508},
      archivePrefix={arXiv},
      primaryClass={astro-ph.HE},
      url={https://arxiv.org/abs/2502.04508}, 
}

@article{Baker_2026,
   title={Explaining the PeV Neutrino Fluxes at KM3NeT and IceCube with Quasiextremal Primordial Black Holes},
   volume={136},
   ISSN={1079-7114},
   url={http://dx.doi.org/10.1103/r793-p7ct},
   DOI={10.1103/r793-p7ct},
   number={6},
   journal={Physical Review Letters},
   publisher={American Physical Society (APS)},
   author={Baker, Michael J. and Juan, Joaquim Iguaz and Symons, Aidan and Thamm, Andrea},
   year={2026},
   month=feb }

@article{dvali2018microscopicmodelholographysurvival,
      title={A Microscopic Model of Holography: Survival by the Burden of Memory}, 
      author={Gia Dvali},
      year={2018},
      eprint={1810.02336},
      archivePrefix={arXiv},
      primaryClass={hep-th},
      url={https://arxiv.org/abs/1810.02336}, 
}

@article{Dvali_2019,
   title={Universe’s primordial quantum memories},
   volume={2019},
   ISSN={1475-7516},
   url={http://dx.doi.org/10.1088/1475-7516/2019/03/010},
   DOI={10.1088/1475-7516/2019/03/010},
   number={03},
   journal={Journal of Cosmology and Astroparticle Physics},
   publisher={IOP Publishing},
   author={Dvali, Gia and Eisemann, Lukas and Michel, Marco and Zell, Sebastian},
   year={2019},
   month=mar, pages={010–010} }

@article{dvali2024memoryburdeneffectblack,
      title={Memory Burden Effect in Black Holes and Solitons: Implications for PBH}, 
      author={Gia Dvali and Juan Sebastián Valbuena-Bermúdez and Michael Zantedeschi},
      year={2024},
      eprint={2405.13117},
      archivePrefix={arXiv},
      primaryClass={hep-th},
      url={https://arxiv.org/abs/2405.13117}, 
}

@article{boccia2025strikeluckkm3230213aevent,
      title={A strike of luck: could the KM3-230213A event be caused by an evaporating primordial black hole?}, 
      author={Andrea Boccia and Fabio Iocco},
      year={2025},
      eprint={2502.19245},
      archivePrefix={arXiv},
      primaryClass={astro-ph.HE},
      url={https://arxiv.org/abs/2502.19245}, 
}

@article{dondarini2025fastslowmergingprobes,
      title={The fast, the slow and the merging: probes of evaporating memory burdened PBHs}, 
      author={Alessandro Dondarini and Giulio Marino and Paolo Panci and Michael Zantedeschi},
      year={2025},
      eprint={2506.13861},
      archivePrefix={arXiv},
      primaryClass={hep-ph},
      url={https://arxiv.org/abs/2506.13861}, 
}

@article{Montero_2023,
   title={The dark dimension and the Swampland},
   volume={2023},
   ISSN={1029-8479},
   url={http://dx.doi.org/10.1007/JHEP02(2023)022},
   DOI={10.1007/jhep02(2023)022},
   number={2},
   journal={Journal of High Energy Physics},
   publisher={Springer Science and Business Media LLC},
   author={Montero, Miguel and Vafa, Cumrun and Valenzuela, Irene},
   year={2023},
   month=feb }

@article{Niikura_2019,
   title={Microlensing constraints on primordial black holes with Subaru/HSC Andromeda observations},
   volume={3},
   ISSN={2397-3366},
   url={http://dx.doi.org/10.1038/s41550-019-0723-1},
   DOI={10.1038/s41550-019-0723-1},
   number={6},
   journal={Nature Astronomy},
   publisher={Springer Science and Business Media LLC},
   author={Niikura, Hiroko and Takada, Masahiro and Yasuda, Naoki and Lupton, Robert H. and Sumi, Takahiro and More, Surhud and Kurita, Toshiki and Sugiyama, Sunao and More, Anupreeta and Oguri, Masamune and Chiba, Masashi},
   year={2019},
   month=apr, pages={524–534} }

@article{Smyth_2020,
   title={Updated constraints on asteroid-mass primordial black holes as dark matter},
   volume={101},
   ISSN={2470-0029},
   url={http://dx.doi.org/10.1103/PhysRevD.101.063005},
   DOI={10.1103/physrevd.101.063005},
   number={6},
   journal={Physical Review D},
   publisher={American Physical Society (APS)},
   author={Smyth, Nolan and Profumo, Stefano and English, Samuel and Jeltema, Tesla and McKinnon, Kevin and Guhathakurta, Puragra},
   year={2020},
   month=mar }

@article{Gould1992,
  title = {Femtolensing of Gamma-Ray Bursters},
  author = {Gould, Andrew},
  journal = {The Astrophysical Journal},
  volume = {386},
  pages = {L5--L7},
  year = {1992},
  month = feb,
  doi = {10.1086/186275},
  url = {https://ui.adsabs.harvard.edu/abs/1992ApJ...386L...5G/abstract}
}

@article{Katz_2018,
   title={Femtolensing by dark matter revisited},
   volume={2018},
   ISSN={1475-7516},
   url={http://dx.doi.org/10.1088/1475-7516/2018/12/005},
   DOI={10.1088/1475-7516/2018/12/005},
   number={12},
   journal={Journal of Cosmology and Astroparticle Physics},
   publisher={IOP Publishing},
   author={Katz, Andrey and Kopp, Joachim and Sibiryakov, Sergey and Xue, Wei},
   year={2018},
   month=dec, pages={005–005} }

@article{Sugiyama_2020,
   title={On the wave optics effect on primordial black hole constraints from optical microlensing search},
   volume={493},
   ISSN={1365-2966},
   url={http://dx.doi.org/10.1093/mnras/staa407},
   DOI={10.1093/mnras/staa407},
   number={3},
   journal={Monthly Notices of the Royal Astronomical Society},
   publisher={Oxford University Press (OUP)},
   author={Sugiyama, Sunao and Kurita, Toshiki and Takada, Masahiro},
   year={2020},
   month=feb, pages={3632–3641} }

@article{LIGOScientific:2019kan,
    author = "Abbott, B. P. and others",
    collaboration = "LIGO Scientific, Virgo",
    title = "{Search for Subsolar Mass Ultracompact Binaries in Advanced LIGO{\textquoteright}s Second Observing Run}",
    eprint = "1904.08976",
    archivePrefix = "arXiv",
    primaryClass = "astro-ph.CO",
    reportNumber = "LIGO-P1900037",
    doi = "10.1103/PhysRevLett.123.161102",
    journal = "Phys. Rev. Lett.",
    volume = "123",
    number = "16",
    pages = "161102",
    year = "2019"
}

@article{Gergely:2006qt,
    author = "Gergely, Laszlo A. and Darazs, Barbara",
    title = "{Weak gravitational lensing in brane-worlds}",
    eprint = "astro-ph/0602427",
    archivePrefix = "arXiv",
    journal = "Publ. Astron. Dep. Eotvos Univ.",
    volume = "17",
    pages = "213--219",
    year = "2006"
}

@article{Whisker_2005,
   title={Strong gravitational lensing by braneworld black holes},
   volume={71},
   ISSN={1550-2368},
   url={http://dx.doi.org/10.1103/PhysRevD.71.064004},
   DOI={10.1103/physrevd.71.064004},
   number={6},
   journal={Physical Review D},
   publisher={American Physical Society (APS)},
   author={Whisker, Richard},
   year={2005},
   month=mar }

@article{Kar:2003gh,
    author = "Kar, S. and Sinha, M.",
    title = "{Bending of light and gravitational signals in certain on-brane and bulk geometries}",
    doi = "10.1023/A:1026057929329",
    journal = "Gen. Rel. Grav.",
    volume = "35",
    pages = "1775--1784",
    year = "2003"
}

@article{Frolov_2003,
   title={Particle and light motion in a space-time of a five-dimensional rotating black hole},
   volume={68},
   ISSN={1089-4918},
   url={http://dx.doi.org/10.1103/PhysRevD.68.064011},
   DOI={10.1103/physrevd.68.064011},
   number={6},
   journal={Physical Review D},
   publisher={American Physical Society (APS)},
   author={Frolov, Valeri and Stojković, Dejan},
   year={2003},
   month=sep }

@article{Pal_2008,
   title={Gravitational lensing in braneworld gravity: formalism and applications},
   volume={25},
   ISSN={1361-6382},
   url={http://dx.doi.org/10.1088/0264-9381/25/4/045003},
   DOI={10.1088/0264-9381/25/4/045003},
   number={4},
   journal={Classical and Quantum Gravity},
   publisher={IOP Publishing},
   author={Pal, Supratik and Kar, Sayan},
   year={2008},
   month=jan, pages={045003} }

@article{Bin_Nun_2010,
   title={Relativistic images in Randall-Sundrum II braneworld lensing},
   volume={81},
   ISSN={1550-2368},
   url={http://dx.doi.org/10.1103/PhysRevD.81.123011},
   DOI={10.1103/physrevd.81.123011},
   number={12},
   journal={Physical Review D},
   publisher={American Physical Society (APS)},
   author={Bin-Nun, Amitai Y.},
   year={2010},
   month=jun }

@article{10.1093/mnras/128.4.295,
    author = {Refsdal, Sjur and Bondi, H.},
    title = { The Gravitational Lens Effect },
    journal = {Monthly Notices of the Royal Astronomical Society},
    volume = {128},
    number = {4},
    pages = {295-306},
    year = {1964},
    month = sep,
    issn = {0035-8711},
    doi = {10.1093/mnras/128.4.295},
    url = {https://doi.org/10.1093/mnras/128.4.295},
    eprint = {https://academic.oup.com/mnras/article-pdf/128/4/295/8073511/mnras128-0295.pdf},
}

@article{Einstein:1936:LLA,
  author  = "Albert Einstein",
  title   = "Lens-Like Action of a Star by the Deviation of Light in the Gravitational Field",
  journal = "Science",
  volume  = "84",
  number  = "2188",
  pages   = "506--507",
  year    = "1936",
  doi     = "10.1126/science.84.2188.506"
}

@ARTICLE{Paczynski:1986,
       author = {{Paczy{\'n}ski}, B.},
        title = "{Gravitational microlensing by the galactic halo}",
      journal = {\text{The Astrophysical Journal}},
         year = 1986,
        month = may,
       volume = {304},
        pages = {1-5},
          doi = {10.1086/164140},
       adsurl = {https://harvard.edu}
}

@article{gould:2000,
  title = {A Natural Formalism for Microlensing},
  author = {Gould, Andrew},
  journal = {The Astrophysical Journal},
  year = {2000},
  month = aug,
  volume = {542},
  number = {2},
  pages = {785--791},
  doi = {10.1086/317037},
  url = {https://iopscience.iop.org/article/10.1086/317037},
  archivePrefix = {arXiv},
  eprint = {astro-ph/0001421},
  primaryClass = {astro-ph}
}

@article{Virbhadra:2000vv,
    author = "Virbhadra, K. S. and Ellis, George F. R.",
    title = "{Schwarzschild black hole lensing}",
    journal = "Phys. Rev. D",
    volume = "62",
    pages = "084003",
    year = "2000",
    eprint = "astro-ph/9904193",
    archivePrefix = "arXiv",
    doi = "10.1103/PhysRevD.62.084003"
}

@ARTICLE{1994ApJ...430..505W,
       author = {{Witt}, Hans J. and {Mao}, Shude},
        title = "{Can Lensed Stars Be Regarded as Pointlike for Microlensing by MACHOs?}",
      journal = {\apj},
         year = 1994,
        month = aug,
       volume = {430},
        pages = {505},
          doi = {10.1086/174426},
       adsurl = {https://ui.adsabs.harvard.edu/abs/1994ApJ...430..505W}
}

@article{Zeldovich:1967lct,
    author = "Zel'dovich, Ya. B. and Novikov, I. D.",
    title = "{The Hypothesis of Cores Retarded during Expansion and the Hot Cosmological Model}",
    journal = "Sov. Astron.",
    volume = "10",
    pages = "602",
    year = "1967"
}

@article{Carr_2020,
   title={Primordial Black Holes as Dark Matter: Recent Developments},
   volume={70},
   ISSN={1545-4134},
   url={http://dx.doi.org/10.1146/annurev-nucl-050520-125911},
   DOI={10.1146/annurev-nucl-050520-125911},
   number={1},
   journal={Annual Review of Nuclear and Particle Science},
   publisher={Annual Reviews},
   author={Carr, Bernard and Kühnel, Florian},
   year={2020},
   month=oct, pages={355–394} }

@article{K_hnel_2016,
   title={Ellipsoidal collapse and primordial black hole formation},
   volume={94},
   ISSN={2470-0029},
   url={http://dx.doi.org/10.1103/PhysRevD.94.063514},
   DOI={10.1103/physrevd.94.063514},
   number={6},
   journal={Physical Review D},
   publisher={American Physical Society (APS)},
   author={Kühnel, Florian and Sandstad, Marit},
   year={2016},
   month=sep }

@article{Kokubu_2018,
   title={Effect of inhomogeneity on primordial black hole formation in the matter dominated era},
   volume={98},
   ISSN={2470-0029},
   url={http://dx.doi.org/10.1103/PhysRevD.98.123024},
   DOI={10.1103/physrevd.98.123024},
   number={12},
   journal={Physical Review D},
   publisher={American Physical Society (APS)},
   author={Kokubu, Takafumi and Kyutoku, Koutarou and Kohri, Kazunori and Harada, Tomohiro},
   year={2018},
   month=dec }

@article{Carr1975,
  author = {Carr, Bernard J.},
  title = {The Primordial black hole mass spectrum},
  journal = {The Astrophysical Journal},
  year = {1975},
  volume = {201},
  pages = {1--19},
  doi = {10.1086/153853},
  url = {https://iop.org}
}

@article{Musco:2008hk,
    author = "Musco, Ilia and Miller, John C. and Polnarev, Alexander G.",
    title = "{Primordial black hole formation in the radiative era: investigation of the critical nature of the collapse}",
    journal = "Class. Quant. Grav.",
    volume = "26",
    pages = "235001",
    year = "2009",
    eprint = "0811.1452",
    archivePrefix = "arXiv",
    primaryClass = "gr-qc",
    doi = "10.1088/0264-9381/26/23/235001"
}

@inbook{Flores:2024eyy,
    author = "Flores, Marcos M. and Kusenko, Alexander",
    title = "{New Ideas on~the~Formation and~Astrophysical Detection of~Primordial Black Holes}",
    eprint = "2404.05430",
    archivePrefix = "arXiv",
    primaryClass = "astro-ph.CO",
    doi = "10.1007/978-981-97-8887-3_9",
    year = "2025"
}

@article{Milligan:2025zbu,
    author = "Milligan, Ethan and Padilla, Luis E. and Mulryne, David J. and Hidalgo, Juan Carlos",
    title = "{Primordial black hole formation in a scalar field dominated universe}",
    eprint = "2504.02600",
    archivePrefix = "arXiv",
    primaryClass = "astro-ph.CO",
    doi = "10.1088/1475-7516/2025/10/025",
    journal = "JCAP",
    volume = "10",
    pages = "025",
    year = "2025"
}
\endgroup

\cleardoublepage
\phantomsection
\addcontentsline{toc}{chapter}{List of Figures}
\begingroup
\hypersetup{linkcolor=black}
\listoffigures
\endgroup


\end{document}